\documentclass[12pt,a4paper]{book}
\usepackage[utf8]{inputenc}
\usepackage[T1]{fontenc} 
\usepackage[english]{babel}
\usepackage[labelfont=bf,textfont=normal]{caption}
\usepackage[font=small, format=hang, margin={1cm, 1cm}]{caption}
\usepackage[a4paper, left=3cm,right=3cm]{geometry}
\usepackage{amsfonts} 
\usepackage{amsmath} 
\usepackage{mathrsfs}  %
\usepackage{amssymb}
\usepackage{mathabx}
\usepackage{slashed}
\usepackage{mathtools}
\usepackage{relsize}
\usepackage{setspace}   
\usepackage{color}  
\usepackage{wasysym}
\usepackage{booktabs}
\usepackage{tensor}
\usepackage{cite}
\usepackage{tikz}
\usetikzlibrary{calc, arrows.meta}
\usepackage{graphicx}
\usepackage{subfig}
\usepackage{caption}
\usepackage[normalem]{ulem}
\usepackage{hyperref}
\usepackage{epstopdf}
\usepackage{comment}
\usepackage{pict2e}
\usepackage{mdframed}
 
\usepackage{tcolorbox} 
\usepackage{wrapfig}

\definecolor{darkblue}{HTML}{004C93} 
\tcbset{
    center title,
		top=1cm,
		right=0cm,
		left=0cm,
    colback=darkblue,
    colframe=white,
    width=13.85cm,
		height=4.62cm,
		halign=right,
		valign=top,
		flush right,
		sharp corners=all
    }

\usepackage{float}
\usepackage{subfig}
\usepackage{tikz}
\usepackage{xcolor}
\definecolor{DarkGreen}{RGB}{0,128,0}

\numberwithin{equation}{section}
\usepackage{dcolumn}
\usepackage{bm}
\usepackage[toc,page]{appendix}

\def\ndelta{\delta\hspace{-0.50em}\slash\hspace{-0.05em} }

\newcommand*\xbar[1]{%
  \hbox{%
    \vbox{%
      \hrule height 0.5pt 
      \kern0.3ex
      \hbox{%
        \kern-0.0em
        \ensuremath{#1}%
        \kern-0.0em
      }%
    }%
  }%
}

\makeatletter
\DeclareRobustCommand{\loplus}{\mathbin{\mathpalette\dog@lsemi{+}}}
\DeclareRobustCommand{\lotimes}{\mathbin{\mathpalette\dog@lsemi{\times}}}
\DeclareRobustCommand{\roplus}{\mathbin{\mathpalette\dog@rsemi{+}}}
\DeclareRobustCommand{\rotimes}{\mathbin{\mathpalette\dog@rsemi{\times}}}

\newcommand{\dog@rsemi}[2]{\dog@semi{#1}{#2}{-90,90}}
\newcommand{\dog@lsemi}[2]{\dog@semi{#1}{#2}{270,90}}
\newcommand{\dog@semi}[3]{%
  \begingroup
  \sbox\z@{$\m@th#1#2$}%
  \setlength{\unitlength}{\dimexpr\ht\z@+\dp\z@\relax}%
  \makebox[\wd\z@]{\raisebox{-\dp\z@}{%
    \begin{picture}(1,1)
    \linethickness{\variable@rule{#1}}
    \roundcap
    \put(0.5,0.5){\makebox(0,0){\raisebox{\dp\z@}{$\m@th#1#2$}}}
    \put(0.5,0.5){\arc[#3]{0.5}}
    \end{picture}%
  }}%
  \endgroup
}
\newcommand{\variable@rule}[1]{%
  \fontdimen8  
  \ifx#1\displaystyle\textfont3\else
    \ifx#1\textstyle\textfont3\else
      \ifx#1\scriptstyle\scriptfont3\else
        \scriptscriptfont3\relax
  \fi\fi\fi
}
\makeatother

\def\bL{\bar{L}}
\def\bP{\bar{P}}
\def\p{\partial}

\def\bp{\bar \p}
\allowdisplaybreaks[1]
\def\bOmega{\bar\Omega}
\def\bomega{\bar\omega}
\def\bY{\bar Y}
\def\bDelta{\bar \Delta}
\newcommand{\bea}{\begin{eqnarray}}
\newcommand{\eea}{\end{eqnarray}}
\newcommand{\be}{\begin{equation}}
\newcommand{\ee}{\end{equation}}

\def\eps{\varepsilon}

\newcommand{\Order}[1]{\mathcal{O}(r^{-#1})}
\newcommand{\pref}{\frac{\sqrt{q}}{16\pi G}}
\newcommand{\prefwg}{\frac{1}{16\pi G}}

\newcommand{\D}{\text{d}}
\newcommand{\beq}{\begin{equation}}
\newcommand{\eeq}{\end{equation}}
\newcommand{\beqn}{\begin{eqnarray}}
\newcommand{\eeqn}{\end{eqnarray}}
\newcommand{\pa}{\partial}

\newcommand{\dd}{\partial}
\renewcommand{\d}{\partial}

\newcommand{\half}{\frac{1}{2}}

\newcommand{\ffrac}[2]{\raisebox{.5pt}%
  {\footnotesize$\displaystyle\frac{#1}{#2}$}\kern1pt}

\newcommand{\dover}[2]{\ffrac{\dd #1}{\dd #2}}

\newcommand{\ddl}[2]{\ffrac{\dd #1}{\dd #2}}

\newcommand{\vddl}[2]{{\ffrac{\delta #1}{\delta #2}}}

\def\cJ{\mathcal{J}}
\def\cK{\mathcal{K}}
\def\cL{\mathcal{L}}
\def\cM{\mathcal{M}}
\def\cN{\mathcal{N}}
\def\cO{\mathcal{O}}

\def\cY{\mathcal{Y}}

\hypersetup{
    colorlinks,%
    citecolor=blue,%
    filecolor=blue,%
    linkcolor=blue,%
   urlcolor=blue,
   linktoc=page
}

\begin{document}

\begin{titlepage}

$ $
\vspace{20pt}

\begin{center}
\centerline{ \LARGE{\bf{On the Various Extensions of the BMS Group}}} 

 \vspace{10mm}

\vspace{2mm} 
\centerline{\large{\bf{Romain Ruzziconi\footnote{e-mail: rruzzico@ulb.ac.be}}}}

\vspace{2mm}
\normalsize
\bigskip\medskip
\textit{Universit\'{e} Libre de Bruxelles and International Solvay Institutes\\
CP 231, B-1050 Brussels, Belgium\\
\vspace{2mm}
}
\vspace{25mm}
\large{Thesis submitted in fulfilment of the requirements of \\ the PhD Degree in Sciences (``Docteur en Sciences'')}

\large{Academic year 2019-2020}


\vspace{10mm}

Supervisor: Prof. Glenn Barnich

\vspace{25mm}

\end{center}
\textbf{Thesis jury:  } \\  \\
Prof. Riccardo Argurio (Université libre de Bruxelles) \\
Prof. Geoffrey Comp\`ere (Université libre de Bruxelles) \\
Prof. Daniel Grumiller (Technische Universität Wien) \\
Prof. Marios Petropoulos (École Polytechnique de Paris) \\

\end{titlepage}


%
%
%
%
%
%
%




\begin{center} 
\textbf{Abstract}
\end{center}      
The Bondi-Metzner-Sachs-van der Burg (BMS) group is the asymptotic symmetry group of radiating asymptotically flat spacetimes. It has recently received renewed interest in the context of the flat holography and the infrared structure of gravity. In this thesis, we investigate the consequences of considering extensions of the BMS group in four dimensions with superrotations. In particular, we apply the covariant phase space methods on a class of first order gauge theories that includes the Cartan formulation of general relativity and specify this analysis to gravity in asymptotically flat spacetime. Furthermore, we renormalize the symplectic structure at null infinity to obtain the generalized BMS charge algebra associated with smooth superrotations. We then study the vacuum structure of the gravitational field, which allows us to relate the so-called superboost transformations to the velocity kick/refraction memory effect. Afterward, we propose a new set of boundary conditions in asymptotically locally (A)dS spacetime that leads to a version of the BMS group in the presence of a non-vanishing cosmological constant, called the $\Lambda$-BMS asymptotic symmetry group. Using the holographic renormalization procedure and a diffeomorphism between Bondi and Fefferman-Graham gauges, we construct the phase space of $\Lambda$-BMS and show that it reduces to the one of the generalized BMS group in the flat limit.
\newpage

\tableofcontents

\chapter*{Acknowledgements}

This thesis could not have been completed without the help and support of several people. First, I am indebted to my supervisor, Glenn Barnich, for having guided my first steps in the marvellous world of asymptotic symmetries, from my Master's studies to my PhD. I am grateful to him for having shared his extensive knowledge on the subject and his enthusiasm about research. He has been an example of rigour and independence that will definitely influence the future of my research. I would also like to thank him for the complete freedom that I had in my research, which allowed me to start new collaborations and complete side-projects. 

Furthermore, I am grateful to Geoffrey Comp\`ere for the many projects and ideas he shared with me. I also thank him for his availability for discussions at all stages of the projects.

I would also like to thank Adrien Fiorucci for the numerous hours of stimulating discussions that were the origin of many of the original ideas presented in this thesis. Our collaboration and friendship are among the best memories I will keep from this PhD experience. 

I thank my other collaborators, Luca Ciambelli, Pujian Mao, Charles Marteau, Marios Petropoulos, for a fruitful exchange of ideas. Collaborating with them was a real enlightenment. 

I would like to thank Francesco Alessio, Laura Donnay, Daniel Grumiller, Yannick Herfray and C\'eline Zwikel for interesting discussions that will set the basis for future collaborations. 

Moreover, I thank all the members of the Mathematical Physics research group for the nice and emulating atmosphere. 

Once again, I would like to thank the members of my PhD jury Riccardo Argurio, Geoffrey Comp\`ere, Daniel Grumiller and Marios Petropoulos for agreeing to be my examiners and for their time dedicated to read this thesis. 

From a personal perspective, I thank my family and friends for their support throughout the whole process. 

This work was supported by the FRIA, F.R.S.-FNRS Belgium (2016-2020).

\chapter{Introduction}
\label{sec:Introduction}

To specify most of the physical theories, one has to consider two ingredients: the kinematics which defines the allowed states and observables of the system, and the dynamics which dictates the evolution of the state through some equations of motion. An essential piece to define the kinematics is the set of boundary conditions that selects, using the equations of motion, the allowed solutions of the theory. Depending on the context, this set of boundary conditions should enable one to determine the exact initial conditions/characteristic initial value problem that one has to provide to select a particular solution in the allowed space. The choice of boundary conditions is dictated by the physical situation one wants to describe. As broadly illustrated in this thesis, several sets of boundary conditions may be relevant to specify the kinematics for the same dynamical part of the theory.

In this work, we are mainly interested by the study of boundary conditions in gauge theories, and especially in general relativity. Indeed, gauge theories are of major importance in physics since they are involved in the description of the four fundamental interactions through the standard model of particle physics and the general relativity theory. Furthermore, as their name suggests, gauge theories exhibit some symmetries of the dynamics called gauge symmetries. Among the gauge symmetries preserving the chosen boundary conditions, several will be trivial and seen as redundancies of the theory, while others will change the physical state of the system by their actions. The latter are called asymptotic symmetries and form a group (or, more generally, a groupoid) known as the asymptotic symmetry group. In particular, different sets of boundary conditions lead to different asymptotic symmetry groups. 

In a series of seminal papers \cite{Bondi:1962px , Sachs:1962wk , Sachs:1962zza}, Bondi, Metzner, Sachs and van der Burg have shown that the asymptotic symmetry group of four-dimensional general relativity in asymptotically flat spacetimes at null infinity is an infinite-dimensional group enhancing the Poincaré group and is called the (global) BMS group. It is given by the the semi-direct product between the Lorentz group and an infinite-dimensional enhancement of the translations, called the supertranslations. This result was very surprising since one could have naively expected to find the symmetry group of flat space by studying the behaviour of the gravitational field in asymptotic regions. However, this infinite-dimensional enhancement was necessary to allow for some radiative spacetime solutions. Furthermore, this analysis led to the Bondi mass loss formula, which states that the the flux of energy-momentum at null infinity is positive. This argument served to resolve the then-controversial debate of whether gravitational waves are physical or a pure gauge artifact of the linearized theory \cite{Ashtekar:2014zsa}.

An extension of the global BMS algebra has recently been proposed, called the extended BMS algebra \cite{Barnich:2009se ,Barnich:2010eb , Barnich:2011ct}. More precisely, the Lorentz part of the semi-direct sum defining the BMS algebra has been enhanced into the infinite-dimensional algebra of conformal transformations in two dimensions. These new symmetries are called superrotations (or super-Lorentz transformations \cite{Compere:2018ylh}). At the level of the group, these superrotations are singular when considering the topology of the sphere as sections of null infinity. Therefore, only the global subgroup of the extended BMS group is globally well defined, which justifies the epithet “global”. As discussed in the following, this singular extension has been shown to be of major importance when considered as symmetries of the $\mathcal{S}$-matrix of quantum gravity \cite{Kapec:2014opa , Strominger:2017zoo , Himwich:2019qmj}. Even more recently, an alternative extension of the BMS group has been considered by replacing the singular superrotations with smooth Diff$(S^2)$ superrotations \cite{Campiglia:2014yka, Campiglia:2015yka}. This new extension, called the generalized BMS group, is made possible by relaxing the definition of asymptotic flatness and allowing a fluctuating induced boundary metric.

It should be noted that the analysis of asymptotic symmetries in general relativity has been purused for other types of asymptotics and other spacetime dimensions, including three- and four-dimensional asymptotically anti-de Sitter (AdS) and asymptotically de Sitter (dS) spacetimes (see e.g. \cite{Brown:1986nw , Henneaux:1985tv , Ashtekar:1996cd, Barnich:2006av, Grumiller:2016pqb , Ashtekar:2014zfa}). Furthermore, it has been performed on other types of gauge theories including Maxwell, Yang-Mills and Chern-Simons theories (see e.g. \cite{Strominger:2013lka, He:2014cra,  Afshar:2018apx , Detournay:2018cbf , Barnich:2013sxa , Capone:2019aiy , Lu:2019jus}). The interests of these investigations are various and depend on the main research question. In section \ref{Overview of the literature}, following \cite{Ruzziconi:2019pzd}, we relate the study of asymptotic symmetries in gauge theories with major research directions in theoretical physics.

\section{Use of asymptotic symmetries}
\label{Overview of the literature}

The study of asymptotic symmetries in gauge theories is an old subject that has recently received renewed interest. A first direction is motivated by the AdS/CFT correspondence where the asymptotic symmetries of the gravity theory in the bulk spacetime correspond to the global symmetries of the dual quantum field theory through the holographic dictionary \cite{tHooft:1993dmi , Susskind:1994vu , Maldacena:1997re , Witten:1998qj , Aharony:1999ti}. A strong control of asymptotic symmetries allows us to investigate new holographic dualities. A second direction is driven by the recently-established connections among asymptotic symmetries, soft theorems and memory effects \cite{Strominger:2017zoo}. These connections furnish crucial information about the infrared structure of quantized gauge theories. In gravity, they may be relevant to solve the long-standing problem of black hole information paradox \cite{Hawking:1976ra , Hawking:2016msc , Hawking:2016sgy , Haco:2018ske , Haco:2019ggi}. 

\subsection{Holography}

The \textit{holographic principle} states that quantum gravity can be described in terms of lower-dimensional dual quantum field theories \cite{tHooft:1993dmi , Susskind:1994vu}. A concrete realization of the holographic principle asserts that the type IIB string theory living in the bulk spacetime AdS$_5$ $\times$ $S^5$ is dual to the $\mathcal{N} = 4$ supersymmetric Yang-Mills theory living on the four-dimensional spacetime boundary \cite{Maldacena:1997re}. The gravitational theory is effectively living in the five-dimensional spacetime AdS$_5$, the five dimensions of the factor $S^5$ being compactified. A first extension of this original holographic duality is the \textit{AdS/CFT correspondence}, which tells us that the gravitational theory living in the $(d+1)$-dimensional asymptotically AdS spacetime is dual to a CFT living on the $d$-dimensional boundary. Other holographic dualities with different types of asymptotics have also been studied. A holographic dictionary enables one to interpret properties of the bulk theory in terms of the dual boundary theory. For example, the dictionary poses the following relationship between the symmetries of the two theories:
\begin{equation}
\left[
\begin{array}{c}
\text{Gauge symmetries in the bulk theory} \\
\Longleftrightarrow \\
\text{Global symmetries in the boundary theory}.
\end{array}
\right]
\label{relation symmetries}
\end{equation} More specifically for us, consider a given bulk solution space with asymptotic symmetries. The correspondence tells us that a set of quantum field theories exist that are associated with the bulk solutions, such that in the UV regime, the \textit{global symmetries} of these theories are exactly the \textit{asymptotic symmetries} of the bulk solution space. Even if the AdS/CFT correspondence has not been proven yet, it has been verified in a number of situations and extended in various directions. 

We now mention a famous hint in favour of this correspondence using the relation \eqref{relation symmetries}. Brown and Henneaux have shown that the asymptotic symmetry group for asymptotically AdS$_3$ spacetime with Dirichlet boundary conditions is given by the infinite-dimensional group of conformal transformations in two dimensions. Furthermore, they have revealed that the associated surface charges are finite, are integrable, and exhibit a non-trivial central extension in their algebra. This \textit{Brown-Henneaux central charge} is given by 
\begin{equation}
c = \frac{3 \ell}{2G} ,
\label{Brown Henneaux}
\end{equation}  where $\ell$ is the AdS$_3$ radius ($\Lambda = -1 / \ell^2$) and $G$ is the gravitational constant. The AdS/CFT correspondence indicates that there is a set of two-dimensional dual conformal field theories. The remarkable fact is that, when inserting the central charge \eqref{Brown Henneaux} into the Cardy entropy formula valid for 2$d$ CFT \cite{Strominger:1997eq}, this reproduces exactly the entropy of three-dimensional BTZ black hole solutions \cite{Banados:1992wn, Banados:1992gq}. 

The holographic principle is believed to hold in all types of asymptotics. In particular, in asymptotically flat spacetimes, from the correspondence \eqref{relation symmetries}, the dual theory would have BMS as the global symmetry. Important steps have been taken in this direction in three and four dimensions (see e.g. \cite{Dappiaggi:2004kv, Bagchi:2010eg, Barnich:2012xq , Riegler:2016hah , Fareghbal:2014qga , Bagchi:2014iea , Basu:2017aqn , Laddha:2020kvp} and references therein). Furthermore, in four-dimensional asymptotically flat spacetimes, traces of two-dimensional CFT seem to appear, enabling the use of well-known techniques of the AdS/CFT correspondence \cite{Barnich:2010eb , deBoer:2003vf , Ball:2019atb , Pasterski:2016qvg , Kapec:2016jld , Cheung:2016iub, Mishra:2017zan , Donnay:2018neh , Puhm:2019zbl , Donnay:2020guq}. The global BMS symmetry can be seen as a conformal Carroll symmetry \cite{Duval:2014uva , Duval:2014lpa , Figueroa-OFarrill:2019sex}, which is especially relevant in the context of the fluid/gravity correspondence \cite{Hubeny:2011hd , Haack:2008cp, Ciambelli:2019lap, Campoleoni:2018ltl, Ciambelli:2017wou, Caldarelli:2012cm}. 

\subsection{Infrared structure of gauge theories}
\label{Infrared structure of gauge theories}

A connection has recently been established among various areas of gauge theories that are \textit{a priori} unrelated, namely \textit{asymptotic symmetries}, \textit{soft theorems} and \textit{memory effects} (see \cite{Strominger:2017zoo} for a review). These fields of research are often referred to as the three corners of the infrared triangle of gauge theories (see figure \ref{Fig:IRSector}).

\tikzset{every node/.style={fill=white, draw=black, text width=2.5cm, align=center, inner sep=10pt}}
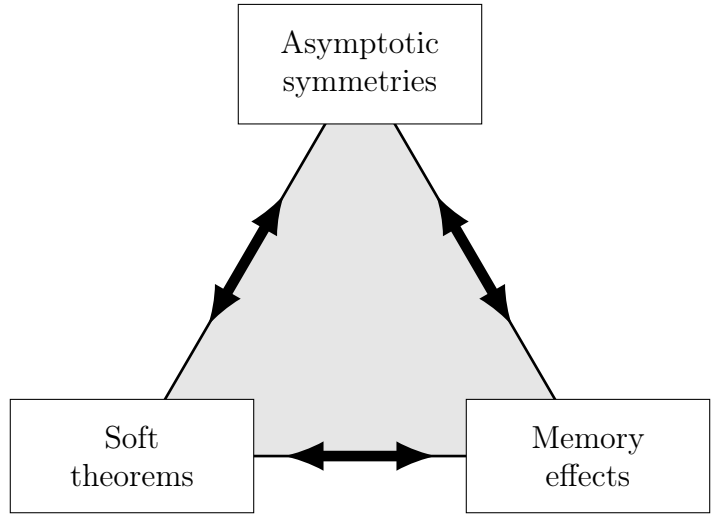
\begin{figure}[h!]
\centering
\begin{tikzpicture}
\draw[white] (-5,-3) -- (-5,4) -- (5,4) -- (5,-3) -- cycle;
\def\a{5.196};
\coordinate (A) at (-3,-2);
\coordinate (B) at (3,-2);
\coordinate (C) at (0,\a-2);
\filldraw[fill=black!10, draw=black, line width=1pt] (A) -- (B) -- (C) -- cycle;
\node[] at (A){Soft \\ theorems};
\node[] at (B){Memory effects};
\node[] at (C){Asymptotic symmetries};
\draw[line width=4pt,latex-latex] (-1,-2) -- (1,-2);
\draw[line width=4pt,latex-latex] (-2,-2+\a/3) -- (-1,-2+2*\a/3);
\draw[line width=4pt,latex-latex] (2,-2+\a/3) -- (1,-2+2*\a/3);
\end{tikzpicture}
\caption{Infrared sector of gauge theories.}
\label{Fig:IRSector}
\end{figure}

The first corner is the area of \textit{asymptotic symmetries}, which is extensively studied in this thesis. The second corner is the topic of \textit{soft theorems} \cite{Low:1954kd , GellMann:1954kc, Yennie:1961ad , Weinberg:1965nx , Weinberg:1995mt}. These theorems state that any $(n+1)$-particles scattering amplitude involving a massless soft particle, namely a particle with momentum $q\to 0$ (that may be a photon, a gluon or a graviton), is equal to the $n$-particles scattering amplitude without the soft particle, multiplied by the soft factor, plus corrections of order $q^0$. We have
\begin{equation}
\mathcal{M}_{n+1}(q, p_1, \ldots p_n) = S^{(0)}  \mathcal{M}_n (p_1, \ldots p_n) + \mathcal{O}(q^0) ,
\end{equation} where $S^{(0)} \sim q^{-1}$ is the soft factor whose precise form depends on the nature of the soft particle involved. Taking as soft particle a photon, gluon or graviton will respectively lead to the soft photon theorem, soft gluon theorem and soft graviton theorem. A remarkable property is that the soft factor is independent of the spin of the $n$ particles involved in the process. Furthermore, some so-called \textit{subleading soft theorems} have been established for the different types of soft particles and they provide some information about the subleading terms in $q$ \cite{Low:1958sn, Burnett:1967km , White:2011yy , Gross:1968in , Jackiw:1968zza}. They take the form
\begin{equation}
\mathcal{M}_{n+1}(q, p_1, \ldots p_n) = (S^{(0)} + S^{(1)}) \mathcal{M}_n (p_1, \ldots p_n) + \mathcal{O}(q) ,
\end{equation} where $S^{(1)} \sim  q^{0}$ is the subleading soft factor. Proposals for sub-subleading soft theorems can also be found \cite{Cachazo:2014fwa , DiVecchia:2016amo, Zlotnikov:2014sva}.

The third corner of the triangle is the topic of memory effects \cite{1974SvA , Braginsky:1986ia , Braginsky1987GravitationalwaveBW , Christodoulou:1991cr , Blanchet:1992br , Thorne:1992sdb , Favata:2010zu , Tolish:2014oda , Winicour:2014ska ,  Pate:2017vwa}. In gravity, the \textit{displacement memory effect} occurs, for example, in the passage of gravitational waves. It can be shown that this produces a permanent shift in the relative positions of a pair of inertial detectors. This shift is controlled by a field in the metric, called the memory field, that is turned on when the gravitational wave is passing through the spacetime region of interest. The analogous memory effects can also be established in electrodynamics (electromagnetic memory effect) \cite{Susskind:2015hpa , Pasterski:2015zua} and in Yang-Mills theory (color memory effect) \cite{Pate:2017vwa}, where a field is turned on as a result of a burst of energy passing through the region of interest, leading to an observable phenomenon. Notice that other memory effects have been identified in gravity \cite{Pasterski:2015tva , Nichols:2018qac , Flanagan:2018yzh , Podolsky:2002sa , Podolsky:2010xh , Compere:2018ylh ,  Donnay:2018ckb , Mao:2018xcw , Nichols:2018qac}, including the \textit{spin memory effect} and the \textit{refraction memory effect}. 

We now briefly discuss the relation between these different topics. It has been shown that if the quantum gravity $\mathcal{S}$-matrix is invariant under the BMS symmetry \cite{Strominger:2013jfa}, then the Ward identity associated with the supertranslations is equivalent to the soft graviton theorem \cite{He:2014laa}. Furthermore, the displacement memory effect is equivalent to performing a supertranslation \cite{Strominger:2014pwa}. More precisely, the action of the supertranslation on the memory field has the same effect as a burst of gravitational waves passing through the region of interest. This can be understood as a vacuum transition process \cite{Compere:2016jwb , Compere:2016hzt , Compere:2016gwf , Scholler:2017uni , Adjei:2019tuj}. Finally, a Fourier transform enables us to relate the soft theorem with the memory effect, which closes the triangle. This triangle controlling the infrared structure of the theory has also been constructed for other gauge theories \cite{Lysov:2014csa , Pate:2017vwa , Mao:2017tey}. Moreover, subleading infrared triangles have been uncovered and discussed \cite{Kapec:2014opa , Lysov:2014csa , Campiglia:2016hvg , Conde:2016csj , Compere:2018ylh , Himwich:2019qmj, Nichols:2018qac}. In particular, the Ward identities of superrotations have been shown to be equivalent to the subleading soft graviton theorem. Furthermore, the spin memory effect and the center-of-mass memory effect have been related to the superrotations.  

Finally, let us mention that this understanding of the infrared structure of quantum gravity is relevant to address the \textit{black hole information paradox} \cite{Hawking:1976ra}. Indeed, an infinite number of soft gravitons are produced in the process of black hole evaporation. Through the above correspondence, these soft gravitons are related with surface charges, called \textit{soft hairs}, that have to be taken into account in the information storage \cite{Hawking:2016msc , Hawking:2016sgy , Haco:2018ske , Haco:2019ggi , Bousso:2017dny , Mirbabayi:2016axw}.

\section{Original results}

The aim of this thesis is to investigate some aspects of the BMS group and its various extensions, including the associated phase spaces, vacuum structures and memory effects. In doing so, we elaborate on the covariant phase space methods for first order gauge theories. In addition, a new version of the BMS symmetry for asymptotically (A)dS$_4$ will be presented. The original results discussed in this thesis are based on the following works:
\begin{itemize}

\item \hypertarget{[A]}{[A]} Conserved currents in the Cartan formulation of general relativity \\
\textit{Glenn Barnich, Pujian Mao, Romain Ruzziconi} \\
Proceedings of the workshop "About various kinds of interactions" (2016) \\
\href{https://arxiv.org/abs/1611.01777}{\texttt{arXiv:1611.01777}}

\item \hypertarget{[B]}{[B]} Superboost transitions, refraction memory and super-Lorentz charge algebra \\
\textit{Geoffrey Comp\`ere, Adrien Fiorucci, Romain Ruzziconi} \\
Journal of High Energy Physics (2018) \\
\href{https://arxiv.org/abs/1810.00377}{\texttt{arXiv:1810.00377}}

\item \hypertarget{[C]}{[C]} The $\Lambda$-BMS$_4$ group of dS$_4$ and new boundary conditions for AdS$_4$ \\
\textit{Geoffrey Comp\`ere, Adrien Fiorucci, Romain Ruzziconi} \\
Classical and Quantum Gravity (2019) \\
\href{https://arxiv.org/abs/1905.00971}{\texttt{arXiv:1905.00971}}

\item \hypertarget{[D]}{[D]} Asymptotic symmetries in the gauge fixing approach and the BMS group \\
\textit{Romain Ruzziconi} \\
Proceedings of Science (2020) \\
\href{https://arxiv.org/abs/1910.08367}{\texttt{arXiv:1910.08367}}

\item \hypertarget{[E]}{[E]} BMS current algebra in the context of the Newman-Penrose formalism \\
\textit{Glenn Barnich, Pujian Mao, Romain Ruzziconi} \\
Classical and Quantum Gravity (2020) \\
\href{https://arxiv.org/abs/1910.14588}{\texttt{arXiv:1910.14588}}

\item \hypertarget{[F]}{[F]} The $\Lambda$-BMS$_4$ charge algebra \\
\textit{Geoffrey Comp\`ere, Adrien Fiorucci, Romain Ruzziconi} \\
Journal of High Energy Physics (2020)\\
\href{https://arxiv.org/abs/2004.10769}{\texttt{arXiv:2004.10769}}

\item \hypertarget{[G]}{[G]} Conserved currents in the Palatini formulation of general relativity \\
\textit{Glenn Barnich, Pujian Mao, Romain Ruzziconi} \\
Proceedings of Science (2020) \\
\href{https://arxiv.org/abs/2004.15002}{\texttt{arXiv:2004.15002}}

\item \hypertarget{[H]}{[H]} Gauges in three-dimensional gravity and holographic fluids \\
\textit{Luca Ciambelli, Charles Marteau, Marios Petropoulos, Romain Ruzziconi} \\
Journal of High Energy Physics (2020) \\
\href{https://arxiv.org/abs/2006.10082}{\texttt{arXiv:2006.10082}} 

\item \hypertarget{[I]}{[I]} Fefferman–Graham and Bondi gauges in the fluid/gravity
correspondence \\
\textit{Luca Ciambelli, Charles Marteau, Marios Petropoulos, Romain Ruzziconi} \\
Proceedings of Science (2020) \\
\href{https://arxiv.org/abs/2006.10083}{\texttt{arXiv:2006.10083}}

\end{itemize} The next subsection briefly summarizes some of the main results and research guidelines of this thesis.

\subsection{First order program}

The covariant phase space methods allowing for constructing meaningful surface charges in gauge theories are based on jet bundles and homotopy operators. This powerful machinery can quickly become complicated for theories of second order derivative or higher. However, for first order theories, namely theories involving at most first order derivatives in the equations of motion and in the transformation of the fields, the computations simplify drastically and do not even require the technology of homotopy operators. Furthermore, for first order theories, the different procedures to construct surface charges, namely Barnich-Brandt \cite{Barnich:2001jy, Barnich:2003xg, Barnich:2007bf} and Iyer-Wald \cite{Wald:1993nt , Iyer:1994ys, Wald:1999wa} procedures, give the same results. 

Another interesting feature is that most of the known gauge theories, including Maxwell, Yang-Mills, general relativity and Chern-Simons, are first order gauge theories, or own a formulation that is of first order using auxiliary fields. For example, Maxwell theory can be formulated as a first order gauge theory by considering the field strength as an auxiliary field (see e.g. \cite{Arnowitt:1962hi}). Similarly, the Cartan formulation of general relativity is a first order theory. 

In this thesis, we discuss covariant phase space formalism in the context of first order gauge theories. More specifically, we consider a class of theories that we call \textit{covariantized Hamiltonian theories}, which includes all the examples cited above. In particular, we investigate the breaking in the conservation of the charges for these theories. We then specify the general results obtained in this context to write the expressions of the surface charges in different first order formulations of gravity, including Cartan formulation and a new Newman-Penrose-type formulation. Finally, we apply these results to the case of four-dimensional gravity in asymptotically flat spacetimes at null infinity and obtain the currents associated with extended BMS. The results of \cite{Barnich:2013axa} are reproduced in a self-consistent way, and enlarged to allow for an arbitrary time-dependent conformal factor for the transverse boundary metric. This discussion is based on the works \hyperlink{[A]}{[A]}, \hyperlink{[E]}{[E]} and \hyperlink{[G]}{[G]}.

\subsection{Extended and generalized BMS}

As discussed in the introduction, two extensions of the global BMS group have been considered. The first is called the extended BMS group and involves singular superrotations (and, consequently, singular supertranslations) on the celestial sphere \cite{Barnich:2009se ,Barnich:2010eb , Barnich:2011ct}. The second one is called the generalized BMS group and involves smooth superrotations on the celestial sphere \cite{Campiglia:2014yka, Campiglia:2015yka}. In this thesis, we investigate both the phase spaces of the first and second extensions, using covariant phase space methods. A common feature between these analyses is that the associated charges are non-integrable but still satisfy an algebra, provided one modifies the Dirac bracket following the prescription of \cite{Barnich:2011mi}. As reviewed in section \ref{Surface charges}, the non-integrability of the charges is related to their non-conservation due to the radiation at null infinity. 

As explained in subsection \ref{Infrared structure of gauge theories}, the BMS symmetries are related to gravitational memory effects. This relation shows up by investigating the vacuum orbit of the theory. Indeed, the vacuum structure of gravity in asymptotically flat spacetime is degenerate. The fields in the metric parametrizing the different vacua, which are turned on when acting with BMS transformations on the Minkowski space, are precisely the memory fields discussed above. In this thesis, we extend the work of \cite{Compere:2016gwf , Compere:2016jwb , Compere:2016hzt} by studying the orbit of Minkowski under generalized BMS transformations. Furthermore, we relate a field that is turned on in the metric under superboost transformations, to the so-called refraction memory/velocity kick \cite{Podolsky:2002sa , Podolsky:2010xh, Podolsky:2016mqg}. We show that this superboost field satisfies a Liouville equation. These results are based on \hyperlink{[B]}{[B]}

\subsection{BMS in asymptotically locally (A)dS$_4$ spacetimes}

The BMS symmetry and its extensions are symmetries of asymptotically flat spacetimes. A legitimate question to ask is if the analogue symmetry also exists in asymptotically locally (A)dS$_4$ spacetimes. Such a generalization would be relevant for the two research guidelines discussed in subsection \ref{Overview of the literature}. Indeed, studying the BMS symmetry in AdS spacetimes, where holography is well controlled, would shed some light on holography in flat space. Furthermore, if the BMS symmetry exists in spacetimes with non-vanishing cosmological constant, it may be related to memory effects and soft theorems in this context \cite{Bieri:2015jwa , Chu:2015yua , Tolish:2016ggo , Hamada:2017gdg , Maldacena:2002vr , Cheung:2007sv , Creminelli:2004yq , Horn:2014rta}.  

In this thesis, we propose a version of the BMS symmetry in presence of a non-vanishing cosmological constant. This new asymptotic symmetry group, called the $\Lambda$-BMS$_4$ group, is obtained by imposing some partial Dirichlet boundary conditions in asymptotically locally (A)dS$_4$ spacetimes. We show that this proposal reduces to the generalized BMS group in the flat limit. Furthermore, we prove that the flat limit also works at the level of the associated phase spaces. This analysis is based on a diffeomorphism between Bondi and Fefferman-Graham gauges that we have explicitly constructed. In particular, the charge algebra of the most general asymptotically locally (A)dS$_4$ spacetime is worked out in the Fefferman-Graham gauge using the covariant phase space methods and the holographic renormalization procedure. Then, imposing the boundary conditions that lead to $\Lambda$-BMS$_4$ symmetry, we translate the symplectic structure into the Bondi gauge, where the flat limit is well defined. Taking the flat limit leads to the symplectic structure of generalized BMS discussed above. This presentation is based on \hyperlink{[C]}{[C]}, \hyperlink{[D]}{[D]} and \hyperlink{[F]}{[F]}.

\section{Plan}

The rest of this thesis is organized as follows. In Chapter \ref{ch:Asymptotic symmetries and surface charges}, we present in a self-consistent way some methods to study asymptotic symmetries in gauge theories and how to construct meaningful surface charges. This chapter is essentially a review of the existing literature, with an attempt to present the different concepts in both a unified and more abstract way. Examples illustrating the general definitions are provided. Some of those are based on results obtained in the framework of this thesis and explained in more detail in the subsequent chapters. 

In chapter \ref{ch:First order formulations and surface charges}, we restrict our study to a particular class of first order gauge theories that we call covariantized Hamiltonian theories. We show that the covariant phase space methods simply drastically in this framework and do not require the technology of jet bundles and homotopy operators discussed in Chapter \ref{ch:Asymptotic symmetries and surface charges} (see also appendix \ref{Useful results}). Furthermore, we provide a discussion on vielbeins and connection by including torsion and non-metricity into the standard discussion. Then we investigate the Cartan formulation of general relativity and its different avatars and derive the expressions of the surface charges. These are particular examples of covariantized Hamiltonian theories. Starting from a Newman-Penrose-adapted variational principle, we derive the BMS current algebra in a self-consistent way for an arbitrary $u$-dependent conformal factor.

In chapter \ref{Generalized BMS and renormalized phase space}, we study the phase space associated with the generalized BMS symmetry. In particular, we show that a renormalization procedure using Iyer-Wald ambiguity is needed to obtain finite symplectic structure. The associated charges are finite, but non-integrable. They satisfy an algebra when using the modified Dirac bracket. 

In chapter \ref{ch:Vacuum structure, superboost transitions and refraction memory}, we act on the Minkowski space with (both extended and generalized) BMS transformations and obtain the orbit of vacua. We then relate the superboosts transformations, which are part of the BMS symmetries, to the velocity kick/refraction memory. Finally, we propose a Wald-Zoupas-like prescription to isolate meaningful finite charges from the infinitesimal non-integrable expressions. Applying this prescription to the generalized BMS charges leads to the finite charges that are used in the Ward identities to establish the equivalence with soft theorems. 

In chapter \ref{ch:LambdaBMS group}, we study the most general solution spaces and the residual gauge diffeomorphisms of Fefferman-Graham and Bondi gauges in three and four dimensions. We relate the results obtained in the two gauges by constructing a diffeomorphism that maps one gauge to the other. We then focus on the four-dimensional case and propose new boundary conditions in asymptotically locally (A)dS$_4$ spacetimes. We show that the associated asymptotic symmetry group, called the $\Lambda$-BMS$_4$ group, is infinite-dimensional and reduces to the generalized BMS group in the flat limit. Then, using the holographic renormalization procedure, we construct the phase space associated with the most general asymptotically locally AdS$_4$ spacetimes in Fefferman-Graham gauge. That allows us to derive the associated charge algebra that we specify for $\Lambda$-BMS$_4$ symmetry. Transforming the $\Lambda$-BMS$_4$ symplectic structure through the diffeomorphism between Fefferman-Graham and Bondi gauges, and taking the flat limit, we prove that it reduces to the generalized BMS symplectic structure. Finally, we study new mixed boundary conditions in asymptotically locally AdS$_4$ spacetime that allow us to have a well-defined Cauchy problem. The associated asymptotic symmetry group is an infinite-dimensional subgroup of $\Lambda$-BMS$_4$ consisting of the area preserving diffeomorphisms and the time translations.  

This thesis also contains several appendices that are referenced in the core of the text.

\chapter{Asymptotic symmetries and surface charges}
\label{ch:Asymptotic symmetries and surface charges}

This chapter is an introduction to asymptotic symmetries in gauge theories, with
a focus on general relativity in four dimensions. We explain how to impose consistent sets of boundary conditions in the gauge fixing approach and how to derive the asymptotic symmetry parameters. The different procedures to obtain the associated charges are presented. As an illustration of these general concepts, the examples of four-dimensional general relativity in asymptotically locally (A)dS$_4$ and asymptotically flat spacetimes are covered. This enables us to discuss the different extensions of the BMS group that will be investigated with more details in the subsequent chapters. 

This chapter essentially reproduces the lecture notes \cite{Ruzziconi:2019pzd}.

\section{Definitions of asymptotics}
\label{Definitions of asymptotics}

Several frameworks exist to impose boundary conditions in gauge theories. Some of them are mentioned next. 

\subsection{Geometric approach}

The geometric approach of boundary conditions was initiated by Penrose, who introduced the techniques of conformal compactification to study general relativity in asymptotically flat spacetimes at null infinity \cite{Penrose:1962ij, Penrose:1964ge}. According to this perspective, the boundary conditions are defined by requiring that certain data on a fixed boundary be preserved. The asymptotic symmetry group $G$ is then defined as the quotient:
\begin{equation}
G = \frac{\text{Gauge transformation preserving the boundary conditions}}{\text{Trivial gauge transformations}} ,
\label{ASG def 1}
\end{equation} where the trivial gauge transformations are the gauge transformations that reduce to the identity on the boundary. In other words, the asymptotic symmetry group is isomorphic to the group of gauge transformations induced on the boundary which preserve the given data. This is the \textit{weak} definition of the asymptotic symmetry group. A \textit{stronger} definition of the asymptotic symmetry group is given by the quotient \eqref{ASG def 1}, where the trivial gauge transformations are now the gauge transformations that have associated vanishing charges. 

The geometric approach was essentially used in gravity theory and led to much progress in the study of symmetries and symplectic structures for asymptotically flat spacetimes at null infinity \cite{Hansen:1978jz, Ashtekar:2014zsa, Ashtekar:1981bq , Dray:1984rfa , Herfray:2020rvq} and spatial infinity \cite{Ashtekar:1978zz , Ashtekar:1991vb}. It was also considered to study asymptotically (A)dS spacetimes \cite{Ashtekar:2014zfa, Ashtekar:2015lla ,Ashtekar:2015lxa  , Ashtekar:2019khv }. Moreover, this framework was recently applied to study boundary conditions and associated phase spaces on null hypersurfaces \cite{Chandrasekaran:2018aop}. 

The advantage of this approach is that it is manifestly gauge invariant, since we do not refer to any particular coordinate system to impose the boundary conditions. Furthermore, the geometric interpretation of the symmetries is transparent. The weak point is that the definition of boundary conditions is rigid. It is a non-trivial task to modify a given set of boundary conditions in this framework to highlight new asymptotic symmetries. It is often \textit{a posteriori} that boundary conditions are defined in this framework, after having obtained the results in coordinates. 

\subsection{Gauge fixing approach}

A gauge theory has redundant degrees of freedom. The gauge fixing approach consists in using the gauge freedom of the theory to impose some constraints on the fields. This enables one to quotient the field space to eliminate some of the unphysical or pure gauge redundancies in the theory. For a given gauge theory, an \textit{appropriate gauge fixing} (where appropriate will be defined below) still allows some redundancy. For example, in electrodynamics, the gauge field $A_\mu$ transforms as $A_\mu \to A_\mu + \partial_\mu \alpha$ ($\alpha$ is a function of the spacetime coordinates) under a gauge transformation. The Lorenz gauge is defined by setting $\partial_\mu A^\mu = 0$. This gauge can always be reached using the gauge redundancy, since $\partial_\mu \partial^\mu \alpha = - \partial_\nu A^\nu$ always admits a solution for $\alpha$, regardless of the exact form of $A_\mu$. However, residual gauge transformations remain that preserve the Lorenz gauge. These are given by $A_\mu \to A_\mu + \partial_\mu \beta$, where $\beta$ is a function of the spacetime coordinates satisfying $\partial_\mu \partial^\mu \beta = 0$ (see, e.g., \cite{Jackson:1998nia}). The same phenomenon occurs in general relativity where spacetime diffeomorphisms can be performed to reach a particular gauge defined by some conditions imposed on the metric $g_{\mu\nu}$. Some explicit examples are discussed below. 

Then, the boundary conditions are imposed on the fields of the theory written in the chosen gauge. The \textit{weak} version of the definition of the asymptotic symmetry group is given by 
\begin{equation}
G_{\text{weak}} = 
\left[
\begin{array}{l}
\text{Residual gauge transformations} \\
\text{preserving the boundary conditions.}
\end{array}
\right]
\label{ASG def 2}
\end{equation} Intuitively, the gauge fixing procedure eliminates part of the pure gauge degrees of freedom, namely, the trivial gauge transformations defined under \eqref{ASG def 1}. Therefore, fixing the gauge is similar to taking the quotient as in equation \eqref{ASG def 1}, and the two definitions of asymptotic symmetry groups coincide in most of the practical situations. As in the geometric approach, a stronger version of the asymptotic symmetry group exists and is given by 
\begin{equation}
G_{\text{strong}} = 
\left[
\begin{array}{l}
\text{Residual gauge transformations preserving the boundary} \\ 
\text{conditions with associated non-vanishing charges.}
\end{array} 
\right]
\label{ASG def 3}
\end{equation} Notice that $G_{\text{strong}} \subseteq G_{\text{weak}}$\footnote{One of the most striking examples of the difference between the weak and the strong definitions of the asymptotic symmetry group is given in gravity by considering Neumann boundary conditions in asymptotically AdS$_{d+1}$ spacetimes. Indeed, in this situation, we have $G_{\text{weak}} = \text{Diff($\mathbb{R} \times S^{d-1}$)}$, and $G_{\text{strong}}$ is trivial \cite{Compere:2008us}.}.

The advantage of the gauge fixing approach is that it is highly flexible to impose boundary conditions, since we are working with explicit expressions in coordinates. For example, the BMS group in four dimensions was first discovered in this framework \cite{Bondi:1962px , Sachs:1962wk , Sachs:1962zza}. Furthermore, a gauge fixing is a local consideration (i.e. it holds in a coordinate patch of the spacetime). Therefore, the global considerations related to the topology are not directly relevant in this analysis, thereby allowing further flexibility. For example, as we will discuss in subsection \ref{Asymptotic symmetry algebra}, this allowed to consider singular extensions of the BMS group: the Witt $\times$ Witt superrotations \cite{Barnich:2009se, Barnich:2011ct}. These new asymptotic symmetries are well-defined locally; however, they have poles on the celestial sphere. In the geometric approach, one would have to modify the topology of the spacetime boundary to allow these superrotations by considering some punctured celestial sphere \cite{Strominger:2016wns , Barnich:2017ubf}. The weakness of this approach is that it is not manifestly gauge invariant. Hence, even if the gauge fixing approach is often preferred to unveil new boundary conditions and symmetries, the geometric approach is complementary and necessary to make the gauge invariance of the results manifest. In section \ref{Asymptotic symmetries in the gauge fixing approach}, we study the gauge fixing approach and provide some examples related to gravity in asymptotically flat and asymptotically (A)dS spacetimes.      

\subsection{Hamiltonian approach}

Some alternative approaches exist that are also powerful in practice. For example, in the Hamiltonian formalism, asymptotically flat \cite{Regge:1974zd} and AdS \cite{Henneaux:1985tv, Brown:1986nw} spacetimes have been studied at spatial infinity. Furthermore, the global BMS group was recently identified at spatial infinity using twisted parity conditions \cite{Henneaux:2018cst, Henneaux:2018gfi , Henneaux:2019yax }. In this framework, the computations are done in a coordinate system making the split between space and time explicit, without performing any gauge fixing. Then, the asymptotic symmetry group is defined as the quotient between the gauge transformations preserving the boundary conditions and the trivial gauge transformations, where trivial means that the associated charges are identically vanishing on the phase space. This definition of the asymptotic symmetry group corresponds to the strong definition in the two first approaches.

\section{Asymptotic symmetries in the gauge fixing approach}
\label{Asymptotic symmetries in the gauge fixing approach}

We now focus on the aforementioned gauge fixing approach of asymptotic symmetries in gauge theories. We illustrate the different definitions and concepts using examples, with a specific focus on asymptotically flat and asymptotically (A)dS spacetimes in four-dimensional general relativity. 

\subsection{Gauge fixing procedure}
\label{sec:Gauge fixing procedure}

\paragraph{Definition \normalfont[Gauge symmetry]} Let us start with a Lagrangian theory in a $n$-dimensional spacetime $M$
\begin{equation}
S[\phi] = \int_M \mathbf{L}[\phi, \partial_\mu \phi, \partial_{\mu}\partial_{\nu} \phi, \ldots] ,
\label{Lagrangian of the theory}
\end{equation} where $\mathbf{L} = L \mathrm{d}^n x$ is the Lagrangian and $\phi = (\phi^i)$ are the fields of the theory. A \textit{gauge transformation} is a transformation acting on the fields, and which depends on parameters $f=(f^\alpha)$ that are taken to be arbitrary functions of the spacetime coordinates. We write
\begin{equation}
\begin{split}
\delta_f \phi^i &= R^i[ f] \\
&=R_\alpha^i f^\alpha + R^{i\mu}_\alpha \partial_\mu f^\alpha + R^{i(\mu\nu)}_\alpha \partial_\mu \partial_\nu f^\alpha + \ldots \\
&= \sum_{k\ge 0} R^{i(\mu_1\ldots\mu_k)}_\alpha \partial_{\mu_1} \ldots \partial_{\mu_k} f^\alpha \\
\end{split}
\label{transformation of the fields}
\end{equation} the infinitesimal gauge transformation of the fields. In this expression, $R^{i(\mu_1\ldots\mu_k)}_\alpha$ are \textit{local functions}, namely functions of the coordinates, the fields, and their derivatives. The gauge transformation is a \textit{symmetry} of the theory if, under \eqref{transformation of the fields}, the Lagrangian transforms as
\begin{equation}
\delta_f \mathbf{L} = \mathrm{d} \mathbf{B}_f,
\label{symmetry}
\end{equation} where $\mathbf{B}_f = B_f^\mu (\mathrm{d}^{n-1}x)_\mu$.  

\paragraph{Examples} We illustrate this definition by providing some examples. First, consider classical vacuum electrodynamics
\begin{equation}
S[A] = \int_M \mathbf{F} \wedge \star \mathbf{F},
\label{electrodynamics}
\end{equation} where $\mathbf{F} = \mathrm{d} \mathbf{A}$ and $\mathbf{A}$ is a $1$-form. It is straightforward to check that the gauge transformation $\delta_\alpha \mathbf{A} = \mathrm{d} \alpha$, where $\alpha$ is an arbitrary function of the coordinates, is a symmetry of the theory.


Now, consider the general relativity theory
\begin{equation}
S[g] = \frac{1}{16\pi G} \int_M (R-2 \Lambda) \sqrt{-g} \mathrm{d}^n x ,
\end{equation} where $R$ and $\sqrt{-g}$ are the scalar curvature and the square root of minus the determinant associated with the metric $g_{\mu \nu}$ respectively, and $\Lambda$ is the cosmological constant. It can be checked that the gauge transformation $\delta_\xi g_{\mu\nu} = \mathcal{L}_\xi g_{\mu \nu} = \xi^\rho \partial_\rho g_{\mu\nu} + g_{\mu\rho} \partial_\nu \xi^\rho +  g_{\rho \nu} \partial_\mu \xi^\rho$, where $\xi^\mu$ is a vector field generating a diffeomorphism, is a symmetry of the theory.

Notice that in these examples, the transformation of the fields \eqref{transformation of the fields} is of the form 
\begin{equation}
\delta_f \phi^i =  R^i_\alpha f^\alpha + R^{i\mu}_\alpha \partial_\mu f^\alpha ,
\label{First order}
\end{equation} namely they involve at most first order derivatives of the parameters. 

\paragraph{Definition \normalfont[Gauge fixing]} Starting from a Lagrangian theory \eqref{Lagrangian of the theory} with gauge symmetry \eqref{transformation of the fields}, the \textit{gauge fixing} procedure involves imposing some algebraic or differential constraints on the fields in order to eliminate (part of) the redundancy in the description of the theory. We write 
\begin{equation}
G[\phi] = 0
\label{generic gauge condition}
\end{equation} a generic gauge fixing condition. This gauge has to satisfy two conditions: 
\begin{itemize}
\item It has to be reachable by a gauge transformation, which means that the number of independent conditions in \eqref{generic gauge condition} is inferior or equal to the number of independent parameters $f = (f^\alpha)$ generating the gauge transformation.
\item It has to use all of the available freedom of the arbitrary functions parametrizing the gauge transformations to reach the gauge\footnote{If the available freedom is not used, we talk about \textit{partial gauge fixing}. In this configuration, there are still some arbitrary functions of the coordinates in the parameters of the residual gauge transformations.}, which means that the number of independent conditions in \eqref{generic gauge condition} is superior or equal to the number of independent parameters $f = (f^\alpha)$ generating the gauge transformations.
\end{itemize}
Considering these two requirements together tells us that the number of independent gauge fixing conditions in \eqref{generic gauge condition} has to be equal to the number of independent gauge parameters $f = (f^\alpha)$ involved in the fields transformation \eqref{transformation of the fields}.

\paragraph{Examples} In electrodynamics, several gauge fixings are commonly used. Let us mention the Lorenz gauge $\partial^\mu A_\mu = 0$, the Coulomb gauge $\partial^i A_i=0$, the temporal gauge $A_0 = 0$, and the axial gauge $A_3 = 0$. As previously discussed, the Lorenz gauge can always be reached by performing a gauge transformation. We can check that the same statement holds for all the other gauge fixings. Notice that these gauge fixing conditions involve only one constraint, as there is only one free parameter $\alpha$ in the gauge transformation. 



In gravity, many gauge fixings are also used in practice. For example, the \textit{De Donder (or harmonic) gauge} requires that the coordinates $x^\mu$ be harmonic functions, namely, $\Box x^\mu = \frac{1}{\sqrt{-g}} \partial_\nu ( \sqrt{-g} \partial^\nu x^\mu)= 0$. Notice that the number of constraints, $n$, is equal to the number of independent gauge parameters $\xi^\mu$. This gauge condition is suitable for studying gravitational waves in perturbation theory (see, e.g., \cite{Blanchet:2013haa}).

Another important gauge fixing in configurations where $\Lambda \neq 0$ is the \textit{Fefferman-Graham gauge} \cite{Starobinsky:1982mr, AST_1985__S131__95_0, Skenderis:2002wp, 2007arXiv0710.0919F, Papadimitriou:2010as}. We write the coordinates as $x^\mu = (\rho, x^a)$, where $a=1, \ldots,n-1$ and $\rho$ is an expansion parameter ($\rho = 0$ is at the spacetime boundary, and $\rho > 0$ is in the bulk). It is defined by the following conditions:
\begin{equation}
g_{\rho\rho} = - \frac{(n-1)(n-2)}{2\Lambda \rho^2}, \quad g_{\rho a} = 0
\label{FG gauge}
\end{equation} ($n$ conditions). The coordinate $\rho$ is spacelike for $\Lambda<0$ and timelike for $\Lambda>0$. The most general metric takes the form
\begin{equation}
ds^2 =- \frac{(n-1)(n-2)}{2\Lambda}\frac{\mathrm{d}\rho^2}{\rho^2} + \gamma_{ab}(\rho,x^c) \mathrm{d}x^a \mathrm{d}x^b. 
\label{fG gauge}
\end{equation}

Finally, the \textit{Bondi gauge} will be relevant for us in the following \cite{Bondi:1962px , Sachs:1962wk , Sachs:1962zza}. This gauge fixing is valid for both $\Lambda = 0$ and $\Lambda \neq 0$ configurations. Writing the coordinates as $(u,r,x^A)$, where $x^A = (\theta_1, \ldots , \theta_{n-2})$ are the transverse angular coordinates on the $(n-2)$-celestial sphere, the Bondi gauge is defined by the following conditions\footnote{Notice that the determinant condition in \eqref{Bondi gauge} is weaker than the historical one considered in \cite{Bondi:1962px , Sachs:1962wk , Sachs:1962zza}. We refer to appendix \ref{app:detcondBondi} for more details on this condition.}:
\begin{equation}
g_{rr} = 0, \quad g_{rA} = 0, \quad \partial_r \left(\frac{\det g_{AB}}{r^{2(n-2)}} \right) = 0
\label{Bondi gauge}
\end{equation}  ($n$ conditions). These conditions tell us that, geometrically, $u$ labels null hypersurfaces in the spacetime, $x^A$ labels null geodesics inside a null hypersurface, and $r$ is the luminosity distance along the null geodesics. The most general metric takes the form
\begin{equation}
ds^2 = e^{2\beta} \frac{V}{r} \mathrm{d}u^2 - 2 e^{2\beta}\mathrm{d}u \mathrm{d}r + g_{AB} (\mathrm{d}x^A - U^A \mathrm{d}u)(dx^B - U^B \mathrm{d}u)
\label{bondi gauge}
\end{equation} where $\beta$, $U^A$ and $\frac{V}{r}$ are arbitrary functions of the coordinates, and the $(n-2)$-dimensional metric $g_{AB}$ satisfies the determinant condition in the third equation of \eqref{Bondi gauge}. Let us mention that the Bondi gauge is closely related to the \textit{Newman-Unti gauge} \cite{Newman:1962cia, Barnich:2011ty} involving only algebraic conditions:
\begin{equation}
g_{rr} = 0, \quad g_{rA} = 0, \quad g_{ru} = -1
\label{NU gauge def}
\end{equation} ($n$ conditions).

\paragraph{Definition \normalfont[Residual gauge transformation]} After having imposed a gauge fixing as in equation \eqref{generic gauge condition}, there usually remain some \textit{residual gauge transformations}, namely gauge transformations preserving the gauge fixing condition. Formally, the residual gauge transformations with generators $F$ have to satisfy $\delta_f G[\phi]  = 0$. They are local functions parametrized as $f = f(a)$, where the parameters $a$ are arbitrary functions of $(n-1)$ coordinates.

\paragraph{Examples} Consider the Lorenz gauge $\partial^\mu A_\mu=0$ in electrodynamics. As we discussed earlier, the residual gauge transformations for the Lorenz gauge are the gauge transformations $\delta_\alpha A_\mu = \partial_\mu \alpha$, where $\partial^\mu\partial_\mu \alpha = 0$. 

Similarly, consider the Fefferman-Graham gauge \eqref{FG gauge} in general relativity with $\Lambda \neq 0$. The residual gauge transformations generated by $\xi^\mu$ have to satisfy $\mathcal{L}_\xi g_{\rho \rho} = 0$ and $\mathcal{L}_\xi g_{\rho a} = 0$. The solutions to these equations are given by
\begin{equation}
\xi^\rho = \sigma (x^a) \rho , \quad \xi^a = \xi^a_0 (x^b) + \frac{(n-1)(n-2)}{2\Lambda} \partial_b \sigma \int^\rho_0 \frac{\mathrm{d} \rho'}{\rho'} \gamma^{ab}(\rho',x^c) .
\label{residual FG}
\end{equation} These solutions are parametrized by $n$ arbitrary functions $\sigma$ and $\xi^a_0$ of $(n-1)$ coordinates $x^a$. 

In the Bondi gauge \eqref{Bondi gauge}, the residual gauge transformations generated by $\xi^\mu$ have to satisfy $\mathcal{L}_\xi g_{rr} = 0$, $\mathcal{L}_\xi g_{rA} = 0$ and $g^{AB} \mathcal{L}_\xi g_{AB} = 2(n-2) \omega$, where $\omega$ is an arbitrary function of $(u,x^A)$ (see appendix \ref{app:detcondBondi}). The solutions to these equations are given by
\begin{equation}
\begin{split}
\xi^u &= f, \\
\xi^A &= Y^A + I^A, \quad I^A = -\partial_B f \int_r^\infty \mathrm{d}r'  (e^{2 \beta} g^{AB}),\\
\xi^r &= - \frac{r}{n-2} (\mathcal{D}_A Y^A - (n-2) \omega + \mathcal{D}_A I^A - \partial_B f U^B + \frac{1}{2} f g^{-1} \partial_u g) ,\\
\end{split}
\label{eq:xir}
\end{equation} 
where $\partial_r f = 0 = \partial_r Y^A$, and $g= \det (g_{AB})$ \cite{Barnich:2013sxa}. The covariant derivative $\mathcal{D}_A$ is associated with the $(n-2)$-dimensional metric $g_{AB}$. The residual gauge transformations are parametrized by the $n$ functions $\omega$, $f$ and $Y^A$ of $(n-1)$ coordinates $(u,x^A)$.

\subsection{Boundary conditions}

\paragraph{Definition \normalfont[Boundary conditions]} Once a gauge condition \eqref{generic gauge condition} has been fixed, we can impose \textit{boundary conditions} for the theory by requiering some constraints on the fields in a neighbourhood of a given spacetime region. Most of those boundary conditions are fall-off conditions on the fields in the considered asymptotic region\footnote{Notice that the asymptotic region could be taken not only at (spacelike, null or timelike) infinity, but also in other spacetime regions, such as near a black hole horizon \cite{Hawking:2016msc, Hawking:2016sgy , Haco:2018ske , Haco:2019ggi , Donnay:2015abr , Donnay:2016ejv , Donnay:2019jiz , Donnay:2019zif , Grumiller:2019fmp}.}, or conditions on the leading functions in the expansion. This choice of boundary conditions is motivated by the physical model that we want to consider. A set of boundary conditions is usually considered to be interesting if it provides non-trivial asymptotic symmetry group and solution space, exhibiting interesting properties for the associated charges (finite, generically non-vanishing, integrable and conserved; see below). If the boundary conditions are too strong, the asymptotic symmetry group will be trivial, with vanishing surface charges. Furthermore, the solution space will not contain any solution of interest. If they are too weak, the associated surface charges will be divergent. Consistent and interesting boundary conditions should therefore be located between these two extreme situations.

\paragraph{Examples} Let us give some examples of boundary conditions in general relativity theory. Many examples of boundary conditions for other gauge theories can be found in the literature (see e.g. \cite{Strominger:2013lka , He:2014cra , Afshar:2018apx , Detournay:2018cbf , Barnich:2013sxa , Capone:2019aiy , Lu:2019jus}). 

Let us consider the \textit{Bondi gauge} defined in equation \eqref{Bondi gauge} in dimension $n \ge 3$. There exist several definitions of \textit{asymptotic flatness at null infinity} ($r\to \infty$) in the literature. For all of them, we require the following preliminary boundary conditions on the functions of the metric \eqref{bondi gauge} in the asymptotic region $r\to\infty$:
\begin{equation}
\beta = o(1), \quad \frac{V}{r} = o(r^{2}), \quad U^A = o(1), \quad g_{AB} = r^2 q_{AB} + r C_{AB} + D_{AB} + \mathcal{O}(r^{-1}) ,
\label{asymp flat 1}
\end{equation} where $q_{AB}$, $C_{AB}$ and $D_{AB}$ are $(n-2)$-dimensional symmetric tensors, which are functions of $(u,x^A)$. Notice in particular that $q_{AB}$ is kept free at this stage. 

A first definition of asymptotic flatness at null infinity (AF1) is a sub-case of \eqref{asymp flat 1}. In addition to all these fall-off conditions, we require the transverse boundary metric $q_{AB}$ to have a fixed determinant, namely,
\begin{equation}
\sqrt{q} = \sqrt{\bar{q}} ,
\label{asymp flat 1 prime}
\end{equation} where $\bar{q}$ is a fixed volume element (which may possibly depend on time) on the $(n-2)$-dimensional transverse space \cite{Campiglia:2014yka, Campiglia:2015yka, Compere:2018ylh , Flanagan:2019vbl}. 

A second definition of asymptotic flatness at null infinity (AF2) is another sub-case of the definition \eqref{asymp flat 1}. All the conditions are the same, except that we require that the transverse boundary metric $q_{AB}$ be conformally related to the unit $(n-2)$-sphere metric, namely,
\begin{equation}
q_{AB} = e^{2\varphi} \mathring{q}_{AB} ,
\label{asymp flat 2}
\end{equation} where $\mathring{q}_{AB}$ is the unit $(n-2)$-sphere metric \cite{Barnich:2010eb}. Note that for $n = 4$, this condition can always be reached by a coordinate transformation, since every metric on a two dimensional surface is conformally flat (but even in this case, as we will see below, this restricts the form of the symmetries). 

A third definition of asymptotic flatness at null infinity (AF3), which is the historical one \cite{Bondi:1962px , Sachs:1962wk , Sachs:1962zza}, is a sub-case of the second definition \eqref{asymp flat 2}. We require \eqref{asymp flat 1} and we demand that the transverse boundary metric $q_{AB}$ be the unit $(n-2)$-sphere metric, namely,
\begin{equation}
q_{AB} = \mathring{q}_{AB} .
\label{asymp flat 3}
\end{equation} Note that this definition of asymptotic flatness is the only one that has the property to be \textit{asymptotically Minkowskian}, that is, for $r\to \infty$, the leading orders of the spacetime metric \eqref{bondi gauge} tend to the Minkowski line element $\mathrm{d}s^2 = -\mathrm{d}u^2 -2 \mathrm{d}u \mathrm{d}r + r^2 \mathring{q}_{AB} \mathrm{d}x^A \mathrm{d}x^B$. 

Let us now present several definitions of \textit{asymptotically (A)dS spacetimes} in both the \textit{Fefferman Graham gauge} \eqref{FG gauge} and \textit{Bondi gauge} \eqref{Bondi gauge}. A preliminary boundary condition, usually called the asymptotically locally (A)dS condition, requires the following conditions on the functions of the Fefferman-Graham metric \eqref{fG gauge}:
\begin{equation}
\gamma_{ab} = \mathcal{O}(\rho^{-2})
\label{as ads}
\end{equation} or, equivalently, $\gamma_{ab} = \rho^{-2} g_{ab}^{(0)} + o(\rho^{-2})$. Notice that the $(n-1)$-dimensional boundary metric $g^{(0)}_{ab}$ is kept free in this preliminary set of boundary conditions, thus justifying the adjective ``locally'' \cite{Fischetti:2012rd}. In the Bondi gauge, as we will see below, these fall-off conditions are (on-shell) equivalent to demand that
\begin{equation}
g_{AB} = \mathcal{O}(r^2)
\label{as ads Bondi}
\end{equation} or, equivalently, $g_{AB} = r^2 q_{AB} + o(r^2)$. 

A first definition of asymptotically (A)dS spacetime (AAdS1) is a sub-case of the definition \eqref{as ads}. In addition to these fall-off conditions, we demand the following constraints on the $(n-1)$-dimensional boundary metric $g^{(0)}_{ab}$ :
\begin{equation}
g_{tt}^{(0)} = \frac{2\Lambda}{(n-1)(n-2)}, \quad g_{tA}^{(0)} = 0, \quad \det(g_{ab}^{(0)}) = \frac{2 \Lambda}{(n-1)(n-2)} \bar{q} ,
\label{boundary gauge fixing FG}
\end{equation} where $\bar{q}$ is a fixed volume form for the transverse $(n-2)$-dimensional space (which may possibly depend on $t$) \cite{Compere:2019bua}. In the Bondi gauge, the boundary conditions \eqref{boundary gauge fixing FG} translate into  
\begin{equation}
\beta = o(1), \quad \frac{V}{r} = \frac{2r^2 \Lambda}{(n-1)(n-2)} + o(r^2), \quad U^A = o(1), \quad  \sqrt{q} = \sqrt{\bar{q}} .
\label{boundary gauge fixing Bondi}
\end{equation} Notice the similarity of these conditions to the definition (AF1) (equations \eqref{asymp flat 1} and \eqref{asymp flat 1 prime}) of asymptotically flat spacetime.

A second definition of asymptotically AdS spacetime\footnote{This choice is less relevant for asymptotically dS spacetimes, since it strongly restricts the Cauchy problem and the bulk spacetime dynamics \cite{Ashtekar:2014zfa , Ashtekar:2015lla}.} (AAdS2) is a sub-case of the definition \eqref{as ads}. We require the same conditions as in the preliminary boundary condition \eqref{as ads}, except that we demand that the $(n-1)$-dimensional boundary metric $g^{(0)}_{ab}$ be fixed \cite{Henneaux:1985tv}. These conditions are called \textit{Dirichlet boundary conditions}. One usually chooses the cylinder metric as the boundary metric, namely,
\begin{equation}
g_{ab}^{(0)} \mathrm{d}x^a \mathrm{d}x^b =  \frac{2 \Lambda }{(n-1)(n-2)} \mathrm{d}t^2 + \mathring{q}_{AB} \mathrm{d}x^A \mathrm{d}x^B ,
\label{BC Dirichlet}
\end{equation} where $\mathring{q}_{AB}$ are the components of the unit $(n-2)$-sphere metric (as in the Bondi gauge, the upper case indices $A,B, \ldots$ run from $3$ to $n$, and $x^a = (t,x^A)$). In the Bondi gauge, the boundary conditions \eqref{BC Dirichlet} translate into 
\begin{equation}
\beta = o(1), \quad \frac{V}{r} = \frac{2r^2 \Lambda}{(n-1)(n-2)} + o(r^2), \quad U^A = o (1), \quad q_{AB} = \mathring{q}_{AB} .
\label{BC Dirichlet Bondi}
\end{equation} Notice the similarity of these conditions to the definition (AF3) (equations \eqref{asymp flat 1} and \eqref{asymp flat 3}) of asymptotically flat spacetime. 

\begin{figure}[h!]
	\centering
	\vspace{10pt}
	\begin{tabular}{ccc}
	\includegraphics[width=0.3\textwidth]{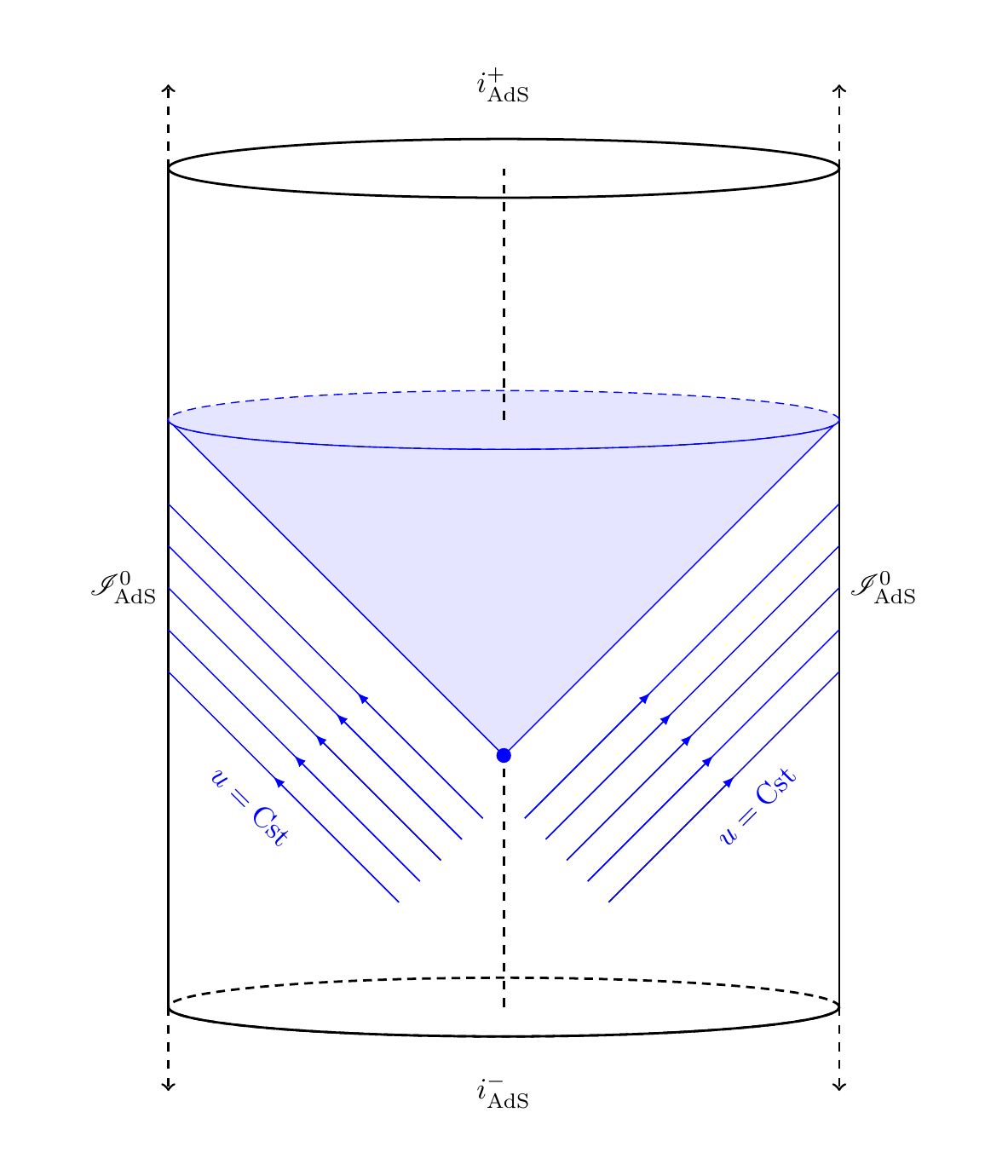}\label{Fig:AdS} & \includegraphics[width=0.3\textwidth]{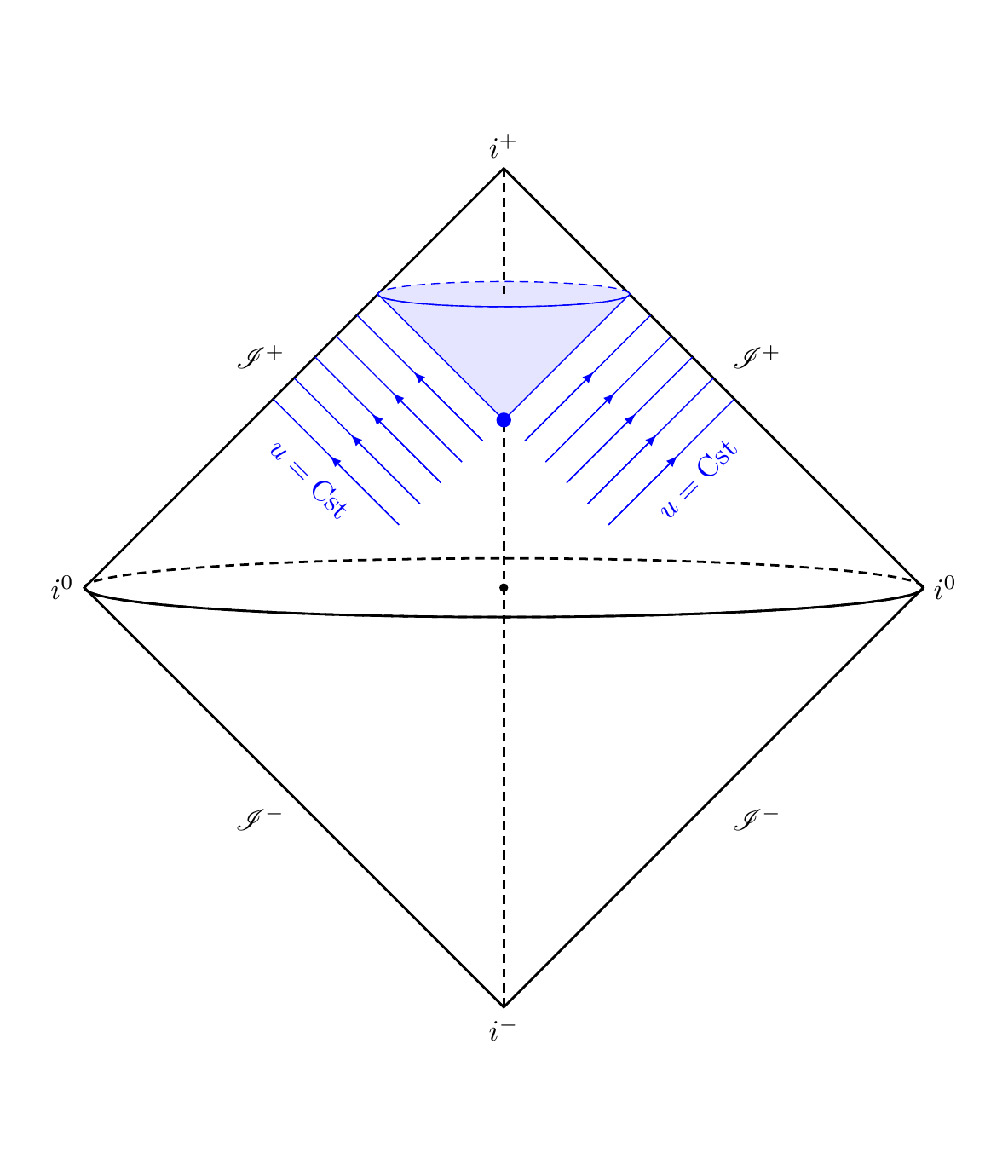}\label{Fig:Flat} & \includegraphics[width=0.3\textwidth]{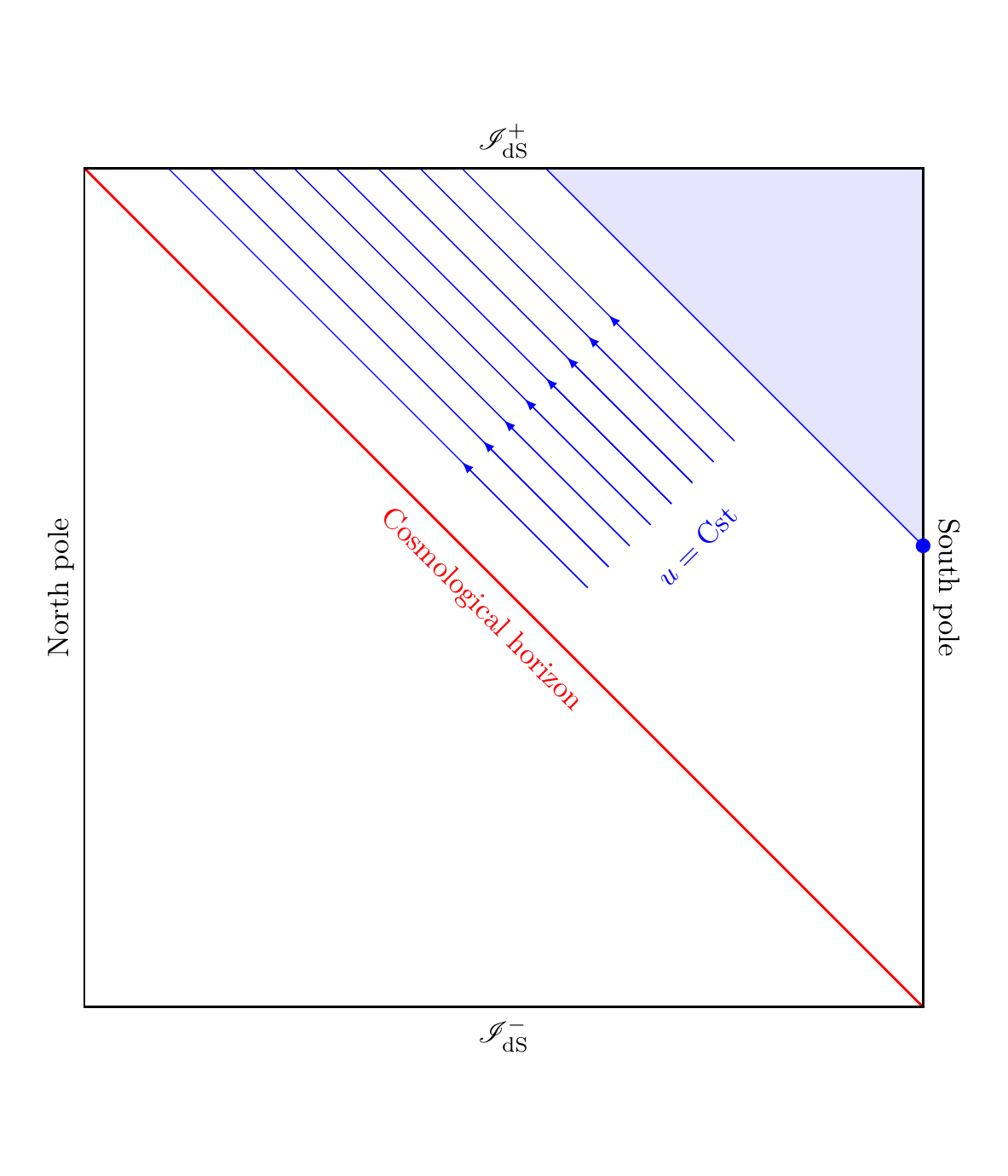}\label{Fig:dS} \\  
	AdS case $(\Lambda<0)$. & Flat case $(\Lambda=0)$. & dS case $(\Lambda>0)$.  
	\end{tabular} 
	\vspace{20pt}
	\caption{Bondi gauge for any $\Lambda$.}
	\label{Fig:FIGUREE}
\end{figure}

As we see it, the Bondi gauge is well-adapted for each type of asymptotics (see figure \ref{Fig:FIGUREE}), while the Fefferman-Graham gauge is only defined in asymptotically (A)dS spacetimes. 


%
%
%


\subsection{Solution space}
\label{Solution space}

\paragraph{Definition \normalfont[Solution space]} Given a gauge fixing \eqref{generic gauge condition} and boundary conditions, a \textit{solution} of the theory is a field configuration $\tilde{\phi}$ satisfying $G[\tilde{\phi}]=0$, the boundary conditions, and the \textit{Euler Lagrange-equations}
\begin{equation}
\left. \frac{\delta \mathbf{L}}{\delta \phi^i}\right\vert_{\tilde{\phi}} = 0 ,
\end{equation} where the \textit{Euler-Lagrange derivative} is defined in equation \eqref{Euler lagrange def}. The set of all solutions of the theory is called the \textit{solution space}. It is parametrized as $\tilde{\phi} = \tilde{\phi}(b)$, where the parameters $b$ are arbitrary functions of $(n-1)$ coordinates.

\paragraph{Examples} We now provide some examples of solution spaces of four-dimensional general relativity in different gauge fixings. These examples will be re-discussed in details in the remaining of the text (see e.g. subsections \ref{subsec:Fefferman-Graham gauge in 4d} and \ref{sec2}). We first consider the \textit{Fefferman-Graham gauge} in asymptotically (A)dS$_4$ spacetimes with preliminary boundary conditions \eqref{as ads}. Solving the Einstein equations
\begin{equation}
G_{\mu\nu} + \Lambda g_{\mu\nu} = 0 ,
\end{equation} we obtain the following analytic fall-offs:
\begin{equation}
\gamma_{ab} = \rho^{-2} g_{ab}^{(0)} + \rho^{-1} g_{ab}^{(1)} + g_{ab}^{(2)} + \rho  g_{ab}^{(3)} + \mathcal{O}(\rho^{2}) ,
\end{equation} where $g^{(i)}_{ab}$ are functions of $x^a$ \cite{Starobinsky:1982mr, AST_1985__S131__95_0, Skenderis:2002wp, 2007arXiv0710.0919F, Papadimitriou:2010as}. The only free data in this expansion are $g_{ab}^{(0)}$ and $g_{ab}^{(3)}$. All the other coefficients are determined in terms of these free data. Following the holographic dictionary, we call $g_{ab}^{(0)}$ the boundary metric and we define 
\begin{equation}
T_{ab} = \frac{\sqrt{3|\Lambda|}}{16\pi G} g_{ab}^{(3)}
\end{equation} as the stress energy tensor. From the Einstein equations, we have
\begin{equation}
g_{ab}^{(0)} T^{ab} = 0, \quad D_a^{(0)} T^{ab} = 0 ,
\label{stress energy tensor cond}
\end{equation} where $D_a^{(0)}$ is the covariant derivative with respect to $g_{ab}^{(0)}$. In summary, the solution space of general relativity in the Fefferman-Graham gauge with the preliminary boundary condition \eqref{as ads} is parametrized by the set of functions 
\begin{equation}
\{g_{ab}^{(0)}, T_{ab} \}_{\Lambda \neq 0} ,
\label{most general sol space fg}
\end{equation} where $T_{ab}$ satisfies \eqref{stress energy tensor cond} ($11$ functions).

Now, for the restricted set of boundary conditions \eqref{boundary gauge fixing FG}, that is, (AAdS1), the solution space reduces to 
\begin{equation}
\{g_{AB}^{(0)}, T_{ab} \}_{\Lambda \neq 0} ,
\end{equation} where $g_{AB}^{(0)}$ has a fixed determinant and $T_{ab}$ satisfies \eqref{stress energy tensor cond} ($7$ functions). Finally, for Dirichlet boundary conditions \eqref{BC Dirichlet} (AAdS2), the solution space reduces to 
\begin{equation}
\{T_{ab} \}_{\Lambda \neq 0} ,
\label{solution space dirichlet}
\end{equation} where $T_{ab}$ satisfies \eqref{stress energy tensor cond} (5 functions).

Let us now consider the \textit{Bondi gauge} in asymptotically (A)dS$_4$ spacetimes with preliminary boundary condition \eqref{as ads Bondi}. From the Fefferman-Graham theorem and the gauge matching between Bondi and Fefferman-Graham that is described in appendix \ref{app:chgt} (see also \cite{Poole:2018koa , Compere:2019bua}), we know that the functions appearing in the metric admit an analytic expansion in powers of $r$. In particular, we can write 
\begin{equation}
g_{AB} = r^2 q_{AB} + r C_{AB} + D_{AB} + \frac{1}{r} E_{AB} + \frac{1}{r^2} F_{AB} +\mathcal{O}(r^{-3}) ,
\label{preliminary boundary condition}
\end{equation} where $q_{AB}$, $C_{AB}$, $D_{AB}$, $E_{AB}$, $F_{AB}$, $\ldots$ are functions of $(u, x^A)$. The determinant condition defining the Bondi gauge and appearing in the third equation of \eqref{Bondi gauge} implies $g^{AB}\partial_r g_{AB}=4/r$, which imposes successively that $\det (g_{AB}) = r^4 \det (q_{AB})$, $q^{AB} C_{AB} = 0$ and
\begin{equation}
\begin{split}
&D_{AB} = \frac{1}{4} q_{AB} C^{CD} C_{CD} + \mathcal{D}_{AB} (u,x^C),  \\
&E_{AB} = \frac{1}{2} q_{AB} \mathcal{D}_{CD}C^{CD} + \mathcal{E}_{AB} (u,x^C), \\
&F_{AB} = \frac{1}{2} q_{AB} \Big[ C^{CD}\mathcal{E}_{CD} + \frac{1}{2} \mathcal{D}^{CD}\mathcal{D}_{CD} - \frac{1}{32} (C^{CD}C_{CD})^2 \Big] + \mathcal{F}_{AB}(u,x^C),
\end{split}
\end{equation}
with $q^{AB} \mathcal{D}_{AB} = q^{AB} \mathcal{E}_{AB} = q^{AB} \mathcal{F}_{AB} = 0$ (indices are lowered and raised with the metric $q_{AB}$ and its inverse). We now sketch the results obtained by solving the Einstein equations
\begin{equation}
G_{\mu\nu} + \Lambda g_{\mu\nu} = 0
\end{equation} for $\Lambda \neq 0$ (we follow \cite{Compere:2019bua, Poole:2018koa}; see also \cite{Mao:2019ahc} for the Newman-Penrose version). The component $(rr)$ gives the following radial constraints on the Bondi functions:
\begin{align}
\beta(u,r,x^A) &= \beta_0 (u,x^A) + \frac{1}{r^2} \Big[ -\frac{1}{32} C^{AB} C_{AB} \Big] + \frac{1}{r^3} \Big[ -\frac{1}{12} C^{AB} \mathcal{D}_{AB} \Big] \label{eq:EOM_beta} \\
&\qquad + \frac{1}{r^4}\Big[ - \frac{3}{32} C^{AB}\mathcal{E}_{AB} - \frac{1}{16} \mathcal{D}^{AB}\mathcal{D}_{AB} + \frac{1}{128} (C^{AB}C_{AB})^2 \Big] + \mathcal{O}(r^{-5}). \nonumber
\end{align} where $\beta_0 (u,x^A)$ is an arbitrary function. The component $(rA)$ yields 
\begin{equation}
\begin{split}
U^A = \,\, & U^A_0(u,x^B) +\overset{(1)}{U^A}(u,x^B) \frac{1}{r} + \overset{(2)}{U^A}(u,x^B) \frac{1}{r^2} \\
&+ \overset{(3)}{U^A}(u,x^B) \frac{1}{r^3} + \overset{(\text{L}3)}{U^A}(u,x^B) \frac{\ln r}{r^3} + o(r^{-3})
\end{split} \label{eq:EOM_UA}
\end{equation}
with
\begin{eqnarray}
\overset{(1)}{U^A}(u,x^B)\hspace{-6pt} &=&\hspace{-6pt} 2 e^{2\beta_0} \partial^A \beta_0 ,\nonumber \\
\overset{(2)}{U^A}(u,x^B)\hspace{-6pt} &=&\hspace{-6pt} - e^{2\beta_0} \Big[ C^{AB} \partial_B \beta_0 + \frac{1}{2} D_B C^{AB} \Big], \nonumber\\
\overset{(3)}{U^A}(u,x^B)\hspace{-6pt} &=& \hspace{-6pt}- \frac{2}{3} e^{2\beta_0} \Big[ N^A - \frac{1}{2} C^{AB} D^C C_{BC} +   (\partial_B \beta_0 - \frac{1}{3} D_B) \mathcal{D}^{AB} - \frac{3}{16} C_{CD}C^{CD} \partial^A \beta_0  \Big], \nonumber\\
\overset{(\text{L}3)}{U^A}(u,x^B) \hspace{-6pt}&=&\hspace{-6pt} -\frac{2}{3}e^{2\beta_0}D_B \mathcal{D}^{AB}. \label{eq:EOM_UA2}
\end{eqnarray}
In these expressions, $U^A_0(u,x^B)$ and $N^A(u,x^B)$ are arbitrary functions. We call $N^A$ the \textit{angular momentum aspect}. Notice that, at this stage, logarithmic terms are appearing in the expansion \eqref{eq:EOM_UA}. However, we will see below that these terms vanish for $\Lambda \neq 0$. The component $(ru)$ leads to
\begin{align}
\frac{V}{r} = &\frac{\Lambda}{3} e^{2\beta_0} r^2 - r (l + D_A U^A_0) \label{eq:EOMVr} \\
&- e^{2\beta_0} \Big[ \frac{1}{2}\Big( R[q] + \frac{\Lambda}{8}C_{AB} C^{AB} \Big) + 2 D_A \partial^A \beta_0 + 4 \partial_A \beta_0 \partial^A \beta_0 \Big] - \frac{2  M}{r} + o(r^{-1}) , \nonumber 
\end{align}
where $l = \partial_u \ln \sqrt{q}$, $R[q]$ is the scalar curvature associated with the metric $q_{AB}$ and $ M(u,x^A)$ is an arbitrary function called the \textit{Bondi mass aspect}. Afterwards, we solve the components $(AB)$ of the Einstein equations order by order, thereby providing us with the constraints imposed on each order of $g_{AB}$. The leading order $\mathcal{O}(r^{-1})$ of that equation yields to
\begin{equation}
\frac{\Lambda}{3} C_{AB} = e^{-2\beta_0} \Big[ (\partial_u - l) q_{AB} + 2 D_{(A} U^0_{B)} - D^C U^0_C q_{AB} \Big].
\label{eq:CAB}
\end{equation} Going to $\mathcal{O}(r^{-2})$, we get
\begin{equation}
\frac{\Lambda}{3} \mathcal{D}_{AB} = 0,\label{eq:DAB}
\end{equation}
which removes the logarithmic term in \eqref{eq:EOM_UA} for $\Lambda \neq 0$ (but not for $\Lambda = 0$). The condition at the next order $\mathcal{O}(r^{-3})$
\begin{equation}
\partial_u \mathcal{D}_{AB} + U_0^C D_C \mathcal{D}_{AB} + 2 \mathcal{D}_{C(A} D_{B)}U_0^C = 0
\end{equation}
is trivial for $\Lambda \neq 0$. Using an iterative argument as in \cite{Poole:2018koa}, we now make the following observation. If we decompose $g_{AB} = r^2 \sum_{n\geq 0} g_{AB}^{(n)} r^{-n}$, we see that the iterative solution of the components $(AB)$ of the Einstein equations organizes itself as $\Lambda g_{AB}^{(n)} = \partial_u g_{AB}^{(n-1)} + (...)$ at order $\mathcal{O}(r^{-n})$, $n\in\mathbb{N}_0$. Accordingly, the form of $\mathcal{E}_{AB}$ should have been fixed by the equation found at $\mathcal{O}(r^{-3})$; however, this is not the case, since both contributions of $\mathcal{E}_{AB}$ cancel between $G_{AB}$ and $\Lambda g_{AB}$. Moreover, the equation $\Lambda g_{AB}^{(4)} = \partial_u g_{AB}^{(3)} + (...)$ at the next order turns out to be a constraint for $g_{AB}^{(4)} \sim \mathcal{F}_{AB}$, determined with other subleading data such as $C_{AB}$ or $\partial_u g_{AB}^{(3)} \sim \partial_u \mathcal{E}_{AB}$. It shows that $\mathcal{E}_{AB}$ is a set of two free data on the boundary, built up from two arbitrary functions of $(u,x^A)$. Morover, it indicates that no more data exist to be uncovered for $\Lambda \neq 0$. Finally, the components $(uu)$ and $(uA)$ of the Einstein equations give some evolution constraints with respect to the $u$ coordinate for the Bondi mass aspect $M$ and the angular momentum aspect $N^A$. We will not describe these equations explicitly here (see \cite{Poole:2018koa , Compere:2019bua} or subsection \ref{sec2}).

In summary, the solution space for general relativity in the Bondi gauge with the preliminary boundary condition \eqref{preliminary boundary condition} and $\Lambda \neq 0$ is parametrized by the set of functions
\begin{equation}
\{ \beta_0, U^A_0, q_{AB}, \mathcal{E}_{AB}, M, N^A \}_{\Lambda \neq 0}
\label{most general sol space bondi}
\end{equation} ($11$ functions), where $M$ and $N^A$ have constrained evolutions with respect to the $u$ coordinate. Therefore, the characteristic initial value problem is well-defined when the following data are given: $\beta_0(u, x^C)$, $U^A_0(u, x^C)$, $\mathcal{E}_{AB}(u, x^C)$, $q_{AB}(u, x^C)$, $M(u_0, x^C)$ and $N^A(u_0, x^C)$, where $u_0$ is a fixed value of $u$.

Notice that for the boundary conditions \eqref{boundary gauge fixing Bondi} (AAdS1), the solution space reduces to 
\begin{equation}
\{q_{AB}, \mathcal{E}_{AB}, M, N^A \}_{\Lambda \neq 0} ,
\end{equation} where $M$ and $N^A$ have constrained evolutions with respect to the $u$ coordinate, and $q_{AB}$ has a fixed determinant \cite{Compere:2019bua} ($7$ functions). Finally, for the Dirichlet boundary conditions \eqref{BC Dirichlet Bondi} (AAdS2), the solution space finally reduces to 
\begin{equation}
\{ \mathcal{E}_{AB}, M, N^A \}_{\Lambda \neq 0} ,
\end{equation} where $M$ and $N^A$ have constrained evolutions with respect to the $u$ coordinate (5 functions). 

Let us finally discuss the \textit{Bondi gauge} in asymptotically flat spacetimes \cite{Bondi:1962px , Sachs:1962wk , Sachs:1962zza , Barnich:2010eb, Flanagan:2015pxa , Compere:2018ylh , Compere:2019bua}. We first consider the preliminary boundary conditions \eqref{asymp flat 1}. From the previous analysis of solution space with $\Lambda \neq 0$, we can readily obtain the solution space with $\Lambda =0$, that is, the solution of
\begin{equation}
G_{\mu\nu} = 0 ,
\end{equation} by taking the flat limit $\Lambda \to 0$. The radial constraints \eqref{eq:EOM_beta}, \eqref{eq:EOM_UA2} and \eqref{eq:EOMVr} are still valid by setting to zero $\beta_0$, $U^A_0$ (see equation \eqref{asymp flat 1}) and all the terms proportional to $\Lambda$. Furthermore, by the same procedure, the constraint equation \eqref{eq:CAB} becomes  
\begin{equation}
(\partial_u - l) q_{AB} = 0 .
\label{time evolution constraint}
\end{equation} Therefore, the asymptotic shear $C_{AB}$ becomes unconstrained, and the metric $q_{AB}$ gets a time evolution constraint. Similarly, the equation \eqref{eq:DAB} becomes trivial and $\mathcal{D}_{AB}$ is not constrained at this order. In particular, this allows for the existence of logarithmic terms in the Bondi expansion (see equation \eqref{eq:EOM_UA}). One has to impose the additional condition $D^A \mathcal{D}_{AB} = 0$ to make these logarithmic terms disappear. Finally, one can see that for $\Lambda = 0$, the subleading orders of the components $(AB)$ of the Einstein equations impose time evolution constraints on $\mathcal{D}_{AB}$, $\mathcal{E}_{AB}$, $\ldots$ , but this infinite tower of functions is otherwise unconstrained and they become free parameters of the solution space. Finally, as for the case $\Lambda \neq 0$, the components $(uu)$ and $(uA)$ of the Einstein equations yield time evolution constraints for the Bondi mass aspect $M$ and the angular momentum aspect $N^A$. 

In summary, the solution space for general relativity in the Bondi gauge with the preliminary boundary condition \eqref{asymp flat 1} is parametrized by the set of functions
\begin{equation}
\{q_{AB}, C_{AB}, M, N^A, \mathcal{D}_{AB}, \mathcal{E}_{AB}, \mathcal{F}_{AB}, \ldots \}_{\Lambda = 0} ,
\label{parametrization solution space}
\end{equation} where $q_{AB}$, $M$, $N^A$, $\mathcal{D}_{AB}$, $\mathcal{E}_{AB}$, $\mathcal{F}_{AB}$, $\ldots$ have constrained time evolutions (infinite tower of independent functions). Therefore, the characteristic initial value problem is well-defined when the following data are given: $C_{AB}(u,c^C)$, $q_{AB}(u_0, x^C)$, $M(u_0, x^C)$, $N^A(u_0, x^C)$, $\mathcal{D}_{AB}(u_0, x^C)$, $\mathcal{E}_{AB}(u_0, x^C)$, $\mathcal{F}_{AB}(u_0, x^C)$, $\ldots$ where $u_0$ is a fixed value of $u$. Notice a subtle point here: by taking the flat limit of the solution space with $\Lambda \neq 0$, we assumed that $g_{AB}$ is analytic in $r$ and can be expanded as \eqref{preliminary boundary condition} (this condition was not restrictive for $\Lambda \neq 0$). This condition is slightly more restrictive than \eqref{asymp flat 1} where analyticity is assumed only up to order $r^{-1}$. Therefore, by this flat limit procedure, we only obtain a subsector of the most general solution space. Writing $g_{AB}(u,r,x^C) = r^2 q_{AB}(u,x^C) + r C_{AB}(u,x^C) + D_{AB}(u,x^C) + \tilde{E}_{AB}  (u,r,x^C)$, where $\tilde{E}_{AB}$ is a function of all the coordinates of order $\mathcal{O}(r^{-1})$ in $r$, the most general solution space can be written as
\begin{equation}
\{q_{AB}, C_{AB}, M, N^A, \mathcal{D}_{AB}, \tilde{\mathcal{E}}_{AB} \}_{\Lambda = 0} ,
\label{parametrization solution space 2}
\end{equation} where $\tilde{\mathcal{E}}_{AB}$ is the trace-free part of $\tilde{{E}}_{AB}$, and $q_{AB}$, $M$, $N^A$, $\mathcal{D}_{AB}$, $\tilde{\mathcal{E}}_{AB}$ obey time evolution constraints. Now, the characteristic initial value problem is well-defined when the following data are given: $C_{AB}(u,c^C)$, $q_{AB}(u_0, x^C)$, $M(u_0, x^C)$, $N^A(u_0, x^C)$, $\mathcal{D}_{AB}(u_0, x^C)$ and $\tilde{\mathcal{E}}_{AB}(u_0,r, x^C)$. 

We complete this set of examples by mentioning the restricted solution spaces in the different definitions of asymptotic flatness introduced above. For boundary conditions (AF1) (equations \eqref{asymp flat 1} with \eqref{asymp flat 1 prime}), we obtain
\begin{equation}
\{q_{AB}, C_{AB}, M, N^A, \mathcal{D}_{AB}, \tilde{\mathcal{E}}_{AB} \}_{\Lambda = 0} ,
\end{equation} where $q_{AB}$, $M$, $N^A$, $\mathcal{D}_{AB}$ and $\tilde{\mathcal{E}}_{AB}$ obey time evolution constraints, and $\sqrt{q}$ is fixed. In particular, if we choose a branch where $\sqrt{q}$ is time-independent, from \eqref{time evolution constraint}, we immediately see that $\partial_u q_{AB} = 0$. For boundary conditions (AF2) (equations \eqref{asymp flat 1} with \eqref{asymp flat 2}), the solution space reduces to 
\begin{equation}
\{\varphi, C_{AB}, M, N^A, \mathcal{D}_{AB}, \tilde{\mathcal{E}}_{AB} \}_{\Lambda = 0} ,
\end{equation} where $M$, $N^A$, $\mathcal{D}_{AB}$ and $\tilde{\mathcal{E}}_{AB}$ obey time evolution equations. Notice that the metric $q_{AB}$ of the form \eqref{asymp flat 2} automatically satisfies \eqref{time evolution constraint}. This agrees with results of \cite{Barnich:2010eb}. Finally, taking the boundary conditions (AF3) (equations \eqref{asymp flat 1} with \eqref{asymp flat 3}) yields the solution space
\begin{equation}
\{C_{AB}, M, N^A, \mathcal{D}_{AB}, \tilde{\mathcal{E}}_{AB} \}_{\Lambda = 0} ,
\label{solution space Bondi dirich}
\end{equation} where $M$, $N^A$, $\mathcal{D}_{AB}$ and $\tilde{\mathcal{E}}_{AB}$ obey time evolution equations. This agrees with the historical results of \cite{Bondi:1962px , Sachs:1962wk , Sachs:1962zza}.

\subsection{Asymptotic symmetry algebra}
\label{Asymptotic symmetry algebra}

\paragraph{Definition \normalfont[Asymptotic symmetry]} Given boundary conditions imposed in a chosen gauge, the \textit{asymptotic symmetries} are defined as the residual gauge transformations preserving the boundary conditions\footnote{This is the weak definition of asymptotic symmetry, in the sense of \eqref{ASG def 2}.}. In other words, the asymptotic symmetries considered on-shell are the gauge transformations $R[f]$ tangent to the solution space. In practice, the requirement to preserve the boundary conditions gives some constraints on the functions parametrizing the residual gauge transformations. In gravity, the generators of asymptotic symmetries are often called \textit{asymptotic Killing vectors}.

\paragraph{Definition \normalfont[Asymptotic symmetry algebra]} Once the asymptotic symmetries are known, we have
\begin{equation}
\begin{split}
[R[f_1], R[f_2]] &= \delta_{f_1} R[ f_2] - \delta_{f_2} R[ f_1] \\ 
&\approx R [[f_1, f_2]_A] ,
\end{split}
\label{representation on the solution space}
\end{equation} where $\approx$ means that this equality holds on-shell, i.e. on the solution space. In this expression, the bracket $[f_1, f_2]_A$ of gauge symmetry generators is given by
\begin{equation}
[f_1, f_2]_A = C(f_1, f_2) - \delta_{f_1} f_2 + \delta_{f_2} f_1 ,
\label{modified lie bracket}
\end{equation} where $C(f_1, f_2)$ is a skew-symmetric bi-differential operator \cite{Barnich:2010xq , Barnich:2018gdh}
\begin{equation}
C(f_1, f_2) = \sum_{k,l \ge 0} C^{(\mu_1 \cdots \mu_k)(\nu_1 \cdots \nu_l)}_{[\alpha\beta]} \partial_{\mu_1} \ldots \partial_{\mu_k}f^\alpha_1 \partial_{\nu_1} \ldots \partial_{\nu_l} f_2^\beta .
\end{equation} The presence of the terms $- \delta_{f_1} f_2 + \delta_{f_2} f_1$ in \eqref{representation on the solution space} is due to the possible field-dependence of the asymptotic symmetry generators. We can verify that \eqref{modified lie bracket} satisfies the Jacobi identity, i.e. the asymptotic symmetry generators form a (solution space-dependent) Lie algebra for this bracket. It is called the \textit{asymptotic symmetry algebra}. The statement \eqref{representation on the solution space} means that the infinitesimal action of the gauge symmetries on the fields forms a representation of the Lie algebra of asymptotic symmetry generators: $[\delta_{f_1}, \delta_{f_2}] \phi = \delta_{[f_1,f_2]_A} \phi$. Let us mention that a Lie algebroid structure is showing up at this stage \cite{Crainic , Barnich:2010xq , Barnich:2017ubf}. The base manifold is given by the solution space, the field-dependent Lie algebra is the Lie algebra of asymptotic symmetry generators introduced above and the anchor is the map $f \to R[f]$.

\paragraph{Examples} The examples that we present here will be re-discussed in much details in the remaining of the text. Let us start by considering \textit{asymptotically AdS$_4$ spacetimes} in the \textit{Fefferman-Graham} and \textit{Bondi gauge}. The preliminary boundary condition \eqref{as ads} does not impose any constraint on the generators of the residual gauge diffeomorphisms of the Fefferman-Graham gauge given in \eqref{residual FG}. Similarly, the generators of the residual gauge diffeomorphisms in Bondi gauge given in \eqref{eq:xir} do not get further constraints with \eqref{as ads Bondi}. 

Now, let us consider the boundary conditions (AAdS1) (equation \eqref{as ads} together with \eqref{boundary gauge fixing FG}) in the Fefferman-Graham gauge. The asymptotic symmetries are generated by the vectors fields $\xi^\mu$ given in \eqref{residual FG} preserving the boundary conditions, namely, satisfying $\mathcal{L}_\xi g_{tt}^{(0)} =0$, $\mathcal{L}_\xi g_{tA}^{(0)} = 0$ and $g^{ab}_{(0)} \mathcal{L}_\xi g_{ab}^{(0)}= 0$. This leads to the following constraints on the parameters:
\begin{equation}
\left(\partial_u - \frac{1}{2} l  \right) \xi_0^t = \frac{1}{2} D_A^{(0)} \xi^A_0 , \quad \partial_u Y^A  = - \frac{\Lambda}{3} g^{AB}_{(0)} \partial_B \xi_0^t, \quad \sigma = \frac{1}{2} (D_A^{(0)} \xi^A_0 + \xi^t_0 l )  ,
\label{constraint equations FG param}
\end{equation} where $l = \partial_u \ln \sqrt{\bar{q}}$. In this case, the Lie bracket \eqref{modified lie bracket} is given by 
\begin{equation}
[\xi_1, \xi_2]_A = \mathcal{L}_{\xi_1} \xi_2 - \delta_{\xi_1} \xi_2 + \delta_{\xi_2} \xi_1
\label{modified lie bracket gravity}
\end{equation}  and is referred as the \textit{modified Lie bracket} \cite{Barnich:2010eb}. Therefore, the asymptotic symmetry algebra can be worked out and is given explicitly by $[\xi (\xi_{0,1}^t, \xi^A_{0,1}), \xi (\xi_{0,2}^t, \xi^A_{0,2})]_A = \xi({\hat{\xi}}_0^t, {\hat{\xi}}_0^A)$, where\footnote{The terms $\delta_{\xi (\xi_{0,1}^t, \xi^A_{0,1}) } \xi_{0,2}^t$ and $\delta_{\xi (\xi_{0,1}^t, \xi^A_{0,1}) } \xi_{0,2}^A$ in \eqref{ASG FG gen} take into account the possible field-dependence of the parameters $(\xi_{0,2}^t, \xi^A_{0,2})$.}
\begin{equation}
\begin{split}
{\hat{\xi}}_0^t &= \xi^A_{0,1} \partial_A \xi_{0,2}^t + \frac{1}{2} \xi_{0,1}^t  D_A^{(0)} \xi^A_{0,2} - \delta_{\xi (\xi_{0,1}^t, \xi^A_{0,1}) } \xi_{0,2}^t - (1 \leftrightarrow 2 ), \\
{\hat{\xi}}_0^A &= {\xi}_{0,1}^B \partial_B \xi^A_{0,2} - \frac{\Lambda}{3} \xi_{0,1}^t g^{AB}_{(0)} \partial_B \xi_{0,2}^t  - \delta_{\xi (\xi_{0,1}^t, \xi^A_{0,1}) } \xi_{0,2}^A - (1 \leftrightarrow 2 ) .\\
\end{split}
\label{ASG FG gen}
\end{equation} In the Bondi gauge with corresponding boundary conditions \eqref{boundary gauge fixing Bondi}, the constraints on the parameters are given by
\begin{equation}
\left(\partial_u - \frac{1}{2} l  \right) f = \frac{1}{2} D_A Y^A , \quad \partial_u Y^A  = - \frac{\Lambda}{3} q^{AB} \partial_B f , \quad \omega = 0
\label{constraint equations Bondi param}
\end{equation} and the asymptotic symmetry algebra is written as $[\xi (f_1, Y^A_{1}), \xi (f_{2}, Y^A_{2})]_A = \xi({\hat{f}}, {\hat{Y}}^A)$,
where 
\begin{equation}
\begin{split}
{\hat{f}} &= Y^A_{1} \partial_A f_{2} + \frac{1}{2} f_{1}  D_A Y^A_2 - \delta_{\xi (f_1, Y^A_1 )} f_2  - (1 \leftrightarrow 2 ), \\
{\hat{Y}}^A &= {Y}_{1}^B \partial_B Y^A_2 - \frac{\Lambda}{3} f_{1} q^{AB} \partial_B f_{2} - \delta_{\xi (f_1, Y^A_1 )} Y^A_2 - (1 \leftrightarrow 2 ). \\
\end{split}
\label{ASG Bondi gen}
\end{equation} This asymptotic symmetry algebra is infinite-dimensional (in particular, it contains the area-preserving diffeomorphisms as a subgroup) and field-dependent. It is called the $\Lambda$-BMS$_4$ algebra \cite{Compere:2019bua} and is denoted as $\mathfrak{bms}_4^\Lambda$. The parameters $f$ are called the supertranslation generators, while the parameters $Y^A$ are called the superrotation generators. These names will be justified below when studying the flat limit of this asymptotic symmetry algebra $\mathfrak{bms}_4^\Lambda$. The computation of the modified Lie bracket \eqref{modified lie bracket gravity} in the Bondi gauge for these boundary conditions\footnote{This completes the results obtained in \cite{Compere:2019bua} where the asymptotic symmetry algebra was obtained by pullback methods.} follows closely \cite{Barnich:2010eb}. 

Let us consider the Fefferman-Graham gauge with Dirichlet boundary conditions\\(AAdS2), that is, \eqref{as ads} together with \eqref{BC Dirichlet}. Compared to the above situation, the equations \eqref{constraint equations FG param} reduce to
\begin{equation}
\partial_u  \xi_0^t = \frac{1}{2} D_A^{(0)} \xi^A_0 , \quad \partial_u Y^A  = - \frac{\Lambda}{3} \mathring{q}^{AB}_{(0)} \partial_B \xi_0^t, \quad \sigma = \frac{1}{2} D_A^{(0)} \xi^A_0  ,
\end{equation} where $D_A^{(0)}$ is the covariant derivative associated with the fixed unit sphere metric $\mathring{q}_{AB}$. Furthermore, there is an additional constraint: $\mathcal{L}_\xi g_{AB}^{(0)} = o(\rho^{-2})$, which indicates that $\xi^A_0$ is a conformal Killing vector of $\mathring{q}_{AB}$, namely,
\begin{equation}
D_A^{(0)} \xi^0_B + D_B^{(0)} \xi^0_A = D_C^{(0)} \xi^C_0 \mathring{q}_{AB} .
\end{equation} The asymptotic symmetry algebra remains of the same form as \eqref{ASG FG gen}, with $\delta_{\xi (\xi_{0,1}^t, \xi^A_{0,1}) } \xi_{0,2}^t = 0 = \delta_{\xi (\xi_{0,1}^t, \xi^A_{0,1}) } \xi_{0,2}^A$. In the Bondi gauge, Dirichlet boundary conditions are given by \eqref{as ads Bondi} together with \eqref{BC Dirichlet Bondi}. The equations \eqref{constraint equations Bondi param} become
\begin{equation}
\partial_u  f = \frac{1}{2} D_A Y^A , \quad \partial_u Y^A  = - \frac{\Lambda}{3} \mathring{q}^{AB} \partial_B \xi_0^t , \quad \omega = 0 ,
\end{equation} where $D_{A}$ is the covariant derivative with respect to $\mathring{q}_{AB}$, while the additional constraint $\mathcal{L}_\xi g_{AB} = o(r^2)$ gives
\begin{equation}
D_A Y_B + D_B Y_A = D_C Y^C \mathring{q}_{AB} .
\end{equation} This means that $Y^A$ is a conformal Killing vector of $\mathring{q}_{AB}$. The asymptotic symmetry algebra \eqref{ASG Bondi gen} remains of the same form, with $\delta_{\xi (f_1, Y^A_1 )} f_2 = 0 = \delta_{\xi (f_1, Y^A_1 )} Y^A_2$. It can be shown that the asymptotic symmetry algebra corresponds to $\mathfrak{so}(3,2)$ algebra for $\Lambda<0$ and $\mathfrak{so}(1,4)$ algebra for $\Lambda>0$ \cite{Barnich:2013sxa} (see also appendix A of \cite{Compere:2019bua}). Therefore, we see how the infinite-dimensional asymptotic symmetry algebra $\mathfrak{bms}_4^\Lambda$ reduces to these finite-dimensional algebras, which are the symmetry algebras of global AdS$_4$ and global dS$_4$, respectively.  

Let us now consider \textit{four-dimensional asymptotically flat spacetimes} in the \textit{Bondi gauge}. The asymptotic Killing vectors can be derived in a similar way to that in the previous examples. Another way in which to proceed is to take the flat limit of the previous results obtained in the Bondi gauge. We sketch the expressions obtained by following these two equivalent procedures. First, consider the preliminary boundary conditions \eqref{asymp flat 1}. The \textit{asymptotic Killing vectors} $\xi^\mu$ are the residual gauge diffeomorphisms \eqref{eq:xir} with the following constraints on the parameters:
\begin{equation}
\left(\partial_u - \frac{1}{2} l  \right) f = \frac{1}{2} D_A Y^A-  \omega , \quad \partial_u Y^A  = 0  ,
\end{equation} where $l =\partial_u \ln \sqrt{q}$. These equations can be readily solved and the solutions are given by
\begin{equation}
f = q^{\frac{1}{4}} \left[ T(x^A) + \frac{1}{2} \int^u_0 \mathrm{d}u' [ q^{-\frac{1}{4}} (D_A Y^A - 2 \omega) ] \right], \quad Y^A = Y^A(x^B) ,
\end{equation} where $T$ are called supertranslation generators and $Y^A$ superrotation generators. Notice that there is no additional constraint on $Y^A$ at this stage. Computing the modified Lie bracket \eqref{modified lie bracket gravity}, we obtain $[\xi (f_1, Y^A_{1},\omega_1), \xi (f_{2}, Y^A_{2}, \omega_2)]_A = \xi({\hat{f}}, {\hat{Y}}^A, \hat{\omega})$ where
\begin{equation}
\begin{split}
{\hat{f}} &= Y^A_{1} \partial_A f_{2} + \frac{1}{2} f_{1}  D_A Y^A_2 - (1 \leftrightarrow 2 ) , \\
{\hat{Y}}^A &= {Y}_{1}^B \partial_B Y^A_2 - (1 \leftrightarrow 2 ), \\
\hat{\omega} &= 0 .
\end{split}
\label{asg flat}
\end{equation} 

Now, we discuss the two relevant sub-cases of boundary conditions in asymptotically flat spacetimes. Adding the condition \eqref{asymp flat 1 prime} to the preliminary condition \eqref{asymp flat 1}, i.e. considering (AF1), gives the additional constraint 
\begin{equation}
\omega = 0
\end{equation} Note that this case corresponds exactly to the flat limit of the (AAdS1) case (equations \eqref{as ads} and \eqref{boundary gauge fixing FG}). The asymptotic symmetry algebra reduces to the semi-direct product
\begin{equation}
\mathfrak{bms}_4^{\text{gen}} = \mathfrak{diff}(S^2) \loplus \mathfrak{s} ,
\label{generalized BMS group}
\end{equation} where $\mathfrak{diff}(S^2)$ are the smooth superrotations generated by $Y^A$ and $\mathfrak{s}$ are the smooth supertranslations generated by $T$. This extension of the original global BMS$_4$ algebra (see below) is called the \textit{generalized BMS$_4$ algebra} \cite{Campiglia:2014yka , Campiglia:2015yka , Compere:2018ylh , Flanagan:2019vbl}. Therefore, the $\Lambda$-BMS$_4$ algebra reduces in the flat limit to the smooth extension \eqref{generalized BMS group} of the BMS$_4$ algebra.  

The other sub-case of boundary conditions for asymptotically flat spacetimes (AF2) is given by adding condition \eqref{asymp flat 2} to the preliminary boundary condition \eqref{asymp flat 1}. The additional constraint on the parameters is now given by
\begin{equation}
D_A Y_B + D_B Y_A = D_C Y^C \mathring{q}_{AB} ,
\end{equation} i.e. $Y^A$ is a conformal Killing vector of the unit round sphere metric $\mathring{q}_{AB}$. If we allow $Y^A$ to not be globally well-defined on the $2$-sphere, then the asymptotic symmetry algebra has the structure
\begin{equation}
[(\mathfrak{diff}(S^1) \oplus \mathfrak{diff}(S^1) )\loplus \mathfrak{s}^*] \oplus \mathbb{R} .
\end{equation}  Here, $\mathfrak{diff}(S^1) \oplus \mathfrak{diff}(S^1)$ is the direct product of two copies of the Witt algebra, parametrized by $Y^A$. Furthermore, $\mathfrak{s}^*$ are the supertranslations, parametrized by $T$, and $\mathbb{R}$ are the abelian Weyl rescalings of $\mathring{q}_{AB}$, parametrized by $\omega$. Note that the supetranslations also contain singular elements since they are related to the singular superrotations through the algebra \eqref{asg flat}. This extension of the global BMS$_4$ algebra is called the \textit{extended BMS$_4$ algebra} \cite{Barnich:2010eb} and is denoted as $\mathfrak{bms}_4^{\text{ext}} \oplus \mathbb{R}$. Finally, as a sub-case of this one, considering the more restrictive constraints \eqref{asymp flat 3}, i.e. (AF3), and allowing only globally well-defined $Y^A$, we recover the \textit{global BMS$_4$ algebra} \cite{Bondi:1962px , Sachs:1962wk , Sachs:1962zza}, which is given by
\begin{equation}
\mathfrak{bms}_4^{\text{glob}} =  \mathfrak{so}(3,1) \loplus \mathfrak{s} ,
\end{equation} where $\mathfrak{s}$ are the supertranslations and $\mathfrak{so}(3,1)$ is the algebra of the globally well-defined conformal Killing vectors of the unit $2$-sphere metric, which is isomorphic to the proper orthocronous Lorentz group in four dimensions. 

\paragraph{Definition \normalfont[Action on the solution space]} Given boundary conditions imposed in a chosen gauge, there is a natural \textit{action of the asymptotic symmetry algebra}, with generators $f=f(a)$, \textit{on the solution space} $\tilde{\phi}= \tilde{\phi}(b)$. The form of this action can be deduced from \eqref{transformation of the fields} by inserting the solution space and the explicit form of the asymptotic symmetry generators\footnote{This action is usually not linear. However, in three-dimensional general relativity, this action is precisely the coadjoint representation of the asymptotic symmetry algebra \cite{Barnich:2012rz, Barnich:2014zoa , Barnich:2015uva , Barnich:2017jgw , Oblak:2017ect}.}.

\paragraph{Examples} Again, the examples given here will be discussed in more details in the text. In the \textit{Fefferman-Graham gauge} with Dirichlet boundary conditions for \textit{asymptotically AdS$_4$ spacetimes} (AAdS2) (\eqref{as ads} with \eqref{BC Dirichlet}), the asymptotic symmetry algebra $\mathfrak{so}(3,2)$ acts on the solution space \eqref{solution space dirichlet} as 
\begin{equation}
\delta_{\xi^c_0} T_{ab} = \left( \mathcal{L}_{\xi^c_0} + \frac{1}{3} D^{(0)}_c \xi^c_0 \right) T_{ab}.
\end{equation} In the \textit{Bondi gauge} with definition (AF3) (\eqref{asymp flat 1} with \eqref{asymp flat 3}) of \textit{asymptotically flat spacetime}, the global BMS$_4$ algebra $\mathfrak{bms}_4^{\text{glob}}$ acts on the leading functions of the solution space \eqref{solution space Bondi dirich} as 
\begin{equation}
\begin{split}
\delta_{(f,Y)} C_{AB} &= [f \partial_u + \mathcal{L}_Y - \frac{1}{2} D_C Y^C ] C_{AB} - 2 D_A D_B f + \mathring{q}_{AB} D_C D^C f,\\
\delta_{(f,Y)} M &= [f \partial_u + \mathcal{L}_Y + \frac{3}{2} D_C Y^C] M  + \frac{1}{4} N^{AB} D_A D_B f \\
&\qquad\qquad+ \frac{1}{2} D_A f D_B N^{AB} + \frac{1}{8} D_C D_B D_A Y^A C^{BC},\\
\delta_{(f,Y)} N_A &= [f\partial_u + \mathcal{L}_Y + D_C Y^C] N_A + 3 M D_A f - \frac{3}{16} D_A f N_{BC} C^{BC}  \\
&\quad - \frac{1}{32} D_A D_B Y^B C_{CD}C^{CD} + \frac{1}{4} (2 D^B f  + D^B D_C D^C f) C_{AB}\\
&\quad - \frac{3}{4} D_B f (D^B D^C C_{AC} - D_A D_C C^{BC}) + \frac{3}{8} D_A (D_C D_B f C^{BC})  \\
&\quad + \frac{1}{2} (D_A D_B f - \frac{1}{2} D_C D^C f \mathring{q}_{AB}) D_C C^{BC} + \frac{1}{2} D_B f N^{BC} C_{AC},
\end{split}
\end{equation} where $N_{AB} = \partial_u C_{AB}$ \cite{Barnich:2010eb}. For the action of the associated asymptotic symmetry group on these solution spaces, see \cite{Barnich:2016lyg}.

\section{Surface charges}
\label{Surface charges}

In this section, we review how to construct the surface charges associated with gauge symmetries. After recalling some results about global symmetries and Noether currents, the Barnich-Brandt prescription to obtain the surface charges in the context of asymptotic symmetries is discussed. We illustrate this construction with the example of general relativity in asymptotically (A)dS and asymptotically flat spacetimes. The relation between this prescription and the Iyer-Wald construction is established.

\subsection{Global symmetries and Noether's first theorem}

\paragraph{Definition \normalfont[Global symmetry]} Let us consider a Lagrangian theory with Lagrangian density $\mathbf{L}[\phi, \partial_\mu \phi, \ldots]$ and a transformation $\delta_Q \phi = Q$ of the fields, where $Q$ is a local function. In agreement with the above definition \eqref{symmetry}, this transformation is said to be a \textit{symmetry} of the theory if
\begin{equation}
\delta_Q \mathbf{L} = \mathrm{d} \mathbf{B}_Q ,
\label{symmetry def}
\end{equation} where $\mathbf{B}_Q = B^\mu_Q (\mathrm{d}^{n-1}x)_\mu$. Then, as defined in \eqref{transformation of the fields}, a \textit{gauge symmetry} is just a symmetry that depends on arbitrary spacetime functions $f = (f^\alpha)$, i.e. $Q = R[f]$. We define an on-shell equivalence relation $\sim$ between the symmetries of the theory as
\begin{equation}
Q \sim Q + R[f] ,
\end{equation} i.e. two symmetries are equivalent if they differ, on-shell, by a gauge transformation $R[f]$. The equivalence classes $[Q]$ for this equivalence relation are called the \textit{global symmetries}. In particular, a gauge symmetry is a trivial global symmetry. 

\paragraph{Definition \normalfont[Noether current]} A \textit{conserved current} $\mathbf{j}$ is an on-shell closed $(n-1)$-form, i.e. $\mathrm{d} \mathbf{j} \approx 0$. We define an on-shell equivalence relation $\sim$ between the currents as
\begin{equation}
\mathbf{j} \sim \mathbf{j} + \mathrm{d} \mathbf{K} ,
\label{Noether current def}
\end{equation} where $\mathbf{K}$ is a $(n-2)$-form. A \textit{Noether current} is an equivalence class $[\mathbf{j}]$ for this equivalence relation.

\paragraph{Theorem \normalfont[Noether's first theorem]} A one-to-one correspondence exists between \textit{global symmetries} $Q$ and \textit{Noether currents} $[\mathbf{j}]$, which can be written as
\begin{equation}
[Q] \stackrel{\text{\scriptsize 1-1}}{\longleftrightarrow} [\mathbf{j}] .
\label{one to one}
\end{equation} In particular, Noether currents associated with gauge symmetries are trivial. Recent demonstrations of this theorem can for example be found in \cite{Barnich:2018gdh , Barnich:2001jy}. 

\paragraph{Remark} This theorem also enables us to construct explicit representatives of the Noether current for a given global symmetry. We have
\begin{equation}
\delta_Q \mathbf{L} = \mathrm{d}\mathbf{B}_Q = (\partial_\mu B^\mu_Q) \mathrm{d}^n x .
\label{stage 1}
\end{equation} Furthermore, writing $\mathbf{L} = L \mathrm{d}^n x$, we obtain 
\begin{equation}
\begin{split}
\delta_Q L &= \delta_Q \phi^i \frac{\partial L}{\partial  \phi^i} + \delta_Q \partial_\mu \phi^i \frac{\partial L}{\partial (\partial_\mu \phi^i)} + \ldots \\
&= Q^i \frac{\partial L}{\partial \phi^i} + \partial_\mu Q^i \frac{\partial L}{\partial (\partial_\mu \phi^i)} + \ldots \\
&= Q^i \left( \frac{\partial L}{\partial \phi^i} - \partial_\mu \frac{\partial L}{\partial (\partial_\mu \phi^i)} + \ldots \right) + \partial_\mu \left( Q^i \frac{\partial L}{\partial (\partial_\mu \phi^i)} + \ldots \right) \\
&= Q^i \frac{\delta L}{\delta \phi^i} + \partial_\mu \left( Q^i \frac{\partial L}{\partial (\partial_\mu \phi^i)} + \ldots \right) ,
\end{split}
\label{stage 2}
\end{equation} where, in the second line, we used 
\begin{equation}
[\delta_Q, \partial_\mu] = 0 
\label{commutator}
\end{equation} and, in the last equality, we used \eqref{Euler lagrange def}. Putting \eqref{stage 1} and \eqref{stage 2} together, we obtain
\begin{equation}
Q^i \frac{\delta L}{\delta \phi^i} = \partial_\mu \left( B^\mu_Q - Q^i \frac{\partial L}{\partial (\partial_\mu \phi^i)} + \ldots  \right) \equiv \partial_\mu j^\mu_Q
\end{equation} or, equivalently
\begin{equation}
Q^i \frac{\delta \mathbf{L}}{\delta \phi^i} = \mathrm{d} \mathbf{j}_Q ,
\label{equality important}
\end{equation}  where $\mathbf{j}_Q = j^\mu_Q (\mathrm{d}^{n-1}x)_\mu$. In particular, $\mathrm{d} \mathbf{j}_Q \approx 0$ holds on-shell. Hence, we have obtained a representative of the Noether current associated with the global symmetry $Q$ through the correspondence \eqref{one to one}. 

\paragraph{Theorem \normalfont[Noether representation theorem]} Defining the bracket as
\begin{equation}
\{ \mathbf{j}_{Q_1} , \mathbf{j}_{Q_2} \} = \delta_{Q_1} \mathbf{j}_{Q_2} ,
\end{equation} we have
\begin{equation}
\{ \mathbf{j}_{Q_1} , \mathbf{j}_{Q_2} \} \approx \mathbf{j}_{[Q_1, Q_2]} 
\label{algebra current Noether}
\end{equation} ($n>1$), where $[Q_1,Q_2] = \delta_{Q_1} Q_2 - \delta_{Q_2} Q_1$. In other words, the Noether currents form a representation of the symmetries. 

To prove this theorem, we apply $\delta_{Q_1}$ on the left-hand side and the right-hand side of \eqref{equality important}, where $Q$ is replaced by $Q_2$. On the right-hand side, using the first equation of \eqref{commutation relation delta Q}, we obtain
\begin{equation}
\delta_{Q_1}\mathrm{d} \mathbf{j}_{Q_2} \approx \mathrm{d} \delta_{Q_1} \mathbf{j}_{Q_2} .
\label{equa 1}
\end{equation} On the left-hand side, we have
\begin{equation}
\begin{split}
\delta_{Q_1} \left( Q_2^i \frac{\delta \mathbf{L}}{\delta \phi^i} \right) &= \delta_{Q_1} Q_2^i \frac{\delta \mathbf{L}}{\delta \phi^i} + Q_2^i \delta_{Q_1} \frac{\delta \mathbf{L}}{\delta \phi^i}  \\
&=\delta_{Q_1} Q_2^i \frac{\delta \mathbf{L}}{\delta \phi^i} +  Q_2^i \frac{\delta }{\delta \phi^i} (\delta_{Q_1} \mathbf{L} ) -  Q_2^i \sum_{k \ge 0} (-1)^k \partial_{\mu_1} \ldots \partial_{\mu_k} \left( \frac{\partial Q_1^j}{\partial \phi_{\mu_1\ldots\mu_k}^i} \frac{\delta \mathbf{L}}{\delta \Phi^j} \right) \\
&= \delta_{Q_1} Q_2^i \frac{\delta \mathbf{L}}{\delta \phi^i}-  Q_2^i \sum_{k \ge 0} (-1)^k \partial_{\mu_1} \ldots \partial_{\mu_k} \left( \frac{\partial Q_1^j}{\partial \phi_{\mu_1\ldots\mu_k}^i} \frac{\delta \mathbf{L}}{\delta \phi^j} \right) , \\
\end{split}
\end{equation} where, to obtain the second equality, we used \eqref{variation and euler lagrange}. In the last equality, we used \eqref{symmetry def} together with \eqref{euler lagrange and total}. Now, using Leibniz rules in the second term of the right-hand side, we get 
\begin{equation}
\begin{split}
\delta_{Q_1} \left( Q_2^i \frac{\delta \mathbf{L}}{\delta \phi^i} \right) &= \delta_{Q_1} Q_2^i \frac{\delta \mathbf{L}}{\delta \phi^i}- \sum_{k \ge 0}  \partial_{\mu_1} \ldots \partial_{\mu_k} Q_2^j  \left( \frac{\partial Q_1^i}{\partial \phi_{\mu_1\ldots\mu_k}^j} \frac{\delta \mathbf{L}}{\delta \phi^i} \right) + \partial_\mu T^\mu_{Q_1} \left( Q_2,  \frac{\delta L}{\delta \phi} \right) \mathrm{d}^n x \\
&= ( \delta_{Q_1} Q_2^i - \delta_{Q_2} Q_1^i ) \frac{\delta \mathbf{L}}{\delta \phi^i} + \partial_\mu T^\mu_{Q_1} \left( Q_2,  \frac{\delta L}{\delta \phi} \right)  \mathrm{d}^n x \\
&= [Q_1, Q_2]^i  \frac{\delta \mathbf{L}}{\delta \phi^i} + \partial_\mu T^\mu_{Q_1} \left( Q_2,  \frac{\delta L}{\delta \phi} \right) d^n x \\
&= \mathrm{d} \mathbf{j}_{[Q_1, Q_2]} + \mathrm{d} \mathbf{T}_{Q_1} \left( Q_2,  \frac{\delta L}{\delta \phi} \right) ,
\end{split}
\label{equa 2}
\end{equation} where $T^\mu_{Q_1} \left( Q_2,  \frac{\delta L}{\delta \phi} \right)$ is an expression vanishing on-shell. In the second equality, we used \eqref{def variation}, and in the last equality, we used \eqref{equality important}. Putting \eqref{equa 1} and \eqref{equa 2} together results in 
\begin{equation}
\mathrm{d} \left[ \delta_{Q_1} \mathbf{j}_{Q_2} - \mathbf{j}_{[Q_1, Q_2]} - \mathbf{T}_{Q_1} \left( Q_2,  \frac{\delta L}{\delta \phi} \right) \right] = 0 .
\label{closed}
\end{equation} We know from Poincar\'e lemma that locally, every closed form is exact\footnote{The Poincar\'e lemma states that in a star-shaped open subset, the de Rham cohomology class $H^p_{dR}$ is given by \[H^p_{dR} = \left\{
    \begin{array}{ll}
         0 & \mbox{if } 0<p \le n \\
        \mathbb{R} & \mbox{if } p =0
    \end{array} .
\right.\]}. However, this cannot be the case in Lagrangian field theories. In fact, this would imply that every $n$-form is exact, and therefore, there would not be any possibility of non-trivial dynamics. Let us remark that the operator $\mathrm{d}$ that we are using is not the standard exterior derivative, but a horizontal derivative in the jet bundle (see definition \eqref{horizontal}) that takes into account the field-dependence. In this context, we have to use the \textit{algebraic Poincar\'e lemma}.

\paragraph{Lemma \normalfont[Algebraic Poincar\'e lemma]} The cohomology class $H^p(\mathrm{d})$ for the operator $\mathrm{d}$ defined in \eqref{horizontal} is given by
\begin{equation}
H^p(\mathrm{d}) = \left\{
    \begin{array}{lll}
        [\boldsymbol{\alpha}^n] & \mbox{if } p = n \\
         0 & \mbox{if } 0<p<n \\
        \mathbb{R} & \mbox{if } p =0
    \end{array}
\right.
\label{Poincare lemma}
\end{equation} where $[\boldsymbol{\alpha}^n]$ designates the equivalence classes of $n$-forms for the equivalence relation $\boldsymbol{\alpha}^n \sim {\boldsymbol{\alpha}'}^n$ if $\boldsymbol{\alpha}^n = {\boldsymbol{\alpha}'}^n + \mathrm{d} \boldsymbol{\beta}^{n-1}$ \cite{Barnich:2018gdh}.

Le us go back to the proof of \eqref{algebra current Noether}. Applying the algebraic Poincar\'e lemma to \eqref{closed} yields 
\begin{equation}
\delta_{Q_1} \mathbf{j}_{Q_2} = \mathbf{j}_{[Q_1, Q_2]} + \mathbf{T}_{Q_1} \left( Q_2,  \frac{\delta L}{\delta \phi} \right) + \mathrm{d}\boldsymbol{\eta} ,
\end{equation} where $\boldsymbol{\eta}$ is a $(n-2)$-form.  Therefore, on-shell, since $\mathbf{T}_{Q_1} \left( Q_2,  \frac{\delta L}{\delta \phi} \right) \approx 0$ and because Noether currents are defined up to exact $(n-1)$-forms, we obtain the result \eqref{algebra current Noether}. Notice that in classical mechanics (i.e. $n=1$), from \eqref{Poincare lemma}, constant central extensions may appear in the current algebra.

\paragraph{Definition \normalfont[Noether charge]} Given a Noether current $[\mathbf{j}]$, we can construct a \textit{Noether charge} by integrating it on a $(n-1)$-dimensional spacelike surface $\Sigma$, with boundary $\partial \Sigma$, as
\begin{equation}
H_Q[\phi] = \int_\Sigma \mathbf{j} .
\label{Noether charge}
\end{equation} If we assume that the currents and their ambiguities vanish at infinity, this definition does not depend on the representative of the Noether current. Indeed, 
\begin{equation}
H'_Q[\phi] = \int_\Sigma (\mathbf{j} + \mathrm{d} \mathbf{K}) = H_Q[\phi] + \int_{\partial \Sigma} \mathbf{K} ,
\end{equation} where we used the Stokes theorem. Since $\int_{\partial \Sigma} \mathbf{K} =0$, we have $H'_Q[\phi] = H_Q[\phi]$.

\paragraph{Remark \normalfont[Conservation and algebra of Noether charges]} The Noether charge \eqref{Noether charge} is conserved in time, that is,
\begin{equation}
\frac{d}{dt} H_Q[\phi] \approx 0 .
\end{equation} In fact, consider two spacelike hypersurfaces $\Sigma_1 \equiv t_1 = 0$ and $\Sigma_2 \equiv t_2 = 0$. We have
\begin{equation}
H^{t_2}_Q[\phi] - H^{t_1}_Q[\phi] = \int_{\Sigma_2} \mathbf{j}_Q  - \int_{\Sigma_1} \mathbf{j}_Q = \int_{\Sigma_2 - \Sigma_1} \mathrm{d} \mathbf{j}_Q \approx 0 ,
\end{equation} where $\Sigma_2 - \Sigma_1$ is the spacetime volume encompassed between $\Sigma_1$ and $\Sigma_2$. In the second equality, we used the hypothesis that currents vanish at infinity and the Stokes theorem. 

The Noether charges \eqref{Noether charge} form a representation of the algebra of global symmetries, i.e.
\begin{equation}
\{ H_{Q_1}, H_{Q_2} \} \approx H_{[Q_1, Q_2]} ,
\end{equation} where the bracket of Noether charges is defined as 
\begin{equation}
\{ H_{Q_1}, H_{Q_2} \} = \delta_{Q_1} H_{Q_2} = \int_\Sigma \delta_{Q_1} \mathbf{j}_{Q_2} .
\end{equation} This is a direct consequence of \eqref{algebra current Noether}.

\subsection{Gauge symmetries and lower degree conservation law}

\paragraph{Definition \normalfont[Noether identities]} Consider the relation \eqref{equality important} for a gauge symmetry:
\begin{equation}
R^i[f] \frac{\delta L}{\delta \phi^i} = \partial_\mu j^\mu_f .
\label{equation importante gauge}
\end{equation} The left-hand side can be worked out as
\begin{equation}
\begin{split}
R^i[f] \frac{\delta L}{\delta \phi^i} =& \left( R^i_\alpha f^\alpha + R^{i\mu}_\alpha  \partial_\mu f^\alpha + R^{i(\mu\nu)}_\alpha \partial_\mu \partial_\nu f^\alpha + \ldots \right) \frac{\delta L}{\delta \phi^i} \\
=& f^\alpha \left[ R^i_\alpha  \frac{\delta L}{\delta \phi^i} - \partial_\mu \left( R^{i\mu}_\alpha   \frac{\delta L}{\delta \phi^i} \right) +  \partial_\mu \partial_\nu \left( R^{i(\mu\nu)}_\alpha \frac{\delta L}{\delta \phi^i} \right) + \ldots \right] \\
&+ \partial_\mu \LaTeXunderbrace{\left[ R^{i\mu}_\alpha f^\alpha  \frac{\delta L}{\delta \phi^i} - f^\alpha \partial_\nu \left( R^{i(\mu\nu)}_\alpha \frac{\delta L}{\delta \phi^i}  \right) + \ldots \right]}_{\equiv S^\mu_f} .
\end{split}
\label{Noether identity proof}
\end{equation} Therefore, the equation \eqref{equation importante gauge} can be rewritten as
\begin{equation}
f^\alpha R^\dagger_\alpha \left( \frac{\delta L}{\delta \phi}  \right) = \partial_\mu (j^\mu_f - S^\mu_f )  ,
\label{Noether proof 2}
\end{equation} where $R^\dagger_\alpha \left( \frac{\delta L}{\delta \phi^i}  \right) = R_\alpha^i  \frac{\delta L}{\delta \phi^i} - \partial_\mu \left( R^{i\mu}_\alpha   \frac{\delta L}{\delta \phi^i} \right) +  \partial_\mu \partial_\nu \left( R^{i(\mu\nu)}_\alpha \frac{\delta L}{\delta \phi^i} \right) + \ldots$ Since $f$ is a set of arbitrary functions, we can apply the Euler-Lagrange derivative \eqref{Euler lagrange def} with respect to $f^\alpha$ on this equation. Since the right-hand side is a total derivative, it vanishes under the action of the Euler-Lagrange derivative (see \eqref{euler lagrange and total}) and we obtain
\begin{equation}
R^\dagger_\alpha \left( \frac{\delta \mathbf{L}}{\delta \phi}  \right) = 0 .
\label{Noether identity}
\end{equation} This identity is called a \textit{Noether identity}. There is one identity for each independent generator $f^\alpha$. Notice that these identities are satisfied off-shell.

\paragraph{Theorem \normalfont[Noether's second theorem]} We have 
\begin{equation}
R^i[f] \frac{\delta \mathbf{L}}{\delta \phi^i} = \mathrm{d} \mathbf{S}_f \left[\frac{\delta L}{\delta \phi} \right] ,
\label{second Noether theorem}
\end{equation} where $\mathbf{S}_f = S^\mu_f (\mathrm{d}^{n-1} x)_\mu$ is the \textit{weakly vanishing Noether current} (i.e. $\mathbf{S}_f \approx 0$) that was defined in the last line of \eqref{Noether identity proof}. This is a direct consequence of \eqref{Noether identity proof}, taking the Noether identity \eqref{Noether identity} into account.

\paragraph{Example} Consider the theory of general relativity $\mathbf{L} = (16\pi G)^{-1} (R - 2 \Lambda) \sqrt{-g} \mathrm{d}^n x$. The Euler-Lagrange derivative of the Lagrangian is given by 
\begin{equation}
\frac{\delta \mathbf{L}}{\delta g_{\mu\nu}} = -(16\pi G)^{-1}(G^{\mu\nu} + g^{\mu\nu} \Lambda)   \sqrt{-g} \mathrm{d}^n x .
\end{equation} The Noether identity associated with the diffeomorphism generated by $\xi^\mu$ is obtained by following the lines of \eqref{Noether identity proof}:
\begin{equation}
\begin{split}
(16 \pi G)\delta_\xi g_{\mu\nu} \frac{\delta \mathbf{L}}{\delta g_{\mu\nu}} &= -2 \nabla_\mu \xi_\nu (G^{\mu\nu} + g^{\mu\nu} \Lambda)   \sqrt{-g} \mathrm{d}^n x \\
&= 2\xi_\nu \nabla_\mu G^{\mu\nu} \sqrt{-g} \mathrm{d}^n x - \partial_\mu [ 2 \xi_\nu (G^{\mu\nu} + g^{\mu\nu} \Lambda)\sqrt{-g}] \mathrm{d}^n x .
\end{split}
\end{equation} Therefore, the Noether identity is the Bianchi identity for the Einstein tensor
\begin{equation}
\nabla_\mu G^{\mu \nu} = 0
\end{equation} and the weakly vanishing Noether current of Noether's second theorem \eqref{second Noether theorem} is given by
\begin{equation}
\mathbf{S}_\xi = -\frac{\sqrt{-g}}{8\pi G}\xi_\nu (G^{\mu\nu} + g^{\mu\nu} \Lambda) (\mathrm{d}^{n-1}x)_\mu .
\label{weakly vanishin noether current gravity}
\end{equation}

\paragraph{Remark} From \eqref{equation importante gauge} and \eqref{second Noether theorem}, we have $\mathrm{d} (\mathbf{j}_f - \mathbf{S}_f ) = 0$, and hence, from the algebraic Poincar\'e lemma \eqref{Poincare lemma},
\begin{equation}
\mathbf{j}_f =  \mathbf{S}_f + \mathrm{d} \mathbf{K}_f ,
\label{ambiguity}
\end{equation} where $\mathbf{K}_f$ is a $(n-2)$-form. Therefore, as already stated in Noether's first theorem \eqref{one to one}, the Noether current associated with a gauge symmetry is trivial, i.e. vanishing on-shell, up to an exact $(n-1)$-form. A natural question arises at this stage: is it possible to define a notion of conserved quantity for gauge symmetries? Naively, following the definition \eqref{Noether charge}, one may propose the following definition for conserved charge:
\begin{equation}
H_f = \int_\Sigma \mathbf{j}_f \approx \int_{\partial \Sigma} \mathbf{K}_f
\label{bad noether charge}
\end{equation} where, in the second equality, we used \eqref{ambiguity} and Stokes' theorem. This charge will be conserved on-shell since $\mathrm{d}\mathbf{j}_f \approx 0$. The problem is that the $(n-2)$-form $\mathbf{K}_f$ appearing in \eqref{bad noether charge} is completely arbitrary. Indeed, the Noether currents are equivalence classes of currents (see equation \eqref{Noether current def}). Therefore, we have to find an appropriate procedure to isolate a particular $\mathbf{K}_f$. 


\paragraph{Definition \normalfont[Reducibility parameter]} \textit{Reducibility parameters} $\bar{f}$ are parameters of gauge transformations satisfying
\begin{equation}
R[\bar{f}] \approx 0 .
\end{equation} Two reducibility parameters $\bar{f}$ and $\bar{f}'$ are said to be equivalent, i.e. $\bar{f}\sim \bar{f}'$, if $\bar{f} \approx \bar{f}'$. Note that for a large class of gauge theories (including electrodynamics, Yang-Mills and general relativity in dimensions superior or equal to three \cite{Barnich:2001jy , Barnich:2018gdh}), these equivalence classes of asymptotic reducibility parameters are determined by field-independent ordinary functions $\bar{f}(x)$ satisfying the off-shell condition
\begin{equation}
R[\bar{f}] = 0 .
\label{exact red param}
\end{equation} We will call them \textit{exact reducibility parameters}.

\paragraph{Theorem \normalfont[Generalized Noether's theorem]} A one-to-one correspondence exists between \textit{equivalence classes of reducibility parameters} and \textit{equivalence classes of on-shell conserved $(n-2)$-forms} $[\mathbf{K}]$, which can be written as
\begin{equation}
[\bar{f}] \stackrel{\text{\scriptsize 1-1}}{\longleftrightarrow} [\mathbf{K}] .
\label{one to one gen}
\end{equation} In this statement, two conserved $(n-2)$-forms $\mathbf{K}$ and $\mathbf{K}'$ are said to be equivalent, i.e. $\mathbf{K}\sim \mathbf{K}'$, if $\mathbf{K} \approx \mathbf{K}' + \mathrm{d}\mathbf{l}$ where $\mathbf{l}$ is a $(n-3)$-form \cite{Barnich:1994db, Barnich:1995ap}. 

\paragraph{Remark} The Barnich-Brandt procedure allows for the construction of explicit representatives of the conserved $(n-2)$-forms for given exact reducibility parameters $\bar{f}$ \cite{Barnich:2001jy , Barnich:2003xg}. From Noether's second theorem \eqref{second Noether theorem} and \eqref{exact red param}, we have
\begin{equation}
\mathrm{d} \mathbf{S}_{\bar{f}} = 0 .
\label{weakly vanishing is conserved}
\end{equation} From the algebraic Poincar\'e Lemma \eqref{Poincare lemma}, we get\footnote{The minus sign on the left-hand side of \eqref{weakly vanishing is conserved conseq} is a matter of convention.}
\begin{equation}
-\mathrm{d} \mathbf{K}_{\bar{f}} = \mathbf{S}_{\bar{f}} \approx 0 .
\label{weakly vanishing is conserved conseq}
\end{equation} Using the homotopy operator \eqref{homotopy operator}, we define 
\begin{equation}
\mathbf{k}_{\bar{f}}[\phi; \delta \phi] = - I_{\delta \phi}^{n-1} \mathbf{S}_{\bar{f}} .
\label{n-2 in case reduc}
\end{equation} This $\mathbf{k}_{\bar{f}}[\phi; \delta \phi]$ is an element of $\Omega^{n-2,1}$ (see appendix \ref{Useful results}) and is defined up to an exact $(n-2)$-form. This enables us to find an explicit expression for the conserved $(n-2)$-form $\mathbf{K}_{\bar{f}}[\phi]$ as
\begin{equation}
\mathbf{K}_{\bar{f}}[\phi] = \int_\gamma \mathbf{k}_{\bar{f}} [\phi; \delta \phi] ,
\label{DEF OF K}
\end{equation} where $\gamma$ is a path on the solution space relating $\bar{\phi}$ such that $S_{\bar{f}}[\bar{\phi}] = 0$ to the solution $\phi$ of interest. Applying the operator $\mathrm{d}$ on \eqref{DEF OF K} gives back \eqref{weakly vanishing is conserved conseq}, using the property \eqref{commutation homotopy op} of the homotopy operator. Notice that the expression \eqref{DEF OF K} of $\mathbf{K}_{\bar{f}}[\phi]$ generically depends on the chosen path $\gamma$. Therefore, in practice, we consider the $(n-2)$-form $\mathbf{k}_{\bar{f}}[\phi; \delta \phi]$ defined in \eqref{n-2 in case reduc} as the fundamental object, rather than $\mathbf{K}_{\bar{f}}[\phi]$.  

%

\paragraph{Example} Let us return to our example of general relativity. The exact reducibility parameters of the theory are the diffeomorphism generators $\bar{\xi}$, which satisfy
\begin{equation}
\delta_{\bar{\xi}} g_{\mu\nu} = \mathcal{L}_{\bar{\xi}} g_{\mu\nu} = 0 ,
\end{equation} i.e. they are the Killing vectors of $g_{\mu\nu}$. Note that for a generic metric, this equation does not admit any solution. Hence, the previous construction is irrelevant for this general case. Now, consider linearized general relativity around a background $g_{\mu\nu} = \bar{g}_{\mu\nu} + h_{\mu\nu}$. We can show that
\begin{equation}
\delta_{\bar{\xi}} h_{\mu\nu} = \mathcal{L}_{\bar{\xi}}\bar{g}_{\mu\nu} = 0 ,
\label{killing eq background}
\end{equation} i.e. the exact reducibility parameters of the linearized theory are the Killing vectors of the background $\bar{g}_{\mu\nu}$. If $\bar{g}_{\mu\nu}$ is taken to be the Minkowski metric, then the solutions of \eqref{killing eq background} are the generators of the Poincar\'e transformations. The $(n-2)$-form \eqref{DEF OF K} can be constructed explicitly and integrated on a $(n-2)$-sphere at infinity. This gives the ADM charges of linearized gravity \cite{Barnich:2001jy}.

%
%
%

\subsection{Asymptotic symmetries and surface charges}
\label{subsec:Asymptotic symmetries and surface charges}

We now come to the case of main interest, where we are dealing with asymptotic symmetries in the sense of the definition in subsection \ref{Asymptotic symmetry algebra}. The prescription to construct the $(n-2)$-form $\mathbf{k}_f [\phi, \delta\phi]$ associated with generators of asymptotic symmetries $f$ is essentially the same as the one introduced above for exact reducibility parameters. However, this $(n-2)$-form will not be conserved on-shell. Indeed, for a generic asymptotic symmetry, \eqref{weakly vanishing is conserved} does not hold; therefore, the equation \eqref{weakly vanishing is conserved conseq} is not valid anymore. Nonetheless, as we will see below, we still have a control on the \textit{breaking} in the conservation law.  

\paragraph{Definition \normalfont[Barnich-Brandt $(n-2)$-form for asymptotic symmetries]} The $(n-2)$-form $\mathbf{k}_f$ associated with asymptotic symmetries generated by $f$ is defined as
\begin{equation}
\mathbf{k}_f [\phi ; \delta \phi] = -I^{n-1}_{\delta \phi} \mathbf{S}_f ,
\label{expression k}
\end{equation} where $I^{n-1}_{\delta \phi}$ is the homotopy operator \eqref{homotopy operator} and $\mathbf{S}_f$ is the weakly vanishing Noether current defined in the last line of \eqref{Noether identity proof}. For a \textit{first order gauge theory}, namely a gauge theory involving only first order derivatives of the gauge parameters $f=(f^\alpha)$ and the fields $\phi = (\phi^i)$ in the gauge transformations, and first order equations of motion for the fields, the $(n-2)$-form \eqref{expression k} becomes
\begin{equation}
\mathbf{k}_f [\phi ; \delta \phi] = -\frac{1}{2} \delta \phi^i \frac{\partial}{\partial (\partial_\mu \phi^i)} \left( \frac{\partial}{\partial \mathrm{d}x^\mu} \mathbf{S}_f  \right) ,
\end{equation} where 
\begin{equation}
\mathbf{S}_f = R^{i\mu}_\alpha f^\alpha \frac{\delta L}{\delta \phi^i} (\mathrm{d}^{n-1}x)_\mu .
\end{equation} The simplicity of these expressions motivates the study of first order formulations of gauge theories in this context \cite{Barnich:2016rwk , Barnich:2019vzx , Frodden:2019ylc , Oliveri:2019gvm} (see chapter \ref{ch:First order formulations and surface charges}).

\paragraph{Example} Let us consider the theory of general relativity. Applying the homotopy operator on the weakly vanishing Noether current $\mathbf{S}_\xi$ obtained in equation \eqref{weakly vanishin noether current gravity}, we deduce the explicit expression
\begin{equation}
\begin{split}
\mathbf{k}_\xi [g;h ] = \frac{\sqrt{-g} }{8\pi G} (\mathrm{d}^{n-2}x)_{\mu\nu} [&\xi^\nu \nabla^\mu h + \xi^\mu \nabla_\sigma h^{\sigma \nu} + \xi_\sigma \nabla^\nu h^{\sigma \mu}  \\
&+ \frac{1}{2} ( h \nabla^\nu \xi^\mu + h^{\mu \sigma} \nabla_\sigma \xi^\nu + h^{\nu\sigma} \nabla^\mu \xi_\sigma) ] ,
\end{split}
\label{Barnich-Brandt}
\end{equation} where $h_{\mu\nu} = \delta g_{\mu \nu}$. Indices are lowered and raised by $g_{\mu\nu}$ and its inverse, and $h = {h^{\mu}}_\mu$ \cite{Barnich:2001jy}. Notice that this expression has also been derived both in the first order Cartan formulation and in the Palatini formulation of general relativity \cite{Barnich:2016rwk, newGlenn}.

\paragraph{Theorem \normalfont[Conservation law]} Define the \textit{invariant presymplectic current as}
\begin{equation}
\mathbf{W}[\phi;\delta \phi , \delta \phi ] = \frac{1}{2} I^{n}_{\delta \phi} \left( \delta \phi^i \frac{\delta \mathbf{L}}{\delta \phi^i} \right) .
\label{def of invariant presympletcic potential}
\end{equation} We have the following \textit{conservation law}
\begin{equation}
\mathrm{d} \mathbf{k}_f[\phi;\delta \phi] \approx \mathbf{W}[\phi; R[f] , \delta \phi] ,
\label{breaking in the conservation}
\end{equation} where, in the equality $\approx$, it is implied that $\phi$ is a solution of the Euler-Lagrange equations and $\delta \phi$ is a solution of the linearized Euler-Lagrange equations. Furthermore, we use the notation $\mathbf{W} [\phi; R[f] ,  \delta \phi]  = i_{R[f]} \mathbf{W}[\phi; \delta\phi, \delta \phi]$.

The proof of this proposition involves the properties of the operators introduced in appendix \ref{Useful results}. We have
\begin{equation}
\begin{split}
\mathrm{d} \mathbf{k}_f [\phi;\delta\phi] &= -\mathrm{d} I^{n-1}_{\delta \phi} \mathbf{S}_f \\
&= \delta \mathbf{S}_f -  I^n_{\delta\phi}\mathrm{d} \mathbf{S}_f \\
&\approx  -I^n_{\delta\phi} \mathrm{d} \mathbf{S}_f \\
&\approx -I^n_{\delta\phi} \left( R^i[f] \frac{\delta \mathbf{L}}{\delta \phi^i} \right) \\
&\approx  \frac{1}{2} i_{R[f]} I^{n}_{\delta \phi} \left( \delta \phi^i \frac{\delta \mathbf{L}}{\delta \phi^i} \right) \\
&\approx  i_{R[f]} \mathbf{W}[\phi;\delta \phi , \delta \phi ] \\
&\approx \mathbf{W}[\phi; R[f] ,  \delta \phi] .
\end{split}
\end{equation} In the second equality, we used \eqref{commutation homotopy op}. In the third equality, we used the fact that $\delta \mathbf{S}_f \approx 0$, since $\delta \phi$ is a solution of the linearized Euler-Lagrange equations. In the fourth equality, we used Noether's second theorem \eqref{second Noether theorem}. In the fifth equality, we used 
\begin{equation}
i_{R[f]} \mathbf{W}[\phi;\delta\phi, \delta\phi]=  I^{n}_{R[f]} \left( \delta \phi^i \frac{\delta \mathbf{L}}{\delta \phi^i} \right) = - I^{n}_{\delta \phi} \left( R^i[f] \frac{\delta \mathbf{L}}{\delta \phi^i} \right) .
\end{equation} The proof of this statement can be found in appendix A.5 of \cite{Barnich:2007bf}. Finally, in the sixth equality, we used the definition \eqref{def of invariant presympletcic potential}.


\paragraph{Definition \normalfont[Surface charges]} Let $\Sigma$ be a $(n-1)$-surface and $\partial\Sigma$ its $(n-2)$-dimensional boundary. We define the \textit{infinitesimal surface charge} $\ndelta H_f[\phi]$ as
\begin{equation}
\ndelta H_f[\phi] = \int_{\partial\Sigma} \mathbf{k}_f[\phi;\delta\phi] \approx \int_\Sigma \mathbf{W}[\phi; R[f] ,  \delta \phi] .
\label{inifnitesimal surf charge}
\end{equation} The infinitesimal surface charge $\ndelta H_f[\phi]$ is said to be \textit{integrable} if it is $\delta$-exact, i.e. if\\$\ndelta H_f[\phi] = \delta H_f [\phi]$. The symbol $\ndelta$ in \eqref{inifnitesimal surf charge} emphasizes that the infinitesimal surface charge is not necessarily integrable. If it is actually integrable, then we can define the \textit{integrated surface charge} $H_f[\phi]$ as
\begin{equation}
H_f [\phi] = \int_\gamma \delta H_f [\phi]  + N[\bar{\phi}] = \int_\gamma
 \int_{\partial\Sigma} \mathbf{k}_f[\phi;\delta\phi]+ N[\bar{\phi}] ,
 \label{integrated charge}
\end{equation} where $\gamma$ is a path in the solution space, going from a reference solution $\bar{\phi}$ to the solution $\phi$. $N[\bar{\phi}]$ is a chosen value of the charge for this reference solution, which is not fixed by the formalism. Notice that for integrable infinitesimal charge, the integrated charge $H_f [\phi]$ is independent from the chosen path $\gamma$ \cite{Compere:2019qed}. 

\paragraph{Theorem \normalfont[Charge representation theorem]} Assuming integrability\footnote{One also has to assume $\mathbf{E}[\phi; \delta \phi, \delta \phi] = 0$, where $\mathbf{E}[\phi; \delta \phi, \delta \phi]$ is defined in \eqref{definition}.}, the integrated surface charges satisfy the algebra
\begin{equation}
\{ H_{f_1} , H_{f_2} \} \approx H_{[f_1, f_2]_A} + K_{f_1; f_2}[\bar{\phi}] .
\label{charge algebra integrable}
\end{equation} In this expression, the integrated charges bracket is defined as
\begin{equation}
\{ H_{f_1} , H_{f_2} \} = \delta_{f_2} H_{f_1} =\int_{\partial \Sigma} \mathbf{k}_{f_1} [\phi;\delta_{f_2} \phi] .
\end{equation} Furthermore, the central extension $K_{f_1;f_2}[\bar{\phi}]$, which depends only on the reference solution $\bar{\phi}$, is antisymmetric with respect to $f_1$ and $f_2$, i.e. $K_{f_1;f_2}[\bar{\phi}] = - K_{f_2;f_1}[\bar{\phi}]$. It satisfies the $2$-cocycle condition
\begin{equation}
K_{[f_1, f_2]_A ;f_3}[\bar{\phi}] + K_{[f_2, f_3]_A ;f_1}[\bar{\phi}] + K_{[f_3, f_1]_A ;f_2}[\bar{\phi}] \approx 0 .
\end{equation} Therefore, the integrated charges form a representation of the asymptotic symmetry algebra, up to a central extension \cite{Barnich:2001jy, Barnich:2007bf}. 

For the proof of this theorem, see e.g. section 1.4 of \cite{Compere:2019qed}.

\paragraph{Remark} In the literature, there are several criteria based on properties of the surface charges, that make a choice of boundary conditions interesting. The main properties are the following:
\begin{itemize}
\item The charges are usually required to be \textit{finite}. Two types of divergences may occur: divergences in the expansion parameter defining asymptotics, say $r$, and divergences when performing the integration on the $(n-2)$-surface $\partial \Sigma$.
\item The charges have to be \textit{integrable}. As explained above, this criterion enables us to define integrated surface charges as in \eqref{integrated charge}. Integrability implies that the charges form a representation of the asymptotic symmetry algebra, up to a central extension (see \eqref{charge algebra integrable}). 
\item The charges have to be \textit{generically non-vanishing}. Indeed, the asymptotic symmetries for which associated integrated charges identically vanish are considered as trivial in the strong definition of asymptotic symmetry group \eqref{ASG def 3}.
\item The charges have to be \textit{conserved} in time when the integration is performed on a spacelike $(n-2)$-dimensional surface $\partial \Sigma$ at infinity. This statement is not guaranteed a priori because of the breaking in the conservation law \eqref{breaking in the conservation}. 
\end{itemize}

However, even if these requirements seem reasonable, in practice, some of them may not be satisfied. Indeed, as we will see below, the BMS charges in four dimensions are not always finite, neither integrable, nor conserved \cite{Barnich:2011mi}. We now discuss the violation of some of the above requirements:
\begin{itemize}
\item The fact that the charges may \textit{not be finite} in terms of the expansion parameter $r$ can be expected when the asymptotic region is taken to be at infinity. Indeed, consider $r$ as a cut-off. It makes sense to integrate on a surface $\partial\Sigma$ at a constant finite value of $r$, encircling a finite volume. Then, taking the limit $r\to \infty$ leads to an infinite volume; therefore, it does not come as a surprise that quantities diverge. Furthermore, it has recently been shown in \cite{Godazgar:2018vmm} that subleading orders in $r$ in the $(n-2)$-form $\mathbf{k}_{f}[\phi;\delta \phi]$ contain some interesting physical information, such as the $10$ conserved Newman-Penrose charges \cite{Newman:1968uj}. Therefore, it seems reasonable to think that overleading orders in $r$ may also contain relevant information (see e.g. \cite{Compere:2017wrj , Campiglia:2016jdj , Compere:2019odm}). 
\item The \textit{non-integrability} of the charges may be circumvented by different procedures to isolate an integrable part in the expression of the charges (see e.g. \cite{Wald:1999wa} and \cite{Compere:2018ylh}). However, the final integrated surface charges obtained by these procedures do not have all the properties that integrable charges would have. In particular, the representation theorem does not generically hold. Another philosophy is to keep working with non-integrable expressions, without making any specific choice for the integrable part of the charges. In some situations, it is still possible to define a modified bracket for the charges, leading to a representation of the asymptotic symmetry algebra, up to a $2$-cocycle which may depend on fields \cite{Barnich:2011mi, Compere:2018ylh }. However, no general representation theorem exists in this context, even if some progress has been made \cite{Troessaert:2015nia}.
\item Finally, the \textit{non-conservation} of the charges contains some important information on the physics. For example, in asymptotically flat spacetimes at null infinity, the non-conservation in time of the charges associated with time translations is known as the \textit{Bondi mass loss}. This tells us that the mass decreases in time at future null infinity because of a flux of radiation through the boundary. Hence, the non-conservation of the charges contains important information on the dynamics of the system.      
\end{itemize} Even if the charges have these pathologies, they still offer important insights on the system. They could be seen as interesting combinations of the elements of the solution space that enjoy some properties in their transformation (see e.g. \cite{Barnich:2016lyg, Barnich:2015uva}). 

\paragraph{Examples} We now provide explicit examples of surface charge constructions in four-dimensional general relativity. First, consider asymptotically AdS$_4$ spacetimes with Dirichlet boundary conditions (AAdS2) (condition \eqref{as ads} together with \eqref{BC Dirichlet}), the associated solution derived in subsection \ref{Solution space} (equation \eqref{solution space dirichlet}), and the associated asymptotic Killing vectors derived in subsection \ref{Asymptotic symmetry algebra}. Inserting this solution space and these asymptotic Killing vectors into the $(n-2)$-form \eqref{Barnich-Brandt} results in an integrable expression at order $\rho^0$. Therefore, we can construct an integrated surface charge \eqref{integrated charge} where the $2$-surface $\partial \Sigma$ is taken to be the $2$-sphere at infinity, written $S^2_\infty$. We have the explicit expression
\begin{equation}
H_\xi [g] = \int_{S^2_\infty} \mathrm{d}^2\Omega ~(\xi_0^a {T_a}^t  ) ,
\end{equation} where $\mathrm{d}^2\Omega$ is the integration measure on the $2$-sphere (see e.g. \cite{Compere:2008us}). These charges are finite and generically non-vanishing. Furthermore, we can easily show that they are conserved in time, i.e.
\begin{equation}
\frac{{d}}{{d}t} H_\xi [g] \approx 0 .
\end{equation}

Now, we consider definition \eqref{asymp flat 1} with \eqref{asymp flat 3} of asymptotically flat spacetimes in four dimensions (AF3). The surface charges are obtained by inserting the corresponding solution space derived in subsection \ref{Solution space} (see equation \eqref{solution space Bondi dirich}) and the asymptotic Killing vectors discussed in subsection \ref{Asymptotic symmetry algebra} into the expression \eqref{Barnich-Brandt}, and then integrating over $S^2_\infty$. The result is given by
\begin{equation}
\ndelta H_\xi [g;\delta g] \approx \delta J_\xi [g] + \Theta_\xi [g; \delta g] ,
\label{BMS charges}
\end{equation} where
\begin{equation}
\begin{split}
&J_\xi [g] = \frac{1}{16\pi G} \int_{S^2_\infty} \mathrm{d}^2 \Omega ~\left[4 f M + Y^A (2N_A + \frac{1}{16} \partial_A (C^{CB} C_{CB}) ) \right] \\
&\Theta_\xi [g;\delta g] = \frac{1}{16 \pi G} \int_{S^2_\infty} \mathrm{d}^2 \Omega ~\left[\frac{f}{2} N_{AB} \delta C^{AB} \right]
\end{split}
\label{integrable and non integrable parts}
\end{equation} and where $N_{AB} = \partial_u C_{AB}$ \cite{Barnich:2011mi}. As mentioned above, the infinitesimal surface charges are not integrable. Therefore, we cannot unambiguously define an integrated surface charge as in \eqref{integrated charge} (see, however, \cite{Wald:1999wa , Compere:2018ylh}). In particular, the representation theorem \eqref{charge algebra integrable} does not hold. Nevertheless, we can define the following modified bracket \cite{Barnich:2011mi}:
\begin{equation}
\{ J_{\xi_1} , J_{\xi_2} \}_* = \delta_{\xi_2} J_{\xi_1}[g] + \Theta_{\xi_2} [g;\delta_{\xi_1}g ] .
\label{modified bracket charges}
\end{equation} We can show that
\begin{equation}
\{ J_{\xi_1} , J_{\xi_2} \}_* \approx J_{[\xi_1, \xi_2]_A} [g] +  K_{\xi_1;\xi_2}[g] ,
\label{algebra modified}
\end{equation} where $ K_{\xi_1;\xi_2}[g]$ is a \textit{field-dependent} $2$-cocycle given explicitly by\footnote{Notice that this $2$-cocycle is zero for globally well-defined conformal transformations on the $2$-sphere. It becomes non-trivial when considering the extended BMS$_4$ group with $\mathfrak{diff}(S^1) \oplus  \mathfrak{diff}(S^1)$ superrotations.}
\begin{equation}
K_{\xi_1;\xi_2}[g] = \frac{1}{32 \pi G} \int_{S^2_\infty} \mathrm{d}^2 \Omega~ [C^{BC} (f_1 D_B D_C D_A Y^A_2 - f_2 D_B D_C D_A Y_1^A)] .
\end{equation} It satisfies the generalized $2$-cocycle condition 
\begin{equation}
K_{[\xi_1, \xi_2]_A, \xi_3} + \delta_{\xi_3} K_{\xi_1, \xi_2} + \text{cyclic (1,2,3)} \approx 0 .
\label{generalized 2 cocycle condition}
\end{equation} For the algebra \eqref{algebra modified} to make sense, its form should not depend on the particular choice of integrable part $J_\xi[g]$ in \eqref{integrable and non integrable parts}. In particular, defining $J' = J - N$ and $\Theta' = \Theta + \delta N$ for some $N = N_\xi[g]$, we obtain
\begin{equation}
\{ J'_{\xi_1} , J'_{\xi_2} \}_* = J'_{[\xi_1, \xi_2]_A} [g] +  K'_{\xi_1;\xi_2}[g] ,
\end{equation} where $\{ J'_{\xi_1} , J'_{\xi_2} \}_*= \delta_{\xi_2} J'_{\xi_1}[g] + \Theta'_{\xi_2}[g;\delta_{\xi_1} g]$ and 
\begin{equation}
K'_{\xi_1;\xi_2} = K_{\xi_1, \xi_2} - \delta_{\xi_2} N_{\xi_1} + \delta_{\xi_1} N_{\xi_2} + N_{[\xi_1, \xi_2]_A} .
\label{modif central charge}
\end{equation} Notice that $- \delta_{\xi_2} N_{\xi_1} + \delta_{\xi_1} N_{\xi_2} + N_{[\xi_1, \xi_2]_A}$ automatically satisfies the generalized $2$-cocycle condition \eqref{generalized 2 cocycle condition} \cite{Barnich:2011mi}. Another property of the surface charges \eqref{BMS charges} and \eqref{integrable and non integrable parts} is that they are not conserved. Indeed, 
\begin{equation}
\frac{d}{du} \ndelta H_{\xi}[g] = \int_{S^2_\infty} \mathbf{W}[g;\delta_\xi g , \delta g] ,
\label{bondi mass loss}
\end{equation} where $\mathbf{W}[g;\delta g , \delta g]$ was computed\footnote{More precisely, in \cite{Compere:2018ylh}, we computed the presymplectic form $\boldsymbol{\omega}[g;\delta g , \delta g ]$ introduced below. However, as we will see, this is equal to the invariant presymplectic current in the Bondi gauge.} in \cite{Compere:2018ylh}. We have
\begin{equation}
\int_{S^2_\infty}  \mathbf{W}[g;\delta g , \delta g] = -\frac{1}{32\pi G}\int_{S^2_\infty} \mathrm{d}^2 \Omega~ [\delta N^{AB} \wedge \delta C_{AB} ] .
\label{non conservation BMS}
\end{equation} Notice that taking $f = 1$ and $Y^A =0$ in \eqref{bondi mass loss}, we recover the famous \textit{Bondi mass loss formula} \cite{Bondi:1962px , Sachs:1962wk , Sachs:1962zza}. This formula indicates that the mass is decreasing in time because of the leak of radiation through $\mathscr{I}^+$. This was a striking argument for the existence of gravitational waves at the non-linear level of the theory. Finally, despite the BMS charges \eqref{BMS charges} and \eqref{integrable and non integrable parts} not being divergent in $r$\footnote{As explained in \cite{Compere:2018ylh, newCompere , Donnay:2020guq}, when taking into account the contact terms due to the meromorphic poles on the celestial sphere, divergences in $r$ actually appear in the expressions \eqref{integrable and non integrable parts}.}, we can show that some of the supertranslation charges diverge for the Kerr solution \cite{Barnich:2011mi}.

\paragraph{Remark} A non-trivial relation seems to exist between conservation and integrability of the surface charges. For example, in the case of Dirichlet boundary conditions in asymptotically AdS$_4$ spacetimes (AAdS2) considered above, we see that the surface charges are both integrable and conserved. Reciprocally, there is a relation between non-conservation and non-integrability of the surface charges. For example, in the asymptotically flat case (AF3), we see that the source of non-integrability is contained in the asymptotic shear $C_{AB}$ and the news function $N_{AB} = \partial_u C_{AB}$. These are precisely the functions involved in the right-hand side of \eqref{non conservation BMS}. We can consider many other examples where this phenomenon appears. Therefore, non-integrability is related to non-conservation of the charges. We will see below that for diffeomorphism-invariant theories, the relation between non-conservation and integrability is transparent in the covariant phase space formalism.

\subsection{Relation between Barnich-Brandt and Iyer-Wald procedures}
\label{Relation between Barnich-Brandt and Iyer-Wald procedures}

In this subsection, we briefly discuss the covariant phase space formalism leading to the Iyer-Wald prescription for surface charges \cite{Wald:1999wa, Iyer:1994ys , Wald:1993nt , Harlow:2019yfa}. Notice that this method is valid only for diffeomorphism-invariant theories (including general relativity), and not for any gauge theories. In practice, this means that the parameters of the asymptotic symmetries are diffeomorphisms generators, i.e. $f \equiv \xi$ and $\delta_f \phi \equiv \mathcal{L}_\xi \phi$. Finally, we relate this prescription to the Barnich-Brandt prescription presented in detail in the previous section.

\paragraph{Definition \normalfont[Presymplectic form]} Consider a diffeomorphism-invariant theory with Lagrangian $\mathbf{L} = L \mathrm{d}^n x$. Let us perform an arbitrary variation of the Lagrangian. Using a similar procedure as in \eqref{stage 2}, we obtain
\begin{equation}
\begin{split}
\delta L &= \delta \phi^i \frac{\partial L}{\partial \phi^i} + \delta \partial_\mu \phi^i   \frac{\partial L}{\partial (\partial_\mu \phi^i)} + \ldots \\
&= \delta \phi^i \frac{\delta L}{\delta \phi^i} + \partial_\mu \left( \delta \phi^i \frac{\partial L}{\partial (\partial_\mu \phi^i ) } + \ldots \right) \\
&= \delta \phi^i \frac{\delta L}{\delta \phi^i} + \partial_\mu  \theta^\mu [\phi; \delta \phi] ,
\end{split}
\label{eq:presymp construct}
\end{equation} where 
\begin{equation}
\boldsymbol{\theta} [ \phi; \delta \phi ] = \theta^\mu [ \phi; \delta \phi ] (\mathrm{d}^{n-1}x)_\mu = \left( \delta \phi^i \frac{\partial L}{\partial (\partial_\mu \phi^i ) } + \ldots \right) (\mathrm{d}^{n-1}x)_\mu = I^n_{\delta \phi} \mathbf{L}
\label{presymplectic potential}
\end{equation} is the \textit{presymplectic potential}. Taking into account that $\delta$ is Grassmann odd, the equation \eqref{eq:presymp construct} can be rewritten as
\begin{equation}
\delta \mathbf{L} = \delta \phi^i \frac{\delta \mathbf{L}}{\delta \phi^i} - \mathrm{d} \boldsymbol{\theta} [\phi; \delta \phi] .
\label{var lagrangien presymp}
\end{equation} Now, the \textit{presymplectic form} $\boldsymbol{\omega}$ is defined as
\begin{equation}
\boldsymbol{\omega}[\phi;\delta \phi, \delta \phi] = \delta \boldsymbol{\theta} [\phi, \delta \phi ] .
\label{presymplectic form}
\end{equation}

\paragraph{Definition \normalfont[Iyer-Wald $(n-2)$-form for asymptotic symmetries]} The \textit{Iyer-Wald $(n-2)$-form} $\mathbf{k}^{IW}_\xi$ associated with asymptotic symmetries generated by $\xi$ is defined as
\begin{equation}
\mathbf{k}_\xi^{IW} [\phi; \delta \phi] = -\delta \mathbf{Q}_\xi [\phi] + \iota_{\xi} \boldsymbol{\theta}[\phi ; \delta \phi ] ,
\label{Iyer-Wald}
\end{equation} up to an exact $(n-2)$-form\footnote{\label{useful footnote} In the definition \eqref{Iyer-Wald}, we assumed that the variational operator $\delta$ in front of the Noether-Wald charge does not see the possible field-dependence of the asymptotic Killing vectors $\xi^\mu$. Strictly speaking, one should write $\mathbf{k}_\xi^{IW} [\phi; \delta \phi] = -\delta \mathbf{Q}_\xi [\phi] +  \mathbf{Q}_{\delta \xi} [\phi] + \iota_{\xi} \boldsymbol{\theta}[\phi ; \delta \phi ]$.}. In this expression, $\mathbf{Q}_\xi [\phi]  = - I_\xi^{n-1} \boldsymbol{\theta} [\phi; \mathcal{L}_\xi \phi]$ is called the \textit{Noether-Wald} surface charge.

\paragraph{Example} For general relativity theory, the (canonical) presymplectic potential \eqref{presymplectic potential} is given by 
\begin{equation}
\boldsymbol{\theta}[g;h] = \frac{\sqrt{-g}}{16\pi G}  (\nabla_\nu h^{\mu\nu} - \nabla^\mu h )(\mathrm{d}^{n-1}x)_\mu ,
\label{canonical presymplectic potential}
\end{equation} where $h_{\mu\nu} = \delta g_{\mu \nu}$. Indices are lowered and raised by $g_{\mu\nu}$ and its inverse, and $h = {h^{\mu}}_\mu$. From this expression, the Noether-Wald charge can be computed; we obtain
\begin{equation}
\mathbf{Q}_\xi [g]  = - I_\xi^{n-1} \boldsymbol{\theta} [g; \mathcal{L}_\xi g] = \frac{\sqrt{-g}}{8\pi G}  \nabla^\mu \xi^\nu (\mathrm{d}^{n-2} x)_{\mu\nu}
\end{equation} and we recognize the \textit{Komar charge}. Finally, inserting these expression into \eqref{Iyer-Wald} yields
\begin{equation}
\mathbf{k}_\xi^{IW} [g;h] = \frac{\sqrt{-g}}{8 \pi G}  \left(   \xi^\mu \nabla_\sigma h^{\nu\sigma} - \xi^\mu \nabla^\nu h + \xi_\sigma \nabla^\nu h^{\mu\sigma} + \frac{1}{2} h \nabla^\nu \xi^\mu - h^{\rho\nu} \nabla_\rho \xi^\mu \right) (\mathrm{d}^{n-2} x)_{\mu\nu} .
\label{IW for GR}
\end{equation}

\paragraph{Theorem \normalfont[Conservation law]} We have the following \textit{conservation law}:
\begin{equation}
\mathrm{d} \mathbf{k}^{IW}_\xi [\phi;\delta \phi] \approx \boldsymbol{\omega} [\phi;  \mathcal{L}_\xi \phi, \delta \phi] ,
\label{breaking in the conservation 2}
\end{equation} where, in the equality $\approx$, it is implied that $\phi$ is a solution of the Euler-Lagrange equations and $\delta \phi$ is a solution of the linearized Euler-Lagrange equations. Furthermore, $\boldsymbol{\omega} [\phi; \mathcal{L}_\xi \phi,  \delta \phi ]  =  i_{\mathcal{L}_\xi \phi} \boldsymbol{\omega}[\phi; \delta\phi, \delta \phi]=- \boldsymbol{\omega}[\phi; \delta \phi ,  \mathcal{L}_\xi \phi]$. 

This can be proved using Noether's second theorem \eqref{second Noether theorem} (see e.g. \cite{Compere:2019qed} for a detailed proof). 

\paragraph{Remark} In the covariant phase space formalism, the relation between non-integrability and non-conservation mentioned in the previous subsection is clear. Indeed, 
\begin{equation}
\begin{split}
\delta \ndelta H_\xi [\phi] &=  \int_{\partial \Sigma} \delta \textbf{k}^{IW}_\xi [\phi, \delta \phi] \\
&= \int_{\partial \Sigma}+ \delta \iota_\xi \boldsymbol{\theta} [g, \delta g] \\
&= - \int_{\partial \Sigma} \iota_\xi  \delta \boldsymbol{\theta} [g, \delta g] \\
&= -\int_{\partial \Sigma} \iota_\xi \boldsymbol{\omega} [g;\delta g, \delta g ] ,
\end{split}
\end{equation} where we used \eqref{Iyer-Wald} and \eqref{presymplectic form} in the second and the fourth equality, respectively. The surface charge $\ndelta H_\xi [\phi]$ is integrable only if $\delta \ndelta H_\xi [\phi] =0 $, if and only if
\begin{equation}
\int_{\partial \Sigma} \iota_\xi \boldsymbol{\omega} [g;\delta g, \delta g ] = 0
\end{equation} Therefore, from 
\begin{equation}
\mathrm{d} \ndelta H_\xi [\phi] =  \int_{\partial \Sigma} \mathrm{d} \mathbf{k}^{IW}_\xi [g,\delta g] \approx \int_{\partial \Sigma} \boldsymbol{\omega} [\phi;  \mathcal{L}_\xi \phi, \delta \phi] ,
\end{equation} the non-conservation is controlled by $\boldsymbol{\omega}[g,\delta g , \delta g]$ and is an obstruction for the integrability.

\paragraph{Remark} As in the Barnich-Brandt procedure, the Iyer-Wald $(n-2)$-form \eqref{Iyer-Wald} is defined up to an exact $(n-2)$-form. However, there is another source of ambiguity here coming from the definition of the presymplectic potential \eqref{presymplectic potential}. In fact, we have the freedom to shift $\boldsymbol{\theta}$ by an exact $(n-1)$-form as
\begin{equation}
\boldsymbol{\theta}[\phi; \delta \phi]\to \boldsymbol{\theta}[\phi; \delta \phi] - \mathrm{d} \mathbf{Y}[\phi; \delta \phi] ,
\label{ambiguity presymplectic potential IW}
\end{equation} where $\mathbf{Y}[\phi; \delta \phi]$ is a $(n-2)$-form. This implies that the presymplectic form \eqref{presymplectic form} is modified as
\begin{equation}
\boldsymbol{\omega}[\phi; \delta \phi, \delta \phi] \to \boldsymbol{\omega} [\phi; \delta \phi, \delta \phi] + \mathrm{d} \delta \mathbf{Y}[\phi; \delta \phi] ,
\end{equation} where we used the fact that both $\mathrm{d}$ and $\delta$ are Grassmann odd. The Noether-Wald charge becomes
\begin{equation}
\mathbf{Q}_\xi [\phi] \to \mathbf{Q}_\xi[\phi] +  \mathbf{Y}[\phi; \mathcal{L}_\xi \phi] ,
\end{equation} up to an exact $(n-2)$-form which can be reabsorbed in the $(n-2)$-form ambiguity for $\mathbf{k}^{IW}_\xi$ discussed above. Therefore, this ambiguity modifies $\mathbf{k}^{IW}_{F}$ given in \eqref{Iyer-Wald} by
\begin{equation}
\mathbf{k}^{IW}_\xi [\phi; \delta \phi] \to \mathbf{k}^{IW}_\xi [\phi; \delta \phi] - \delta \mathbf{Y}[\phi; \mathcal{L}_\xi \phi] -\iota_\xi \mathrm{d} \mathbf{Y}[\phi; \delta \phi] .
\label{modif on n-2}
\end{equation} 

%

\paragraph{Definition} Let us introduce an important $(n-2)$-form which is involved in the relation between the Barnich-Brandt and Iyer-Wald prescriptions discussed in the remark below. We define
\begin{equation}
\mathbf{E}[\phi; \delta \phi , \delta \phi] = - \frac{1}{2} I^{n-1}_{\delta \phi} \boldsymbol{\theta} = - \frac{1}{2} I^{n-1}_{\delta \phi} I^n_{\delta \phi} \mathbf{L} .
\label{definition}
\end{equation} 

\paragraph{Remark} We now relate the Barnich-Brandt and the Iyer-Wald prescriptions to construct the $(n-2)$-form. Let us start from the expression \eqref{var lagrangien presymp} of the variation of the Lagrangian. We apply the homotopy operator on each side of the equality. We have
\begin{equation}
\begin{split}
I_{\delta \phi}^n \delta \mathbf{L} &= I^n_{\delta \phi} \left( \delta \phi \frac{\delta \mathbf{L}}{\delta \phi} \right) - I^n_{\delta \phi} \mathrm{d} \boldsymbol{\theta} \\
&= I^n_{\delta \phi} \left( \delta \phi \frac{\delta \mathbf{L}}{\delta \phi} \right) - \delta \boldsymbol{\theta} - \mathrm{d} I^{n-1}_{\delta \phi} \boldsymbol{\theta} .
\end{split}
\end{equation} Therefore,
\begin{equation}
I_{\delta \phi}^n \delta \mathbf{L} + \delta \boldsymbol{\theta}= I^n_{\delta \phi} \left( \delta \phi \frac{\delta \mathbf{L}}{\delta \phi} \right) - \mathrm{d} I^{n-1}_{\delta \phi} \boldsymbol{\theta} .
\end{equation} 
Since $[\delta , I_{\delta \phi}^n] = 0$ because $\delta^2 = 0$, the left-hand side of the last equality can be rewritten as $\delta I_{\delta \phi}^n  \mathbf{L} + \delta \boldsymbol{\theta} = 2 \delta \boldsymbol{\theta} = 2 \boldsymbol{\omega}$ where we used \eqref{presymplectic potential}. Now, using \eqref{def of invariant presympletcic potential} and \eqref{definition}, we obtain the relation between the presymplectic form $\boldsymbol{\omega}$ and the invariant presymplectic current $\mathbf{W}$ as
\begin{equation}
\boldsymbol{\omega}[\phi;\delta \phi, \delta \phi] = \mathbf{W}[\phi;\delta \phi, \delta \phi] + \mathrm{d} \mathbf{E}[\phi;\delta \phi, \delta \phi] .
\end{equation} Contracting this relation with $i_{\mathcal{L}_\xi \phi}$ results in
\begin{equation}
\boldsymbol{\omega}[\phi;\mathcal{L}_\xi \phi, \delta \phi] = \mathbf{W}[\phi;\mathcal{L}_\xi \phi, \delta \phi] + \mathrm{d} \mathbf{E}[\phi;\delta \phi,\mathcal{L}_\xi \phi] .
\end{equation} Finally, using the on-shell conservation laws \eqref{breaking in the conservation} and \eqref{breaking in the conservation 2}, we obtain
\begin{equation}
\mathbf{k}^{IW}_\xi [\phi;\delta \phi] \approx \mathbf{k}_\xi [\phi;\delta \phi] + \mathbf{E}[\phi;\delta \phi, \mathcal{L}_\xi \phi] , 
\label{ling IW BB}
\end{equation} up to an exact $(n-2)$-form. Therefore, the Barnich-Brandt $(n-2)$-form $\mathbf{k}_\xi [\phi;\delta \phi]$ differs from the Iyer-Wald $(n-2)$-form $\mathbf{k}_\xi^{IW} [\phi;\delta \phi]$ by the term $\mathbf{E}[\phi;\delta \phi, \mathcal{L}_\xi \phi]$. 

\paragraph{Examples} We illustrate these concepts with the case of general relativity. The $(n-2)$-form $\mathbf{E}[\phi;\delta \phi, \delta \phi]$ can be computed using \eqref{definition}. We obtain 
\begin{equation}
\mathbf{E}[g;\delta g, \delta g] = \frac{\sqrt{-g}}{32\pi G} {(\delta g)^\mu}_\sigma \wedge (\delta g)^{\sigma \nu} (\mathrm{d}^{n-2} x)_{\mu\nu} .
\label{def en gr}
\end{equation} When contracted with $i_{\mathcal{L}_\xi g}$, this leads to
\begin{equation}
\mathbf{E}[g; \delta g , \mathcal{L}_\xi g] =- \frac{\sqrt{-g}}{16\pi G} (\nabla^\mu \xi_\sigma + \nabla_\sigma \xi^\mu) (\delta g)^{\sigma\nu} (\mathrm{d}^{n-2}x)_{\mu\nu} ,
\label{difference BB IW}
\end{equation} up to an exact $(n-2)$-form. This expression can also be obtained from \eqref{ling IW BB} by comparing the explicit expressions \eqref{Barnich-Brandt} and \eqref{IW for GR}. Notice that the difference between the Barnich-Brandt and the Iyer-Wald definitions \eqref{difference BB IW} vanishes for a Killing vectors $\xi^\mu$. Furthermore, a simple computation shows that the relevant components of the $(n-2)$-form \eqref{def en gr} involved in the computation of the surface charges vanish in both the Fefferman-Graham gauge \eqref{FG gauge} and the Bondi gauge \eqref{Bondi gauge}. Therefore, the Barnich-Brandt and the Iyer-Wald prescriptions lead to the same surface charges in these gauges. For an example where the two prescriptions do not coincide, see for instance, \cite{Azeyanagi:2009wf}.

\chapter{First order formulations and surface charges}
\label{ch:First order formulations and surface charges}

As mentioned in the previous chapter, the formalism to construct the co-dimension 2 forms containing the information on the surface charges is particularly well-adapted for first order gauge theories. In this chapter, we review some first order formulations of general relativity and apply the techniques of the covariant phase space formalism in this context. 

In section \ref{Generalities}, we study a class of theories that encompasses most of the first order gauge theories, including general relativity in Cartan and Newman-Penrose formulations, first order Maxwell theory, first order Yang-Mills theory and Chern-Simons theory. We also discuss vielbeins and connections in presence torsion and non-metricity. In section \ref{First order formulations of general relativity}, we review important first order formulations of general relativity, Cartan and Newman-Penrose formulations, and apply the surface charges formalism. For each case, we relate the obtained results to the standard second order metric formulation of general relativity discussed in the examples in section \ref{Surface charges}. 

This chapter essentially reproduces \cite{Barnich:2016rwk , Barnich:2019vzx , newGlenn}.

\section{Generalities}
\label{Generalities}

\subsection{Covariantized Hamiltonian formulations}
\label{sec:covar-hamilt-form}

In this subsection, we study an important class of first order gauge theories that is particularly well-adapted to the application of the surface charges formalism presented in section \ref{Surface charges}. Let us consider a first order
theory that depends at most linearly on the derivatives of the
fields,
\begin{equation}
  \label{eq:146}
  L=a^\mu_j\d_\mu\phi^j-h,
\end{equation}
with a generating set of gauge
transformations that depends at most on first order derivatives of the
gauge parameters,
\begin{equation}
  \label{eq:147}
  \delta_f\phi^i=R^i[f] = R^i_\alpha f^\alpha +R^{i\mu}_\alpha\partial_\mu
  f^\alpha,
\end{equation}
and where the derivatives of the fields occur at most linearly in the
term that does not contain derivatives of gauge parameters,
\begin{equation}
  \label{eq:94}
  R^i_\alpha=R^{i0}_\alpha+R^{i\nu}_{j\alpha}\d_\nu\phi^j.
\end{equation}
We thus assume that
$a^\mu_j[x,\phi],
h[x,\phi],R^{i0}_\alpha[x,\phi],R^{i\nu}_{j\alpha}[x,\phi],
R^{i\mu}_\alpha[x,\phi]$ do not depend on derivatives of the
fields.

As the notation indicates, this is a covariantized version of first
order Hamiltonian actions, where $\phi^i$ contains both the canonical
variables and the Lagrange multipliers, while $h$ includes both the
canonical Hamiltonian and the constraints. For instance, for a first
class Hamiltonian system, we have
\begin{equation}
  \label{eq:153}
  L[z,u]=a_A(z)\dot z^A -H(z)-u^a\gamma_a(z).
\end{equation}
Here $z^A$ are the phase-space variables and $a_A(z)$ are
  the components of
  the symplectic potential. In the case of
  Darboux coordinates for instance, $z^A=(q^i,p_j)$ and
  $a_A=(p_1,\dots p_n,0\dots,0)$.  Furthermore, $H$ is the
  Hamiltonian, $\gamma_a$ are the first-class constraints and
  $u^a$ are the associated Lagrange multipliers. The symplectic $2$-form
$\sigma_{AB}=\d_A a_B-\d_Ba_A$ is assumed to be invertible,
$\sigma^{CA}\sigma_{AB}=\delta^C_B$ with associated Poisson bracket
$\{F,G\}=\dover{F}{z^A}\sigma^{AB}\dover{G}{z^B}$ and
\begin{equation}
  \label{eq:154}
  \{\gamma_a,\gamma_b\}=C^c_{ab}(z)\gamma_c,\quad
  \{H,\gamma_a\}=V^b_a(z)\gamma_b.
\end{equation}
For such systems, a generating set of gauge symmetries is given by
\begin{equation}
  \label{eq:155}
  \delta_fz^A=\{z^A,\gamma_a\}f^a,\quad
\delta_fu^a=\dot f^a-C^a_{bc}u^bf^c-V^a_bf^b,
\end{equation}
see e.g.~\cite{Henneaux:1992ig} for more details.

By using suitable sets of \textit{auxiliary fields}, namely fields whose equations of motion can be solved algebraically in terms of the other fields and their derivatives \cite{Henneaux:1992ig}, the class of theories \eqref{eq:146} is relevant for gravity in
the standard Cartan formulation or the one adapted to the
Newman-Penrose formalism, as discussed below. Chern-Simons theory is
directly of this type, while Yang-Mills theories are of this type when
using the curvatures as auxiliary fields (see
e.g.~\cite{Arnowitt:1962hi} for the case of Maxwell's theory).

For a Lagrangian of the form \eqref{eq:146}, the Euler-Lagrange derivative of the Lagrangian reduces to $\vddl{L}{\phi^i}=\dover{L}{\phi^i}-\d_\mu \left(\dover{L}{\d_\mu\phi^i} \right)$ and is
explicitly given by
\begin{equation}
  \label{eq:148}
  \vddl{L}{\phi^i}=\sigma^{\mu}_{ij}\d_\mu\phi^j-\d_ih
  -\dover{}{x^\mu} a^\mu_i,\quad \sigma^\mu_{ij}=\d_i a^\mu_j-\d_j
  a^\mu_i \Longrightarrow \d_{[i} \sigma^\mu_{jk]}=0,
\end{equation}
where $\d_i=\dover{}{\phi^i}$.

It is instructive to repeat the reasoning of section \ref{Surface charges} leading to \eqref{Noether identity} and \eqref{second Noether theorem} for the class of theories at hand. Starting from 
\begin{equation}
\delta_f\phi^i\vddl{L}{\phi^i}=\partial_\mu j^\mu_f ,
\end{equation} where $j^\mu_f$ is a representative of the Noether current (see equation \eqref{equation importante gauge}),
and integrating by parts on the left hand side so as to make the
undifferentiated gauge parameters appear, one obtains
\begin{equation}
  \label{eq:60}
  f^\alpha [R^i_\alpha \vddl{L}{\phi^i}-\partial_\mu (R^{i\mu}_\alpha
  \vddl{L}{\phi^i})]=\partial_\mu(j^\mu_f-S^\mu_f),\quad S^\mu_f=f^\alpha R^{i\mu}_\alpha
  \vddl{L}{\phi^i}.
\end{equation}
Since this is an off-shell identity that has to hold for all
$f^\alpha[x]$, one concludes not only that the Noether
identities
\begin{equation}
  \label{eq:149}
  R^i_\alpha\vddl{L}{\phi^i}-\d_\mu(R^{i\mu}_\alpha\vddl{L}{\phi^i})=0
\end{equation}
hold, but also that $\d_\mu(j^\mu_f-S^\mu_f)=0$, which is the second Noether theorem. This implies in
particular that $S^\mu_f$ is a representative for the Noether current
associated with gauge symmetries that is trivial in the sense
that it vanishes on-shell. Furthermore, every other representative
$j^\mu_f$ differs from $S^\mu_f$ at most by the divergence of
an arbitrary superpotential $\d_\nu \eta^{[\mu\nu]}_f$.

Since the fields and their derivatives can be seen as independent coordinates on the jet space (see appendix \ref{Jet bundles}), the Noether identities \eqref{eq:149} can be separated into terms involving
$\d_\mu\d_\nu\phi^j$, $\d_\mu\phi^k\d_\nu\phi^j$, $\d_\mu\phi^j$ or no
derivatives. The vanishing of the coefficients of these terms yields
\begin{equation}
  \label{eq:150}
  \begin{split}
    & R^{i(\mu}_\alpha \sigma_{ij}^{\nu)}=0,\\
    & \d_k(R^{i
      \mu}_\alpha\sigma^\nu_{ij})+\d_j(R^{i\nu}_\alpha\sigma^\mu_{ik})
    -R^{i\mu}_{k\alpha}\sigma^\nu_{ij}-R^{i\nu}_{j\alpha}\sigma^\mu_{ik}=0,\\
    &R^{i0}_\alpha\sigma^\mu_{ij}+\d_j[R^{i\mu}_\alpha(\d_ih+\dover{}{x^\nu}
    a^\nu_i)]-R^{k\mu}_{j\alpha}(\d_k h+\dover{}{x^\nu}
    a^\nu_k)-\dover{}{x^\nu} (R^{i\nu}_\alpha\sigma_{ij}^\mu)=0,\\
    & R^{i0}_\alpha(\d_i h+\dover{}{x^\nu}
    a^\nu_i)-\dover{}{x^\mu}[R^{i\mu}_\alpha(\d_i h+\dover{}{x^\nu}
    a^\nu_i)]=0.
  \end{split}
\end{equation}

The construction of the co-dimension 2 form deeply relies on the linearized theory. Writing $\varphi^i$ the Grassmann even variation of $\phi^i$, the Lagrangian $L^{(2)}[\phi ; \varphi ]$ of the linearized theory is obtained by collecting the quadratic terms in the expansion of $L[\phi +\varphi]$ in $\varphi^i$ and their derivatives around a solution $\phi$. We obtain 
  \begin{equation}
    \label{eq:152}
    L^{(2)}[\phi ; \varphi]=\d_i a_j^\mu \varphi^i\d_\mu\varphi^j +\half
    \d_i\d_j a^\mu_k\varphi^i \varphi^j\d_\mu\phi^k-\half \d_i\d_j
    h \varphi^i \varphi^j.
  \end{equation}
  The linearized equations of motion are then given by
\begin{equation}
  \label{eq:151}
\vddl{L^{(2)}[\phi ; \varphi]}{\varphi^i}=  [\sigma^\mu_{ij}\d_\mu+\d_j\sigma^\mu_{ik}\d_\mu\phi^k-\d_j(\d_i h+ \dover{}{x^\nu}
    a^\nu_i)]\varphi^j = 0.
  \end{equation}

Consider now the co-dimension 2 form,
\begin{equation}
    \label{eq:81}
k^{[\mu\nu]}_f [\phi; \delta \phi] =
  R^{i[\mu}_\alpha\sigma^{\nu]}_{ij}\delta \phi^jf^\alpha,
\end{equation}
and the invariant presymplectic current
\begin{equation}
  \label{eq:83}
  W^\mu [\phi ; \delta \phi_1,\delta \phi_2]=\frac{1}{2}\sigma_{ij}^\mu\delta \phi^i_1 \wedge \delta \phi^j_2.
\end{equation}
By using the equations of motion, the linearized equations of motion
and the Noether identities in the form of \eqref{eq:150}, one may then
check by a direct computation that
\begin{equation}
  \label{eq:85}
  \d_\nu
  k^{[\mu\nu]}_f[\phi; \delta \phi]= - W^\mu [\phi;R[f],\delta \phi] \quad {\rm when}\quad
    \vddl{L}{\phi^i}=0=\vddl{L^{(2)}[\phi ; \varphi]}{\varphi^i}.
\end{equation}
This means that this co-dimension 2 form is conserved on all solutions
of the linearized equations of motion around a given background
solution $\phi$ when using reducibility parameters $\bar f^\alpha$,
which satisfy
\begin{equation}
  \label{eq:86}
  R[\bar{f}] = R^i_\alpha \bar f^\alpha+ R^{i\mu}_\alpha\d_\mu \bar f^\alpha=0.
\end{equation}

In terms of forms, we can write
\begin{equation}
  \label{eq:157}
 \mathbf{k}_f [\phi; \delta \phi]=
  R^{i\mu}_\alpha\sigma^{\nu}_{ij}\delta \phi^jf^\alpha (\mathrm{d}^{n-2}x)_{\mu\nu},
\end{equation}
where
\begin{equation}
    \label{eq:159}
    \mathrm{d} \mathbf{k}_f [\phi; \delta \phi] = \mathbf{W}[\phi ; R[f],\delta \phi]\quad {\rm when}\quad
    \vddl{L}{\phi^i}=0=\vddl{L^{(2)}[\phi; \varphi]}{\varphi^i},
  \end{equation}
  with
  \begin{equation} 
    \mathbf{W}[\phi; \delta\phi , \delta\phi] =\half
  \sigma^\mu_{ij}\delta\phi^i \wedge \delta\phi^j
  (\mathrm{d}^{n-1} x)_\mu .
  \label{presymplectic current first order}
  \end{equation} We see that we have re-derived some key results of section \ref{Surface charges} (in particular, the conservation law \eqref{breaking in the conservation}) without using the properties of the homotopy operator \eqref{homotopy operator}. Therefore, the class of covariantized Hamiltonian theories considered here drastically simplifies the computations in the Barnich-Brandt formalism. 

These results can be related to the Iyer-Wald formalism introduced in subsection \ref{Relation between Barnich-Brandt and Iyer-Wald procedures}. The presymplectic potential is given by
\begin{equation}
  \label{eq:158}
 \boldsymbol{\theta}[\phi; \delta \phi] =a^\mu_i \delta\phi^i 
  (\mathrm{d}^{n-1}x)_\mu,
\end{equation}
and its associated presymplectic form reads as
\begin{equation} 
  \label{eq:92}
 \boldsymbol{\omega}[\phi; \delta \phi, \delta \phi] =  \delta \boldsymbol{\theta}[\phi; \delta \phi] = \mathbf{W}[\phi; \delta \phi, \delta \phi]. 
\end{equation}
In particular, we see that the $(n-2)$-form defined in \eqref{definition} and that controls the difference between Barnich-Brandt and Iyer-Wald procedures vanishes for covariantized Hamiltonian theories, i.e. $\mathbf{E}[\phi ; \delta \phi , \delta \phi] = 0$. Finally, equation \eqref{eq:157} can be expressed in terms of $\boldsymbol{\theta}[\phi; \delta \phi]$ as
\begin{equation}
  \label{eq:161}
 \mathbf{k}_f [\phi; \delta \phi] =- \half \left(f^\alpha
 R^{i\mu}_\alpha\frac{\d}{\d \delta \phi^i} \right)\frac{\d}{\d \D x^\mu}\delta \boldsymbol{\theta}[\phi; \delta \phi].
\end{equation}

\subsection{Vielbeins and connection}
\label{sec:first}

Now, we recall several notions of vielbeins and connections by including torsion and non-metricity into the standard discussion. This formalism is useful in section \ref{First order formulations of general relativity} when discussing the first order formulations of general relativity. 

\subsubsection{General case}
\label{sec:general-case}

Consider an $n$-dimensional spacetime with a
\textit{moving frame} (or \textit{vielbein})
\begin{equation}
e_a={e_a}^\mu\ddl{}{x^\mu},\quad e^a={e^a}_\mu \D x^\mu, \label{eq:2}
\end{equation}
where ${e_a}^\mu{e^a}_\nu=\delta^\mu_\nu$,
${e_a}^\mu{e^b}_\mu=\delta_a^b$, and $\d_a f=e_a(f)$.  Under a
combined frame and coordinate transformation, we have
${e'_a}^\mu(x')={\Lambda_a}^b(x){e_b}^\nu(x){\Lambda^\mu}_\nu(x)$,
where ${\Lambda^a}_b(x)$ denotes a local $GL(n,\mathbb R)$ element
with ${\Lambda_a}^b={(\Lambda^{-1})^b}_a$ while
${\Lambda^\mu}_\nu=\ddl{{x'}^\mu}{x^\nu}$ is the Jacobian matrix
  of the coordinate transformation, with
${\Lambda_\mu}^\nu=\ddl{x^\nu}{{x'}^\mu}$. The Lie algebra generators
of $\mathfrak{gl}(n,\mathbb R)$ are denoted by ${\Delta_a}^b$,
${({\Delta_a}^b)^c}_d=\delta_a^c\delta_d^b$, with generators for the
vector representation denoted by $t_a$,
\begin{equation}
  \label{eq:22}
  [{\Delta_a}^b,{\Delta_c}^d]=\delta^b_c{\Delta_a}^d-\delta^d_a{\Delta_c}^b,
  \quad {\Delta_a}^b t_c=\delta^b_c t_a.
\end{equation}
The structure functions are defined by
\begin{equation}
[e_a,e_b]={D^c}_{ab}e_c \iff \D e^a=-\half {D^a}_{bc}e^be^c.\label{eq:13}
\end{equation}
For further use, note that if ${\mathbf e}={\rm det}\,{e^a}_\mu$, then
\begin{equation}
  \label{eq:82}
  \d_\mu(\mathbf{e}\,{e^\mu}_a)=\mathbf{e}\, {D^b}_{ba},
\end{equation}
and, if we define, 
\begin{equation}
  \label{eq:49}
   {d^a}_{bc}={e^a}_\lambda \d_b {e_c}^\lambda,
\end{equation}
then
\begin{equation}
{d^\sigma}_{\rho\mu}=-{e_d}^\sigma\d_\rho {e^d}_\mu,\quad
{D^a}_{bc}=2{d^a}_{[bc]}, \label{eq:100}
\end{equation}
where it is understood that tangent space indices $a,b,\dots$ and
world-indices $\mu,\nu,\dots$ are transformed into each other by using
the vielbeins and their inverse.

In addition, we assume that there is an \textit{affine connection}
\begin{equation}
D_a e_b={\Gamma^c}_{ba}e_c\iff D_b v^a=\d_b v^a+{\Gamma^a}_{cb} v^c. \label{eq:12}
\end{equation}
If ${\Gamma^a}_b={\Gamma^a}_{bc} e^c$,
${\Gamma}={\Gamma^a}_b{\Delta_a}^b$, and ${e}=e^a t_a$, the \textit{torsion}
tensor and \textit{curvature} tensors are defined by
\begin{equation}
{\mathcal T}=T^at_a=\D {e} +{\Gamma} \wedge {e},\quad {\mathcal R}={R^a}_b{\Delta_a}^b=\D {\Gamma}
+\half[{\Gamma},{\Gamma}]\label{eq:23},
\end{equation}
where the bracket is the graded
commutator. More explicitly,
$T^a=\half {T^a}_{bc} e^b \wedge e^c=\D  e^a+{\Gamma^a}_b \wedge e^b$, so that
\begin{equation}
{T^a}_{\mu\nu}=\d_\mu {e^a}_\nu-\d_\nu
{e^a}_\mu+{\Gamma^a}_{b\mu}
{e^b}_\nu-{\Gamma^a}_{b\nu}{e^b}_\mu,\label{eq:70}
\end{equation}
\begin{equation}
{T^c}_{ab}=2{\Gamma^c}_{[ba]}+{D^c}_{ba}=2({\Gamma^c}_{[ba]}+{d^c}_{[ba]}),\label{eq:14}
\end{equation} 
and ${{R}^a}_b=\half {R^{a}}_{bcd} e^c \wedge e^d=\D 
{\Gamma^a}_b+{\Gamma^a}_c \wedge {\Gamma^c}_b$, so that
\begin{equation}
{R^f}_{c\mu\nu}=\d_\mu
  {\Gamma^f}_{c\nu}-\d_\nu {\Gamma^f}_{c\mu}
  +{\Gamma^f}_{d\mu}{\Gamma^d}_{c\nu}-{\Gamma^f}_{d\nu}{\Gamma^d}_{c\mu},\label{eq:71}
\end{equation}
\begin{equation}
{R^f}_{cab}=\d_a
  {\Gamma^f}_{cb}-\d_b {\Gamma^f}_{ca}
  +{\Gamma^f}_{da}{\Gamma^d}_{cb}-{\Gamma^f}_{db}{\Gamma^d}_{ca}-{D^d}_{ab}{\Gamma^f}_{cd}.
\label{eq:15}
\end{equation}
Furthermore, 
\begin{equation}
  \label{eq:20}
  [D_a,D_b]v_c=-{R^d}_{cab}v_d-{T^d}_{ab}D_dv_c.
\end{equation}

Under a local frame transformation, we have
\begin{equation}
  \label{eq:25}
  {e}'=\Lambda {e},\quad {\Gamma'}=\Lambda
  {\Gamma}\Lambda^{-1}+\Lambda \D  \Lambda^{-1},
\end{equation}
so that
\begin{equation}
  \label{eq:26}
  {\mathcal T'}=\Lambda{\mathcal T},\quad {\mathcal
    R'}=\Lambda{\mathcal R}\Lambda^{-1}.
\end{equation}
Defining $\Lambda=\mathbf 1+{\omega}+O(\omega^2)$, with
$\omega= {\omega^a}_b{\Delta_a}^b$ and also
${\omega_b}^a=-{\omega^a}_b$, we have
\begin{equation}
  \label{eq:35}
  \delta_\omega \Gamma=-(\D \omega+[\Gamma,\omega])\iff \delta_\omega
  {\Gamma^a}_b=\D {\omega_b}^a+{\Gamma^a}_c{\omega_b}^c-{\Gamma^c}_b{\omega_c}^a,
\end{equation}
and also
\begin{equation}
\delta_\omega e=[\omega,e]\iff \delta_\omega e^a={\omega^a}_be^b. \label{eq:49a}
\end{equation}
Under a coordinate transformation, we have
\begin{equation}
  \label{eq:58}
  {{e'}^a}_\mu={\Lambda_\mu}^\nu{e^a}_\nu,\quad
  {{\Gamma'}^a}_{b\mu}={\Lambda_\mu}^\nu{{\Gamma}^a}_{b\nu},
\end{equation}
and for ${x'}^\mu=x^\mu-\xi^\mu+ O(\xi^2)$,
${\Lambda^\mu}_\nu=\delta^\mu_\nu-\d_\nu\xi^\mu+ O(\xi^2)$, so that
${\omega_\nu}^\mu=\d_\nu\xi^\mu$ and
\begin{equation}
  \label{eq:59}
  \delta_\xi {e^a}_\mu={\mathcal L}_\xi e^{a}_\mu,\quad  \delta_\xi
  {\Gamma^a}_{b\mu}={\mathcal L}_\xi {\Gamma^a}_{b\mu},
\end{equation}
where ${\mathcal L}_\xi$ denotes the Lie derivative.

The \textit{Bianchi
identities} are
\begin{equation}
  \label{eq:21}
 \D {\mathcal T}+{{\Gamma}} \wedge {\mathcal T}={\mathcal R}\wedge {e},\quad \D 
  {\mathcal R}+[\Gamma,\mathcal R]=0.
\end{equation}
Explicitly,
\begin{equation}
  \label{eq:24}
  {R^a}_{[bcd]}=D_{[b}{T^a}_{cd]}+{T^a}_{f[b}{T^f}_{cd]},\quad
D_{[f}{R^a}_{|b|cd]}=-{R^a}_{bg[f}{T^g}_{cd]},
\end{equation}
where a bar encloses indices that are not involved in
the (anti) symmetrization. The Ricci tensor is defined by
${R}_{ab}={R^{c}}_{acb}$, while
$S_{ab}={R^c}_{cab}$. Contracting the Bianchi identities gives
\begin{equation}
  \label{eq:27}
  {R}_{ab}-{R}_{ba}=S_{ab}-D_c {T^c}_{ab}
-2D_{[a} {T^c}_{b]c}-{T^c}_{dc}{T^d}_{ab},
\end{equation}
\begin{equation}
2D_{[f}{R}_{|b|d]}+D_c{R^c}_{bdf}={R}_{bg}{T^g}_{df}
-2{R^c}_{b[f|g|}{T^g}_{d]c}, \label{eq:28a}
\end{equation}
\begin{equation}
  \label{eq:28b}
  D_{[f}S_{cd]}=-S_{g[f}{T^g}_{cd]}.
\end{equation}

Assume now that there is a \textit{pseudo-Riemannian metric},
\begin{equation}
g_{\mu\nu}={e^a}_\mu g_{ab} {e^b}_\nu\label{eq:11},
\end{equation}
i.e., a symmetric, non-degenerate $2$-tensor . 
As usual, tangent space indices $a,b,\dots$ and world indices
$\mu,\nu,\dots$ are lowered and raised with $g_{ab}$, $g_{\mu\nu}$,
and their inverses.

The \textit{non-metricity tensor} is defined as
$\Xi^{ab}=dg^{ab}+2\Gamma^{(ab)}$. The associated Bianchi identities
are given by
$d\Xi^{ab}+{\Gamma^{a}}_c\Xi^{cb}+{\Gamma^b}_c\Xi^{ac}=2R^{(ab)}$.
More explicitly,
\begin{equation}
  \label{eq:95}
  {\Xi^{ab}}_c=D_c g^{ab}, \quad 2D_{[c}
  {\Xi^{ab}}_{d]}=-{\Xi^{ab}}_f{T^f}_{cd}+2{R^{(ab)}}_{cd}.
\end{equation}
It should also be noted that from $g^{ab}g_{bc}=\delta^a_c$, it follows
\begin{equation}
  \label{eq:40}
  D_c g_{ab}=-\Xi_{abc}. 
\end{equation}
Contracting the last of \eqref{eq:95} with $g_{ab}$
gives
\begin{equation}
  \label{eq:86}
  S_{cd}=g_{ab}D_{[c} {\Xi^{ab}}_{d]}+\half
  {\Xi^a}_{af}{T^f}_{cd},
\end{equation}
while \eqref{eq:28a} with $g^{bf}$ gives
\begin{multline}
  \label{eq:34}
  D^b {R}_{ba}-\half D_a R=\half
  {R^{bc}}_{da}{T^{d}}_{bc}+{{R}^b}_c{T^c}_{ab}\\- \half({\Xi^{bc}}_c{
    R}_{ba}+{\Xi^{cd}}_b{R^b}_{cda}+{\Xi^{bc}}_a{
    R}_{bc})\\+D_c(D_{[b}{\Xi^{bc}}_{a]}+\half{\Xi^{bc}}_d{T^d}_{ba})+
  (D_{[b}{\Xi^{bc}}_{d]}+\half{\Xi^{bc}}_d{T^d}_{bd}){T^d}_{ac}.
\end{multline}

The curvature scalar is defined by
${R}=g^{ab}{R}_{ab}$, the Einstein tensor by
\begin{equation}
  \label{eq:33}
  G_{ab}={R}_{(ab)}-\half g_{ab} {R}.
\end{equation} 
When combining with \eqref{eq:27}, the contracted Bianchi identity
\eqref{eq:34} written in terms of the Einstein tensor is 
\begin{multline}
  \label{eq:34a}
D^b {G}_{ba}=\half
  {R^{bc}}_{da}{T^{d}}_{bc}+{{R}^b}_c{T^c}_{ab} - \half {\Xi_{ab}}^b R  \\
 +\half D^b(S_{ab}-D_c {T^c}_{ab}
-2D_{[a} {T^c}_{b]c}-{T^c}_{dc}{T^d}_{ab})\\ - \half({\Xi^{bc}}_c{
    R}_{ba}+{\Xi^{cd}}_b{R^b}_{cda}+{\Xi^{bc}}_a{
    R}_{bc})\\+D_c(D_{[b}{\Xi^{bc}}_{a]}+\half{\Xi^{bc}}_d{T^d}_{ba})+
  (D_{[b}{\Xi^{bc}}_{d]}+\half{\Xi^{bc}}_d{T^d}_{bd}){T^d}_{ac} .
\end{multline}

By the usual manipulations, one may show that the existence of the metric
implies that the most general connection can be written as
\begin{equation}
  \label{eq:17}
  \Gamma_{abc}=
  \{{}_{abc}\}+M_{abc}+K_{abc}+r_{abc}
\end{equation}
where the \textit{Christoffel symbols}, the \textit{conmetricity}, the \textit{contorsion
tensor}, and the \textit{co-structure functions} are given by
\begin{equation}
\{{}_{abc}\}=\half(g_{ab,c}+g_{ac,b}-g_{bc,a})=\{{}_{acb}\}, \label{eq:97}
\end{equation}
\begin{equation}
M_{abc}=\half(\Xi_{abc}+\Xi_{acb}-\Xi_{bca})=M_{acb}, \label{eq:97}
\end{equation}
\begin{equation}
  \label{eq:99c}
K_{abc}=\half(T_{bac}+T_{cab}-T_{abc})=-K_{bac},
\end{equation}
\begin{equation}
  r_{abc}=\half(D_{bac}+D_{cab}-D_{abc})=-r_{bac}.\label{eq:96}
\end{equation}
Furthermore, one can directly show that
\begin{equation}
  \label{eq:105}
  {\Gamma^a}_{b\mu}={e^a}_\nu(\d_\mu
  {e_b}^\nu+{\Gamma^\nu}_{\rho\mu}{e^\rho}_b)\iff
{\Gamma}_{abc}=e_{a\nu}\d_c{e_b}^\nu+{e_a}^\mu{e_b}^\nu{e_c}^\rho\Gamma_{\mu\nu\rho}.
\end{equation}

Finally, we need the following variations,
\begin{equation}
  \label{eq:47}
  \delta
  {R^a}_{b\mu\nu}=D_\mu\delta{\Gamma^a}_{b\nu}-D_\nu\delta{\Gamma^a}_{b\mu},
\end{equation}
\begin{multline}
  \label{eq:50}
  \delta
  {R^a}_{bcd}=D_c\delta{\Gamma^a}_{bd}-D_d\delta{\Gamma^a}_{bc}+{T^f}_{cd}\delta{\Gamma^a}_{bf}\\
+{e^f}_\sigma\big[{\Gamma^a}_{bf}{\mathcal D}_d
+\d_f{\Gamma^a}_{bd}-{D^g}_{fd}{\Gamma^a}_{bg}\big]\delta{e_c}^\sigma\\
-{e^f}_\sigma\big[{\Gamma^a}_{bf}{\mathcal D}_c
+\d_f{\Gamma^a}_{bc}-{D^g}_{fc}{\Gamma^a}_{bg}\big]\delta{e_d}^\sigma.
\end{multline}
To write the latter variation, we introduced the derivative operator $\mathcal{D}$ defined through
\begin{equation}
  \label{eq:48}
  {\mathcal D}_\mu\delta{e_b}^\sigma=
\d_\mu\delta{e_b}^\sigma+{d^\sigma}_{\mu\rho}\delta{e_b}^\rho-{d^c}_{\mu
b}\delta{e_c}^\sigma.
\end{equation}

\subsubsection{Metricity and Lorentz metric}
\label{sec:metricity}

When requiring metricity, $D_a g^{bc}=0={\Xi^{bc}}_a$, the connection is given by
\begin{equation}
  \label{eq:17c}
  \Gamma_{abc}=
  \{{}_{abc}\}+K_{abc}+r_{abc},
\end{equation}
From $D_{[a}D_{b]}g_{cd}=0$ it also follows that
\begin{equation}
  \label{eq:29}
  R_{abcd}=-R_{bacd}, \quad  S_{ab}=0.
\end{equation}
This can be used to show that
\begin{multline}
  \label{eq:30}
  R_{abcd}-R_{cdab}=\frac{3}{2}(D_{[b}T_{|a|cd]}+T_{af[b}{T^f}_{cd]}
-D_{[a}T_{|b|cd]}-T_{bf[a}{T^f}_{cd]}\\-D_{[d}T_{|c|ab]}-T_{cf[d}{T^f}_{ab]}
+D_{[c}T_{|d|ab]}+T_{df[c}{T^f}_{ab]}),
\end{multline}
while the contracted Bianchi identities \eqref{eq:34a} become
\begin{equation}
  \label{eq:34b}
D^b {G}_{ba}=\half
  {R^{bc}}_{da}{T^{d}}_{bc}+{{R}^b}_c{T^c}_{ab}  -\half D^b(D_c {T^c}_{ab}
+2D_{[a} {T^c}_{b]c}+{T^c}_{dc}{T^d}_{ab}) .
\end{equation}

If metricity holds and we assume a constant Lorentz metric
$g_{ab}=\eta_{ab}$, we have $\{abc\}=0$ and
\begin{equation}
\Gamma_{abc}=-\Gamma_{bac}\label{eq:32}.
\end{equation}
Local Lorentz transformations are denoted by ${\Lambda_a}^b(x)$ with
${\Lambda_a}^b\eta_{bc}{\Lambda_d}^c=\eta_{ad}$, or equivalently,
${\Lambda^d}_b {\Lambda_a}^b=\delta_a^d$.  In terms of the Poincar\'e
algebra,
\begin{equation}
  \label{eq:16}
  [J_{ab},J_{cd}]=\eta_{bc}J_{ad}-\eta_{ac}J_{bd}-\eta_{bd}J_{ac}+\eta_{ad}J_{bc},
  \quad [J_{ab},P_c]=\eta_{bc} P_a-\eta_{ac}P_b,
\end{equation}
one defines the Lorentz connection $\Gamma=\half \Gamma^{ab}J_{ab}$,
$e=e^a P_a$, $R=\half R^{ab} J_{ab}$, $T=T^a P_a$, so that
$R=\D \Gamma+\half[\Gamma,\Gamma]$, $T=\D e+[\Gamma,e]$. In this case,
\begin{equation}
  \label{eq:36}
  \d_\mu({\mathbf e}\, v^\mu)={\mathbf e}\,
  (D_\mu+{e_b}^\nu\d_\mu{e^b}_\nu)v^\mu=D_\mu({\mathbf
    e} v^\mu),
\end{equation}
with $D_\mu v^\mu=\d_\mu v^\mu$ for the Lorentz connection and the
definition
\begin{equation}
  \label{eq:73}
  D_\mu{\mathbf e}={\mathbf e}\,({e_b}^\nu\d_\mu{e^b}_\nu).
\end{equation}
In particular, this implies that
\begin{equation}
  \label{eq:94}
  D_\mu(\mathbf{e}\,{e^\mu}_a)=\mathbf{e}\,{T^b}_{ab}.
\end{equation}
The connection reduces to
\begin{equation}
  \label{eq:37}
  \Gamma_{abc}=K_{abc}+r_{abc}.
\end{equation}

Finally, if one imposes, in addition, vanishing of torsion, the connection is reduced further to
\begin{equation}
  \label{eq:19}
  \Gamma_{abc}=r_{abc},
\end{equation}
and the contracted Bianchi identities \eqref{eq:34b} reduce to
\begin{equation}
  \label{eq:38}
  D_b {G^b}_a=0.
\end{equation}

\subsubsection{Coordinate basis, torsionless connection}
\label{sec:coordinate-basis}

In a coordinate basis, ${e_a}^\mu={\delta_a}^\mu$,
${D^\lambda}_{\mu\nu}=0$ and
${T^\lambda}_{\mu\nu}={\Gamma^\lambda}_{\nu\mu}-{\Gamma^\lambda}_{\mu\nu}$. Imposing
vanishing of torsion,
${\Gamma^\lambda}_{\mu\nu}= {\Gamma^\lambda}_{\nu\mu}$, \eqref{eq:27}
implies $S_{\mu\nu}=R_{\mu\nu}-R_{\nu\mu}$ and the contracted Bianchi
identities \eqref{eq:34a} become
\begin{multline}
  \label{eq:34c}
D^\nu {G}_{\nu\mu}=
 D^\nu R_{[\mu\nu]} +D_\lambda
 {R^{(\lambda\nu)}}_{\nu\mu}\\
 - \half(D_\nu g^{\nu\lambda}{
    R}_{\lambda\mu}+D_\nu g^{\lambda\rho} {R^\nu}_{\lambda\rho\mu}+
  D_\mu g^{\nu\lambda} {
    R}_{\nu\lambda}+ D^\nu g_{\nu \mu} R),
\end{multline}
while the variation \eqref{eq:47} simplifies to
\begin{equation}
  \label{eq:51}
   \delta
  {R^\alpha}_{\beta\mu\nu}=D_\mu\delta{\Gamma^\alpha}_{\beta\nu}
-D_\nu\delta{\Gamma^\alpha}_{\beta\mu}.
\end{equation}
We also have
\begin{equation}
\d_\mu(\sqrt{|g|} v^\mu)=\sqrt{|g|}
(D_\mu-{\Gamma^\nu}_{\mu\nu} +\half g^{\nu\lambda}\d_\mu
g_{\nu\lambda}) v^\mu=D_{\mu}(\sqrt{|g|}v^\mu),  \label{eq:41}
\end{equation}
where we defined
\begin{equation}
  \label{eq:53}
  D_\mu\sqrt{|g|}=\sqrt{|g|}(\half g^{\nu\lambda}\d_\mu
g_{\nu\lambda}-{\Gamma^\nu}_{\mu\nu})
\end{equation}
to write the last equality. Under an infinitesimal coordinate
transformation, besides $\delta_\xi g_{\mu\nu}={\mathcal L}_\xi g_{\mu\nu}$, we
have
\begin{equation}
\delta_\xi
  {\Gamma^\mu}_{\nu\rho}=\d_\rho\d_\nu\xi^\mu+\xi^\sigma\partial_\sigma
  {\Gamma^\mu}_{\nu\rho}-\d_\sigma\xi^\mu
  {\Gamma^\sigma}_{\nu\rho}+\d_\nu\xi^\sigma
  {\Gamma^\mu}_{\sigma\rho}+\d_\rho\xi^\sigma
  {\Gamma^\mu}_{\nu\sigma}.\label{eq:61}
\end{equation}

Requiring in addition metricity, this leads to the Levi-Civita connection ($D_\mu \equiv \nabla_\mu$)
\begin{equation}
  \label{eq:18}
  \Gamma_{\lambda\mu\nu}=\half(\d_\nu g_{\lambda\mu}+\d_\mu
  g_{\lambda\nu}-\d_\lambda g_{\mu\nu}),
\end{equation}
while the contracted Bianchi identities \eqref{eq:34c}
reduce to
\begin{equation}
  \label{eq:42}
  \nabla^\nu {G_{\nu\mu}}=0,
\end{equation}
and \eqref{eq:41} to
\begin{equation}
  \label{eq:43}
  \d_\mu(\sqrt{|g|} v^\mu)=\sqrt{|g|} \nabla_\mu v^\mu.
\end{equation}

\section{First order formulations of general relativity}
\label{First order formulations of general relativity}

In this section, we review some first order formulations of general relativity including Cartan and Newman-Penrose formulations, and apply the surface charges formalism. In particular, we see that these are first order formulation in the sense of the covariantized Hamiltonian theories of subsection \ref{sec:covar-hamilt-form}. For each case, we relate the obtained results to the standard second order metric formulation of general relativity discussed in the examples in section \ref{Surface charges}. We use the compact notation
\begin{equation}
\kappa = \frac{1}{16 \pi G} .
\end{equation}

\subsection{Cartan formulation I}
\label{sec:cartan-formulation}

\subsubsection{Variational principle}
\label{sec:vari-princ}

In the standard Cartan formulation, the variables of the variational
principle are the components of the vielbein ${e_a}^\mu$ and a Lorentz
connection 1-form in the coordinate basis, ${\Gamma^a}_{b\mu}$ in
terms of which the action is
\begin{equation}
  \label{eq:69}
  S^C[{e_a}^\mu,{\Gamma^b}_{c\nu}]=\kappa \int \D^nx\, L^C= \kappa\int \D^nx\, {\mathbf e}\,
  ({R^{ab}}_{\mu\nu}{e_a}^\mu{e_b}^\nu-2\Lambda).
\end{equation}
Using \eqref{eq:47},
the variation of the action is given by
\begin{equation}
  \label{eq:74}
  \delta S^C=\kappa \int \D^nx\, {\mathbf e}\,\big[2({G^a}_\mu
+\Lambda{e^a}_\mu)\delta {e_a}^\mu+{e_a}^\mu{e_b}^\nu
(D_\mu\delta {\Gamma^{ab}}_\nu-D_\nu\delta{\Gamma^{ab}_\mu})\big].
\end{equation}
Now, using \eqref{eq:36} and neglecting boundary terms, this gives
\begin{equation}
  \label{eq:75}
  \delta S^C= \kappa \int \D^nx\, \big[2{\mathbf e}\,({G^a}_\mu
  +\Lambda{e^a}_\mu)\delta {e_a}^\mu+2 D_\nu({\mathbf e}\,
{e_a}^\mu{e_b}^\nu)\delta {\Gamma^{ab}}_\mu\big],
\end{equation}
so that
\begin{equation}
  \label{eq:76}
  \vddl{L^C}{{e_a}^\mu}=2{\mathbf e}\,({G^a}_\mu
  +\Lambda{e^a}_\mu),
\end{equation}
\begin{equation}
  \label{eq:77}
  \vddl{L^C}{{\Gamma^{ab}}_\mu}=2 D_\nu({\mathbf e}\,
{e_{[a}}^\mu{e_{b]}}^\nu)={\mathbf e}\, ({T^\mu}_{ab}+2e^\mu_{[a}{T^c}_{b]c}).
\end{equation}
Contracting the equations of motions associated to \eqref{eq:77} with
${e_\mu}^b$ gives ${T^b}_{ab}=0$. When re-injecting, this implies
${T^a}_{bc}=0$. It follows that when the equations of motion for
${\Gamma^{ab}}_\mu$ hold, the connection is torsionless and thus given
by $\Gamma_{abc}=r_{abc}$. The fields ${\Gamma^{ab}}_\mu$ are thus
entirely determined by ${e_a}^\mu$ so that
${\Gamma^{ab}}_\mu$ are auxiliary fields.

\subsubsection{Symmetries}

The gauge symmetries of the action \eqref{eq:69} are the diffeomorphisms and the Lorentz gauge transformations. The infinitesimal transformations of the fields under these symmetries are given by
\begin{equation}
\begin{split}
\delta_{\xi, \omega} {e_a}^\mu &= \xi^\nu \p_\nu {e_a}^\mu - \partial_\nu \xi^\mu  {e_a}^\nu + {\omega_a}^b {e_b}^\mu , \\
\delta_{\xi, \omega} {\Gamma^{ab}}_{\mu} &= \xi^\nu \p_\nu {\Gamma^{ab}}_{\mu} + \partial_\mu \xi^\nu {\Gamma^{ab}}_{\nu}  - D_\mu \omega^{ab}.
\end{split}
\end{equation}

Following the lines of \eqref{Noether identity proof}, we consider
\begin{equation}
  \label{eq:78}
  \vddl{L^C}{{e_a}^\mu}
\delta_{\xi,\omega} {e_a}^\mu
+\vddl{L^C}{{\Gamma^{ab}}_\mu}\delta_{\xi,\omega} {\Gamma^{ab}}_\mu 
\end{equation}
and integrate by parts in order to isolate the
undifferentiated gauge parameters $\omega^{ab}$ and $\xi^\rho$. Discarding the boundary terms, this leads to the Noether identities
\begin{equation}
  \label{eq:79}
  \vddl{L^C}{e^{[a|\mu|}}{e_{b]}}^\mu+D_\mu\vddl{L^C}{{\Gamma^{ab}}_\mu}=0,
\end{equation}
\begin{equation}
  \label{eq:80}
  \vddl{L^C}{{e_a}^\mu}\d_\rho {e_a}^\mu+\vddl{L^C}{{\Gamma^{ab}}_\mu}
\d_\rho{\Gamma^{ab}_\mu}+\d_\mu \left(\vddl{L^C}{{e_a}^\rho}{e_a}^\mu
-\vddl{L^C}{{\Gamma^{ab}}_\mu}{\Gamma^{ab}}_\rho \right)=0.
\end{equation}
Identity \eqref{eq:79}, associated with Lorentz gauge symmetries, can be shown to be equivalent to
\eqref{eq:27}. Using \eqref{eq:79}, identity \eqref{eq:80}, associated with diffeomorphisms,
can be written as
\begin{equation}
  \label{eq:83}
  \d_\mu \left(\vddl{L^C}{{e_a}^\rho}{e_a}^\mu \right)+\vddl{L^C}{{e_a}^\mu}D_\rho
  {e_a}^\mu
  +\vddl{L^C}{{\Gamma^{ab}}_\mu}{R^{ab}}_{\rho\mu}=0,
\end{equation}
and can then be shown to be equivalent to \eqref{eq:34b}.

\subsubsection{Construction of the co-dimension 2 form}
\label{sec:constr-co-dimens}

When keeping the boundary terms, one finds the weakly vanishing Noether
current associated with the gauge symmetries as
\begin{equation}
  \label{eq:81}
  \kappa^{-1} S^\mu_{\xi,\omega}=\vddl{L^C}{{\Gamma^{ab}}_\mu}(-{\omega^{ab}}
  +{\Gamma^{ab}}_\rho\xi^\rho)
  -\vddl{L^C}{{e_a}^\rho}{e_a}^\mu\xi^\rho.
\end{equation}
The associated co-dimension 2 form
$k_{\xi,\omega}=k^{\mu\nu}_{\xi,\omega}(\D^{n-2}x)_{\mu\nu}$ computed
through \eqref{eq:157} is
given by
\begin{multline}
  \label{eq:84}
\kappa^{-1}  k^{\mu\nu}_{\xi,\omega}={\mathbf e}\,\big[(2\delta{e_a}^\mu
  {e_b}^\nu+{e^c}_\lambda\delta
  {e_c}^\lambda{e_a}^\nu{e_b}^\mu)(-\omega^{ab}+{\Gamma^{ab}}_\rho\xi^\rho)\\+
\delta
  {\Gamma^{ab}}_\rho (\xi^\rho {e_a}^\mu{e_b}^\nu
+2\xi^{\mu} {e_a}^{\nu} {e_b}^\rho) -(\mu\longleftrightarrow\nu)\big].
\end{multline}
This can also be written as
\begin{equation}
  \label{eq:102}
 \mathbf{k}_{\xi,\omega}=-\delta
  \mathbf{K}^K_{\xi,\omega}+ \mathbf{K}^K_{\delta\xi,\delta\omega}-\xi^\nu\ddl{}{\D x^\nu}\boldsymbol{\Theta},
\end{equation}
where
\begin{equation}
  \label{eq:87}
 \mathbf{K}^K_{\xi,\omega}=2\kappa\mathbf{e}\,{e_a}^\nu {e_b}^\mu
  (-\omega^{ab}+{\Gamma^{ab}}_\rho\xi^\rho)(\D^{n-2}x)_{\mu\nu},\quad
\boldsymbol{\Theta}= 2 \kappa \mathbf{e}\,
\delta{\Gamma^{ab}}_\rho{e_a}^\mu{e_b}^\rho (\D^{n-1}x)_\mu.
\end{equation}

According to the general results reviewed in section
\ref{Surface charges}, the co-dimension 2 form is closed, $\D
\mathbf k_{\xi,\omega}= 0$, or, equivalently,
$\d_\nu k^{\mu\nu}_{\xi,\omega}=0$, if ${e_a}^\mu,{\Gamma^{ab}}_\mu$
are solutions to the Euler-Lagrange equations of motion, and thus to
the Einstein equations, $\delta {e_a}^\mu,\delta {\Gamma^{ab}}_\mu$
solutions to the linearized equations and $\omega^{ab},\xi^\rho$
satisfy
\begin{equation}
  {\mathcal L}_\xi {e_a}^\mu+{\omega_a}^b{e_b}^\mu\approx 0,\quad
 {\mathcal L}_\xi {\Gamma^{ab}}_\mu\approx D_\mu
  \omega^{ab}\label{eq:85},
\end{equation}
where $\approx$ now denotes on-shell for the background solution and
is relevant in case the parameters
$\omega^{ab},\xi^\rho$ explicitly depend on the background solution
${e_a}^\mu,{\Gamma^{ab}}_\mu$ around which one linearizes. The first equation also implies in particular that $\xi^\rho$ is a
possibly field dependent Killing vector of the background solution
$g_{\mu\nu}$,
\begin{equation}
  \label{eq:88}
  {\mathcal L}_\xi g_{\mu\nu}\approx 0,
\end{equation}
and that
\begin{equation}
  \label{eq:106}
\omega^{ab}\approx -{e^{b}}_\mu{\mathcal L}_\xi
{e^{a\mu}}\approx -{e^{[b}}_\mu{\mathcal L}_\xi
{e^{a]\mu}}.
\end{equation}

\subsubsection{Reduction to the metric formulation}
\label{sec:reduct-metr-form}

To compare with the results in the metric formulation, let us
go on-shell for the auxiliary fields ${\Gamma^{ab}}_\mu$ and eliminate
$\omega^{ab}$ using \eqref{eq:106}. The former implies that we are in
the torsionless case with the Lorentz connection simplified to
${\Gamma^{ab}}_\mu={r^{ab}}_\mu$, while \eqref{eq:105} reduces to
\begin{equation}
  \label{eq:103}
  {\Gamma^{ab}}_\mu={e^a}_\nu\nabla_\mu
  e^{b\nu}={e^{[a}}_\nu\nabla_\mu
  e^{b]\nu},
\end{equation}
with $\nabla_\mu v^\nu=\d_\mu v^\nu+\{{{}^\nu}_{\rho\mu}\}v^{\rho}$.
Note also that the Killing equation can be written as
$\nabla_\mu\xi_\nu+\nabla_\nu\xi_\mu\approx 0$.
Together with
\eqref{eq:103}, we have
\begin{equation}
-\omega^{ab}+{\Gamma^{ab}}_\rho\xi^\rho\approx
-{e^{[a}}_\rho{e^{b]}}_\sigma\nabla^\rho\xi^\sigma,\label{eq:90}
\end{equation}
\begin{equation}
  \label{eq:91}
  \delta{\Gamma^{ab}}_\rho=\delta {e^{[a}}_\sigma \nabla_\rho
    {e^{b]\sigma}}+ {e^{[a}}_\sigma \delta \{{{}^\sigma}_{\tau\rho}\}
      {e^{b]\tau}}+{e^{[a}}_\sigma \nabla_\rho \delta {e^{b]\sigma}},
\end{equation}
with
\begin{equation}
  \label{eq:92}
  \delta \{{{}^\sigma}_{\tau\rho}\}=\half g^{\sigma\delta}(
  \nabla_\rho\delta g_{\delta\tau}+\nabla_\tau\delta
  g_{\delta\rho}
  -\nabla_\delta\delta g_{\tau\rho}).
\end{equation}
Using that
\begin{equation}
  \label{eq:107}
  \delta {e^a}_\mu{e_a}_\nu=\half h_{\mu\nu}+\delta {e^{a}}_{[\mu}e_{|a|\nu]},
\end{equation}
with $h_{\mu\nu}=\delta g_{\mu\nu}$, indices being lowered and raised
with $g_{\mu\nu}$ and its inverse, and $h=h^\mu_\mu$, substitution
into \eqref{eq:84} gives
\begin{equation}
6\sqrt{|g|}\nabla_\rho(\delta {e_a}^{[\mu} e^{|a|\nu}\xi^{\rho]})+
  k^{\mu\nu}_\xi,\label{eq:3}
\end{equation}
where the first term can be dropped since it is trivial in the sense
that it corresponds to the exterior derivative of an $n-3$ form, while
\begin{multline}
  \label{eq:108}
  k^{\mu\nu}_\xi=\sqrt{|g|}\,\Big[\xi^\nu\nabla^\mu h+\xi^\mu\nabla_\sigma
  h^{\sigma\nu}+\xi_\sigma \nabla^\nu h^{\sigma\mu}\\+\half
  h\nabla^\nu\xi^\mu+\half h^{\mu\sigma}\nabla_\sigma\xi^\nu+\half
  h^{\nu\sigma}\nabla^\mu\xi_\sigma-(\mu\longleftrightarrow \nu)\Big].
\end{multline}
We have thus recovered the results of the metric formulation since the
last expression agrees with the one given in \eqref{Barnich-Brandt}.

\subsection{Cartan formulation II}
\label{Cartan formulation with non-holonomic connection}

\subsubsection{Variational principle}

This version of the Cartan formulation is an intermediate between the Cartan formulation of subsection \ref{sec:cartan-formulation} and the Newman-Penrose formulation discussed in subsection \ref{Newman-Penrose formulation}. Here, the variables of the variational principle are the components
$\Gamma_{abc}=\Gamma_{[ab]c}$ of a Lorentz connection in the non-holonomic frame and the vielbein components ${e_a}^\mu$. In term of these variables, the action reads
\begin{equation}
S^{CNH}[\Gamma_{abc},{e_a}^\mu] = \kappa \int \D^nx L^{CNH}= \kappa\int \D^nx \mathbf{e} (R_{abcd} \eta^{ac} \eta^{bd} - 2 \Lambda) .
\label{CNH action}
\end{equation} Varying the action by using \eqref{eq:50} and dropping the boundary terms, one obtains 
\begin{equation}
\label{Einstein EOM CNH}
\frac{\delta L^{CNH}}{\delta e^\tau_h} = 2 \mathbf{e}({G^h}_c e^c_\tau + \Lambda e_\tau^h) + \mathbf{e} (2 {T^g}_{cg} {\Gamma^{ch}}_\tau - {T^h}_{cd}{\Gamma^{cd}}_\tau) ,
\end{equation}
\begin{equation}
\label{zero torsion CNH}
\begin{split}
\frac{\delta L^{CNH}}{\delta \Gamma_{abf}} &=  \mathbf{e} {T^f}_{cd} \eta^{ac} \eta^{bd} + 2 D_\mu [\mathbf{e}e^\mu_c (\eta^{[a|f|} \eta^{b]c})] \\
&= \mathbf{e}(T^{fab} + 2 \eta^{f[a}{T^{|c|b]}}_c ).   \\
\end{split}
\end{equation} Contracting the equations of motion associated with \eqref{zero torsion CNH} with $\eta_{fa}$ gives ${T^{cb}}_c = 0$. When re-injecting, this implies ${T^f}_{ab} = 0$. This torsionless condition for on-shell connection is the analogue of the one encountered in \eqref{eq:77} above. The fields $\Gamma_{abf}$ are thus auxiliary fields in this formulation. Taking these fields on-shell in \eqref{Einstein EOM CNH}, one obtains the Einstein equations, as expected.

\subsubsection{Symmetries}

The gauge symmetries of the action \eqref{CNH action} are the diffeomorphisms and the Lorentz gauge transformations. The infinitesimal transformations of the fields under these symmetries are given by
\begin{align}
\delta_{\xi, \omega} {e_a}^\mu &= \xi^\nu \p_\nu {e_a}^\mu - \partial_\nu \xi^\mu  {e_a}^\nu + {\omega_a}^b {e_b}^\mu , \\
\delta_{\xi, \omega} \Gamma_{a b c} &= \xi^\rho \partial_\rho \Gamma_{a b c} - D_c \omega_{ab} + \Gamma_{abd} \omega_c^{\;d} .
\label{variation NH connection}
\end{align}

Following the lines of \eqref{Noether identity proof}, we consider
\begin{equation}
  \label{eq:78}
  \vddl{L^C}{{e_a}^\mu}
\delta_{\xi,\omega} {e_a}^\mu
+\vddl{L^C}{{\Gamma_{abc}} }\delta_{\xi,\omega} {\Gamma_{abc}} 
\end{equation}
and integrate by parts to isolate the
undifferentiated gauge parameters $\omega^{ab}$ and $\xi^\rho$. Discarding the boundary terms, this leads to the Noether identities
\begin{equation}
\frac{\delta L^{CNH}}{\delta e^{[a|\tau}}e^\tau_{b]} + D_\mu \left(e^\mu_f \frac{\delta L^{CNH}}{\delta {\Gamma^{ab}}_f} \right) + \frac{\delta L^{CNH}}{\delta {\Gamma_{cf}}^{[a}} \Gamma_{|cf|b]} = 0 , 
\label{Noeth1}
\end{equation}
\begin{equation}
\frac{\delta L^{CNH}}{\delta e^\tau_h} \partial_\rho e^\tau_h + \frac{\delta L^{CNH}}{\delta \Gamma_{abf}} \partial_\rho \Gamma_{abf} + \partial_\mu \left(\frac{\delta L^{CNH}}{\delta e^{\rho}_h}e^\mu_h \right) =0 . 
\label{Noeth2}
\end{equation} The first identity corresponds to Lorentz gauge symmetry and can be shown to be exactly the same as the one found in \eqref{eq:79} in the first Cartan formulation. The second identity corresponds to diffeomorphism symmetry and can be shown to be the same as \eqref{eq:80}. Then, as above, the Noether identities are equivalent to the Bianchi identities \eqref{eq:27} and \eqref{eq:34b}.

\subsubsection{Co-dimension 2 form and equivalence with the other formulations}

When keeping the boundary terms, one finds the weakly vanishing Noether current
\begin{equation}
 \kappa^{-1} S_{\xi, \omega}^\mu = - \frac{\delta L^{CNH}}{\delta \Gamma_{abf}} e^\mu_f \omega_{ab} - \frac{\delta L^{CNH}}{\delta e^\tau_h} e^\mu_h \xi^\tau.
\label{weakly vanishing CNH}
\end{equation} Then the co-dimension 2 form is given by
\begin{multline}
  \label{2form CNH}
   \kappa^{-1} k^{\mu\nu}_{\xi,\omega}={\mathbf e}\,\big[(2\delta{e_a}^\mu
  {e_b}^\nu+{e^c}_\lambda\delta
  {e_c}^\lambda{e_a}^\nu{e_b}^\mu)(-\omega^{ab}+{\Gamma^{ab}}_d e^d_\rho \xi^\rho)\\+
\delta ({\Gamma^{ab}}_d e^d_\rho) (\xi^\rho {e_a}^\mu{e_b}^\nu
+2\xi^{\mu} {e_a}^{\nu} {e_b}^\rho) -(\mu\longleftrightarrow\nu)\big], 
\end{multline} which is obviously the same as \eqref{eq:84} by performing the field redefinition ${\Gamma^{ab}}_\rho = {\Gamma^{ab}}_c e^c_\rho$.

\subsection{Newman-Penrose formulation}
\label{Newman-Penrose formulation}

\subsubsection{Variational principle}

The Newman-Penrose (NP) equations are a set of first order equations involving the spin coefficients, the vielbein and the curvature components at the same footage \cite{Newman:1961qr , Geroch:1973am}. The NP formulation that we introduce here leads to Euler-Lagrange equations that impose vanishing of torsion together with all NP equations. This is achieved by introducing additional auxiliary fields in the Cartan formulation II \eqref{CNH action}. It involves as dynamical variables the vielbein components ${e_a}^\mu$, the Lorentz
connection components in the non-holonomic frame $\Gamma_{abc}$,
and a suitable set of auxiliary fields
${\mathbf R}_{abcd}=\mathbf{R}_{[ab][cd]},
\lambda^{abcd}=\lambda^{[ab][cd]}$,
\begin{equation}
\begin{split}
S[\Gamma_{abc},{e_a}^\mu, {\mathbf R}_{abcd}, \lambda^{abcd} ] &=  \kappa\int \mathrm{d}^n x L^{NP} \\
&= \kappa\int \mathrm{d}^n x \mathbf{e}
[\mathbf{R}_{abcd}(\eta^{ac}\eta^{bd} - \lambda^{abcd})
   + \lambda^{abcd} R_{abcd}-2\Lambda],\label{eq:1}
   \end{split}
\end{equation} where $R_{abcd}=\eta_{ae}{R^e}_{bcd}$ is explicitly given in
\eqref{eq:15} as a function of the variables
${e_a}^\mu,\Gamma_{abc}$ and their first order derivatives.

The equations of motion for the auxiliary fields follow from equating
to zero the Euler-Lagrange derivatives of $L^{NP}$
\begin{equation}
  \label{eq:11a}
  \begin{split}
   \frac{\delta L^{NP}}{\delta \mathbf R_{abcd}} & = -
  \mathbf{e} \left[\lambda^{abcd}-\half (\eta^{ac}\eta^{bd}-\eta^{ad}\eta^{bc})
    \right],\\
 \frac{\delta L^{NP}}{\delta {\mathbf
     \lambda^{abcd}}} & =-\mathbf{e} \left[\mathbf
   R_{abcd}-R_{abcd}\right].
  \end{split}
\end{equation}
They thus fix the auxiliary $\lambda$ fields in terms of the Minkowski
metric,
\begin{equation}
\lambda^{abcd}= \half
(\eta^{ac}\eta^{bd}-\eta^{ad}\eta^{bc})\equiv\lambda^{abcd}_\eta,\label{eq:119}
\end{equation}
and impose the definition of the Riemann tensor in terms of vielbein
and connection components as on-shell relations,
$\mathbf R_{abcd}=R_{abcd}$, which is desirable from the viewpoint of
the NP formalism. They can be eliminated by solving inside the
action. The resulting reduced action coincides with the
action associated with the Cartan formulation II \eqref{CNH action}. 

The next equations of motion follow from the vanishing of
\begin{equation}
  \frac{\delta L^{NP}}{\delta \Gamma_{abc}}=2
  \mathbf{e} \left[D_f\lambda^{abcf}+\lambda^{abdf} ({T^h}_{fh}\delta^c_d+\half
    {T^c}_{df})\right]. \label{2bis}
\end{equation}
When putting $\lambda^{abcd}$ on-shell, they are equivalent to
vanishing of torsion, ${T^a}_{bc}=0$. It follows that
$\Gamma_{abc}=r_{abc}$ or, equivalently, that
${\Gamma^a}_{bc} ={e^a}_\nu {e_c}^\mu\nabla_\mu {e_b}^\nu$, where
$\nabla_\mu$ denotes the Christoffel connection. In other words, the
connection components are also auxiliary fields that can be expressed
in terms of vielbein components and eliminated by their own equations
of motion.

The last equations of motion follow from the vanishing of
\begin{equation}
  \frac{\delta L^{NP}}{\delta
    {e_a}^\mu}={e^b}_\mu\left[2 \mathbf{e}(\lambda^{cdfa}R_{cdfb})-\frac{\delta
      L^{NP}}{\delta \Gamma_{cda}}\Gamma_{cdb}\right]
  - {e^a}_\mu\left[\mathbf{e} (\mathbf R -2\Lambda)+
    \lambda^{bcdf}\frac{\delta L^{NP}}{\delta\lambda^{bcdf}}\right].
    \label{eom NP e}
\end{equation}
On-shell for the auxiliary fields, we have
\begin{equation}
  \label{eq:14}
  \frac{\delta \mathcal L}{\delta
    {e_a}^\mu}|_{\rm aux\ on-shell}=2\mathbf{e} {e^b}_\mu ({G^a}_b+\Lambda\delta^a_b),
\end{equation}
which imply the standard Einstein equations.

Finally, it should be noted that the equations of motion associated with \eqref{2bis} and \eqref{eom NP e} consistently reduce to \eqref{Einstein EOM CNH} and \eqref{zero torsion CNH} when taking the auxiliary fields $\lambda^{abcd}$ and $\mathbf{R}_{abcd}$ on-shell.

%
%

\subsubsection{Improved gauge transformations and Noether identities}

Diffeomorphisms and local Lorentz transformations are extended in
a natural way to the auxiliary fields. If ${\xi}^\mu,{{\omega'}^{a}}_b=-{\omega_b}^a$
denote parameters for the infinitesimal transformations, they act on
the fields as
\begin{equation}
  \label{eq:15num2}
  \begin{split}
    \delta_{\xi,\omega}{e_a}^\mu &={\xi}^\nu\partial_\nu
    {e_a}^\mu-\d_\nu{\xi}^\mu{e_a}^\nu +{\omega_a}^b{e_b}^\mu,\\
\delta_{\xi, \omega} \Gamma_{a b c} &= {\xi}^\nu \partial_\nu \Gamma_{a
  b c} - D_c {\omega}_{a b} + {\omega_c}^{d}\Gamma_{abd} ,\\
\delta_{\xi, \omega} \mathbf R_{abcd} &={\xi}^\nu \partial_\nu \mathbf R_{abcd}
+ {\omega}\indices{_a^f}\mathbf R_{fbcd} + {\omega}\indices{_b^f}\mathbf R_{afcd}
+ {\omega}\indices{_c^f}\mathbf R_{abfd} + {\omega}\indices{_d^f} \mathbf R_{abcf}, \\
\delta_{\xi, \omega} \lambda^{abcd} &={\xi}^\nu \partial_\nu
\lambda^{abcd}
+ {\omega}^a_{\;\;f} \lambda^{fbcd}  + {\omega}^b_{\;\;f} \lambda^{afcd}
+ {\omega}^c_{\;\;f} \lambda^{abfd}  + {\omega}^d_{\;\;f}\lambda^{abcf}.
\end{split}
\end{equation}
In terms of the redefined gauge parameters, which are spacetime
scalars and thus in agreement with the general strategy of the NP approach,
\begin{equation}
{\xi'}^a={e^a}_\mu{\xi}^\mu,\quad
{\omega'_a}^b={\omega_a}^b+{\xi}^\mu{\Gamma^b}_{ac}{e^c}_\mu,\label{eq:70}
\end{equation}
these gauge transformations become
\begin{equation}
  \label{eq:23}
  \begin{split}
    \delta_{\xi',\omega'}{e_a}^\mu &=({\xi'}^c {T^b}_{ac}
    -D_a{\xi'}^b+{\omega'_a}^b){e_b}^\mu,\\
    \delta_{\xi',\omega'} \Gamma_{a b c} &= -{\xi'}^d
    R_{abcd}+ ({\xi'}^f{T^d}_{cf}-D_c
    {\xi'}^d+{\omega'_c}^d)\Gamma_{abd}
    -D_c\omega'_{ab},\\
    \delta_{\xi',\omega'} \mathbf R_{abcd} &= {\xi'}^f D_f \mathbf R_{abcd} +
    {\omega'_a}^f\mathbf R_{fbcd} +
    {\omega'_b}^f\mathbf R_{afcd}
    + {\omega'_c}^f \mathbf R_{abfd} +
    {\omega'_d}^f \mathbf R_{abcf} , \\
    \delta_{\xi',\omega'} \lambda^{abcd} &={\xi'}^f D_f
    \lambda^{abcd} + {{\omega'}^a}_f \lambda^{fbcd} +
    {{\omega'}^b}_f \lambda^{afcd} + {{\omega'}^c}_f \lambda^{abfd}
    + {{\omega'}^d}_f\lambda^{abcf}.
\end{split}
\end{equation}

Isolating the undifferentiated gauge parameters by dropping the
exterior derivative of an $n-1$ form, the invariance of action
\eqref{eq:1} under these transformations leads to the Noether
identities. Since the change of gauge parameters is invertible, the
identities associated with both sets are equivalent. We can thus
concentrate on this second set. For later use, note that
\begin{equation}
\delta_{\xi',\omega'}\Gamma_{abc}-(\delta_{\xi',\omega'}{e_c}^\mu)
{e^d}_\mu\Gamma_{abd}=-{\xi'}^d
R_{abcd}-D_c\omega'_{ab}.\label{eq:91}
\end{equation}

When using \eqref{eq:73}, the Noether identities associated with the
Lorentz parameters $\omega'_{ab}$ become
\begin{multline}
  2 \frac{\delta  L^{NP}}{\delta \mathbf R_{[a|cdf|}} {\mathbf
    R^{b]}}_{cdf} + 2 \frac{\delta L^{NP}}{\delta \mathbf R_{cd[a|f|}}
  {{\mathbf R_{cd}}^{b]}}_f + 2 \frac{\delta L^{NP}}{\delta
    \lambda^{fhcd}} \eta^{f[a}\lambda^{b]hcd}
  + 2 \frac{\delta L^{NP}}{\delta \lambda^{cdfh}} \eta^{f[a}\lambda^{|cd|b]h} \\
  +\frac{\delta L^{NP}}{\delta {e_{[a}}^\mu} e^{b]\mu} + \frac{\delta
    L^{NP}}{\delta \Gamma_{cd[a}} {\Gamma_{cd}}^{b]} + \mathbf{e}
  \big[(D_c +{T^c}_{cf})(\mathbf{e}^{-1} \frac{\delta L^{NP}}{\delta \Gamma_{abc}})\big]
  =0. \label{eq:NL}
\end{multline}
while the Noether identities for the vector fields $ {\xi'}^f$ read
\begin{multline}
  \frac{\delta  L^{NP}}{\delta \mathbf R_{abcd}} D_f \mathbf R_{abcd}
  + \frac{\delta  L^{NP}}{\delta \lambda^{abcd}} D_f \lambda^{abcd} +
  \frac{\delta  L^{NP}}{\delta {e_a}^\mu} {T^b}_{af}{e_b}^\mu
  +\frac{\delta  L^{NP}}{\delta \Gamma_{abc}}({T^d}_{cf}\Gamma_{abd} -R_{abcf})\\
  + \mathbf{e} \big[(D_c+{T^h}_{ch})\mathbf{e}^{-1}( \frac{\delta  L^{NP}}{\delta {e_c}^{\mu}}{e_f}^\mu +\frac{\delta  L^{NP}}{\delta \Gamma_{abc}}\Gamma_{abf})\big] =0.\label{eq:ND}
\end{multline}

It follows from general results on auxiliary fields (see e.g. \cite{Henneaux:1992ig}) that these Noether
identities are equivalent to those of the Cartan formulation II (see equations \eqref{Noeth1} and \eqref{Noeth2}),
which have been investigated and related to the Bianchi identities. More explicitly, we have $L^{NP}=L^{CNH}+A$ with
$A=[(\mathbf
R_{abcd}-R_{abcd})(\eta^{ac}\eta^{bd}-\lambda^{abcd})]$. Identity
\eqref{eq:NL} for $L^{NP}$ replaced by $A$ is equivalent to
\eqref{eq:27}. This then implies that \eqref{eq:NL} reduces to
\begin{equation}
  \label{eq:16}
  \frac{\delta L^{CNH}}{\delta {e_{[a}}^\mu} e^{b]\mu}
+ \frac{\delta L^{CNH}}{\delta \Gamma_{cd[a}} {\Gamma_{cd}}^{b]}
+ \mathbf{e} \big[(D_c+{T^f}_{cf}) (\mathbf{e}^{-1} \frac{\delta 
  L^{CNH}}{\delta \Gamma_{abc}})\big] =0,
\end{equation}
which in turn is also equivalent to \eqref{Noeth1} and so to \eqref{eq:27}.

Identity \eqref{eq:ND} for $L^{NP}$ replaced by $A$ is equivalent to the
second identity of \eqref{eq:24}. This then implies that \eqref{eq:ND} reduces
to
\begin{multline}
  \label{eq:28}
 \frac{\delta L^{CNH}}{\delta {e_a}^\mu} {T^b}_{af}{e_b}^\mu
  +\frac{\delta L^{CNH}}{\delta \Gamma_{abc}}({T^d}_{fc}\Gamma_{abd} -R_{abcf})
  \\+ \mathbf{e} \big[ (D_c+{T^h}_{ch})\mathbf{e}^{-1} (\frac{\delta L^{CNH}}{\delta {e_c}^{\mu}}{e_f}^\mu
  +\frac{\delta L^{CNH}}{\delta \Gamma_{abc}}\Gamma_{abf})\big] =0,
\end{multline}
which is equivalent to \eqref{Noeth2} and so to \eqref{eq:34b}.

\subsubsection{Co-dimension 2 form and breaking}
\label{sec:appl-non-holon}

Writing $\phi^i=(\mathbf R_{abcd},\lambda^{abcd},\Gamma_{abc},{e_a}^\mu)$, the presymplectic potential associated with the action
\eqref{eq:1} is given by
\begin{equation}
  \label{eq:93}
  \boldsymbol{\theta}[\phi ; \delta\phi]=2\kappa \mathbf{e}\lambda^{abcd}{e_c}^\mu \delta
  \Gamma_{ab\nu}{e^\nu}_d{e_c}^\mu (\D^{n-1} x)_\mu,
\end{equation}
where $\delta
  \Gamma_{ab\nu}{e^\nu}_d=\delta\Gamma_{abd}-\Gamma_{abf}{e^f}_\nu \delta
  {e_d}^\nu$.

Writing the gauge parameters as $f^\alpha=(\omega_{ab},\xi^a)$, the weakly vanishing
Noether current is given by
\begin{equation}
\mathbf{S}_{\xi', \omega'}[\phi] = -\kappa\Big[ \frac{\delta L^{NP}}{\delta \Gamma_{abc}}  (\omega'_{ab}+ \Gamma_{abf}{\xi'}^f) + \frac{\delta L^{NP}}{\delta e^\tau_c} e^\tau_f {\xi'}^f \Big] e^\mu_c (\D^{n-1}x)_\mu.
\label{weakly NP}
\end{equation}  The co-dimension 2 form can be obtained from \eqref{eq:157} or alternatively from \eqref{eq:161}. Using the Euler-Lagrange equations for the auxiliary fields, we obtain
\begin{multline}
  \label{2form NP}
  \mathbf{k}_{\xi',\omega'}[\phi ; \delta \phi]=2 \kappa{\mathbf e}\,\big[-(2\delta{e_a}^\mu
  {e_b}^\nu+{e^c}_\lambda\delta
  {e_c}^\lambda{e_a}^\nu{e_b}^\mu){\omega'}^{ab}\\+
\delta ({\Gamma^{ab}}_d e^d_\rho) ({\xi'}^c {e_c}^\rho {e_a}^\mu{e_b}^\nu
+2{\xi'}^{c} {e_c}^\mu {e_a}^{\nu} {e_b}^\rho) \big] (\D^{n-2}x)_{\mu\nu}. 
\end{multline} Notice that this last expression obtained from the NP formulation is exactly the same as \eqref{2form CNH} obtained from the Cartan formulation II, up to the parameters redefinition \eqref{eq:70}.   

The breaking in the conservation law of $\mathbf{k}[\phi,\delta \phi]$ (see equation \eqref{breaking in the conservation}) is given by the invariant presymplectic current \eqref{presymplectic current first order} with an evaluated variation. For the present case, using the equations of motion for the auxiliary fields, we explicitly find 
\begin{multline}
\mathbf{W}[\phi ; \delta_{\xi',\omega'} \phi , \delta \phi] = 2 \kappa \mathbf{e}\big[\delta_{\xi',\omega'}
      {e_b}^\mu\delta{\Gamma^{ab}}_\nu{e_a}^\nu{e^c}_\mu+\delta_{\xi',\omega'}
      {e_a}^\mu\delta{\Gamma^{ac}}_\mu \\
      -\delta_{\xi',\omega'}\ln {\mathbf
        e}\,\delta{\Gamma^{cb}}_\nu{e_b}^\nu
     -(\delta_{\xi',\omega'}\longleftrightarrow\delta)\big] e_c^\rho (\D^{n-1}x)_\rho.
     \label{breaking NP formula}
\end{multline}

\paragraph{Exact reducibility parameters}

General considerations on auxiliary fields imply that,
on-shell, reducibility parameters should be given by Killing vectors
$\bar {\xi'}^a$ of the metric (see e.g. \cite{Henneaux:1992ig}). Let us see how this comes about here.

Merely the first of \eqref{eq:23} encodes gauge
transformations of fields that are not auxiliary. The associated
equation
$\delta_{\bar\omega',\bar\xi'}{e_a}^\mu\approx 0$ is equivalent to
\begin{equation}
  \label{eq:8}
  D_{(a}\bar\xi'_{b)}-\bar\xi'^cT_{(ba)c}\approx 0,\quad
  \bar\omega'_{ab}\approx D_{[a}\bar\xi'_{b]}-\bar{\xi'}^cT_{[ba]c}.
\end{equation}
On-shell when torsion vanishes, the first indeed requires $\bar{\xi'}^a$
to be a Killing vector, while the second uniquely fixes the Lorentz
parameters in terms of it. In particular,
\begin{equation}
\bar\omega'_{ab}\approx
D_a\bar\xi'_b\approx -D_b\bar\xi'_a. \label{eq:76}
\end{equation}

The other equations impose no additional constraints. Indeed,
$\delta_{\bar\omega',\bar\xi'}\lambda^{abcd}\approx 0$ is satisfied
identically on account of the skew-symmetry of
$\bar\omega^{\prime ab}$. Instead of
$\delta_{\bar\xi',\bar\omega'}\Gamma_{abc}\approx 0$ we can consider the
combination \eqref{eq:91}. Requiring this to vanish on-shell amounts
to
\begin{equation}
  \label{eq:78}
  D_c\omega'_{ab}\approx -{\bar\xi}^{\prime d} R_{abcd},
\end{equation}
which holds as a consequence of the second equation in \eqref{eq:8},
when using that
\begin{equation}
D_aD_b\bar\xi'_c\approx
{R^d}_{abc}\bar\xi'_d\label{eq:77num2},
\end{equation}
which can be shown as in \cite{Wald:1984rg} appendix C.3, and when
also using \eqref{eq:30}. Finally,
$\delta_{\bar\xi',\bar\omega'}{\mathbf R}_{abcd}\approx 0$, reduces
on-shell to
\begin{equation}
  \label{eq:31num2}
  \bar{\xi}^{\prime f} D_f  R_{abcd} +
    {{\bar{\omega}}^{\prime~f}_a} R_{fbcd} +
    {\bar\omega}^{\prime~f}_b R_{afcd}
    + {\bar\omega}^{\prime~f}_c R_{abfd} +
    {\bar\omega}^{\prime ~f}_d R_{abcf}\approx 0.
\end{equation}
This equation holds because one can show that, on-shell, the left-hand
side is equal to its opposite when using the previous relations
\eqref{eq:76}, \eqref{eq:77} together with the Bianchi identities
\eqref{eq:24} and the on-shell symmetries of the Riemann tensor.

\section{Application to asymptotically flat 4d gravity}
\label{Application to asymptotically flat 4d gravity}

In this section, we illustrate the general results obtained in the first order formulations presented above into a concrete situation. More precisely, starting from the solution space of general relativity in asymptotically flat spacetime in NP formalism, we compute the asymptotic symmetries, the currents and the breaking in their conservation laws. This derivation is done in a self-consistent way, without resorting to the metric formulation. This contrasts with the approach used in \cite{Barnich:2011ty , Barnich:2013axa}, where the expressions of the currents in the NP formulation were obtained by translation from the metric formalism. Our direct road allows us to extend the previous results for a time-dependent (but non-dynamical) conformal factor $P = P (u, \zeta, \bar{\zeta})$. 

In this section (and this section only), we follow the conventions of \cite{Barnich:2016lyg} and work with a metric of signature $(+--~-)$, which is more adapted to the NP literature. In particular, one has to adapt some signs in the equations established in the previous section before applying them here. We refer to our article \cite{Barnich:2019vzx} where the convention $(+--~-)$ was chosen from the very beginning.

\subsection{Newman-Penrose notations}
\label{sec:relat-newm-penr}

Following the literature on NP formalism \cite{Newman:1961qr , Geroch:1973am , Penrose:1987uia}, we assign some notations to the fields of the NP formulation introduced in subsection \ref{Newman-Penrose formulation}. In four spacetime dimensions, the tetrads
$e_1=l,\,e_2=n,\,e_3=m,\,e_4=\bar m$ are chosen
as null vectors, $e_a\cdot e_b=\eta_{ab}$ with
\begin{equation}
\eta_{ab}=\eta^{ab} = \begin{pmatrix}
0 & 1 & 0 & 0 \\
1 & 0 & 0 & 0 \\
0 & 0 & 0 &-1 \\
0 & 0 & -1 &0 \\
\end{pmatrix}.
\end{equation}
The components of the Lorentz connection are traded for the spin
coefficients,
\begin{equation}
  \label{spinconnection}
  \begin{split}
    &\kappa=\Gamma_{311},\;\;\pi=-\Gamma_{421},\;\;\epsilon=\half(\Gamma_{211} - \Gamma_{431}),\\
    &\tau=\Gamma_{312},\;\;\nu=-\Gamma_{422},\;\;\gamma=\half(\Gamma_{212} - \Gamma_{432}),\\
    &\sigma=\Gamma_{313},\;\;\mu=-\Gamma_{423},\;\;\beta=\half(\Gamma_{213} -\Gamma_{433}),\\
    &\rho=\Gamma_{314},\;\;\lambda=-\Gamma_{424},\;\;\alpha=\half(\Gamma_{214}
    - \Gamma_{434}).
\end{split}
\end{equation}
The other half of the spin coefficients are denoted with a bar on the
symbols in the left-hand sides and obtained by exchanging the index $3$
and $4$ on the right-hand sides. The Weyl tensor $C_{abcd}$ is encoded
in terms of
\begin{equation}
  \label{weyl}
\Psi_0=-C_{1313},\;\;\Psi_1=-C_{1213},\;\;\Psi_2=-C_{1342},\;\;\Psi_3=-C_{1242},\;\;\Psi_4=-C_{2324},
\end{equation}
with the same rule as above for $\bar\Psi_i$, $i=0,\dots, 4$, while the
Ricci tensor is organized as
\begin{equation}
\begin{array}{lll}
\Phi_{00}=-\half R_{11}, & \Phi_{11}=-\dfrac{1}{4}(R_{12}+R_{34}), & \Phi_{22}=-\half R_{22},\\
\Phi_{02}=-\half R_{33}, & \Phi_{01}=-\half R_{13}, & \Phi_{12}=-\half R_{23},\\
\Phi_{20}=-\half R_{44}, & \Phi_{10}=-\half
                           R_{14},& \Phi_{21}=-\half R_{24},\\
  & \tilde{\Lambda}=\dfrac{1}{24}R=\dfrac{1}{12} (R_{12}-R_{34}). &
  \end{array}
\end{equation}
There is no torsion in the NP approach, ${T^a}_{bc}=0$. In this case,
the vacuum Einstein equations in flat space are equivalent to the
vanishing of the $\Phi$'s. The equations governing the NP quantities
can then be interpreted as follows: (i) The metric equations express
commutators of tetrads in terms of spin coefficients. This is the
first of \eqref{eq:13} when taking into account that
${D^a}_{bc}=2{\Gamma^a}_{[cb]}$ in the absence of torsion. (ii) The
spin coefficient equations express directional derivatives of spin
coefficients in terms of spin coefficients and the Weyl and Ricci
tensors. In the torsion-free case, they are equivalent to the
definition of $R_{abcd}$ in \eqref{eq:15}. (iii) The
Bianchi identities express directional derivatives of the $\Psi's$ and
$\Phi$'s in terms of spin coefficents and $\Psi$'s and $\Phi$'s. They
are equivalent to the second of \eqref{eq:24} in the absence of
torsion.

\subsection{Solution space}
\label{sec:solution-space-1}

Four-dimensional asymptotically flat spacetimes at null infinity in
the NP formalism have been studied in
\cite{Newman:1961qr,Newman:1962cia,Exton:1969im} (see
\cite{Barnich:2016lyg} for a summary and conventions appropriate to the
current context). In terms of coordinates $x^\mu = (u, r, x^A)$,
$x^A = (\zeta, \bar{\zeta})$ and using the notations of section
\ref{sec:relat-newm-penr}, the Newman-Unti solution space is entirely
determined by the conditions
\begin{equation}
\begin{split}
  &\kappa = \epsilon = \pi = 0,\quad \rho=\overline{\rho}, \quad \tau=\bar\alpha+\beta,\\
  &l = \frac{\partial}{\partial r}, \quad n = \frac{\partial}{\partial
    u} + U \frac{\partial}{\partial r} + X^A \frac{\partial}{\partial
    x^A} , \quad m = \omega \frac{\partial}{\partial r} + L^A
  \frac{\partial}{\partial x^A},
\end{split}
\label{gauge conditions}
\end{equation}
where $U$, $X^A$, $\omega$ and $L^A$ are arbitrary functions of the
coordinates, together with the fall-off conditions
\begin{equation}
\begin{split}
&X^A = \cO(r^{-1}), \quad \Psi_0 = \Psi^0_0 r^{-5} +
\mathcal{O}(r^{-6}), \quad \rho = -\dfrac{1}{r}+\cO(r^{-3}) , \quad \tau = \cO (r^{-2}),\\
&g_{AB} \mathrm{d}x^A \mathrm{d}x^B = -2 r^2 \frac{\mathrm{d} \zeta \mathrm{d}\bar{\zeta} }{P \bar{P}} +
\mathcal{O}(r).
\end{split}
\label{fall-off conditions}
\end{equation}
Here, $\Psi^0_0$ and $P$ are arbitrary complex functions of
$(u,\zeta,\bar{\zeta})$. Below we also use the real function
$\varphi(u,\zeta,\bar\zeta)$ defined by $P\bP=2 e^{-2\varphi}$. The
associated asymptotic expansion of the solution space in terms of
$\Psi_0 (u_0, r, \zeta, \bar{\zeta})$ ,
$(\Psi^0_2 + \bar{\Psi}^0_2)(u_0, \zeta, \bar{\zeta})$,
$\Psi^0_1 (u_0, \zeta, \bar{\zeta})$ at fixed $u_0$ and of the
asymptotic shear $\sigma^0(u, \zeta, \bar{\zeta})$ and the conformal
factor $P(u, \zeta, \bar{\zeta})$ is summarized in appendix \ref{NP
  solution}. These data characterizing  the solution space are collectively
denoted by $\chi$.

On a space-like cut of $\mathscr{I}^+$, we
use coordinates $\zeta, \bar{\zeta}$, and the (rescaled) induced
metric
\begin{equation}
\D\bar{s}^2 = -\bar{\gamma}_{AB} \D x^A \D x^B =  -2 (P \bar{P})^{-1} \D \zeta \D \bar{\zeta},
\end{equation} with $\bar{P}P> 0$. For the unit sphere, we have $\zeta
= \cot \frac{\theta}{2} e^{i\phi}$ in terms of standard spherical
coordinates and
\begin{equation}
P_S (\zeta, \bar{\zeta}) = \frac{1}{\sqrt{2}}(1+ \zeta \bar{\zeta}).\label{unit}
\end{equation} The covariant derivative on the 2 surface is
encoded in the operators
\begin{equation}\begin{split}
    &\eth \eta^s=P\bP^{-s}\bp(\bP^s \eta^s)=P\bp \eta^s
    + s P \bp \ln \bP \eta^s=P\bp \eta^s + 2 s\xbar\alpha^0 \eta^s,\\
    &\xbar\eth \eta^s=\bP P^{s}\p(P^{-s} \eta^s)=\bP\p \eta^s
    - s \bP \p \ln P \eta^s=\bP\p \eta^s -2 s \alpha^0 \eta^s,
\end{split}\end{equation}
where $s$ is the spin weight of the field $\eta$ and
$\p=\p_{\zeta},\bar\p=\p_{\bar\zeta}$. The spin and conformal weights
of relevant fields are listed in Table \ref{t1}.
\begin{table}[h]
\caption{Spin and conformal weights}\label{t1}
\begin{center}\begin{tabular}{c|c|c|c|c|c|c|c|c|c|c|c|c|c|c|c|c|c}
& $\eth$ & $\p_u$ & $\gamma^0$ & $\nu^0$ & $\mu^0$ & $\sigma^0$ &
   $\lambda^0$
  & $\Psi^0_4$ &  $\Psi^0_3$ & $\Psi^0_2$ & $\Psi^0_1$ & $\Psi_0^0$ & $\cY$   \\
\hline
s & $1$& $0$& $0$& $-1$& $0$& $2$& $-2$  &
  $-2 $&  $-1$ & $0$ & $1$ & $2$ & $-1$  \\
\hline
w  & $-1$& $-1$& $-1$& $-2$& $-2$& $-1$& $-2$  &
  $-3 $&  $-3$ & $-3$ & $-3$ & $-3$ & $1$  \\
\end{tabular}\end{center}\end{table}
Complex conjugation transforms the spin weight into its opposite and
leaves the conformal weight unchanged. The operators $\eth$,
$\overline{\eth}$ respectively raise and lower the spin weight by one
unit. The Laplacian is
$\bar{\Delta} = 4 e^{-2 \varphi} \partial \bar{\partial} = 2 \xbar\eth
\eth$. Note that $P$ is of spin weight $1$ and ``holomorphic'',
$\xbar\eth P = 0$ and that
\begin{equation}
[\xbar\eth, \eth] \eta^s = \frac{s}{2} R \eta^s,
\end{equation} with $R= -4\mu^0=\bar{\Delta} \ln (P\bar{P})$, $R_S
=2$. We also have
\begin{equation}
[\partial_u, \eth] \eta^s =-2 (\xbar\gamma^0 \eth + s \eth
\gamma^0)\eta^s,\quad [\p_u,\xbar\eth]\eta^s=-2(\gamma^0\xbar\eth -
s\xbar\eth\xbar\gamma^0)\eta^s.
\end{equation}

The components of the inverse metric associated with the tetrad given in
\eqref{gauge conditions} is
\begin{equation*}
g^{0\mu} = \delta^\mu_1,~ g^{rr} = 2(U - \omega \overline{\omega}),~
g^{rA} = X^A - (\overline{\omega} L^A + \omega \overline{L}^A),~
g^{AB} = - (L^A \overline{L}^B + L^B \overline{L}^A ).
\end{equation*} Note furthermore that if $L_A = g_{AB} L^B$ with
$g_{AB}$ the two dimensional metric inverse to $g^{AB}$, then $L^A
\overline{L}_A = -1$, $L^A L_A = 0 = \overline{L}^A
\overline{L}_A$. The co-tetrad is given by
\begin{equation}
\begin{split}
e^{1} = - [U + X^A (\omega \overline{L}_A + \overline{\omega}
L_A)]\D u + \D r + (\omega \overline{L}_A + \overline{\omega} L_A) \D x^A, &
\\
e^{2} = \D u, \quad e^{3} = X^A \overline{L}_A \D u -
\overline{L}_A \D x^A, \quad e^{4} = X^A L_A \D u - L_A \D x^A. &
\end{split}
\label{cotetrad 4d case}
\end{equation}

\subsection{Residual gauge transformations}
\label{sec:resid-infn-gauge}

The parameters of residual gauge transformations that preserve the
solution space are entirely determined by asking that conditions
\eqref{gauge conditions} and \eqref{fall-off conditions} be preserved
on-shell. This is worked out in detail in appendix \ref{ASG}, where it
is shown that these parameters are given by
\begin{equation}
  f(u,\zeta,\bar\zeta),\quad Y^\zeta=Y(\zeta),\quad
  Y^{\bar\zeta}=\bar Y(\bar\zeta),\quad
  \omega^{34}_0(u,\zeta,\bar\zeta).
  \label{orig}
\end{equation}
The associated residual gauge
transformations are explicitly determined by the gauge parameters,
\begin{equation}
\begin{split}
\xi^u = f(u,\zeta,\bar{\zeta}),\quad
\xi^A= Y^A(x^A) - \p_B f \int^{+\infty}_r \D r[L^A \bL^B +  \bL^AL^B ], \\
\xi^r=-\p_u f r + \frac{1}{2} \bar{\Delta}f - \p_A f
\int^{+\infty}_r \D r[\omega \bL^A + \bomega L^A + X^A], \\
\end{split}
\end{equation}
and
\begin{equation}
\begin{split}
\omega^{12}=& \p_u f + X^A \p_A f, \quad
\omega^{23}= \bL^A \p_A f,\quad
\omega^{24}= L^A \p_A f, \\
\omega^{13}=&(\gamma^0 + \bar{\gamma}^0)\bar{P} {\partial} f -
\bar{P} \partial_u {\partial}f + \p_A f \int^{+\infty}_r \D r[\lambda
L^A + \mu \bL^A], \\
\omega^{14}=&(\gamma^0 + \bar{\gamma}^0)P \bar{\partial} f - P
\partial_u \bar{\partial}f + \p_A f \int^{+\infty}_r \D r[\bar{\lambda}
\bL^A + \bar{\mu} L^A], \\
\omega^{34}=&  \omega^{34}_0(u,\zeta, \bar{\zeta}) - \partial_A f
\int^{+\infty}_{r} \D r [(\bar{\alpha} - \beta ) \bar{L}^A +
(\bar{\beta} - \alpha) L^A].
\end{split}
\end{equation}
For the computations below, the leading orders of their asymptotic
on-shell expansions are also useful,
\be\begin{split}\label{asymmetry1} \xi^u=& f,
  \quad 
\xi^\zeta= Y-\frac{\bP \eth f}{r}+\frac{\sigma^0\bP \xbar \eth
  f}{r^2}+O(r^{-3}),\quad \xi^{\bar\zeta}=\overline{\xi^\zeta},\\
\xi^r=& -r\p_u f+ \half \bDelta f -\frac{\xbar\eth \sigma^0 \xbar\eth
  f + \eth \xbar\sigma^0 \eth f}{r}+O(r^{-2}),
\end{split}\ee
and
\be\begin{split}\label{asymmetry2}
\omega^{12}=& \p_u f+O(r^{-3}),\quad
\omega^{23}= \frac{\xbar \eth f}{r} - \frac{\xbar\sigma^0 \eth
  f}{r^2} + \frac{\sigma^0\xbar\sigma^0 \xbar\eth f}{r^3} +O(r^{-4}),
\\
\omega^{13}=& (\gamma^0+\xbar\gamma^0) \xbar \eth f- \xbar\eth \p_u f
+ \frac{\lambda^0\eth f + \mu^0 \xbar\eth f}{r} \\ &\hspace{3cm}-
\frac{\xbar\sigma^0
  \mu^0 \eth f + \sigma^0\lambda^0\xbar\eth f}{r^2} - \frac{\Psi^0_2
  \xbar\eth f}{2r^2}+ O(r^{-3}),\\
\omega^{34}=& \omega^{34}_0 + \frac{\bar{P}
  \partial \ln P \eth f-P \bar{\partial} \ln \bar{P}
  \overline{\eth} f }{r} \\ &\hspace{3cm}+
\frac{P
  \bar{\partial} \ln \bar{P} \bar{\sigma}^0 \eth f
  -\bar{P} \partial \ln P \sigma^0 \overline{\eth} f }{r^2}  +  O(r^{-3}),
\end{split}\ee
with $\omega^{24}=\overline{\omega^{23}}$,
$\omega^{14}=\overline{\omega^{13}}$,
$\omega^{34}=-\overline{\omega^{34}}$.

\subsection{Residual symmetry algebra}
\label{sec:acti-symm-solut}

The variation of the free
data parametrizing the solution space under residual gauge transformation
in terms of the parametrization provided by \eqref{orig} is given by
\begin{equation}
\delta_{f,Y,\omega_0} P = P \partial_u f + f \partial_u P + Y \partial
P + \bar{Y}\bar{\partial} {P}- P \bar{\partial} \bar{Y} + P
\omega^{34}_0,
\label{conformal factors}
\end{equation}
together with the variation of the rest of the free data and derived
quantities that is written in appendix \ref{sec:oracti}.

To make these variations more transparent, it is useful to
re-parametrize residual gauge symmetries through {\em field-dependent
  redefinitions}. In a first step, one trades the real function
$\d_uf(u, \zeta, \bar{\zeta})$ and the imaginary
$\omega^{34}_0(u, \zeta, \bar{\zeta})$ for a complex
$\Omega(u, \zeta, \bar{\zeta})$ according to
\begin{equation}
\begin{split}
\partial_u f &= \half[\bar\d\bar Y-\bar Y\bar \d\ln (P\bar P) + \d Y
-Y\d \ln (P\bar P)] + f
(\gamma^0+\xbar\gamma^0)+\half(\Omega+\bOmega), \\
\omega^{34}_0 &= \half[\bar\d \bar Y-\bar Y\bar \d\ln P +\bar Y \bar \d
\ln \bar P- \d Y + Y\d \ln \bar P -Y \d \ln P]\\ &\hspace{8cm}+
f(\xbar\gamma^0 - \gamma^0)+ 
\half(\Omega-\bOmega).\\
\end{split}
\label{parameter redef 4d}
\end{equation}
It then follows that the first of \eqref{parameter redef 4d} can be
solved for $f$ in terms of an 
integration function $T_R(\zeta,\bar\zeta)$, (called $\tilde T$ in
\cite{Barnich:2010eb,Barnich:2011mi,Barnich:2011ty})
\begin{equation}
f(u, \zeta, \bar{\zeta}) = \frac{1}{\sqrt{P\bar{P}}} [T_R(\zeta,
\bar{\zeta}) + \frac{\tilde{u}}{2} (\partial Y + \bar{\partial}
\bar{Y}) - Y \partial \tilde{u} - \bar{Y} \bar{\partial} \tilde{u} +
\frac{1}{2} (\tilde{\Omega} + \bar{\tilde{\Omega}})]
\label{explicit f},
\end{equation}
where
\begin{equation}
\tilde{u} = \int^u_{u_0} \D u \sqrt{P\bar{P}}, \quad \tilde{\Omega} =
\int^u_{u_0} \D u \sqrt{P\bar{P}}\,\Omega.
\end{equation}
This redefinition of parameters is such that
\begin{equation}
\delta_{Y,T_R,\Omega} P = \Omega P ,
\label{transfo finale}
\end{equation}
together with the complex conjugate relation
$\delta_{Y,T_R,\Omega} \bar{P} = \bar{\Omega}\bar{P}$.

Denoting by $\phi^\alpha$ the fields
$({e_a}^\mu,\Gamma_{abc})$ (together with the auxiliary fields ${\bf
R}_{abcd},\lambda^{abcd}$ if useful), it follows from \eqref{eq:15num2}
that 
\begin{equation}
  \begin{split}
& [\delta_{\xi_1,\omega_1},\delta_{\xi_2,\omega_2}]\phi^\alpha=
\delta_{\hat\xi,\hat\omega}\phi^\alpha,\\
& \hat\xi^\mu=[\xi_1,\xi_2]^\mu, \quad (\hat{\omega})\indices{_a^b}
  ={\xi_1}^\rho\p_\rho{\omega_{2a}}^b+{\omega_{1a}}^c{\omega_{2c}}^b
  -(1\leftrightarrow
  2),
\end{split}
\end{equation}
when the gauge parameters $\xi,\omega$ are field-independent. In
case gauge parameters do depend on the fields, one finds instead
\begin{equation}
  \begin{split}
& [\delta_{\xi_1,\omega_1},\delta_{\xi_2,\omega_2}]\phi^\alpha=
\delta_{\hat\xi_M,\hat\omega_M}\phi^\alpha,\\
& \hat\xi^\mu_M=[\xi_1,\xi_2]^\mu
-\delta_{\xi_1,\omega_1}{\xi}^\mu_2+\delta_{\xi_2,\omega_2}{\xi}^\mu_1,
  \\
 & (\hat{\omega}_M)\indices{_a^b}
 ={\xi_1}^\rho\p_\rho{\omega_{2a}}^b+{\omega_{1a}}^c{\omega_{2c}}^b
 -\delta_{\xi_1,\omega_1}{\omega_{2a}}^b 
  -(1\leftrightarrow
  2).
\end{split}
\end{equation}
We now have the following result:

The gauge parameters $(\xi[Y,T_R,\Omega],\omega[Y,T_R\Omega])$
equipped with the modified commutator for field dependent gauge
transformations realize the direct sum of the abelian ideal of complex
Weyl rescalings with the (extended) BMS algebra,  $\mathfrak{bms}^{\text{ext}}_4$,
everywhere in the bulk spacetime,
\begin{equation}
  \label{eq:105num2}
  \begin{split}
  &\hat \xi_M=\xi[\hat Y,\hat T_R,\hat \Omega],\quad \hat
  \omega_M=\omega[\hat Y,\hat T_R,\hat \Omega],\\
  & \hat Y^A=Y_1^B\partial_B Y^A_2-Y_2^B\partial_B Y^A_1,\\
  &\hat T_R=Y_1^A\d_A T_{R2}+\half T_{R1}\d_A Y^A_2 -(1\leftrightarrow
  2),\\
  & \hat\Omega=0. 
\end{split}
\end{equation}
The proof follows by adapting the ones provided in 
\cite{Barnich:2010eb,Barnich:2011ty,Barnich:2013sxa} to the current
set-up. 

\subsection{Action of symmetries on solutions}
\label{sec:acti-symm-solut-1}

A further field-dependent redefinition consists in 
\begin{equation}
  \label{eq:95bis}
  Y =\bar P \xbar \cY,\quad \bar Y =P \cY.
\end{equation}
The transformations \eqref{transfo solution space 1} then become
\begin{equation}
\begin{split}
&\delta_s \sigma^0=[\cY \eth +\xbar \cY \xbar\eth +
\frac32 \eth\cY - \frac12 \xbar\eth\xbar\cY + \frac{3}{2}\Omega -
\frac{1}{2}\bar{\Omega}]\sigma^0 + f\bar\lambda^0 - \eth^2 f,\\
&\delta_s \Psi^0_0=[\cY \eth +\xbar \cY \xbar\eth +
\frac52 \eth\cY + \frac12\xbar\eth\xbar\cY +\frac{5}{2}\Omega +
\frac{1}{2}\bar{\Omega}]\Psi^0_0 + f\eth\Psi_1^0 + 3f\sigma^0\Psi_2^0
+ 4 \Psi^0_1 \eth f,\\
&\delta_s \Psi^0_1=[\cY \eth +\xbar \cY \xbar\eth + 2
\eth\cY + \xbar\eth\xbar\cY + 2\Omega + \bar{\Omega}]\Psi^0_1 +
f\eth\Psi_2^0 + 2f\sigma^0\Psi_3^0 + 3 \Psi^0_2 \eth f,\\
&\delta_s \left(\frac{\Psi^0_2 +
    \bar{\Psi}^0_2}{2}\right)=[\cY \eth +\xbar \cY \xbar\eth + \frac32
\eth\cY + \frac32\xbar\eth\xbar\cY  + \frac{3}{2}\Omega +
\frac{3}{2}\bar{\Omega}]\left(\frac{\Psi^0_2 +
    \bar{\Psi}^0_2}{2}\right)\\
&~~~~~~~~~~~~~~~~~~~~~~~~~~~~~~ + \frac{1}{2}( f\eth\Psi_3^0
+f\sigma^0\Psi_4^0+ 2 \Psi^0_3 \eth f + \text({c.c.})),
\end{split}
\label{transfo final 2}
\end{equation}
while \eqref{transfo solution space 2}-\eqref{transfo solution space
  4} read as
\begin{multline}
\delta_s \Psi^1_0 =\big[\cY \eth +\xbar \cY \xbar\eth
+ 3 \eth\cY + \xbar\eth\xbar\cY
+3\Omega + \bar{\Omega}\big]\Psi^1_0\\ - \overline{\eth}\big[5  \eth f
\Psi^0_0 + f \eth \Psi^0_0 + 4 f \Psi^0_1 \sigma^0 \big],
\end{multline}
\begin{multline}
\delta_s \Psi^2_0 = \big[\cY \eth +\xbar \cY \xbar\eth
+\frac{7}{2} \eth\cY +\frac{3}{2} \xbar\eth\xbar\cY +\frac{7}{2}\Omega
+ \frac{3}{2}\bar{\Omega}\big] \Psi^2_0 \\ + \big[-3 \bar{\Delta} f -
\xbar \eth f \eth - 3 \eth f \overline{\eth} -
\frac{1}{2} f \eth \overline{\eth} - \frac{5}{4} f R \big] \Psi^1_0
\\ +\big[-5 f \Psi^0_2 - \frac{5}{2} f \bar{\Psi}^0_2 + \frac{5}{2} f
\sigma^0 \overline{\eth}^2 + 5f \overline{\eth} \sigma^0
\overline{\eth} + 3 f \eth \bar{\sigma}^0 \eth + \frac{1}{2} f
\bar{\sigma}^0 \eth^2 + \frac{5}{2} f \eth^2 \bar{\sigma}^0 +
\frac{5}{2}f \sigma^0
\lambda^0 \\ + 5 \overline{\eth} \sigma^0 \overline{\eth} f + 15 \eth
\bar{\sigma}^0 \eth f + 5 \sigma^0 \overline{\eth} f \overline{\eth} +
3 \bar{\sigma}^0 \eth f \eth \big] \Psi^0_0
\\ +\big[ 5 f \Psi_1^0 + 12 \sigma^0 \bar{\sigma}^0 \eth f  + 12 f \sigma^0
\eth \bar{\sigma}^0 + 2f \eth \sigma^0 \bar{\sigma}^0 + \frac{9}{2}f
\sigma^0 \bar{\sigma}^0 \eth  \big] \Psi^0_1 \\
+ \frac{15}{2} f(\sigma^0)^2 \bar{\sigma}^0 \Psi^0_2,
\end{multline}
\begin{multline}
\delta_s \Psi^n_0 =\big[ \cY \eth +\xbar \cY \xbar\eth +
\frac{5+n}{2} \eth\cY + \frac{1+n}{2}
\xbar\eth\xbar\cY + \frac{5+n}{2} \Omega +
\frac{1+n}{2} \bar{\Omega}\big]\Psi^n_0 \\
+ (\text{inhomogeneous terms}).
\label{transfo final 3}
\end{multline}
Finally, the variations \eqref{transfo
  solution space} are given by
\begin{equation}
\begin{split}
&\delta_s \lambda^0=[\cY \eth +\xbar \cY \xbar\eth +2
\xbar\eth\xbar\cY + 2 \bar{\Omega}]\lambda^0 - f \Psi_4^0 -
\frac12\xbar\eth^2(\eth\cY+\xbar\eth\xbar\cY),\\
&\delta_s \Psi^0_2=[\cY \eth +\xbar \cY \xbar\eth +
\frac32 \eth\cY + \frac32\xbar\eth\xbar\cY  + \frac{3}{2}\Omega +
\frac{3}{2}\bar{\Omega}]\Psi^0_2 \\ &\hspace{8cm} + f\eth\Psi_3^0 +f\sigma^0\Psi_4^0+ 2
\Psi^0_3 \eth f,\\
&\delta_s \Psi^0_3=[\cY \eth +\xbar \cY \xbar\eth +
\eth\cY + 2\xbar\eth\xbar\cY+ \Omega +2\bar{\Omega}]\Psi^0_3 +
f\eth\Psi_4^0 + \Psi^0_4 \eth f,\\
&\delta_s \Psi^0_4=[\cY \eth +\xbar \cY \xbar\eth +
\frac12 \eth\cY + \frac52\xbar\eth\xbar\cY +\frac{1}{2}\Omega +
\frac{5}{2}\bar{\Omega}]\Psi^0_4 \\ &\hspace{8cm} + f \p_u \Psi_4^0 +
2(2\gamma^0+\xbar\gamma^0)\Psi_4^0.
\end{split}
\label{transfo finale v2}
\end{equation}

\subsection{Reduction of solution space}
\label{sec:reduct-solut-space}

Besides conditions \eqref{gauge conditions} and \eqref{fall-off
  conditions}, additional constraints may be imposed on solution
space. A standard choice is to fix the conformal factor $P$ to be
equal to $P_S$ given in \eqref{unit}. We also fix $P$ here,
without committing to a specific value. In other words, we consider
$P$ to be part of the background structure. As a consequence,
infinitesimal complex Weyl rescalings (whose finite counterparts have
been discussed in \cite{Barnich:2016lyg}) are frozen and $\Omega=0$ in
the formulas above, while in the formulas below, $s$ stands for
$(\cY,\xbar\cY,T_R,0)$. The main reason we do not perform the
analysis below while keeping $P(u,\zeta,\bar\zeta)$ arbitrary is
computational simplicity. We plan to return elsewhere to a detailed
discussion of the current algebra and its interpretation when complex
Weyl rescalings are allowed.

\subsection{Breaking and co-dimension 2 form}
\label{sec:break-co-dimens}

Under this additional constraint on the solution space, the invariant presymplectic current can be
computed using equation \eqref{breaking NP formula},
\begin{equation}
\mathbf{W}[\phi, \delta_s \phi , \phi] = - W^r_{s(0)} \D u \D \zeta \D \bar{\zeta} +
\mathcal{O}(r^{-1})\label{eq:98},
\end{equation}
where
\begin{equation}
W^r_{s (0)}=\frac{1}{8\pi G
  P\bP}\left(\delta\sigma^0\delta_s\lambda^0 +
  \delta\xbar\sigma^0\delta_s\xbar\lambda^0 -
  \delta\lambda^0\delta_s\sigma^0 -
  \delta\xbar\lambda^0\delta_s\xbar\sigma^0\right).
\label{breaking 4d}
\end{equation}  The expression containing the information about the non-conservation of the currents, it should not come as a surprise that it involves the news functions encoded in $\lambda^0$ and $\bar{\lambda}^0$.

Furthermore, the co-dimension $2$ form \eqref{2form NP} takes the form
\begin{equation}
\mathbf{k}_s[\phi;\delta \phi] = k^{ur}_{s(0)}  \D\zeta \D\bar{\zeta} - k^{\zeta
  r}_{s(0)} \D u  \D \bar{\zeta} + k^{\bar{\zeta}r}_{s(0)} \D u  \D\zeta +
\mathcal{O}(r^{-1}) ,
\end{equation} where
\begin{multline}
k^{ur}_{s(0)}=-\frac{1}{P\bP 8\pi G}\Big(\delta\big[f(\Psi^0_2 + \sigma^0
\lambda^0) +\cY(\sigma^0\eth\xbar\sigma^0  + \half
\eth(\sigma^0\xbar\sigma^0)+ \Psi^0_1) \\- \half \eth(\cY
\sigma^0\xbar\sigma^0) - r \eth(\xbar\cY \xbar\sigma^0)\big]  - f
\lambda^0 \delta \sigma^0 + c.c.\Big),\label{current-full}
\end{multline}
\begin{multline}
k^{\zeta r}_{s(0)}=-\frac{1}{P 8\pi
  G}\Big(\delta\big[\xbar\cY(\xbar\lambda^0\xbar\sigma^0 - \xbar
\Psi^0_2) - f\xbar\Psi^0_3 + \half \xbar\eth\sigma^0(\eth \cY -
\xbar\eth\xbar\cY)+\half \sigma^0\xbar\eth(\xbar\eth\xbar\cY -
\eth\cY)\\ - \xbar\lambda^0\xbar\eth f + r \cY (\xbar\lambda^0 +
\sigma^0(\gamma^0+\xbar\sigma^0))\big]-\xbar\cY(\xbar\lambda^0\delta\xbar\sigma^0
+ \lambda^0\delta\sigma^0) \Big),
\end{multline}
and $k^{\bar\zeta r}_{s(0)}$ given by the complex conjugate.
By construction
\begin{equation}
\p_u k^{ur}_{s(0)} + \p_\zeta k^{\zeta r}_{s(0)} +
\p_{\bar\zeta} k^{\bar\zeta r}_{s(0)}=-W^r_{s (0)}\label{eq:97},
\end{equation}
which may also be checked by direct computation. Note that
$k^{ur}_{s(0)},k^{\zeta r}_{s(0)},k^{\bar\zeta r}_{s(0)}$ also contain,
in addition to a finite contribution, linearly divergent terms when
$r\to\infty$.  Following \cite{Barnich:2013axa}, the latter can be
removed through an exact $2$-form $\p_\rho
\eta_s^{\mu\nu\rho}$. Defining
\begin{equation}
\bP\eta_s^{[ur\bar{\zeta}]}=\cN_s^u=-r\xbar\cY \xbar\sigma^0 - \half \cY
\sigma^0\xbar\sigma^0,\quad
\eta_s^{[\zeta r\bar{\zeta}]}=\cN^\zeta_s=0,
\end{equation}
and splitting into
an integrable part
\begin{equation}
\cJ^u_s=-\frac{1}{8\pi
  G}\big[f(\Psi^0_2+ \sigma^0\lambda^0) + \cY[\sigma^0\eth\xbar\sigma^0 +
\Psi^0_1 + \half \eth(\sigma^0\xbar\sigma^0)]+ \text{c.c.}\big],
\label{current comp u}
\end{equation}
\begin{multline}
  \cJ^\zeta_s=\frac{1}{8\pi G}\bigg[\xbar\cY \xbar
  \Psi^0_2 + f\xbar\Psi^0_3 +\half \xbar\cY(\lambda^0\sigma^0 -
  \xbar\lambda^0\xbar\sigma^0) + \half
  \xbar\eth\sigma^0(\xbar\eth\xbar\cY - \eth \cY)\\ - \half
  \sigma^0\xbar\eth(\xbar\eth\xbar\cY - \eth\cY) +
  \xbar\lambda^0\xbar\eth f\bigg],
\end{multline}
and a non-integrable one
\begin{equation}
\Theta^u_s(\delta\chi)=\frac{1}{8\pi G}(f\lambda^0\delta\sigma^0 + \text{c.c.}),\quad
\Theta^\zeta_s(\delta\chi)=\frac{1}{8\pi G}\xbar\cY(\lambda^0\delta\sigma^0 +
\xbar\lambda^0\delta\xbar\sigma^0),
\label{non integrable part}
\end{equation}
one finally arrives at
\begin{equation}
  \begin{split}
\delta \cJ^u_s&=P\bP[k^{ur}_{s(0)} - \bp \eta_s^{[ur\bar{\zeta}]} - \p
\eta_s^{[ur\zeta]}] - \Theta^u_s,\\
\delta \cJ_s^\zeta&=P [k^{\zeta r}_{s(0)} + \p_u \eta_s^{[ur\zeta]}
+\bp\eta_s^{[\bar{\zeta} r \zeta]}]- \Theta_s^\zeta,
\end{split}
\end{equation}
where $\cJ_s^{\bar{\zeta}},\Theta_s^{\bar{\zeta}}$ are the complex
conjugates of $\cJ_s^{\zeta},\Theta_s^{\zeta}$. The results of
\cite{Barnich:2013axa} are recovered when taking $P$ to be
$u$-independent, which implies $\gamma^0=\xbar\gamma^0=0$ and
$\lambda^0=\dot{\xbar\sigma^0}$. The associated forms are
given by
\begin{equation}
  \begin{split}
J_s &= (P\bar P)^{-1}\cJ^u_s \D\zeta \D\bar{\zeta} - P^{-1}\cJ^\zeta_s \D u \D\bar{\zeta} +
\bar P^{-1} \cJ^{\bar{\zeta}}_s \D u \D\zeta, \\
\theta_s &= (P\bar P)^{-1}\Theta^u_s \D \zeta \D \bar{\zeta} - P^{-1}\Theta_s^{\zeta} \D u
\D\bar{\zeta} + \bar P^{-1} \Theta_s^{\bar{\zeta}} \D u \D \zeta.
\end{split}\label{forms currents}
\end{equation}

\subsection{Current algebra}
\label{sec:algebra}

Using the relations of appendix \ref{Useful
  relations}, the first independent component of the current algebra can
be written as
\begin{equation}
\delta_{s_2} \cJ^u_{s_1} +
\Theta_{s_2}^u(\delta_{s_1}\chi)\approx \cJ^u_{[s_1,s_2]}+\cK^u_{s_1,s_2}+\eth\cL_{s_1,s_2}+\xbar\eth\xbar{\cL_{s_1,s_2}},
\label{algrebra u}
\end{equation}
where
\begin{equation}
\cK^u_{s_1,s_2}=\frac{1}{8\pi G}\Big[\Big(\frac12\xbar\sigma^0\left[f_1 \eth^2(\eth\cY_2 +
  \xbar\eth\xbar\cY_2)\right] - f_1\eth f_2 \xbar\eth\mu^0
-(1\leftrightarrow 2)\Big)+ \text{c.c.}\Big],
\end{equation}
and
\begin{multline}
\cL_{s_1,s_2}=\cY_2 \cJ^u_{s_1} - f_2 \cJ^{\bar{\zeta}}_{s_1}\\
-\frac{1}{8\pi G}\Big[\big(
\frac12 (\eth\cY_1 + \xbar\eth\xbar\cY_1) \eth f_2 -\frac12
\cY_1\eth^2f_2 - \xbar\cY_1\eth\xbar\eth f_2\big)\xbar\sigma^0 \\ -
\frac12 \cY_1 \xbar\eth^2f_2 \sigma^0 - \cY_1\eth f_2\eth\xbar\sigma^0
+ \xbar\cY_1 \xbar\eth f_2\eth \xbar\sigma^0 - f_1\eth f_2
\lambda^0\Big].
\end{multline}
The second independent component of the current algebra is
\begin{equation}
\delta_{s_2} \cJ^{\bar{\zeta}}_{s_1} +
\Theta_{s_2}^{\bar{\zeta}}(\delta_{s_1}\chi)
\approx \cJ^{\bar{\zeta}}_{[s_1,s_2]}+\cK^{\bar{\zeta}}_{s_1,s_2}
-\p_u\cL_{s_1,s_2}-2\gamma^0\cL_{s_1,s_2}+\xbar\eth\xbar{
\cM_{s_1,s_2}},\label{eq:101}
\end{equation}
where
\begin{multline}
\cK^{\bar{\zeta}}_{s_1,s_2}=-\frac{1}{8\pi G}\Big[f_2\eth f_1 \xbar\eth\nu^0 + \frac12 \eth
f_1\xbar\eth^3\xbar\cY_2 + \cY_1\xbar\eth f_2 \eth \mu^0 + f_1
\cY_2(\sigma^0\xbar\eth\nu^0 + \xbar\sigma^0\eth\xbar\nu^0) \\+
\frac12\cY_2\xbar\eth^2(\eth\cY_1 + \xbar\eth\xbar\cY_1)\sigma^0 +
\frac12\cY_2\eth^2(\eth\cY_1 + \xbar\eth\xbar\cY_1)\xbar\sigma^0 -
(1\leftrightarrow2)\Big],
\end{multline}
and
\begin{equation}
\xbar{\cM_{s_1,s_2}}=\xbar\cY_2 \cJ^{\bar{\zeta}}_{s_1}
-\frac{1}{8\pi G}\Big[\frac12\xbar\eth(\xbar\eth\xbar\cY_1 - \eth\cY_1 )\eth f_2 + \frac12
\eth \cY_1 \eth\xbar\eth f_2\Big] - \text{c.c.}.\label{eq:102}
\end{equation}

\subsection{Cocycle condition}
\label{sec:cocycle-condition}

The components of $\cK_{s_1,s_2}$
satisfy the 2-cocycle conditions
\begin{equation}
\cK^u_{[s_1,s_2],s_3}+\delta_{s_3} \cK^u_{s_1,s_2} +
\text{cyclic}(1,2,3)=\eth\cN_{s_1,s_2,s_3}+\xbar\eth\xbar\cN_{s_1,s_2,s_3},
\end{equation}
where
\begin{equation}
\cN_{s_1,s_2,s_3}=-f_3 \cK^{\bar{\zeta}}_{s_1,s_2}+ \text{cyclic}(1,2,3),
\end{equation}
and
\begin{equation}
\cK^{\bar{\zeta}}_{[s_1,s_2],s_3}+\delta_{s_3}
\cK^{\bar{\zeta}}_{s_1,s_2} + \text{cyclic}(1,2,3)=-\p_u\cN_{s_1,s_2,s_3} - 2
\gamma^0 \cN_{s_1,s_2,s_3} + \xbar\eth\xbar{\cO_{s_1,s_2,s_3}},
\end{equation}
where
\begin{multline}
\xbar{\cO_{s_1,s_2,s_3}}=-\frac{1}{8\pi G}\xbar\cY_3\Big[(f_1\cY_2-f_2\cY_1)\sigma^0\xbar\eth\nu^0 +
\frac12\sigma^0(\cY_2\xbar\eth^3\xbar\cY_1 -
\cY_1\xbar\eth^3\xbar\cY_2) \\+ \frac12(\eth f_1\xbar\eth^3\xbar\cY_2
- \eth f_2\xbar\eth^3\xbar\cY_1) + (f_2\eth f_1 - f_1\eth
f_2)\xbar\eth\nu^0\Big]-\text{c.c.} + \text{cyclic}(1,2,3).
\end{multline}
A situation where this 2-cocycle is relevant is discussed in
\cite{Barnich:2017ubf}. 

\subsection{Discussion}
\label{sec:discussion}

The results obtained in this section generalize the results discussed in one of the examples of subsection \ref{subsec:Asymptotic symmetries and surface charges} for an arbitrary time-dependent non-dynamical conformal factor $P= P(u,\zeta, \bar{\zeta})$ and in the Newman-Penrose formalism. In particular, equations \eqref{breaking 4d}, \eqref{current comp u}, \eqref{non integrable part} and \eqref{algrebra u} can be compared with \eqref{non conservation BMS}, \eqref{integrable and non integrable parts} and \eqref{algebra modified}. 

Let us now give some comments about the results. The BMS current algebra discussed in the previous section not only involves a consistent mathematical structure, but also contains some physical information on the system that would have been lost by considering only an integrable piece in the currents. To illustrate this claim, let us restrict ourselves to globally well-defined quantities on the
sphere, with $P=P_S=\frac{1}{\sqrt{2}}(1+\zeta\bar\zeta)$, there are
no superrotations and $\cK^u_{s_1,s_2}=0=\cK^{\bar\zeta}_{s_1,s_2}$. In
this case, BMS charges
are defined by integrating the forms \eqref{forms currents} at fixed retarded
time over the celestial sphere,
\begin{equation}
Q_s=\int_{u={\rm cte}} J_s=\int_{u={\rm cte}} (P_S\bar
P_S)^{-1}\cJ^u_s \D\zeta \D\bar\zeta \label{eq:86}
\end{equation} (see one of the example in subsection \ref{subsec:Asymptotic symmetries and surface charges}). If one also defines
\begin{equation}
  \label{eq:120}
  \Theta_{s}=\int_{u={\rm cte}} \theta_s=\int_{u={\rm cte}} (P_S\bar
P_S)^{-1}\Theta^u_s \D\zeta \D\bar\zeta,
\end{equation}
and the bracket
\begin{equation}
\{Q_{s_1},Q_{s_2}\}_*=\delta_{s_2}
Q_{s_1}+\Theta_{s_2}[\delta_{s_1}\chi], \label{eq:113}
\end{equation}
the integrated version
of equation \eqref{algrebra u},
becomes
\begin{equation}
  \label{eq:116}
  \{Q_{s_1},Q_{s_2}\}_*=Q_{[s_1,s_2]},
\end{equation}
This charge algebra contains for instance information on
non-conservation of BMS charges. Indeed, let us take $s_2=\d_u$, by which we mean that $T_R=\sqrt{P_S\bar P_S}, Y=0=\bar Y$, so that
  $f=1,\cY=0=\bar\cY$. In this case, equation \eqref{eq:116} together with the
  definition on the left-hand side in \eqref{eq:113}
  becomes
  \begin{equation}
    \label{eq:118}
    \delta_{\d_u} Q_s+\Theta_{\d_u}[\delta_s\chi]=Q_{[s,\d_u]}.
  \end{equation}
  When using that 
  \begin{equation}
    \label{eq:117}
    \frac{\D}{\D u} Q_s= \delta_{\d_u} Q_s+\dover{}{u}Q_s,
  \end{equation}
  and $\dover{}{u}Q_s =Q_{{\d s}/{\d u}}=-Q_{[s,\d_u]}$, it follows
  that
\begin{equation}
  \frac{\D}{\D u} Q_s= -\Theta_{\d_u}[\delta_{s}\chi].
\end{equation}
If one now chooses $s=\d_u$, one recovers the Bondi mass loss
formula.

More generally, equation \eqref{algrebra u} is the local
version of \eqref{eq:116}, where superrotations and arbitrary fixed
$P(u,\zeta,\bar\zeta)$ are allowed. When choosing $s_2=\d_u$ in that
equation, it encodes the non-conservation of BMS currents
(cf.~equation (4.22) in \cite{Barnich:2013axa}). For particular
choices of $s_1$, it controls the time evolution of the Bondi mass and
angular momentum aspects.

Even though we concentrated here on the case of standard Einstein
gravity, all the kinematics is in place to generalize the
constructions to gravitational theories with higher derivatives and/or
dynamical torsion.

For the most part of section \ref{Application to asymptotically flat 4d gravity}, the standard discussion has been extended
to include an arbitrary u-dependent conformal factor $P$. This
has been done so as to manifestly include the Robinson-Trautman
solution \cite{Robinson:1960zzb,Robinson:1962zz} in the solution
space. The application of the current set-up to these solutions
requires the inclusion of a dynamical conformal factor in the
derivation of the current algebra.

\chapter{Generalized BMS$_4$ and renormalized phase space}
\label{Generalized BMS and renormalized phase space}

In section \ref{Application to asymptotically flat 4d gravity}, we discussed the solution space of four-dimensional general relativity in asymptotically flat spacetime (AF2) (see equations \eqref{asymp flat 1} and \eqref{asymp flat 2} for these boundary conditions in metric formalism). Furthermore, we investigated the associated phase space, assuming the conformal factor to be non-dynamical. As stated in subsection \ref{Asymptotic symmetry algebra}, the asymptotic symmetry algebra is given by the extended BMS algebra written as $\mathfrak{bms}_4^{\text{ext}}$, namely the semi-direct sum between the superrotations $\mathfrak{diff}(S^1) \oplus \mathfrak{diff}(S^1)$ and the supertranslations $\mathfrak{s}^*$ \cite{Barnich:2009se,Barnich:2010eb,Barnich:2011mi}. 

In this chapter, we consider another set of boundary conditions  (AF1) (see equations \eqref{asymp flat 1} and \eqref{asymp flat 1 prime}) corresponding to a new definition of asymptotic flatness. We also study the associated asymptotic symmetry algebra and phase space. The former is given by a new extension of the global BMS algebra, called the generalized BMS algebra. This is given by $\mathfrak{bms}_4^{\text{gen}} = \mathfrak{diff}(S^2) \loplus \mathfrak{s}$ \cite{Campiglia:2014yka, Campiglia:2015yka , Compere:2018ylh , Flanagan:2019vbl , Campiglia:2020qvc }. This alternative extension of the global BMS algebra is motivated by two points: $(i)$ it is essential to establish the full equivalence between Ward identities for superrotations and subleading soft graviton theorem, and $(ii)$ $\mathfrak{bms}_4^{\text{gen}}$ is obtained in the flat limit of $\mathfrak{bms}^{\Lambda}_4$, the latter being a version of BMS in asymptotically locally (A)dS$_4$ spacetime. 

In section \ref{sec:Bondi}, we recall the set of boundary conditions that leads to the generalized BMS group and we discuss the associated solution space. In section \ref{sec:phasespace}, we compute the corresponding symplectic structure and notice the presence of divergences in $\sim r$. We renormalize these divergences by using the Iyer-Wald ambiguity and obtain a finite symplectic structure, from which we derive the charge algebra.

This chapter has strong intersections with \cite{Compere:2018ylh}. 

\section{Generalized BMS$_4$ group and solution space}
\label{sec:Bondi}

\subsection{Solution space}
\label{sec:BC1}

We recall that the Bondi gauge \eqref{Bondi gauge} leads to coordinates $(u,r,x^A)$ where $u$ labels null outgoing geodesic congruences, $r$ is a parameter along these geodesics, and $x^A$ are two coordinates on the $2$-sphere. The Bondi metric is parametrized as
\begin{equation}
\text{d}s^2 = \frac{V}{r} e^{2\beta} \text{d}u^2 -2e^{2\beta} \text{d}u \text{d}r + g_{AB} (\text{d}x^A - U^A \text{d}u)(\text{d}x^B - U^B \text{d}u), 
\label{gauge condition 1}
\end{equation} where $g_{AB}$ satisfies the determinant condition
\begin{equation}
\partial_r \left(\frac{\det(g_{AB})}{r^4}\right) = 0.
\label{gauge condition 2}
\end{equation}

We choose the definition (AF1) of asymptotic flatness (equations \eqref{asymp flat 1} and \eqref{asymp flat 1 prime}) and repeat it here:
\begin{equation}
\begin{split}
&\beta = o(1), \quad \frac{V}{r} = o(r^{2}), \quad U^A = o(1), \\ 
&\quad g_{AB} = r^2 q_{AB} + r C_{AB} + D_{AB} + \mathcal{O}(r^{-1}) , \quad \sqrt{q} = \sqrt{\bar{q}}
\label{repeated BC}
\end{split}
\end{equation}  Using these gauge and fall-off conditions, the Einstein equations entirely determine the solution space (see equations \eqref{parametrization solution space} for a parametrization of the solution space). Furthermore, we assume two additional constraints:
\begin{equation}
\mathcal{D}_{AB} = 0 , \qquad \sqrt{\bar{q}} = \sqrt{\mathring{q}}
\label{two additional cond}
\end{equation} where $\mathcal{D}_{AB}$ is the trace-free part of $D_{AB}$ and $\mathring{q}$ is the determinant of the unit $2$-sphere metric. The first condition in \eqref{two additional cond} guarantees that there is no logarithmic term in the expansion of $\beta$ and $U^A$. The second condition ensures that $l = \partial_u \ln \sqrt{\bar{q}} = 0$. The solution space is then explicitly given by 
\begin{equation}
\begin{split}
\frac{V}{r} &= - \frac{R}{2}+ \frac{2M}{r} + \mathcal{O}(r^{-2}) ,\\
\beta &= -\frac{1}{32 r^2}  C^{AB}C_{AB} + \mathcal{O}(r^{-3}), \\
g_{AB} &= r^2 q_{AB} + r C_{AB} +\frac{1}{4}q_{AB}C^{CD}C_{CD}+ \mathcal{O}(r^{-1}) ,\\
U^A &= - \frac{1}{2r^2} D_{B}C^{AB}  -\frac{2}{3} \frac{1}{r^3} \left[ N^A - \frac{1}{2} C^{AB} D^C C_{BC} \right] + \mathcal{O}(r^{-4}) 
\end{split}
\label{fall-off}
\end{equation} 
where all functions appearing in the expansions of $\frac{1}{r}$ depend upon $u$ and $x^A$. All 2-sphere indices in \eqref{fall-off} are raised and lowered with $q_{AB}$, and $D_A$ is the Levi-Civita connection associated with $q_{AB}$. The determinant condition \eqref{gauge condition 2} imposes in particular that $q_{AB} C^{AB}=0$. $C_{AB}$ is otherwise completely arbitrary, and its time derivative $N_{AB} = \partial_u C_{AB}$ is the Bondi news tensor which describes gravitational radiation. 

Let us recall that the Einstein equations impose the following time evolution equations:
\begin{align}
\partial_u q_{AB} &= 0, \label{qu} \\
\partial_u M &= - \frac{1}{8} N_{AB} N^{AB} + \frac{1}{4} D_A D_B N^{AB} + {\color{black}\frac{1}{8} D_A D^A {R}},\label{duM} \\ 
\partial_u N_A &= D_A M + \frac{1}{16} D_A (N_{BC} C^{BC}) - \frac{1}{4} N^{BC} D_A C_{BC} \nonumber \\
&\quad -\frac{1}{4} D_B (C^{BC} N_{AC} - N^{BC} C_{AC}) - \frac{1}{4} D_B D^B D^C C_{AC} \label{EOM1} \\
&\quad + \frac{1}{4} D_B D_A D_C C^{BC} + {\color{black} \frac{1}{4} C_{AB} D^B \mathring{R}}. \nonumber
\end{align}
Here $M(u,x^A)$ is the Bondi mass aspect, $N_A(u,x^B)$ is the angular momentum aspect. Concerning this quantity, our conventions are those of Barnich-Troessaert \cite{Barnich:2010eb,Barnich:2011mi} (also followed by \cite{Distler:2018rwu}), but differ from those of Flanagan-Nichols $(FN)$ \cite{Flanagan:2015pxa} and Hawking-Perry-Strominger $(HPS)$ \cite{Hawking:2016sgy}. Here is the dictionary to match the different conventions:
\begin{align}
N_A^{(FN)} &= N_A + \frac{1}{4} C_{AB} D_C C^{BC} + \frac{3}{32} \partial_A (C_{BC} C^{BC}) ,\label{chgtFN}\\
N_A^{(HPS)} &= N_A^{(FN)} - u D_A M.
\end{align}

\subsection{Asymptotic Killing vectors}

The asymptotic Killing vectors $\xi^\mu$ are obtained by imposing the preservation of the Bondi gauge (equations \eqref{gauge condition 1} and \eqref{gauge condition 2}) and the asymptotic flatness conditions (equations \eqref{repeated BC} and \eqref{two additional cond}). They are explicitly given by 
\begin{equation}
\begin{split}
\xi^u &= f(u,x^A) ,\\
\xi^A &= Y^A(u,x^A) + I^A, \quad I^A = - D_B f \int_r^\infty \text{d}r' (e^{2\beta} g^{AB}), \\
\xi^r &= -\frac{1}{2} r (D_A Y^A + D_A I^A - U^ B D_B f ),
\end{split} \label{eq:BMSvectors}
\end{equation} with $\partial_r f = \partial_r Y^A = 0$. Furthermore, the parameters satisfy
\begin{equation}
\begin{split}
&\partial_u Y^A = 0 \Longleftrightarrow Y^A = Y^A (x^B),\\
&\partial_u f = \frac{1}{2}D_A Y^A \Longleftrightarrow f = T(x^B) + \frac{u}{2}D_A Y^A .
\end{split}
\label{fAndY}
\end{equation} We can perform the radial integration in \eqref{eq:BMSvectors} to get a perturbative expression of the infinitesimal residual diffeomorphisms using the explicit solution space \eqref{fall-off}:
\begin{align}
\xi^u &= f , \label{eq:XiR1}\\
\xi^A &= Y^A - \frac{1}{r} D^A f + \frac{1}{r^2}  \left( \frac{1}{2} C^{AB} D_B f \right) + \frac{1}{r^3} \left( -\frac{1}{16} C_{BC}C^{BC} D^A f \right)  + \mathcal{O}(r^{-4}) ,\label{eq:XiR2}\\
\xi^r &= -\frac{1}{2} r D_A Y^A + \frac{1}{2} D_A D^A f + \frac{1}{r} \left( -\frac{1}{2} D_A C^{AB} D_B f - \frac{1}{4} C^{AB} D_A D_B f \right) + \mathcal{O}(r^{-2}). \label{eq:XiR}
\end{align}
The asymptotic Killing vectors are spanned by $\mathfrak{diff}(S^2)$ \textit{super-Lorentz} transformations generated by $Y^A (x^B)$ and by (smooth) \textit{supertranslations} generated by $T(x^A)$. We therefore denote them as $\xi(T,Y)$. Notice that \textit{in chapters} \ref{Generalized BMS and renormalized phase space} \textit{and} \ref{ch:Vacuum structure, superboost transitions and refraction memory} (and only in these chapters), we find it convenient to call the extension of the Lorentz transformations as the super-Lorentz transformations instead of superrotations. Any 2-vector on the sphere can be decomposed into a divergence-free part and a rotational-free part. A super-Lorentz transformation whose pullback on the celestial sphere is divergence-free is a superrotation. This generalizes the rotations. A super-Lorentz transformation whose pullback on the celestial sphere is rotational-free
is a superboost. This generalizes the boosts.

\subsection{Asymptotic symmetry algebra}

As discussed in subsection \ref{Asymptotic symmetry algebra}, to obtain the asymptotic symmetry algebra, one has to consider the modified Lie bracket 
\begin{equation}
[\xi_1,\xi_2]_A = [\xi_1,\xi_2] - \left( \delta^g_{\xi_1} \xi_2 - \delta^g_{\xi_2} \xi_1 \right),
\end{equation}
where $\delta^g_{\xi_1} \xi_2$ denotes the variation of $\xi_2$ caused by the Lie dragging along $\xi_1$ of the metric contained in the definition of $\xi_2$. We find
\begin{equation}
[\xi (T_1, Y^A_1),\xi (T_2, Y^A_2)]_A = \xi (\hat{T}, \hat{Y} ),
\end{equation} where 
\begin{equation}
\begin{split}
\hat{T} &= Y_1^A D_A T_2 + \frac{1}{2} T_1 D_A Y^A_2 - (1\leftrightarrow 2) , \\
\hat{Y}^A &= Y^B_1 D_B Y_2^A - (1\leftrightarrow 2).
\end{split} \label{struct const gen}
\end{equation}
This defines the \textit{generalized BMS algebra} $\mathfrak{bms}_4^{\text{gen}}$. It consists of the semi-direct sum of the diffeomorphism algebra on the celestial $2$-sphere $\mathfrak{diff}(S^2)$ and the abelian ideal $\mathfrak{s}$ of supertranslations, consisting of arbitrary smooth densities (of weight $-1/2$) on the $2$-sphere:
\begin{equation}
\mathfrak{bms}_4^{\text{gen}} = \mathfrak{diff}(S^2) \loplus \mathfrak{s}.
\end{equation}

\subsection{Action on the solution space}

The vectors \eqref{eq:BMSvectors} preserve the solution space in the sense that infinitesimally
\begin{equation}
\mathcal{L}_{\xi(T,Y)} g_{\mu\nu} [\phi^i] = g_{\mu\nu} [ \phi^i + \delta_{(T,Y)} \phi^i ] - g_{\mu\nu} [ \phi^i ] 
\end{equation}
where $\phi^i = \{ q_{AB}, C_{AB}, M, N_A \}$ denotes the collection of relevant fields that describe the metric in Bondi gauge. The action of the vectors preserve the form of the metric but modify the fields $\phi^i$, in such a way that the above equation is verified. We can show that
\begin{align}
\delta_{(T,Y)} q_{AB} &= 2 D_{(A} Y_{B)} - D_C Y^C q_{AB},\label{dqAB} \\
\delta_{(T,Y)} C_{AB} &= [f \partial_u + \mathcal{L}_Y - \frac{1}{2} D_C Y^C ] C_{AB} - 2 D_A D_B f + q_{AB} D_C D^C f,\label{dCAB}\\
\delta_{(T,Y)} N_{AB} &= [f\partial_u + \mathcal{L}_Y] N_{AB} - (D_A D_B D_C Y^C - \frac{1}{2} q_{AB} D_C D^C D_D Y^D),\label{dNAB}\\
\delta_{(T,Y)} M &= [f \partial_u + \mathcal{L}_Y + \frac{3}{2} D_C Y^C] M + \frac{1}{4} D_A f D^A R \\
& + \frac{1}{8} D_C D_B D_A Y^A C^{BC} + \frac{1}{4} N^{AB} D_A D_B f + \frac{1}{2} D_A f D_B N^{AB},\\
\delta_{(T,Y)} N_A &= [f\partial_u + \mathcal{L}_Y + D_C Y^C] N_A + 3 M D_A f - \frac{3}{16} D_A f N_{BC} C^{BC}  \nonumber \\
&\quad - \frac{1}{32} D_A D_B Y^B C_{CD}C^{CD} + \frac{1}{4} (D^B f \mathring{R} + D^B D_C D^C f) C_{AB} \nonumber \\
&\quad - \frac{3}{4} D_B f (D^B D^C C_{AC} - D_A D_C C^{BC}) + \frac{3}{8} D_A (D_C D_B f C^{BC}) \nonumber \\
&\quad + \frac{1}{2} (D_A D_B f - \frac{1}{2} D_C D^C f q_{AB}) D_C C^{BC} + \frac{1}{2} D_B f N^{BC} C_{AC}.   \label{dNA}
\end{align}
Note that the boundary Ricci scalar ${R}$ transforms as
\begin{equation}
\delta_{(T,Y)} {R} = Y^A D_A {R} + D_A Y^A{R} + D^2  D_B Y^B.
\end{equation}

\section{Renormalized phase space}
\label{sec:phasespace}


In this section, we define an extended phase space invariant under the action of Diff$(S^2)$ super-Lorentz transformations and supertranslations. Super-Lorentz transformations are overleading in the sense that they change the boundary metric, which is usually fixed in standard asymptotically flat spacetimes. We can therefore expect that a renormalization procedure will be required. In this context, we use the Iyer-Wald procedure described in subsection \ref{Relation between Barnich-Brandt and Iyer-Wald procedures} that allows us to remove the divergences by utilizing the ambiguity in the definition of the presymplectic potential.

\subsection{Presymplectic potential} 

In the metric formalism, the canonical presymplectic potential is given in equation \eqref{canonical presymplectic potential}. Plugging the solution space \eqref{fall-off} into this expression, we obtain 
\begin{eqnarray}
\theta^u [g;\delta g]&=& r \,\theta^u_{(div)} + \theta^u_{(0)} + r^{-1} \theta^u_{(1)} + \Order{2}\label{th1}, \\
\theta^r[g;\delta g] &=& r \,\theta^r_{(div)} + \theta^r_{(0)} + \Order{1}\label{th2}.
\end{eqnarray}
We have $\theta^u_{(div)} \propto \delta \sqrt{q}$ and therefore  $\theta^u_{(div)} =0$ as a result of the boundary condition \eqref{repeated BC}. Furthermore, we find $\theta^u_{(1)}  =0$. The other components are
\begin{eqnarray}
\theta^u_{(0)} &=& \pref \frac{1}{2} C_{AB} \delta q^{AB}  ,\\
\theta^r_{(div)} &=& -\pref \delta R - \frac{1}{2} \pref  N_{AB} \delta q^{AB} , \\
\theta^r_{(0)} &=& -\pref \, \delta \left[ \frac{1}{8}N^{AB}C_{AB} - 2 M + \frac{1}{2} D_A D_B C^{AB} \right] \nonumber\\ 
&&+ \bar \theta_{flux} + \pref D_A \left[\frac{1}{2} D^C C_{BC}  \delta q^{AB} \right], 
\end{eqnarray}
where we define with hindsight the important quantity
\bea
\bar \theta_{flux} &\equiv&  \pref \Big[ \frac{1}{2}  N_{AB}\delta C^{AB} - \frac{1}{4}  R C_{AB} \delta q^{AB} - \frac{1}{2}  D^C C_{BC}  D_A \delta q^{AB} \Big].\label{thetaflux}
\eea

We note that one can isolate a total derivative and a total variation as
\bea
\theta^u_{(0)} &=& -\p_r Y^{ur}, \label{d1}\\
r \theta^r_{(div)}&=& -\p_u Y^{ru} - \delta (\sqrt{q} {R}\; r) = -\p_u Y^{ru} -\p_A Y^{rA}\label{d2}
\eea
where $Y^{ur}=-Y^{ru}= -r \frac{1}{2} \pref C_{AB} \delta q^{AB}$ and $Y^{rA} = r \, \prefwg \, \theta^A_{2d}( q ; \delta q)$ is $r$ times the presymplectic potential of the two-dimensional Einstein-Hilbert action, $\p_A \theta^A_{2d} =  \delta (\sqrt{q} {R})$. 

\subsection{Presymplectic form}

The bare Lee-Wald presymplectic form is given by ${\boldsymbol{\omega}}[g;\delta g, \delta g] = \delta \boldsymbol{\theta} [g;\delta g]$ (see equation \eqref{presymplectic form}). We have already obtained that the bare presymplectic potential $\boldsymbol{\theta} [g;\delta g]$ is divergent (see \eqref{th1} and \eqref{th2}). However, it is ambiguous under the change $\boldsymbol{\theta} [g;\delta g] \to \boldsymbol{\theta} [g;\delta g] - \D \mathbf{Y}[g;\delta g]$, where $\mathbf{Y}[g;\delta g]$ is a co-dimension 2 form (see \eqref{ambiguity presymplectic potential IW} and the associated discussion). We have already identified in \eqref{d1} and \eqref{d2} the counterterms required to make the presymplectic potential finite. 

Let us discuss this point in detail, starting from the bare presymplectic form. We have 
\begin{align}
\omega^u &= \pref \left( \frac{1}{2} \delta q_{AB} \wedge \delta C^{AB} \right) + \mathcal{O}(r^{-2}) ,\\
\omega^r &=- r \pref \left( \frac{1}{2} \delta N_{AB} \wedge \delta q^{AB} \right) \\
&\phantom{=} + \pref \left[ \frac{1}{2} \delta \left( N^{AB} + \frac{1}{2} {R} q^{AB} \right)\wedge \delta C_{AB} + \frac{1}{2} \delta (D_A D^C C_{BC} ) \wedge \delta q^{AB} \right] + \mathcal{O}(r^{-1}) . \nonumber
\end{align}
Clearly, such a presymplectic form is divergent. After choosing the boundary term $\mathbf{Y}[g;\delta g]$ as in \eqref{d1} and \eqref{d2}, the presymplectic form becomes well-defined, 
\begin{align}
\omega^u_{\text{ren}} &= \mathcal{O}(r^{-2}) ,\\
\omega^r_{\text{ren}} &= \pref \left[ \frac{1}{2} \delta \left( N^{AB} + \frac{1}{2} {R} q^{AB} \right)\wedge \delta C_{AB} + \frac{1}{2} \delta (D_A D^C C_{BC}) \wedge \delta q^{AB} \right] + \mathcal{O}(r^{-1}) . \label{omegegar}
\end{align}
Since $Y^{Ar}$ is exact, it does not contribute here. This defines the presymplectic structure at $\mathscr{I}^+$
\begin{align}
&\Omega_{\text{ren}}[g;\delta g, \delta g] \nonumber \\
&\quad=  \frac{1}{16 \pi G}\int_{\mathscr{I}^+} \text{d}u \,  \text{d}^2 \Omega  \,\left[ \frac{1}{2} \delta \left( N^{AB} + \frac{1}{2}{R} q^{AB} \right)\wedge \delta C_{AB} + \frac{1}{2} \delta (D_A D_C C^C_{B}) \wedge \delta q^{AB} \right] \nonumber\\
&\quad= \int_{\mathscr{I}^+} \text{d}u \, \text{d}^2 x \left(  \delta \bar \theta_{flux}[g;\delta g]  \right) , \label{Omega}
\end{align}
where $\bar \theta_{flux}$ is defined in \eqref{thetaflux}, after discarding a boundary term. 


\subsection{Infinitesimal surface charges} 

The $(ur)$-component of the bare Iyer-Wald co-dimension $2$ form is
\begin{align}
k^{ur}_\xi [g;\delta g] = -\delta Q^{ur}_\xi [g] + Q^{ur}_{\delta\xi} [g]  + \xi^u \theta^r [g;\delta g]- \xi^r \theta^u [g;\delta g]\label{decom:charge}
\end{align} (we note the presence of the term $Q^{ur}_{\delta\xi} [g]$ compared to the expression \eqref{Iyer-Wald}, as already discussed in the footnote \ref{useful footnote} of subsection \ref{Relation between Barnich-Brandt and Iyer-Wald procedures}).
Expanding in powers of $1/r$, we get
\begin{equation}
k^{ur}_\xi [g;\delta g] = r k^{ur}_{(div)} + k^{ur}_{(0)} + \mathcal{O}(r^{-1}).
\end{equation}
We define $\slashed{\delta}\bar H_\xi = \int_{S^2_\infty} \mathbf{k}_\xi [g;\delta g]$. The divergent term is
\begin{equation}
\slashed{\delta}\bar H_\xi^{(div)} = \prefwg \int_{S^2_\infty} \text{d}^2 \Omega \left[  \delta(Y^A D_B C_{A}^B) - f \delta {R} - \frac{1}{2} f N_{AB} \delta q^{AB} + \frac{1}{4} D_C Y^C q_{AB} \delta C^{AB} \right]\label{divch}
\end{equation}
while the finite term is
\begin{align}
\slashed{\delta}\bar H_\xi^{(0)} &= \prefwg \int_{S^2_\infty} \text{d}^2 \Omega \left[  \delta  \left( 4 f M + 2 Y^A N_A +\frac{1}{16} Y^A D_A (C_{BC}C^{BC}) \right) \right] \label{chi1}\\
&\phantom{=} + \prefwg \int_{S^2_\infty} \text{d}^2 \Omega \left[  \frac{1}{2} f \left(  N^{AB} + \frac{1}{2}  q^{AB}  {R} \right) \delta C_{AB} \right. \nonumber \\
&\phantom{=}\qquad \left. +  D_A f D^C C_{BC} \delta q^{AB} + \frac{1}{2} f   D_A D^C C_{BC} \delta q^{AB} - \frac{1}{4} D^2 f q_{AB} \delta C^{AB}  \right] .\nonumber
\end{align}
Clearly, the charges are neither finite nor integrable. 

For covariant counter-term $\mathbf{Y}[g;\delta g]$, the modification at the level of the co-dimension 2 form is
\begin{equation}
k^{ur}_\xi \to k^{ur}_\xi - \delta Y^{ur}[g ; \mathcal{L}_\xi g ] + Y^{ur}[g ; \mathcal{L}_{\delta\xi} g] + \xi^u \Delta \theta^r - \xi^r \Delta \theta^u .\label{shiftk}
\end{equation}
More specifically, for the $\mathbf{Y}[g; \delta g]$ given in \eqref{d1} and \eqref{d2}, we have
\begin{align}
- \delta Y^{ur}[g ; \mathcal{L}_\xi g ] &= \frac{1}{2} r \frac{\sqrt{q}}{16\pi G}  \delta \left( C_{AB} \delta_{(T,Y)} q^{AB} \right) \label{1} ;\\
Y^{ur}[g ; \mathcal{L}_{\delta\xi} g] &=0 \label{12},
\end{align}
where the last equation follows from the fact that the fields are not modified by $\delta \xi$ at leading order in $r$. Moreover, 
\begin{align}
\xi^u \Delta \theta^r &= f (\partial_u Y^{ru} + \partial_A Y^{rA}) = \frac{1}{2} r \frac{\sqrt{q}}{16\pi G} f N_{AB} \delta q^{AB} + \frac{\sqrt{q}}{16\pi G} r \left( f \delta {R}  \right) ;\label{2} \\
-\xi^r \Delta \theta^u &= -\xi^r(\partial_r Y^{ur} ) = -\frac{1}{4} \frac{\sqrt{q}}{16\pi G} r D_C Y^C C_{AB} \delta q^{AB} + \frac{1}{4} \frac{\sqrt{q}}{16\pi G} D_C D^C f C_{AB} \delta q^{AB} +\mathcal{O}(r^{-1}). \label{3}
\end{align}
Here, the $\mathcal{O}(1)$ part in (\ref{3}) is due to the $\mathcal{O}(1)$ contribution from $\xi^r$, and exactly cancels the last non-integrable term at $\mathcal{O}(1)$ in the charges \eqref{chi1}. 

We see that any divergent term will disappear due to this choice of $Y^{ur}$, and the infinitesimal surface charges reduce to
\begin{align}
\slashed{\delta}H^\text{intermediate}_\xi &= \prefwg \int_{S^2_\infty} \text{d}^2 \Omega \left[  \delta  \left( 4 f M + 2 Y^A N_A + \frac{1}{16} Y^A D_A (C_{BC}C^{BC}) \right) \right] \nonumber \\
&\phantom{=} + \prefwg \int_{S^2_\infty} \text{d}^2 \Omega \left[ \frac{1}{2} f \left( N^{AB} + \frac{1}{2} q^{AB} {R} \right) \delta C_{AB} + \frac{1}{2} D_A (f D_C C^C_{B}) \delta q^{AB}\right] .
\end{align} Before proceeding, let us notice a very subtle point. Using the Iyer-Wald ambiguity, we saw that $\mathbf{Y}[g; \delta g]$ allowed us to cancel the divergent part of the presymplectic form, but did not affect the finite part. However, when reporting this Iyer-Wald modification at the level of the co-dimension 2 form in \eqref{shiftk}, we showed that it affected the finite part of the charges. This is in tension with the fact that the co-dimension 2 form is completely determined by the presymplectic potential through the fundamental relation \eqref{breaking in the conservation 2}. This apparent discrepancy in the formalism is due to an inappropriate use of the algorithm that brings the renormalization at the level of the charge. More precisely, the formula \eqref{shiftk} is valid only for $\mathbf{Y}[g; \delta g]$ covariant with respect to the bulk and thus does not hold in our current context. We therefore used an alternative way to obtain the surface charges and checked that the finite part \eqref{chi1} of the bare charge satisfies the fundamental relation
\begin{equation}
\partial_u \ndelta {\bar{H}}^{(0)}_\xi  = -\Omega_{\text{ren}} [g ; \delta_\xi g,\delta g], 
\end{equation} where $\Omega_{\text{ren}} [g ; \delta_\xi g, g]$ is the renormalized symplectic form given in \eqref{Omega}. The final renormalized infinitesimal charge is finite and given explicitly by 
\begin{equation}
\begin{split}
\slashed{\delta} H_\xi &= \prefwg \int_{S^2_\infty} \text{d}^2 \Omega \left[  \delta  \left( 4 f M + 2 Y^A N_A +\frac{1}{16} Y^A D_A (C_{BC}C^{BC}) \right) \right] \\
&\phantom{=} + \prefwg \int_{S^2_\infty} \text{d}^2 \Omega \left[  \frac{1}{2} f \left(  N^{AB} + \frac{1}{2}  q^{AB}  {R} \right) \delta C_{AB} \right.  \\
&\phantom{=}\qquad \left. +  D_A f D^C C_{BC} \delta q^{AB} + \frac{1}{2} f   D_A D^C C_{BC} \delta q^{AB} - \frac{1}{4} D^2 f q_{AB} \delta C^{AB}  \right] .
\end{split}
\label{infinitesimal charges}
\end{equation} 

When $q_{AB}$ is the fixed unit metric on the sphere, it reproduces the expression of Barnich-Troessaert \cite{Barnich:2011mi} (see equations \eqref{BMS charges} and \eqref{integrable and non integrable parts}).

The infinitesimal surface charges can be written as 
\begin{equation}
\slashed{\delta}H_\xi [g] = \delta H_\xi [g] + \Xi_\xi [g ;\delta g] ,
\label{decomposition charge}
\end{equation} where
\bea
H_\xi [g] = \prefwg \int_{S^2_\infty} \text{d}^2 \Omega  \left( 4 f M + 2 Y^A N_A + \frac{1}{16} Y^A D_A (C_{BC}C^{BC}) \right) \label{FNCharges}
\eea
and 
\begin{equation}
\begin{split}
\Xi_\xi [g; \delta g] &= \prefwg \int_{S^2_\infty} \text{d}^2 \Omega \left[  \frac{1}{2} f \left(  N^{AB} + \frac{1}{2}  q^{AB}  {R} \right) \delta C_{AB} +  D_A f D^C C_{BC} \delta q^{AB}  \right.  \\
&\phantom{=}\qquad\qquad \left. + \frac{1}{2} f   D_A D^C C_{BC} \delta q^{AB} - \frac{1}{4} D^2 f q_{AB} \delta C^{AB}  \right] .
\end{split}
\end{equation} Of course, the canonical Hamiltonian \eqref{FNCharges} cannot be deduced solely from the relation \eqref{decomposition charge} since one can shift $H_\xi$ as
\begin{equation}
\slashed{\delta}H_\xi [g] = \delta (H_\xi [g] +\Delta H_\xi[g])  + \Xi_\xi [g;  \delta g] - \delta \Delta H_\xi[g].\label{shifts}
\end{equation} 
We therefore need additional input to fix the finite Hamiltonian. This is discussed in subsection \ref{subsec: Finte charges}.

\subsection{Charge algebra}

After an involved computation, we get the following charge algebra
\begin{equation}
\delta_{\xi_2} H_{\xi_1} [g] + \Xi_{\xi_2} [g,\delta_{\xi_1} g] = H_{[\xi_1,\xi_2]_A}[g]   + \mathcal{K}_{\xi_1,\xi_2} [g]. \label{eq:Algebra} 
\end{equation} In this relation,
\begin{equation}
\begin{split}
\Xi_{\xi_2} [g; \delta_{\xi_1} g] &= \prefwg \int_{S^2_\infty} \text{d}^2 \Omega \left[  \frac{1}{2} f_2 \left(  N^{AB} + \frac{1}{2}  q^{AB}  {R} \right) \delta_{\xi_1} C_{AB} +  D_A f_2 D^C C_{BC} \delta_{\xi_1} q^{AB} \right.  \\
&\phantom{=}\qquad \qquad \left.  + \frac{1}{2} f_2   D_A D^C C_{BC} \delta_{\xi_1} q^{AB} - \frac{1}{4} D^2 f_2 q_{AB} \delta_{\xi_1} C^{AB}  \right] .
\end{split} 
\label{anomalous}
\end{equation} Furthermore, the 2-cocycle $\mathcal{K}_{\xi_1,\xi_2} [g]$ is antisymmetric and satisfies 
\begin{equation}
\mathcal K_{[\xi_1, \xi_2], \xi_3} + \delta_{\xi_3} \mathcal K_{\xi_1, \xi_2} + \text{cyclic}(1,2,3) =0 .
\label{cocycle condition}
\end{equation} It is given explicitly by
\begin{equation}
\begin{split}
\mathcal{K}_{\xi_1,\xi_2} [g] =& \prefwg \int_{S^2_\infty} \text{d}^2\Omega \, \left[ \frac{1}{2} f_1 D_A f_2 D^A \mathring{R} + \frac{1}{2} C^{BC} f_1 D_B D_C D_D Y^D_2  - (1\leftrightarrow 2) \right] .
\end{split}
\end{equation} 

The result \eqref{eq:Algebra} is the analogue of \eqref{algebra modified} (or \eqref{algrebra u} in NP formalism), but for $\mathfrak{diff}(S^2)$ superrotations. 

\chapter{Vacuum structure, superboost transitions and refraction memory}
\label{ch:Vacuum structure, superboost transitions and refraction memory}
\chaptermark{Vacuum structure and refraction memory}

In section \ref{sec:vac}, we study the vacuum structure of the gravitational field in asymptotically flat spacetimes for both extensions of the BMS group. This analysis shows that one field in the metric that is turned on after acting with superboost transformations exhibits the properties of a Liouville field. Furthermore, in section \ref{sec:transitions}, we argue that this field is precisely the memory field associated with the velocity kick/refraction memory effects. In section \ref{subsec: Finte charges}, using some definitions introduced in the previous sections, we propose a prescription to obtain meaningful finite charges out of \eqref{infinitesimal charges}. Applying this procedure, we find precisely the charges needed to establish the equivalence between Ward identities and soft theorems.

\section{Vacuum structure}
\label{sec:vac}

The orbit of Minkowski spacetime under the BMS group is defined as the class of Riemann-flat metrics obtained by exponentiating a general BMS transformation starting from Minkowski spacetime as a seed. The subset of this orbit where only supertranslations act are the non-equivalent vacua of asymptotically flat spacetimes, which are characterized, contrary to Minkowski spacetime, by non-vanishing super-Lorentz charges, while all Poincar\'e charges remain zero \cite{Compere:2016jwb}. In the global BMS case, the exponentiation leads to a single fundamental field labeling inequivalent vacua: the supertranslation field $C(x^A)$. The displacement memory effect is a transition among vacua mediated by gravitational or other null radiation, which effectively induces a supertranslation of $C$ \cite{Strominger:2014pwa}.

For the extended BMS asymptotic symmetry group, this exponentiation leads to two fundamental fields: the supertranslation field and what we will call the superboost or Liouville field $\Phi$. The corresponding solution in Bondi and Newman-Unti gauges was constructed in \cite{Compere:2016jwb}. Here, we extend the construction to finite \text{Diff}$(S^2)$ super-Lorentz transformations following methods similar to the appendix of \cite{Compere:2016hzt}. The corresponding boundary fields will also be the supertranslation $C$ and superboost $\Phi$ fields, complemented by an additional superrotation field $\Psi$. To understand the memory effects associated with super-Lorentz transformations, we start by deriving the structure of the vacua.

\subsection{Generation of the vacua}
\label{Generation of the vacua}

In this subsection, we construct the finite diffeomorphisms associated with the asymptotic Killing vectors (\ref{eq:XiR1}-\ref{eq:XiR}) acting on the Minkowski space. We start from the Minkowski metric written in complex plane coordinates\footnote{The metric \eqref{complex plane coordinates} can be related to the standard Minkowski metric $ds^2 = - du^2_s -2du_s dr_s + \frac{4 r_s^2}{(1+z_s \bar{z}_s)^2} dz_s d\bar{z}_s$
by performing the following coordinate transformation (see appendix A of \cite{Compere:2016hzt}): $
r_c = \frac{\sqrt{2}r_s}{1+ z_s \bar{z}_s} + \frac{u_s}{\sqrt{2}}$,
$u_c = \frac{1+ z_s \bar{z}_s}{\sqrt{2}} u_s - \frac{z_s\bar{z}_s u^2_s}{2r_c}$,
$z_c = z_s - \frac{z_s u_s}{\sqrt{2}r_c}$, $\bar{z}_c = \bar{z}_s - \frac{\bar{z}_s u_s}{\sqrt{2}r_c}$.}
\begin{equation}
ds^2 = -2\text{d}u_c \text{d}r_c + 2 r_c^2 dz_c d\bar{z}_c. 
\label{complex plane coordinates}
\end{equation}
We define the background structures
\begin{equation}
\gamma_{ab} = \left[ \begin{array}{cc}
0 & 1 \\ 
1 & 0
\end{array} \right], \quad \epsilon_{ab} = \left[ \begin{array}{cc}
0 & 1 \\ 
-1 & 0
\end{array} \right]
\end{equation}
with inverse $\gamma^{ab} = \gamma_{ab}$, $\epsilon^{ab} = \epsilon_{ab}$. The goal is to introduce a diffeomorphism to Bondi gauge $(u_c,r_c,z_c,\bar{z}_c)\to (u,r,z,\bar{z})$ that exponentiates $\text{Diff}(S^2)$ super-Lorentz transformations and supertranslations. Requiring that $(u,r,z,\bar z)$ are Bondi coordinates leads to two sets of conditions: $(i)$ the algebraic conditions $g_{rr} = 0 = g_{rA}$, and $(ii)$ the determinant condition $\partial_r (r^{-4} \det g_{AB}) = 0$.

The first set of conditions yields 
\begin{align}
r_c &= r_c (r,u,z,\bar{z}), \\
u_c &= W(u,z,\bar{z}) - r_c^{-1} \gamma_{ab} H^a (u,z,\bar{z}) H^b (u,z,\bar{z}), \\
z_c^a &= G^a (z,\bar{z}) - r_c^{-1} H^a (u,z,\bar{z}), \quad
H^a (u,z,\bar{z}) = - D_G^{-1} \epsilon^{ab} \gamma_{bc}\epsilon^{AB} \partial_A W \partial_B G^c
\end{align}
where $D_G = \det (\partial_A G^b) = \frac{1}{2!}\epsilon_{ab}\epsilon^{AB}\partial_A G^a \partial_B G^b$. The second condition fixes the functional dependence of $r_c$ as
\begin{align}
r_c (r,u,z,\bar{z}) &= R_0 (u,z,\bar{z}) + \sqrt{\frac{r^2}{(\partial_u W)^2} + R_1 (u,z,\bar{z})}.\label{rc}
\end{align}

For $l = \partial_u \ln \sqrt{\bar{q}} = 0$ to be satisfied (see \eqref{two additional cond}), we have to impose that $\partial^2_u W = 0$, so $W$ is at most linear in $u$. Moreover, regularity implies that $\partial_u W$ is nowhere vanishing. Therefore,
\begin{equation}
W(u,z^c) =  \exp \left[\frac{1}{2}\Phi (z,\bar z)\right] (u + C(z,\bar z)).
\label{Wform}
\end{equation}
Expanding $g_{AB}$ in powers of $r$ as in \eqref{fall-off}, we can read the boundary metric as 
\begin{equation}
q_{AB} =q_{AB}^{\text{vac}} \equiv e^{-\Phi} \partial_A G^a \partial_B G^b \gamma_{ab}.
\label{qABform}
\end{equation} It is indeed the result of a large diffeomorphism and a Weyl transformation. If one is restricted to the transformations that lead to $\sqrt{q} = \sqrt{\mathring{q}}$ (see equation \eqref{two additional cond}), we have the relation
\begin{equation}
|\det (\partial_A G^a )| = \frac{2}{(1+ z \bar{z})^2} e^{\Phi} .
\end{equation}
The shear $C_{AB}$ is found to be the trace-free (TF) part of the following tensor
\begin{equation}
C_{AB} = C_{AB}^{\text{vac}} \equiv \left[ \frac{2}{(\partial_u W)^2} \partial_u \left( D_A W D_B W \right) - \frac{2}{\partial_u W} D_A D_B W \right]^{\text{TF}}.
\label{CABGeneral}
\end{equation}
Introducing \eqref{Wform}, it becomes 
\begin{equation}
C^{\text{vac}}_{AB}[\Phi,C] =  (u +C) N^{\text{vac}}_{AB} + C^{(0)}_{AB}, \quad \left\lbrace \begin{array}{ccl}
N^{\text{vac}}_{AB} & = & \left[\frac{1}{2}D_A \Phi D_B \Phi - D_A D_B \Phi  \right]^{\text{TF}}, \label{PhiL} \\ 
C^{(0)}_{AB} & = &   - 2 D_A D_B C + q_{AB} D^2 C.
\end{array} \right.
\end{equation}
We find that all explicit reference on $\gamma_{ab}$ or $G^a$ disappeared. Moreover, the news tensor of the vacua $N^{\text{vac}}_{AB}$ is only built up with $\Phi$. It can be checked that the boundary Ricci scalar is given in terms of $\Phi$ as 
\begin{equation}
 R = D^2 \Phi,
\label{Liouville}
\end{equation}
which implies 
\bea
D_A N_{\text{vac}}^{AB}=-\frac{1}{2}D^B  R. \label{Li2}
\eea
We can therefore add a trace to $N^{\text{vac}}_{AB}$ to form the conserved stress-tensor
\begin{equation}
T_{AB}[\Phi]= \frac{1}{2}D_A \Phi D_B \Phi - D_A D_B \Phi + \frac{1}{2}q_{AB} \left( 2 D^2 \Phi  -\frac{1}{2} D^C \Phi D_C \Phi \right).
\end{equation}
Its trace is equal to $D^2 \Phi$. The tensor $T_{AB}$ is precisely the stress-tensor of Euclidean Liouville theory 
\bea
L[\Phi ; q_{AB}] = \sqrt{q}\left( \frac{1}{2} D^A \Phi D_A \Phi +\Lambda e^\Phi + R[q] \Phi  \right),
\eea
where the parameter $\Lambda$ is zero in order to satisfy \eqref{Liouville}. To derive the stress-tensor from the Lagrangian, one must set the Liouville field off-shell by not imposing the equation \eqref{Liouville} but considering the metric as a background field. Under a super-Lorentz transformation 
\bea
\delta_Y (D^2 \Phi -  R) = (\mathcal L_Y + D_A Y^A)  (D^2 \Phi -  R).\label{actionRL}
\eea 
Therefore, imposing the Liouville equation is consistent with the action of super-Lorentz transformations. 

Using this boundary metric and shear, one can work out the covariant expressions for $R_0$ and $R_1$ in \eqref{rc}. They are given by
\begin{equation}
R_0 = \frac{1}{2} e^{-\Phi}D^2 W \qquad \text{and} \qquad R_1 = \frac{1}{8} e^{-\Phi} C_{AB} C^{AB}.
\end{equation}
Finally, after some algebra, one can write the full metric as
\begin{equation}
\begin{split}
ds^2 &= - \frac{R}{2}\text{d}u^2 - 2 \text{d}\rho \text{d}u + (\rho^2 q_{AB} + \rho C^{\text{vac}}_{AB} + \frac{1}{8}C^{\text{vac}}_{CD}C_{\text{vac}}^{CD} q_{AB})\text{d}x^A \text{d}x^B \\ &+ D^B C^{\text{vac}}_{AB} \text{d}x^A \text{d}u\label{metf}
\end{split}
\end{equation}
where $\rho = \sqrt{r^2+ \frac{1}{8}C^{\text{vac}}_{CD}C_{\text{vac}}^{CD}}$ is a derived quantity in terms of the Bondi radius $r$. The metric is more natural in Newman-Unti gauge $(u,\rho,z^A)$ where $g_{\rho \mu} = -\delta_\mu^\rho$ (see \eqref{NU gauge def}). 

Let us also comment on the meromorphic extension of the Lorentz group instead of Diff$(S^2)$. When super-Lorentz transformations reduce to local conformal Killing vectors on $S^2$, i.e. $G^z = G(z)$ and $G^{\bar z} \equiv \bar{G}(\bar z) $, the boundary metric after a diffeomorphism is the unit round metric on the sphere 
\bea
\mathring{q}_{AB} \D z^A \D z^B = 2 \gamma_s \D z \D \bar{z},\qquad \gamma_s = \frac{2}{(1+z\bar z)^2}
\eea
(and $\mathring{R} = 2$) \emph{except at the singular points of }$G(z)$. The Liouville field reduces to the sum of a meromorphic and an anti-meromorphic part minus the unit sphere factor 
\bea
\Phi = \phi(z)+\bar \phi(\bar z)- \log \gamma_s.\label{Phim}
\eea
The metric \eqref{metf} then exactly reproduces the expression of \cite{Compere:2016jwb} with the substitution $T^{(\text{there})}_{AB} = 1/2 N^{\text{vac}}_{AB}$. We have therefore found the generalization of the metric of the vacua for arbitrary $\text{Diff}(S^2)$ super-Lorentz transformations together with arbitrary supertranslations.

\subsection{The superboost, superrotation and supertranslation fields}

A general vacuum metric is parametrized by a boundary metric $q^{\text{vac}}_{AB}$, the field $C$ that we call the \emph{supertranslation field} and $\Phi$ that we will call either the \emph{Liouville field} or the \emph{superboost field}. Under a BMS transformation, the bulk metric transforms into itself, with the following transformation law of its boundary fields:
\bea
\delta_{T,Y}q^{\text{vac}}_{AB} &=& D_A Y_B + D_B Y_A - q_{AB}^{\text{vac}} D_C Y^C, \\
\delta_{T,Y}\Phi &=& Y^A \p_A \Phi + D_A Y^A,\\
\delta_{T,Y}C &=& T + Y^A \p_A C - \frac{1}{2}C D_A Y^A. \label{deltaC}
\eea
Only the divergence of a general super-Lorentz transformation sources the Liouville field. Since rotations are divergence-free but boosts are not, we call $\Phi$ the \emph{superboost field}. In general, one can decompose a vector on the 2-sphere as a divergence and a rotational part. For a generic superrotation, there should be a field that is sourced by the rotational of $Y^A$. We call this field the \emph{superrotation field} $\Psi$ and we postulate its transformation law
\bea
\delta_{T,Y}\Psi &=& Y^A \p_A \Psi + \eps^{AB} D_A Y_B.
\eea
Where is that field in \eqref{metf}? In fact, the boundary metric $q^{\text{vac}}_{AB}$ is not a fundamental field. It depends upon the Liouville field $\Phi$ and the background metric $\gamma_{ab}$. Since it transforms under superrotations, the metric \eqref{qABform} should also depend upon the superrotation field $\Psi$. The explicit form $q^{\text{vac}}_{AB}[\gamma_{ab},\Phi,\Psi]$ is not known to us. We call the set of boundary fields $(\Phi,\Psi)$ the super-Lorentz fields.

Under a BMS transformation, the news of the vacua $N^{\text{vac}}_{AB}$ and  the tensor $C^{(0)}_{AB}$ transform inhomogeneously as 
\bea\label{dRN}
\delta_{T,Y} N^{\text{vac}}_{AB} &=& \mathcal L_Y N^{\text{vac}}_{AB} -D_A D_B D_C Y^C + \frac{1}{2}q_{AB}D^2 D_C Y^C,\\\delta_{T,Y}C^{(0)}_{AB}  &=&  \mathcal L_Y C^{(0)}_{AB} - \frac{1}{2}D_C Y^C C^{(0)}_{AB}    -2 D_A D_B T + q_{AB}D^2 T.\label{baret}
\eea

From \eqref{metf}, one can read off the explicit expressions of the Bondi mass and angular momentum aspects of the vacua
\begin{equation}
\begin{split}
M &= -\frac{1}{8} N^{\text{vac}}_{AB} C_{\text{vac}}^{AB} ,\\
N_{A} &= -\frac{3}{32} D_A (C^{\text{vac}}_{BC}C_{\text{vac}}^{BC}) - \frac{1}{4} C^{\text{vac}}_{AB} D_C C_{\text{vac}}^{BC}. 
\end{split}\label{valvac}
\end{equation}
The Bondi mass is time-dependent and its spectrum is not bounded from below because $\p_u M = -\frac{1}{8}N_{AB}^{\text{vac}}N^{AB}_{\text{vac}}$ as observed in \cite{Compere:2016jwb}. Nonetheless, the Weyl tensor is identically zero, so the standard Newtonian potential vanishes. This indicates that the mass is identically zero. The relationship between the Bondi mass and the mass is given below in subsection \ref{sec:chv}.

\section{Superboost transitions}
\label{sec:transitions}

The main interest of the non-trivial vacua lies in the dynamical processes that allow transitions from one vacuum to another. In what follows, we focus on the processes associated with both Diff($S^1$)$\times$Diff($S^1$) and Diff$(S^2)$ super-Lorentz extensions of the BMS$_4$ group. We study several examples of transitions and investigate the related memory effects at null infinity.

\subsection{The impulsive Robinson-Trautman metric as a vacuum transition}

The impulsive limit of the Robinson-Trautman type N of positive 2-curvature ($M= 0$, $ R = 2$)  can be rewritten after a coordinate transformation as the metric of the impulsive gravitational waves of Penrose \cite{Penrose:1972aa,Nutku:1992aa}, as shown in \cite{Podolsky:1999zr,Griffiths:2002gm}\footnote{It is exactly the solution (2.10) of \cite{Podolsky:2002sa} with $\eps = +1$ upon substituting $U \rightarrow u/\sqrt{2}$, $V \rightarrow -\sqrt{2}\rho$, $H \rightarrow -1/2 N^{\text{vac}}_{zz}$. Strictly speaking, $g_{uu} = -1-\frac{D^2 \phi}{2}$ at the poles of the meromorphic function $\phi(z)$, but $g_{uu}=  -1$ otherwise.}
\begin{equation}
\begin{split}
ds^2 =& -\text{d}u^2 - 2 \text{d}\rho \text{d}u \\
&+ (\rho^2 q_{AB} + u \rho \Theta(u) N^{\text{vac}}_{AB} + \frac{u^2}{8}\Theta(u) N^{\text{vac}}_{CD}N_{\text{vac}}^{CD} q_{AB})\text{d}x^A \text{d}x^B ,\label{imp}
\end{split}
\end{equation}
where $N^{\text{vac}}_{AB}  = \left[\frac{1}{2}D_A \phi_f D_B \phi_f - D_A D_B \phi_f  \right]^{\text{TF}}$. The vacuum news coincides with \eqref{PhiL} after substituting $\Phi = -\log\gamma_s + \phi_f$ as in \eqref{Phim}. This metric is in Newman-Unti gauge, not in Bondi gauge. It represents the transition between two vacua labelled by distinct meromorphic superboost fields\footnote{The singular impulsive limit requires considering singular diffeomorphisms transitions which turn out to reduce to meromorphic superboost transitions.} (initial $\phi_i = 0$ for $u< 0$ and final $\phi_f = \phi(z)+\bar\phi(\bar z)$ for $u > 0$). The metric $q_{AB}$ is the unit sphere metric globally for $u<0$ and locally for $u > 0$ but it contains singularities at isolated points for $u > 0$. These singularities can be understood as cosmic string decays \cite{Griffiths:2002hj,Griffiths:2002gm,Strominger:2016wns}. 

\subsection{General impulsive gravitational wave transitions}

In general, both the supertranslation field $C$ and the superboost field $\Phi$ can change with hard (finite energy) processes involving null radiation reaching $\mathcal I^+$. Such processes induce vacuum transitions among initial $(C_-,\Phi_-)$ and final $(C_+,\Phi_+)$ boundary fields. The difference between these fields can be expressed in terms of components of the matter stress-tensor and metric potentials reaching $\mathcal I^+$. The simplest possible transition between vacua are shockwaves that carry a matter stress-tensor proportional to a $\delta(u)$ function, as in the original Penrose construction \cite{Penrose:1972aa}. A distinct vacuum lies on each side of the shockwave and the transition between the boundary fields is dictated by the matter stress-tensor. Such a general shockwave takes the form 
\begin{equation}
\begin{split}
ds^2 =& - \frac{\mathring R}{2}\text{d}u^2 - 2 \text{d}\rho \text{d}u + (\rho^2 q_{AB} + \rho C_{AB} + \frac{1}{8}C_{CD}C^{CD} q_{AB})\text{d}x^A \text{d}x^B \\
&+ D^B C_{AB} \text{d}x^A \text{d}u,\label{metf2}
\end{split}
\end{equation}
where 
\bea
q_{AB}&=&\Theta(-u) q_{AB}^{\text{vac}}[\Phi_-] + \Theta(u) q_{AB}^{\text{vac}}[\Phi_+],\\
C_{AB}&=& \Theta(-u) C_{AB}^{\text{vac}}[\Phi_-, C_-] + \Theta(u) C_{AB}^{\text{vac}}[\Phi_+,C_+],
\eea
where $q_{AB}^{\text{vac}}[\Phi]$ and $C_{AB}^{\text{vac}}[\Phi,C]$ are given in \eqref{qABform} and \eqref{PhiL}. The metric \eqref{imp} is recovered for $\Phi_- = -\log\gamma_s$, $\Phi_+ = -\log\gamma_s + \phi(z)+\bar\phi(\bar z)$ as in \eqref{Phim} and $C_+ = C_- = 0$.

\subsection{Conservation of the Bondi mass aspect and the center-of-mass}

In the absence of superboost transitions and for the standard case of the unit round celestial sphere, the integral between initial $u_i$ and final retarded times $u_f$ of the conservation equation for the Bondi mass aspect \eqref{duM} can be reexpressed as the differential equation determining the difference of supertranslation field $\Delta C = C_+ - C_-$ between initial and final retarded times after assuming suitable fall-off conditions \cite{Strominger:2014pwa}
\bea
-\frac{1}{4} D^2 (D^2 + 2)  \Delta C  =  \Delta M + \int_{u_-}^{u_+} \text{d}u \: T_{uu},\label{eq:98}
\eea
where $T_{uu}=\frac{1}{8}N_{AB}N^{AB}$ and $\Delta M$ is the difference between the Bondi mass aspects after and before the burst. The four lowest spherical harmonics $\ell=0,1$ are zero modes of the differential operator appearing on the left-hand side of \eqref{eq:98}. Recall that translations precisely shift the supertranslation field as \eqref{deltaC}. The four lowest harmonics of $C$ can thus be interpreted as the center-of-mass of the asymptotically flat system. This center-of-mass is not constrained by the conservation law \eqref{eq:98}.

A new feature arises in the presence of a superboost transition. The four zero modes of the supertranslation field $C$ are now determined by the conservation equation. This can be seen in the context of impulsive transitions \eqref{metf2}. For simplicity, we take $C_- = 0$ and $\Phi_- = -\log\gamma_s$ ($q_{AB}[\Phi_-] = \mathring q_{AB}$ the unit round sphere metric). Given that the Bondi mass aspect and the Bondi news of the vacua are non-zero \eqref{valvac}, we first define the renormalized Bondi mass aspect and Bondi news as
\bea
\hat M &=& M + \frac{1}{8} C_{AB} N^{AB}_{\text{vac}}[\Phi_+],\\
\hat N_{AB} &=& N_{AB} - \Theta(u) N_{AB}^{\text{vac}}[\Phi_+],
\eea
which are zero for the vacua \eqref{metf}. This mass is obtained in section \ref{subsec: Finte charges} in \eqref{mf}. 

After integration over $u$ of \eqref{duM} and using the corollary of the Liouville equation \eqref{Li2} we obtain 
\begin{multline}
-\frac{1}{4} D^2 (D^2 + R)  C_+ +\frac{1}{4}N^{AB}_{\text{vac}}[\Phi_+] D_A D_B C_+  + \frac{1}{8} C_+ D^2  R  \\ 
= \Delta \hat M +  \int_{u_-}^{u_+} \text{d}u \: T_{uu}, \label{eq4}
\end{multline}
where $T_{uu}=\frac{1}{8}\hat N_{AB}\hat N^{AB}$ and $\Delta \hat M$ act as sources for $C_+$ and all quantities are evaluated on the final metric $q_{AB}[\Phi_+]$. We have that $\Delta \hat M=0$ for transitions between vacua but we included it for making the comparison with \eqref{eq:98} more manifest.

The lowest $\ell=0,1$ spherical harmonics of $C$ are not zero modes of the quartic differential operator on the left-hand side of \eqref{eq4} for any inhomogeously curved boundary metric. Therefore, the center-of-mass is also determined by the conservation law of the Bondi mass aspect.

\subsection{Refraction/Velocity kick memory}

We now consider the simplified case where the change of the boundary metric is localized at individual points. This happens for impulsive gravitational wave transitions that relate the initial and final boundary metric by a meromorphic super-Lorentz transformation (which is a combination of superboosts and superrotations). One example is the original Penrose construction \cite{Penrose:1972aa}. In these cases we consider observers away from these singular points so that we can ignore these singularities. 

We can consider either timelike or null geodesics leading respectively to the velocity kick and refraction memory. Let us first discuss a congruence of timelike geodesics that evolve at finite large radius $r$ in the impulsive gravitational wave spacetime \eqref{imp}. Such observers have a velocity $v^\mu \p_\mu = \p_u + \mathcal{O}(\rho^{-1})$. The deviation vector $s^\mu$ between two neighboring geodesics obeys $\nabla_v \nabla_v s^\mu = R^\mu_{\;\; \alpha\beta\gamma}v^\alpha v^\beta s^\gamma$, where the directional derivative is defined as $\nabla_v = v^\mu \nabla_\mu$. We have $R_{uA uB}= -\frac{\rho}{2} \p^2_u C_{AB} + \mathcal{O}(\rho^0)$ where $C_{AB} = u \Theta(u) N_{AB}^{\text{vac}}$ and therefore 
\bea
q_{AB} \p_u^2 s^B &=& \frac{1}{2\rho} \delta(u) N_{AB}^{\text{vac}}s^B + \mathcal{O}(\rho^{-2}). 
\eea
We deduce that $s^A =s^A_{lead}(x^A)+ \frac{1}{\rho}s^A_{sub}(u,x^A)+\mathcal{O}(\rho^{-2})$ and after two integrations in $u$,  
\bea
s^A_{sub} = \frac{u}{2} \Theta(u) q^{AB } N_{BC}^{\text{vac}}s^C_{lead}. \label{devG}
\eea
Before the shockwave, there is no relative angular velocity between observers. After the shockwave, there will be a relative angular velocity at order $\propto \rho^{-1}$. This is the velocity kick between two such neighboring geodesics due to the shockwave \cite{Podolsky:2002sa,Podolsky:2010xh,Podolsky:2016mqg}. This is a qualitatively distinct effect from the displacement memory effect \cite{Zeldovich:1974aa,Blanchet:1987wq,Christodoulou:1991cr,Blanchet:1992br,0264-9381-9-6-018} and the spin memory effect \cite{Pasterski:2015tva , Himwich:2019qmj , Pasterski:2019msg}.

Analogously, one can consider a congruence of null geodesics that admits a constant leading angular velocity $\Omega^A(x^B)\p_A$, with total 4-velocity 
\bea
v^\mu \p_\mu = (\sqrt{\Omega^A q_{AB} \Omega^B} + \mathcal{O}(\rho^{-1}))\p_u + \mathcal{O}(\rho^{-1})\p_\rho + \frac{1}{\rho}( \Omega^A + \mathcal{O}(\rho^{-1}))\p_A . 
\eea
We consider again a deviation vector of the form $s^A =s^A_{lead}(x^A)+ \frac{1}{\rho}s^A_{sub}(u,x^A)+\mathcal{O}(\rho^{-2})$. The deviation vector obeys again \eqref{devG}. Null geodesics are refracted by the shockwave. This is the refraction memory effect usually described in the bulk of spacetime \cite{Podolsky:2002sa,Podolsky:2010xh,Podolsky:2016mqg}. We identified here the class of null geodesics that displays the refraction memory effect close to null infinity.  

Let us now shortly discuss the case where the change of boundary metric is not localized at individual points. The main point is that timelike geodesics will now admit non-trivial deviation vector already at leading order $\propto\, \rho^0$, $s^A =s^A_{lead}(u,x^A) +\mathcal{O}(\rho^{-1})$, with  
\bea
\frac{1}{2} q_{AB} \p_u^2 s_{lead}^B + \frac{1}{2} \p_u^2 (q_{AB} s^B_{lead}) &=&- \frac{1}{2} \p_u^2 q_{AB} s_{lead}^B . 
\eea
A velocity kick will therefore already occur at order $\rho^0$.

\subsection{A new non-linear displacement memory}
\label{sec:memory}

It should also be mentioned that there is a non-linear displacement memory induced by a superboost transition, when it is accompanied by a supertranslation transition. This case was not considered in \cite{Podolsky:2002sa,Podolsky:2010xh,Podolsky:2016mqg}, where all supertranslation transitions were vanishing. In order to describe the effect, we can consider either timelike or null geodesics. For definiteness, we consider a congruence of timelike geodesics that evolve at finite large radius $\rho$ in the general impulsive gravitational wave spacetime \eqref{metf2}. For simplicity we assume global Minkowski in the far past and we only consider the simplified case where the change of the boundary metric is localized at individual points. In other words, we assume $\Phi_- = -\log\gamma_s$ ($q^{\text{vac}}_{AB}[\Phi_-]$ is the unit sphere metric), $C_- = 0$, $\Phi_+ = -\log\gamma_s + \phi(z)+\bar \phi(\bar z)$ and $C_+=C_+(z,\bar z)$ arbitrary. The velocity is now $v^\mu \p_\mu = \sqrt{\frac{2}{\mathring R}}\p_u +\mathcal{O}(\rho^{-1})$. We have  $R_{uA uB}= -\frac{\rho}{2} \p^2_u C_{AB} + \mathcal{O}(\rho^0)$. Following the same procedure as above, we obtain  $s^A =s^A_{lead}(x^A)+ \frac{1}{\rho}s^A_{sub}(u,x^A)+\mathcal{O}(\rho^{-2})$ and away from the singular points on the sphere, 
\bea
s^A_{sub} &=& \frac{1}{2} q^{AB}C_{BC} s^C_{lead}. \\
&=& \frac{1}{2} \Theta(u)  q^{AB}C_{BC}^{\text{vac}} s^C_{lead}.\\
&=& \frac{1}{2} q^{AB} (u \Theta(u) N_{BC}^{\text{vac}} +  \Theta(u) C^{(0)}_{BC}  + \Theta(u) C N_{BC}^{\text{vac}}  )s^C_{lead}.
\eea
The first term $\propto\, u \Theta(u)$ leads to the velocity kick memory effect. The second term $\propto\, \Theta(u) C^{(0)}_{BC} $ leads to the displacement memory effect due to a change of supertranslation field $C$ between the final and initial states \cite{Strominger:2014pwa}.  The third and last term $\propto\, \Theta(u) C N^{\text{vac}}_{BC} $ is a new type of non-linear displacement memory effect due to changes of both the superboost field $\Phi$ and the supertranslation field $C$. The four lowest spherical harmonics $\ell=0,1$ of $C$, interpreted as the center-of-mass, do not contribute to the standard displacement memory effect because they are zero modes of the differential operator $C_{AB}^{(0)}$. Here, they do contribute to the non-linear displacement memory effect. The transition  of the supertranslation field and in particular of the center-of-mass are determined by \eqref{eq4}, as discussed earlier.

\section{Finite charges and soft theorems}
\label{subsec: Finte charges}

In this section, we present a prescription to extract a meaningful integrable charge from the non-integrable infinitesimal charge expression obtained in equation \eqref{infinitesimal charges} of the previous chapter. This procedure is based on ingredients introduced in section \ref{sec:vac} and is inspired by the Wald-Zoupas procedure\footnote{Notice that the Wald-Zoupas procedure discussed in \cite{Wald:1999wa} could not be readily applied to our case due to the non-vanishing contribution of the term $Q^{ur}_{\delta \xi}[g]$ in \eqref{decom:charge} to the non-integrable part.} \cite{Wald:1999wa}. We then relate our associated finite charge to the existing literature and show that this is consistent with the soft graviton theorems, the action of asymptotic symmetries and the vanishing energy of the vacua.

\subsection{Finite surface charges}

In this analysis, we assume that the Liouville equation 
\begin{equation}
R = D^2 \Phi ,
\label{Liouville assumption}
\end{equation} which was derived for the vacua orbit in \eqref{Liouville}, holds in our phase space. This is not guaranteed a priori for the general phase space studied in chapter \ref{Generalized BMS and renormalized phase space} and is therefore an additional restriction. In particular, equation \eqref{Li2} will be satisfied.

Starting from \eqref{infinitesimal charges}, we want to define the finite charges $H_\xi$ associated with $\xi$. Following the Wald-Zoupas procedure \cite{Wald:1999wa}, it would be natural to request that the flux $\p_u H_\xi [g] $ is identically zero in the absence of news. However, the news tensor transforms inhomogeneously under (both Diff$(S^1) \times$Diff$(S^1)$ and Diff$(S^2)$) super-Lorentz transformations so this condition is not invariant under the action of the asymptotic symmetry group. Instead, we request that the flux $\p_u H_\xi [g] $ is identically zero in the absence of shifted news $\hat N_{AB}$, 
\bea
\hat N_{AB} = N_{AB} - N_{AB}^{\text{vac}}. \label{cond1}
\eea
Since the latter transforms homogeneously under super-Lorentz transformations, this prescription is \textit{invariant} under the action of all asymptotic symmetries. For future use, we define the shifted $\hat C_{AB}$ tensor
\bea
\hat C_{AB} = C_{AB} - u N_{AB}^{\text{vac}}. \label{cond2}
\eea
such that $\p_u \hat C_{AB} = \hat N_{AB}$. To obtain our ansatz, let us start with the charge \eqref{FNCharges}. The flux associated with \eqref{FNCharges} reads as
\begin{align}
\partial_u H^{int}_\xi [g] = -\frac{1}{32\pi G} \int_{S^2_\infty}& \text{d}^2 \Omega \left[  f N_{AB} N^{AB} -2 f D_A D_B N^{AB} - f D_A D^A {R}  - Y^A\mathcal{H}_A (N,C) \right. \nonumber \\
&\left.  \quad + Y^A D_B D^B D^C C_{AC} - Y^A D_B D_A D_C C^{BC} - Y^A C_{AB} D^B {R} \right] 
\end{align}
Here, we defined for later convenience the bilinear operator on rank-2 spherical traceless tensors $P_{AB}$ and $Q_{AB}$:
\begin{equation}
\mathcal{H}_A (P,Q) \equiv \frac{1}{2} \partial_A (P_{BC} Q^{BC}) - P^{BC} D_A Q_{BC} + D_B (P^{BC} Q_{AC} - Q^{BC} P_{AC})
\end{equation}
which enjoys the property $\mathcal{H}_A(P,P)=0$. When $\hat N_{AB} = 0$, we are left with 
\begin{equation}
\begin{split}
\p_u H^{int}_\xi |_{\hat N_{AB}=0} = -\frac{1}{32\pi G}\int_{S^2_\infty} \text{d}^2\Omega &\left[ f N_{AB}^{\text{vac}} N^{AB}_{\text{vac}} - Y^A\mathcal{H}_A (N^{\text{vac}},C) - Y^A C_{AB} D^B {R} \right.   \\
&\left. \quad + Y^A D_B D^B D^C C_{AC} - Y^A D_B D_A D_C C^{BC} \right]
\end{split}
\label{eq:FluxNABphys}
\end{equation}
after using the relation \eqref{Li2}, which follows from our additional boundary condition \eqref{Liouville assumption}. We now want to define a counterterm that is only built out of the fields at $\mathscr{I}^+$ ($q_{AB},C_{AB},N_{AB}$) and out of $N_{AB}^{\text{vac}}$, which is the only boundary field that appears in the condition \eqref{cond1}. Our prescription that cancels the right-hand side of \eqref{eq:FluxNABphys} is 
\bea
\Delta H_\xi[g ; \delta g] &\equiv& \prefwg \int_{S^2_\infty} \text{d}^2 \Omega \Big[ \frac{u}{2} Y^A D_B D^B D^C C_{AC} - \frac{u}{2} Y^A D_B D_A D_C C^{BC}    \nonumber \\
  &&- \frac{u}{2} Y^A C_{AB} D^B {R} + \frac{1}{2} T C_{AB} N_{\text{vac}}^{AB} - \frac{u}{2} Y^A\mathcal{H}_A(N^{\text{vac}},C) \nonumber \\
  &&+ \frac{u^2}{8} D_C Y^C N^{\text{vac}}_{AB} N_{\text{vac}}^{AB} + \frac{u^2}{4} Y^A N^{\text{vac}}_{AB} D^B {R} \Big].
\label{DeltaH}
\eea
This is the minimal ansatz that cancels the right-hand side of \eqref{eq:FluxNABphys}. Of course, there is considerable ambiguity in defining $\Delta H_\xi[g ; \delta g]$. We will justify our minimal choice in \eqref{DeltaH} by showing consistency with the leading and subleading soft theorems, and for defining the charges of the vacua. 

Our final prescription for the canonical charges is $H_\xi [g] = H^{int}_\xi [g] +\Delta H_\xi[g]$. The charges are conveniently written as
\begin{equation}
 H_\xi[g] = \frac{1}{16 \pi G}\int_{S^2_\infty} \text{d}^2 \Omega \left[  4 T \hat M  + 2 Y^A  \hat N_A \right]\label{Hhat}
\end{equation}
where the final mass and angular momentum aspects are given by 
\bea
\hat M &=& M + \frac{1}{8} C_{AB} N^{AB}_{\text{vac}}; \label{mf}\\
\hat N_A &=& N_A -u \p_A M + \frac{1}{32} \partial_A (\hat C_{CD}\hat C^{CD}) + \frac{u}{16} \p_A (\hat C^{CD} N_{CD}^{\text{vac}})  - \frac{1}{32} u^2 \p_A (N_{BC}^{\text{vac}} N^{BC}_{\text{vac}}) \label{NAf}\\
&& - \frac{u}{4} \mathcal{H}_A (N^{\text{vac}},\hat C) - \frac{u}{4} \hat C_{A}^B D_B \mathring{R}  
+ \frac{u}{4} D_B D^B D_C \hat C_{A}^C - \frac{u}{4} D_B D_A D^C \hat C^{B}_C -  \frac{u^2}{8} N_{AB}^{\text{vac}} D^B \mathring{R}.\nonumber
\eea
This is a new prescription for the charges. In the standard asymptotically flat spacetimes where the boundary metric is the round sphere ($q_{AB} = \mathring q_{AB}$ with $\mathring R =2$), our expressions reduce to 
\begin{equation}
\begin{split}
\hat M &= M ;\\
\hat N_A &= N_A -u \p_A M + \frac{1}{32} \partial_A (C_{CD} C^{CD})+ \frac{u}{4} D_B D^B D^C C_{AC} - \frac{u}{4} D_B D_A D_C C^{BC}.\label{finalNA}
\end{split}
\end{equation}
The Lorentz charges differ from the existing prescriptions \cite{Barnich:2011mi,Flanagan:2015pxa,Hawking:2016sgy} since the angular momentum aspect is now enhanced with the two soft terms linear in $u$. We will show that our prescription correctly reproduces the fluxes needed for the subleading soft theorem. Furthermore, our expressions are exactly those needed for the BMS flux balance laws discussed in \cite{Compere:2019gft}. 

Let us mention that another interesting prescription for $\Delta H_\xi[g ; \delta g]$ has been proposed in \cite{newCompere}. The finite charges considered in that reference have two interesting properties: $(i)$ the charges of the vacua are all vanishing, and $(ii)$ the charges represent $\mathfrak{bms}_4^{\text{gen}}$ without central extension at the corners of null infinity under the standard Dirac bracket. We refer to \cite{newCompere} for more details about this prescription.

\subsection{Flux formulae for the soft Ward identities}

Let us show that our expressions for the fluxes reproduce the expressions of the literature used in the Ward identities displaying the equivalence to the leading \cite{Weinberg:1965nx} and subleading \cite{Cachazo:2014fwa} soft graviton theorems. The final flux can be decomposed in soft and hard parts, where the soft terms (resp. hard terms) are linear (resp. quadratic) in $\hat C_{AB}$ or its time variation $\hat N_{AB}$. We have
\begin{equation}
\int_{\mathscr{I}^+} \text{d}u \, \p_u H_\xi[g] = Q_S[T] + Q_H[T] + Q_S[Y] + Q_H[Y]
\label{eq:FinalFlux}
\end{equation}
where 
\begin{align}
Q_S[T] &= \frac{1}{16\pi G}\int_{\mathscr{I}^+} \text{d}u \, \text{d}^2\Omega\; \p_u \left( T D_A D_B \hat C^{AB} \right), \label{SoftT} \\
Q_H[T] &= \frac{1}{16\pi G}\int_{\mathscr{I}^+} \text{d}u \, \text{d}^2\Omega \left( - \frac{1}{2} T \hat N_{AB}  N^{AB}  \right), \label{HardT} \\
Q_S[Y] &= \frac{1}{16\pi G}\int_{\mathscr{I}^+} \text{d}u \, \text{d}^2\Omega \;u \,\p_u \left(  \hat C^{AB} s_{AB} \right), \label{SoftY} \\
Q_H[Y] &= \frac{1}{16\pi G}\int_{\mathscr{I}^+} \text{d}u \, \text{d}^2\Omega \left(\frac{1}{2} Y^A \mathcal{H}_A(\hat N ,\hat C) + \frac{u}{2} Y^A N^{D}_C D_A \hat N_{D}^C + \frac{u}{2}  N^{CD}_{\text{vac}} Y^A D_A \hat N_{CD} \right) \label{HardY}
\end{align}
and 
\begin{equation}
s_{AB} = \left[  D_A D_B D_C Y^C + \frac{\mathring{R}}{2} D_{(A} Y_{B)} - \frac{1}{2}D_{(A} (D^2 + \frac{\mathring{R}}{2}) Y_{B)} \right]^{\text{TF}}
\label{sAB}
\end{equation}
after integrations by parts on the sphere. 

In the standard case where $N_{AB}^{\text{vac}} = 0$, the flux of supermomenta reproduces (2.11) of \cite{He:2014laa} up to a conventional overall sign, which itself agrees with previous results \cite{Ashtekar:1978zz}. After one imposes the antipodal matching condition on $\hat M$ at spatial infinity, one can equate the flux on $\mathscr{I}^+$ with the antipodally related flux on $\mathscr{I}^-$. The result of \cite{He:2014laa} is precisely that the quantum version of this identity is the Ward identity of the leading soft graviton theorem. We have now obtained a generalization in the presence of superboost background flux.

We now consider the hard terms for super-Lorentz transformations. Using the identity
\begin{equation}
\frac{1}{2} ( \hat N^{AC} \hat C_{BC} + \hat C^{AB} \hat N_{BC}) = \frac{1}{2} (\hat N^{BC} \hat C_{BC}) \delta^A_B
\end{equation} and integrating by parts, it can be shown that \eqref{HardY} can be rewritten as 
\begin{align}
Q_H[Y] &= \prefwg \int_{\mathscr{I}^+} \text{d}u \, \text{d}^2\Omega \, \left[- \frac{1}{2} \hat N^{AB} \left(\mathcal{L}_Y \hat C_{AB} - \frac{1}{2} D_C Y^C \hat C_{AB} + \frac{u}{2} D_C Y^C \hat N_{AB} \right) \right. \nonumber\\ 
&\left. \qquad\qquad\qquad\qquad\qquad\qquad+ u N^{AB}_{\text{vac}} Y^C D_C \hat N_{AB} \right] \nonumber \\
&= -\frac{1}{32 \pi  G}\int_{\mathscr{I}^+} \text{d}u\, \text{d}^2\Omega \, \left[ \hat N^{AB} \delta^H_Y \hat C_{AB} - 2u N^{AB}_{\text{vac}} Y^C D_C \hat N_{AB} \right]
\end{align}
where $\delta^H_Y$ is the homogeneous part of the transformation of $\hat C_{AB}$. After restricting to standard configurations where $N_{AB}^{\text{vac}} = 0$, the expression matches (up to the overall conventional sign) with equation (40) of \cite{Campiglia:2015yka}. 

Next, we consider the soft terms for super-Lorentz transformations. Noting that $D^C \delta_Y q_{AC} = D_C D^C Y_A + \frac{R}{2} Y_A$ we can rewrite \eqref{sAB} as
\begin{equation}
s_{AB} = \left[  D_A D_B D_C Y^C + \frac{R}{2} D_{(A} Y_{B)} - \frac{1}{2}D_{(A} D^C \delta_Y q_{B)C} \right]^{\text{TF}}.
\end{equation}
The tensor $s_{AB}$ is recognized as the generalization of equation (47) of \cite{Campiglia:2015yka} in the presence of non-trivial boundary curvature. After some algebra, we can rewrite it in terms of the inhomogeneous part $\delta^I_Y C_{AB}$ of the transformation law of $C_{AB}$ \eqref{dCAB}:
 \bea
-u s_{AB} = \delta_Y^I C_{AB} \equiv -u (D_A D_B D_C Y^C + \frac{1}{2}q_{AB} D_C D^C D_E Y^E). 
\eea
Now that we identified our expressions with the ones of \cite{Campiglia:2015yka}, we can use their results. After imposing the antipodal matching condition on $\hat N_A$ at spatial infinity, one can equate the flux of super-Lorentz charge on $\mathscr{I}^+$ with the antipodally related flux on $\mathscr{I}^-$ as originally proposed in \cite{Hawking:2016sgy} (but where the expression for $\hat N_A$ should be modified to \eqref{finalNA}). The result of \cite{Campiglia:2015yka} is precisely that this identity is the Ward identity of the subleading soft graviton theorem \cite{Cachazo:2014fwa}.

We end up with two further comments. Note that the soft charges for super-Lorentz transformations agree with equation (41) of \cite{Campiglia:2015yka} (up to an overall conventional sign) after an integration by parts on $u$ and after using the restrictive boundary condition $\hat C_{AB} = o(u^{-1})$,
\begin{equation}
\begin{split}
\pref \int \text{d}u \, \hat C^{AB}  s_{AB} &= \pref \left[u \hat C^{AB} s_{AB}\right]_{\mathcal{I}_-^+}^{\mathcal{I}_+^+} - \pref \int \text{d}u \, (u \hat N^{AB}  s_{AB})\\
&= - \pref \int \text{d}u \, (u \hat N^{AB} s_{AB}) = -Q_S [Y].
\end{split}
\end{equation}
However, the boundary condition $\hat C_{AB} = o(u^{-1})$ is not justified since displacement memory effects lead to a shift of $C$, e.g. in binary black hole mergers. Therefore, using more general boundary conditions, the valid expression for the soft charge is only given by \eqref{SoftY}. 

Considering only the background Minkowski spacetime ($q_{AB} =\mathring {q}_{AB}$ the unit metric on the 2-sphere and $N_{AB}^{\text{vac}} = 0$), one can check that in stereographic coordinates one has $s_{zz} =  \partial_z^3 Y^z =  D_z^3 Y^z$. The soft charge then reads as 
\begin{equation}
Q_S [Y] = \frac{1}{16\pi G}\int_{\mathscr{I}^+} \text{d}u  \text{d}^2 z \, \gamma_{s} \, (u \hat N^{zz} D_{{z}}^3 Y^{{z}} + u \hat N^{\bar{z}\bar{z}} D_{\bar{z}}^3 Y^{\bar{z}} )
\end{equation}
where we keep $Y^A \partial_A = Y^z (z,\bar{z}) \partial_z + Y^{\bar{z}} (z,\bar{z}) \partial_{\bar{z}}$ arbitrary. In the case of meromorphic super-Lorentz transformations, this reproduces equation (5.3.17) of \cite{Strominger:2017zoo} (up to a conventional global sign). It shows that the Ward identities of supertranslations and super-Lorentz transformations are equivalent to the leading and subleading soft graviton theorems following the arguments of \cite{He:2014laa,Campiglia:2014yka}.

\subsection{Charges of the vacua}
\label{sec:chv}

Using the values \eqref{valvac} in our prescription \eqref{Hhat} we deduce the mass and angular momenta of the vacua
\begin{equation}
H^{\text{vac}}_\xi [\Phi,C]= \frac{1}{8\pi G} \int_{S^2_\infty} \text{d}^2\Omega [2 T \hat{M}^{\text{vac}} +  Y^A \hat{N}_A^{\text{vac}} ]
\end{equation} where
\begin{equation}
\begin{split}
\hat{M}^{\text{vac}} &=0, \\
\hat N_A^{\text{vac}}  &= -\frac{1}{4} \hat C_{AB} D_C \hat C^{BC} - \frac{1}{16} \partial_A (\hat C_{CD} \hat C^{CD}),
\end{split}
\end{equation}
and $\hat C_{AB}= C N_{AB}^{\text{vac}} - 2 D_A D_B C + q_{AB} D^C D_C C$ in this case. 

The supermomenta are all identically vanishing. Remember that the Lorentz generators are uniquely defined as the six global solutions $Y^A$ to the conformal Killing equation $D_A Y_B + D_B Y_A = q_{AB} D_C Y^C$. 
In general, the Lorentz charges as well as the super-Lorentz charges are non-vanishing. 

For the round sphere metric $q_{AB} = \mathring q_{AB}$ ($\Phi = -\log \gamma_s$), we have $\mathring{R}=2$, $N_{AB}^{\text{vac}} = 0$ and $D^B \hat C_{AB} = D^B C_{AB}^{(0)} =-D_A (D^2+2)C$. The charges then reduce to 
\begin{equation}
H^{\text{vac}}_\xi[C] = \frac{1}{8\pi G} \int_{S^2_\infty}  \text{d}^2\Omega \left[ T \times 0 +  Y^A \Big( -\frac{1}{4} C^{(0)}_{AB} D_C C_{(0)}^{BC} - \frac{1}{16} \partial_A (C_{CD}^{(0)} C^{CD}_{(0)})\Big)\right].
\end{equation} 
As shown in appendix A.3 of \cite{Compere:2016jwb}, the Lorentz charges are identically zero. The difference of charges between our prescription and the one of \cite{Compere:2016jwb} are the last two terms of \eqref{finalNA}, which exactly cancel for the vacua with a round sphere boundary metric. Therefore, we confirm that the vacua with only the supertranslation field turned on do not carry Lorentz charges. The super-Lorentz charges are conserved and non-vanishing in general, which allows us to distinguish the vacua.

\section{Discussion}

Supertranslation BMS symmetry, the leading soft graviton theorem and the displacement memory effect form three corners of a triangle describing the leading infrared structure of asymptotically flat spacetimes at null infinity \cite{Strominger:2017zoo}. The three edges of the triangle can be described in the language of vacuum transitions, Ward identities and Fourier transforms. In the case of super-Lorentz BMS symmetry, it seems that this network of relations is more subtle. Indeed, while the connection among super-Lorentz symmetry, subleading soft theorem and spin memory effect has been established \cite{Kapec:2014opa,Campiglia:2014yka,Campiglia:2015yka,  Pasterski:2015tva , Himwich:2019qmj}, we have shown in this chapter that another memory effect associated with superboosts appeared at the leading order metric at null infinity. More precisely, we clarified how the superboost transitions lead to the refraction or velocity kick memory effect at null infinity. We also described a non-linear displacement memory effect that occurs in the case of combined superboost and supertranslation transitions. Finally, we obtained a new definition of the angular momentum for standard asymptotically flat spacetimes that is consistent with the fluxes required for the subleading soft graviton theorem.

\chapter{$\Lambda$-BMS$_4$ group}
\label{ch:LambdaBMS group}

In this chapter, we investigate a new set of boundary conditions in asymptotically locally (A)dS$_4$ spacetime which is such that the associated asymptotic symmetry algebra is infinite-dimensional and reduces to the generalized BMS algebra $\mathfrak{bms}^{\text{gen}}_4$ in the flat limit ($\Lambda \to 0$). For this reason, we call this new asymptotic symmetry algebra the $\Lambda$-BMS$_4$ algebra\footnote{As we discuss in this chapter, $\mathfrak{bms}^\Lambda_4$ is not strictly speaking a Lie algebra, but a Lie algebroid.} and we write it as $\mathfrak{bms}^\Lambda_4$. 

In section \ref{Relation between Bondi and Fefferman-Graham gauges in three dimensions}, we investigate the most general solution spaces of three-dimensional general relativity in both Fefferman-Graham and Bondi gauges in asymptotically locally (A)dS$_3$ spacetime. We construct the explicit diffeomorphism between the two gauges and identify their solution space. Imposing the Dirichlet boundary conditions, we show how the associated asymptotic symmetry group $\mathfrak{diff}(S^1)\oplus \mathfrak{diff}(S^1)$ reduces to $\mathfrak{bms}_3 = \mathfrak{diff}(S^1) \loplus_{\text{ad}} \mathfrak{vect}(S^1)$ in the flat limit. 

After this warm-up, in section \ref{Relation between Bondi and Fefferman-Graham gauges in four dimensions}, we repeat the analysis in four-dimensional asymptotically locally (A)dS$_4$ spacetime. We derive the most general solution spaces in Fefferman-Graham and Bondi gauges and construct the explicit diffeomorphism that maps one to the other. Then we propose a new set of boundary conditions that leads to the $\mathfrak{bms}^\Lambda_4$ asymptotic symmetry algebra and we show how it reduces to $\mathfrak{bms}^{\text{gen}}_4$ in the flat limit.

In section \ref{Holographic renormalization}, repeating the holographic renormalization procedure, we construct the phase space and derive the associated co-dimension 2 form for the most general solution space in Fefferman-Graham gauge. In section \ref{phase space and its flat limit}, imposing the new set of partial Dirichlet boundary conditions, we find the $\mathfrak{bms}^\Lambda_4$ phase space. In the flat limit, we exactly recover the regularized phase space of section \ref{sec:phasespace}. 

Finally, in section \ref{New boundary conditions for asymptotically locally}, we restrict the analysis to the case $\Lambda < 0$ and require that the symplectic flux vanishes at infinity to have a globally hyperbolic spacetime. This is done by imposing Neumann-type boundary conditions, in addition to the partial Dirichlet boundary conditions already imposed. The associated asymptotic symmetry algebra is an infinite-dimensional subalgebra of $\mathfrak{bms}^\Lambda_4$ formed by the direct sum between the area preserving diffeomorphisms and the abelian time translations. We show that the phase space contains interesting solutions, including a new stationary and axisymmetric solution different from Kerr-AdS$_4$.   

This chapter has strong intersections with \cite{Compere:2019bua , newCompere}, except for section \ref{Relation between Bondi and Fefferman-Graham gauges in three dimensions}, which partially reproduces \cite{newMarios, newMarios2}.

\section{Bondi and Fefferman-Graham gauges in three dimensions}
\label{Relation between Bondi and Fefferman-Graham gauges in three dimensions}
\chaptermark{Bondi and Fefferman-Graham gauges in 3d}

In this section, we present the Fefferman-Graham and Bondi gauges in three dimensions following this general pattern: off-shell definition of the gauge, residual gauge diffeomorphisms, solution space and on-shell variation of the solution space. This analysis follows the logic discussed in section \ref{Asymptotic symmetries in the gauge fixing approach} and particularizes it to the three-dimensional case. In Bondi gauge, the results that we obtain generalize previous considerations by allowing an arbitrary boundary structure encoding the notion of asymptotically locally (A)dS$_3$ spacetime. Furthermore, we construct the explicit diffeomorphism that maps one gauge to the other. We finish this section with a discussion on the asymptotic symmetries aspects and investigate the flat limit in the Bondi gauge. 

\subsection{Fefferman-Graham gauge in 3d}

\subsubsection*{Definition}
In the Fefferman-Graham gauge \eqref{FG gauge}, the metric is given by 
\begin{equation}
\D s^2 = -\frac{1}{\Lambda \rho^2} \D\rho^2 + \gamma_{ab}(\rho, x) \D x^a \D x^b,
\label{gfFG}
\end{equation}
with coordinates $(\rho, x^a)$, $x^a = (t, \phi)$, and the boundary located at $\rho=0$. The three gauge fixing conditions are
\begin{equation}
g_{\rho \rho} = -\frac{1}{\Lambda\rho^2}, \quad g_{\rho a} = 0.
\label{gfFG fixing}
\end{equation}

Residual gauge diffeomorphisms $\xi$, namely diffeomorphisms that preserve the gauge fixing conditions \eqref{gfFG fixing}, satisfy
\begin{equation}
\mathcal{L}_\xi g_{\rho \rho} = 0, \quad \mathcal{L}_\xi g_{\rho a} = 0.
\end{equation}
The explicit solution of these equations is given by 
\begin{equation}
\xi^\rho =  \rho \sigma (x), \quad \xi^a = \xi^a_0 (x) + \frac{1}{\Lambda} \partial_b \sigma \int^\rho_0 \frac{\D \rho'}{\rho'} \gamma^{ab} (\rho', x), 
\label{residual gauge diffeomorphisms FG}
\end{equation}
where $\sigma(x)$ and $\xi^a_0(x)$ are arbitrary functions of $x^a$.

\subsubsection*{Solution space}
\label{Solution space FG}

We impose the preliminary boundary condition $\gamma_{ab} = \mathcal{O}(\rho^{-2})$. Solving the three-dimensional Einstein equations leads to the analytic finite expansion
\begin{equation}
\gamma_{ab}(\rho , x)= \rho^{-2} g_{ab}^{(0)}(x) +  g_{ab}^{(2)}(x)  + \rho^2  g_{ab}^{(4)}(x),
\end{equation} where $g_{ab}^{(4)}$ is determined by $g_{ab}^{(0)}$ and $g_{ab}^{(2)}$ as
\begin{equation}
 g_{ab}^{(4)} = \frac{1}{4} g^{(2)}_{ac}g_{(0)}^{cd}g^{(2)}_{db}.
\end{equation} 
On the other hand, the Einstein equations leave $g^{(2)}_{ab}$ unspecified up to its trace $\text{Tr}\left[ g^{(2)}\right]=\frac{1}{2\Lambda }R^{(0)}$ and the dynamical constraint $D_{(0)}^ag^{(2)}_{ab}=\frac{1}{2 \Lambda}g^{(0)}_{ab}D^a_{(0)}R^{(0)}$. Here, $D^a_{(0)}$ is the covariant derivative with respect to $g^{(0)}_{ab}$ and indices are lowered and raised by $g^{(0)}_{ab}$ and its inverse. Motivated by the holographic dictionary \cite{Balasubramanian1999d, Skenderis2001}, we define the holographic energy-momentum tensor
\beqn
T_{ab} &=& \frac{\sqrt{|\Lambda|}}{8 \pi G} \left(g^{(2)}_{ab} - g_{ab}^{(0)} \text{Tr} \left[g^{(2)}\right]\right) \nonumber\\
&=& \frac{\sqrt{|\Lambda|}}{8 \pi G} \left(g^{(2)}_{ab} - \frac{1}{2 \Lambda} g_{ab}^{(0)}R^{(0)} \right).
\eeqn 
Therefore the Einstein equations infer 
\begin{equation}
{T_a}^a = \eta\frac{c}{24 \pi} R^{(0)} , \quad D^a_{(0)} T_{ab} = 0,
\label{trace condition}
\end{equation} where $\eta =- \text{sgn}(\Lambda)$ and $c = \frac{3}{2 G |\Lambda|}$ is the three-dimensional Brown--Henneaux central charge \cite{Brown:1986nw, Henningson:1998gx}.  

The solution space is thus characterized by five arbitrary functions of $x^a$. Three are in the symmetric tensor $g_{ab}^{(0)}$ and two in the symmetric tensor $T_{ab}$ with constrained trace. These data are subject to two dynamical equations given by $D_{(0)}^a T_{ab} = 0$.

\subsubsection*{Variation of the solution space}

The residual gauge diffeomorphisms \eqref{residual gauge diffeomorphisms FG} evaluated on-shell are given by
\begin{equation}
\xi^\rho = \sigma \rho , \quad \xi^a = \xi^a_0 + \frac{\rho^2}{2 \Lambda} g_{(0)}^{ab} \partial_b \sigma - \frac{\rho^4}{4\Lambda} g_{(0)}^{ac}g_{cd}^{(2)}g_{(0)}^{db} \partial_b \sigma + \mathcal{O}(\rho^6) .
\end{equation}
Under these residual gauge diffeomorphisms, the unconstrained part of the solution space transforms as 
\begin{equation}
\delta_\xi g_{ab}^{(0)} = \mathcal{L}_{\xi_0} g_{ab}^{(0)} - 2 \sigma g_{ab}^{(0)}
\end{equation} while the constrained part transforms as
\begin{equation}
\delta_\xi g_{ab}^{(2)} = \mathcal{L}_{\xi_0} g_{ab}^{(2)} + \frac{1}{2 \Lambda} \mathcal{L}_{\partial \sigma} g_{ab}^{(0)} ,
\end{equation}
from which one can extract the variation of $T_{ab}$.

\subsection{Bondi gauge in 3d}
\label{sec:Bondi gauge in 3dnew}

We now repeat this analysis for the Bondi gauge and extend the results of \cite{Barnich:2006av, Barnich:2010eb} to asymptotically locally (A)dS$_3$ space-times by including the boundary metric in the solution space.

\subsubsection*{Definition}

In the Bondi gauge \eqref{Bondi gauge}, the metric is given by
\begin{equation}
\D s^2 = \frac{V}{r} e^{2\beta} \D u^2 - 2 e^{2 \beta} \D u \D r + r^2 e^{2 \varphi} (\D \phi - U \D u)^2 ,
\label{Bondi metric}
\end{equation} with coordinates $(u, r, \phi)$. In this expression, $V$, $\beta$ and $U$ are functions of $(u, r, \phi)$, and $\varphi$ is a function of $(u,\phi)$. The three gauge fixing conditions are 
\begin{equation}
g_{rr} = 0, \quad g_{r \phi} = 0, \quad g_{\phi \phi} = r^2  e^{2 \varphi} .
\label{Bondi gauge fixing}
\end{equation} Note that $g_{\phi\phi} =  r^2 e^{2 \varphi}$ is the unique solution of the determinant condition
\begin{equation}
\partial_r \left(\frac{g_{\phi\phi}}{r^2} \right) =0,
\end{equation} which can be generalized to define the Bondi gauge in higher dimensions (see \eqref{Bondi gauge} and appendix \ref{app:detcondBondi}). 

The residual gauge diffeomorphisms $\xi$ preserving the Bondi gauge fixing \eqref{Bondi gauge fixing} have to satisfy the three conditions
\begin{equation}
\mathcal{L}_\xi g_{rr} = 0, \quad \mathcal{L}_\xi g_{r \phi} = 0, \quad g^{\phi\phi} \mathcal{L}_\xi g_{\phi \phi} = 2 \omega (u,\phi) .
\end{equation} The explicit solution of these equations is given by 
\beqn
\xi^u &=&  f, \label{resB1}\\
\xi^\phi &=& Y - \partial_\phi f\, e^{-2 \varphi} \int_r^{+\infty} \frac{\D r'}{{r'}^2} e^{2\beta} ,\label{resB2}\\
\xi^r &=& - r [ \partial_\phi \xi^\phi - \omega - U \partial_\phi f + \xi^\phi \partial_\phi \varphi + f \partial_u \varphi ],
\label{resB3}
\eeqn where $f(u,\phi)$, $Y(u,\phi)$ and $\omega (u,\phi)$ are arbitrary functions of $(u,\phi)$. 

\subsubsection*{Solution space}

In this section, we discuss the most general solution space for three-dimensional general relativity in Bondi gauge. This analysis is new and generalizes the results of \cite{Barnich:2010eb}. Interestingly, we do not have to impose any preliminary boundary condition here. This is in contrast with the procedure followed in the Fefferman-Graham gauge. Therefore, in three dimensions, the gauge conditions \eqref{Bondi gauge fixing} are to some extent stronger than those imposed to define the Fefferman-Graham gauge \eqref{gfFG fixing}. 

First we impose the Einstein equations leading to the metric radial constraints. Solving $G_{rr} + \Lambda g_{rr} = R_{rr} = 0$ gives
\begin{equation}
\beta = \beta_0 (u, \phi).
\end{equation} The equation $G_{r\phi} + \Lambda g_{r\phi} = R_{r\phi} = 0$ leads to
\begin{equation}
U= U_0(u, \phi) + \frac{1}{r}  2 e^{2\beta_0} e^{-2 \varphi} \partial_\phi \beta_0  - \frac{1}{r^2} e^{2\beta_0} e^{-2 \varphi} N(u, \phi).
\end{equation} Eventually, $G_{u r} + \Lambda g_{ur} = 0$ gives
\begin{equation}
\frac{V}{r} = \Lambda r^2 e^{2 \beta_0} - 2 r (\partial_u \varphi + D_\phi U_0 ) + M (u, \phi) + \frac{1}{r} 4 e^{2 \beta_0} e^{-2 \varphi} N \partial_\phi \beta_0 - \frac{1}{r^2} e^{2 \beta_0} e^{-2 \varphi} N^2,
\end{equation} where $D_\phi U_0 = \partial_\phi U_0 + \partial_\phi \varphi U_0$. Taking into account the previous results, the Einstein equation $G_{\phi\phi} + \Lambda g_{\phi \phi} =0$ is automatically satisfied at all orders. 

We now solve the Einstein equations to get the time evolution constraints on $M$ and $N$. The equation $G_{u\phi} + \Lambda g_{u\phi} =0$ returns
\begin{align}
(\partial_u + \partial_u \varphi ) N =& \left(\frac{1}{2} \partial_\phi + \partial_\phi \beta_0 \right) M -2 N \partial_\phi U_0 - U_0 (\partial_\phi N + N \partial_\phi \varphi )\nonumber\\
& + 4 e^{2 \beta_0-2 \varphi} [2 (\partial_\phi \beta_0)^3 - (\partial_\phi \varphi) (\partial_\phi \beta_0)^2 + (\partial_\phi \beta_0 ) (\partial_\phi^2 \beta_0) ].
\end{align}

Moreover, $G_{uu} + \Lambda g_{uu} = 0$ imposes
\begin{align}
\partial_u M =~& (- 2 \partial_u \varphi + 2 \partial_u \beta_0-2\pa_\phi U_0+U_0 2\partial_\phi \beta_0- U_0 2\partial_\phi \varphi-U_0\partial_\phi)  M \nonumber \\
&- 2 \Lambda e^{4 \beta_0-2 \varphi} [\partial_\phi N + N (4 \partial_\phi \beta_0 - \partial_\phi \varphi ) ] \nonumber\\
&-2 e^{2 \beta_0-2 \varphi} \{\partial_\phi U_0 [8 (\partial_\phi \beta_0)^2 - 4 \partial_\phi \beta_0 \partial_\phi \varphi + (\partial_\phi \varphi)^2 + 4 \partial_\phi^2 \beta_0 - 2 \partial_\phi^2 \varphi ] -\partial_\phi^3 U_0 \nonumber\\
& + U_0 [ \partial_\phi \beta_0 (8 \partial_\phi^2 \beta_0 - 2 \partial_\phi^2 \varphi ) + \partial_\phi \varphi (- 2 \partial_\phi^2 \beta_0 + \partial_\phi^2 \varphi ) + 2 \partial_\phi^3 \beta_0 - \partial_\phi^3 \varphi ] \nonumber\\
&+ 2 \partial_u \partial_\phi \beta_0 (4 \partial_\phi \beta_0 - \partial_\phi \varphi ) + \partial_u \partial_\phi \varphi (-2 \partial_\phi \beta_0 + \partial_\phi \varphi ) + 2 \partial_u \partial_\phi^2 \beta_0 - \partial_u \partial_\phi^2 \varphi \}
\end{align}

The solution space is thus characterized by five arbitrary functions of $(u, \phi)$, given by $\beta_0$, $U_0$, $M$, $N$, $\varphi$, with two dynamical constraints expressing the time evolution of $M$ and $N$. This counting argument is in agreement with the results obtained by solving the Einstein equations in the Fefferman-Graham gauge. The precise matching between the two solution spaces is established in section \ref{Fefferman-Graham and Bondi}. 

\subsubsection*{Variation of the solution space}

The residual gauge diffeomorphisms (\ref{resB1}--\ref{resB3}) evaluated on-shell are given by
\begin{align}
\xi^u =&~ f,\label{resBON1}\\
\xi^\phi =&~ Y - \frac{1}{r} \partial_\phi f\, e^{2\beta_0-2\varphi} ,\label{resBON2}\\
\xi^r =&~ - r [ \partial_\phi Y - \omega - U_0 \partial_\phi f + Y \partial_\phi \varphi + f \partial_u \varphi ] \nonumber\\
&+ e^{2\beta_0 - 2 \varphi} (\partial_\phi^2 f - \partial_\phi f \partial_\phi \varphi + 4 \partial_\phi f \partial_\phi \beta_0) - \frac{1}{r} e^{2\beta_0 - 2 \varphi}  \partial_\phi f\, N .\label{resBON3}
\end{align}
Under these residual gauge diffeomorphisms, the unconstrained part of the solution space transforms as
\begin{align}
\delta_\xi \varphi =&~ \omega, \label{var1}\\
\delta_\xi \beta_0 =&~ (f \partial_u + Y \partial_\phi)\beta_0 + \left(\frac{1}{2}\partial_u - \frac{1}{2} \partial_u \varphi + U_0 \partial_\phi \right) f - \frac{1}{2}(\partial_\phi Y + Y \partial_\phi \varphi - \omega ), \\
\delta_\xi U_0 =&~ (f \partial_u + Y \partial_\phi - \partial_\phi Y ) U_0 - \left(\partial_u Y + \Lambda e^{4 \beta_0} e^{-2 \varphi} \partial_\phi f\right) + U_0 (\partial_u f +  U_0 \partial_\phi f) ,
\end{align}
while the constrained part transforms as
\begin{align}
\delta_\xi N =&~  (f \partial_u + Y \partial_\phi + 2 \partial_\phi Y +f \partial_u \varphi+  Y \partial_\phi \varphi -\omega-2 U_0 \partial_\phi f )N \nonumber \\ &+ M \partial_\phi f -e^{2\beta_0-2\varphi}[3\partial_\phi^2 f (2\partial_\phi \beta_0 - \partial_\phi \varphi)+ \partial_\phi^3 f\nonumber \\
& + \partial_\phi f( 4 (\partial_\phi \beta_0)^2 - 8 \partial_\phi \beta_0 \partial_\phi \varphi + 2 (\partial_\phi \varphi)^2 + 2 \partial_\phi^2 \beta_0 -\partial_\phi^2 \varphi)  ], \\
\delta_\xi M =&~ -4\Lambda \pa_\phi f e^{4\beta_0-2 \varphi}N+(f\pa_u+ \pa_u f+f \pa_u\varphi+ Y \partial_\phi  +\pa_\phi Y+Y\pa_\phi\varphi-\omega)M \nonumber  \\ &-2e^{2\beta_0-2\varphi}\Big[2\pa_\phi^2f\pa_u\beta_0+4\pa_u\pa_\phi f\pa_\phi\beta_0+\pa_u\pa_\phi^2 f +\pa_\phi^2f\pa_\phi U_0+8\pa_\phi^2 f\pa_\phi \beta_0 U_0 \nonumber \\
&+\pa_\phi f\Big((4\pa_\phi\beta_0-\pa_\phi\varphi)(2\pa_u \beta_0-\pa_u \varphi)+4\pa_u\pa_\phi\beta_0+\pa_\phi U_0(8\pa_\phi\beta_0-2\pa_\phi\varphi)\nonumber\\&-\pa_\phi^2 U_0-2\pa_u\pa_\phi\varphi
+U_0(-4\pa_\phi\beta_0\pa_\phi\varphi+8(\pa_\phi\beta_0)^2+4\pa_\phi^2\beta_0+(\pa_\phi\varphi)^2-2\pa_\phi^2\varphi)\Big) \nonumber \\ &-2\pa_\phi^2 f U_0\pa_\phi\varphi+\pa_\phi^3 fU_0 -\pa_u\pa_\phi f\pa_\phi\varphi-\pa_\phi^2 f\pa_u\varphi-2f\pa_\phi\beta_0\pa_u\pa_\phi\varphi \nonumber\\&+2\pa_\phi\beta_0\pa_\phi\omega-2\pa_\phi\beta_0\pa_\phi Y\pa_\phi\varphi-2\pa_\phi\beta_0\pa_\phi^2 Y-2\pa_\phi \beta_0 Y\pa_\phi^2\varphi\Big]. \label{var2}
\end{align}
These are the most general variations of the solution space in the Bondi gauge. They are key ingredients in the computation of the asymptotic charge algebra.

\subsection{Gauge matching}
\label{Fefferman-Graham and Bondi}

In this section, we perform an explicit diffeomorphism to go from the Bondi gauge to the Fefferman-Graham gauge (we refer to appendix \ref{app:chgt} for the four-dimensional analogous discussion). This will enable us to identify the most general solution spaces obtained separately in the two gauges. We proceed in two stages.

First, we pass from Bondi to tortoise coordinates $(u,r,\phi) \to (t_\star,r_\star,\phi_\star)$, where
\begin{equation}
\begin{split}
u &\to t_\star -r_\star, \quad \phi \to \phi_\star, \\
r &\to     \left\{
    \begin{array}{ll}
        -\frac{1}{\sqrt{-\Lambda}} \cot \left( r_\star \sqrt{-\Lambda}\right) & \mbox{if } \Lambda < 0 \\
       \frac{1}{\sqrt{\Lambda}} \text{coth} \left( r_\star \sqrt{\Lambda} \right)  & \mbox{if } \Lambda > 0
    \end{array}
\right.     .
\end{split}
\end{equation}
Second, we go from tortoise to Fefferman-Graham performing the coordinates transformation $(t_\star,r_\star,\phi_\star) \to (\rho,t,\phi)$, with
\beqn
t_\star &=& t + T_1(t,\phi) \rho +  T_2(t,\phi) \rho^2 + T_3 (t, \phi) \rho^3 + \mathcal{O}(\rho^4), \label{diffeo Bondi FG 1} \\
r_\star &=& R_1(t,\phi) \rho + R_2(t,\phi)\rho^2 + R_3(t, \phi) \rho^3 + \mathcal{O}(\rho^4),\\
\phi_\star &=& \phi + Z_1(t,\phi) \rho + Z_2 (t,\phi) \rho^2 + Z_3 (t, \phi) \rho^3 + \mathcal{O}(\rho^4).
\label{diffeo Bondi FG 2}
\eeqn 
The explicit form of the functions $T_i (t,\phi)$, $R_i(t,\phi)$ and $Z_i (t,\phi)$ ($i=1,2,3$) can be worked out explicitly. For the sake of brevity, we report here only the leading orders
\begin{align}
R_1(t, \phi) =&~ \frac{1}{\Lambda},\\
R_2 (t, \phi) =&~ - e^{-2\beta_0}\frac{1}{\Lambda^2}(\partial_\phi U_0 + U_0 \partial_\phi \varphi + \partial_t \varphi ), \\
& \nonumber \\
T_1(t,\phi) =&~ -\frac{1}{\Lambda} (1 - e^{-2\beta_0}), \\
T_2(t,\phi) =&~ -e^{-4\beta_0} \frac{1}{\Lambda^2} [e^{2\beta_0}\partial_\phi U_0 + U_0 (\partial_\phi\beta_0 + e^{2\beta_0} \partial_\phi \varphi) + \partial_t \beta_0 + e^{2\beta_0} \partial_t \varphi ], \\
& \nonumber \\
Z_1 (t,\phi) =&~  \frac{1}{\Lambda} e^{-2\beta_0} U_0, \\
Z_2 (t,\phi) =&~ -\frac{1}{2 \Lambda} \Big[2 e^{-2\varphi} \partial_\phi \beta_0 + \frac{2}{\Lambda} e^{-4\beta_0} U_0^2 \partial_\phi \beta_0 \nonumber \\ &\qquad\qquad\qquad- \frac{1}{\Lambda} e^{-4\beta_0} \partial_t U_0- \frac{1}{\Lambda} e^{-4\beta_0}  U_0 (\partial_\phi U_0 - 2 \partial_t \beta_0)\Big].
\end{align}

\subsubsection*{Solution space matching}

In this subsection, we use the notation $\Lambda = -1/ \ell^2$ for compactness of the expressions ($\ell \in \mathbb{R}$ if $\Lambda <0$ and $i \ell \in \mathbb{R}$ if $\Lambda > 0$). Using the diffeomorphism (\ref{diffeo Bondi FG 1}-\ref{diffeo Bondi FG 2}), the solution space of the Fefferman-Graham gauge (left-hand side) is related to the solution space of the Bondi gauge (right-hand side) through 
\begin{equation}
g^{(0)}_{ab} = \begin{pmatrix}
-\frac{e^{4\beta_0}}{\ell^2}+ e^{2\varphi} U_0^2 &-e^{2\varphi} U_0 \\
-e^{2\varphi} U_0 &e^{2\varphi} 
\end{pmatrix} 
\label{boundary metric in 3d}
\end{equation} and 
\begin{align}
T_{tt} =&~ \frac{1}{16\pi G \ell} e^{-4\beta_0 - 2 \varphi} \{4e^{8 \beta_0} [2 (\partial_\phi \beta_0)^2 - \partial_\phi \beta_0 \partial_\phi \varphi + \partial_\phi^2 \beta_0 ] + e^{4 \beta_0 + 2 \varphi}[ e^{2 \beta_0} (M-4 N U_0 )\nonumber\\
&- \ell^2 ( (\partial_\phi U_0)^2 + U_0^2 (-8 \partial_\phi \beta_0 \partial_\phi \varphi + (\partial_\phi \varphi)^2 + 4 \partial_\phi^2 \varphi)+ (\partial_t \varphi)^2 \nonumber \\ 
&+ 2 \partial_\phi U_0 (U_0 (-4 \partial_\phi \beta_0 + 3 \partial_\phi \varphi )+ \partial_t \varphi) + 2 U_0 (2 \partial_\phi^2 U_0 + (-4 \partial_\phi\beta_0 + \partial_\phi \varphi) \partial_t \varphi \nonumber\\
&+ 2 \partial_t \partial_\phi \varphi ))]+ e^{4 \varphi} \ell^2 U_0^2 [ e^{2\beta_0} M + \ell^2 ( (\partial_\phi U_0)^2 +U_0^2 (- 4 \partial_\phi \beta_0 \partial_\phi \varphi + (\partial_\phi \varphi)^2 + 2 \partial_\phi^2 \varphi ) \nonumber\\
&+2 \partial_\phi \varphi \partial_t U_0 + \partial_t \varphi (-4 \partial_t \beta_0  + \partial_t \varphi) + 2 \partial_\phi U_0 (2 U_0 (- \partial_\phi \beta_0 + \partial_\phi \varphi ) \nonumber \\
&- 2 \partial_t \beta_0+ \partial_t \varphi ) + 2 U_0 (\partial_\phi^2 U_0 - 2 \partial_\phi \beta_0 \partial_t \varphi + \partial_\phi \varphi ( -2 \partial_t \beta_0 + \partial_t \varphi ) + 2 \partial_t \partial_\phi \varphi) \nonumber \\ &+ 2 (\partial_t \partial_\phi U_0 + \partial_t^2 \varphi ) )] \} ,\\
T_{t \phi} =&~ \frac{1}{16\pi G \ell}e^{-4\beta_0}\{ 2e^{6 \beta_0} N - 2 e^{4\beta_0} \ell^2 [\partial_\phi U_0 (2 \partial_\phi \beta_0 - \partial_\phi \varphi) - \partial_\phi^2 U_0 \nonumber \\
&+ U_0 (2 \partial_\phi \beta_0 \partial_\phi \varphi - \partial_\phi^2 \varphi ) + 2 \partial_\phi \beta_0  \partial_t \varphi - \partial_t \partial_\phi \varphi ]   \nonumber \\
&+ e^{2 \varphi} \ell^2 U_0 [ - e^{2 \beta_0} M - \ell^2 ( (\partial_\phi U_0)^2 + U_0^2 (-4\partial_\phi \beta_0 \partial_\phi \varphi + (\partial_\phi \varphi )^2 + 2 \partial^2_\phi \varphi) \nonumber\\
&+ 2 \partial_\phi \varphi \partial_t U_0 + \partial_t \varphi (-4 \partial_t \beta_0 + \partial_t \varphi ) + 2 \partial_\phi U_0 ( 2 U_0 (-\partial_\phi \beta_0 + \partial_\phi \varphi ) -2 \partial_t \beta_0 + \partial_t \varphi) \nonumber\\
&+ 2 U_0 (\partial_\phi^2 U_0 - 2 \partial_\phi \beta_0 \partial_t \varphi + \partial_\phi \varphi ( - 2 \partial_t \beta_0 + \partial_t \varphi) +2 \partial_t \partial_\phi \varphi) + 2 (\partial_t \partial_\phi U_0 + \partial_t^2 \varphi ))] \} ,\\
T_{\phi \phi} =&~ \frac{1}{16\pi G\ell} e^{-4 \beta_0+ 2 \varphi} \{ e^{2 \beta_0} \ell^2 M + \ell^4 [ (\partial_\phi U_0)^2 + U_0^2 (-4 \partial_\phi \beta_0 \partial_\phi \varphi + (\partial_\phi \varphi)^2 + 2 \partial_\phi^2 \varphi )\nonumber \\
&+ 2 \partial_\phi \varphi \partial_t U_0 + \partial_t \varphi (-4 \partial_t \beta_0 + \partial_t \varphi ) + 2 \partial_\phi U_0 (2 U_0 (-\partial_\phi \beta_0 + \partial_\phi \varphi) - 2 \partial_t \beta_0 + \partial_t \varphi )  \nonumber\\
&+ 2 U_0 (\partial_\phi^2 U_0 - 2 \partial_\phi \beta_0 \partial_t \varphi+\partial_\phi \varphi (-2 \partial_t \beta_0 + \partial_t \varphi) + 2 \partial_t \partial_\phi \varphi ) + 2 (\partial_t \partial_\phi U_0 + \partial_t^2 \varphi )] \}.
\end{align}
One can check on the right-hand side expressions that the trace condition given by the first equation of \eqref{trace condition} is satisfied.

Taking $U_0 = 0$, $\beta_0 = 0$ and $\varphi = \bar{\varphi}$ (three-dimensional analogue of the boundary gauge fixing of \cite{Compere:2019bua}), $T_{ab}$ reduces to\footnote{In term of $\Lambda$, the pre-factor in the right-hand side of \eqref{Tab reduced 3d} is given by $\sqrt{|\Lambda|}/16 \pi G$.}
\begin{equation}
T_{ab} = \frac{1}{16\pi G \ell} \begin{pmatrix}
M - \ell^2 (\partial_t \bar{\varphi})^2 &2N + 2 \ell^2 \partial_t \partial_\phi \bar{\varphi}\\
2N + 2 \ell^2 \partial_t \partial_\phi \bar{\varphi} &e^{2\bar{\varphi}} \ell^2 [M + \ell^2 ((\partial_t \bar{\varphi})^2 + 2 \partial_t^2 \bar{\varphi} )]  
\end{pmatrix} .
\label{Tab reduced 3d}
\end{equation}

\subsubsection*{Residual gauge parameters matching}

Using the diffeomorphism (\ref{diffeo Bondi FG 1}-\ref{diffeo Bondi FG 2}), the parameters of the residual gauge diffeomorphisms of the Fefferman-Graham gauge \eqref{residual gauge diffeomorphisms FG} are related to those of the Bondi gauge (\ref{resB1}--\ref{resB3}) as
\beqn
\xi_0^t &=& f, \\
\xi_0^\phi &=& Y, \\
\sigma &=& \partial_\phi Y - \omega  - U_0 \partial_\phi f + Y \partial_\phi \varphi + f \partial_t \varphi .
\eeqn

\subsection{Flat limit in the Bondi gauge}

In this subsection, we investigate the flat limit of the above results. Notice that the flat limit is well defined in the Bondi gauge, but not in the Fefferman-Graham gauge. This is an illustration of our general statement: the Fefferman-Graham gauge is well adapted for computations in asymptotically (locally) (A)dS spacetime due to its covariance with respect to the boundary structure. However, to consider the flat limit, the results have to be translated into the Bondi gauge, where the computations are more involved but the limit is perfectly defined.  

Let us first study the solution space in the Bondi gauge for vanishing cosmological constant. In three dimensions, the full solution space in the Bondi gauge for vanishing cosmological constant can be readily obtained by taking the flat limit of the solution space obtained in section \ref{sec:Bondi gauge in 3dnew} for non-vanishing cosmological constant. This contrasts with the four-dimensional case where only the analytic part of the solution space is recovered (see the difference between \eqref{parametrization solution space} and \eqref{parametrization solution space 2} and the associated discussion). In practice, we take $\Lambda \to 0$ in the equations. The equation $G_{rr} = 0$ gives
\begin{equation}
\beta = \beta_0 (u, \phi)
\end{equation} Solving $G_{r\phi} = 0$ leads to
\begin{equation}
U= U_0(u, \phi) + \frac{1}{r}  2 e^{2\beta_0} e^{-2 \varphi} \partial_\phi \beta_0  - \frac{1}{r^2} e^{2\beta_0} e^{-2 \varphi} N(u, \phi) .
\end{equation} Solving $G_{u r} = 0$ gives
\begin{equation}
\frac{V}{r} =  - 2 r (\partial_u \varphi + D_\phi U_0 ) + M (u, \phi) + \frac{1}{r} 4 e^{2 \beta_0} e^{-2 \varphi} N \partial_\phi \beta_0 - \frac{1}{r^2} e^{2 \beta_0} e^{-2 \varphi} N^2 ,
\end{equation} where $D_\phi U_0 = \partial_\phi U_0 + \partial_\phi \varphi U_0$. Taking the previous results into account, the Einstein equation $G_{\phi\phi} =0$ is satisfied at all orders. Finally, we solve the Einstein equations giving the time evolution constraints on $M$ and $N$. The equation $G_{u\phi} =0$ gives
\begin{align}
(\partial_u + \partial_u \varphi ) N =& \left(\frac{1}{2} \partial_\phi + \partial_\phi \beta_0 \right) M -2 N \partial_\phi U_0 - U_0 (\partial_\phi N + N \partial_\phi \varphi )\nonumber \\
&+ 4 e^{2 \beta_0} e^{-2 \varphi} [2 (\partial_\phi \beta_0)^3 - (\partial_\phi \varphi) (\partial_\phi \beta_0)^2 + (\partial_\phi \beta_0 ) (\partial_\phi^2 \beta_0) ] ,
\end{align}
whereas $G_{uu}= 0$ results in 
\begin{align}
\partial_u M =~& (- 2 \partial_u \varphi + 2 \partial_u \beta_0-2\pa_\phi U_0+U_0 2\partial_\phi \beta_0- U_0 2\partial_\phi \varphi-U_0\partial_\phi)  M \nonumber \\
&-2 e^{2 \beta_0-2 \varphi} \{\partial_\phi U_0 [8 (\partial_\phi \beta_0)^2 - 4 \partial_\phi \beta_0 \partial_\phi \varphi + (\partial_\phi \varphi)^2 + 4 \partial_\phi^2 \beta_0 - 2 \partial_\phi^2 \varphi ] -\partial_\phi^3 U_0 \nonumber\\
& + U_0 [ \partial_\phi \beta_0 (8 \partial_\phi^2 \beta_0 - 2 \partial_\phi^2 \varphi ) + \partial_\phi \varphi (- 2 \partial_\phi^2 \beta_0 + \partial_\phi^2 \varphi ) + 2 \partial_\phi^3 \beta_0 - \partial_\phi^3 \varphi ] \nonumber\\
&+ 2 \partial_u \partial_\phi \beta_0 (4 \partial_\phi \beta_0 - \partial_\phi \varphi ) + \partial_u \partial_\phi \varphi (-2 \partial_\phi \beta_0 + \partial_\phi \varphi ) + 2 \partial_u \partial_\phi^2 \beta_0 - \partial_u \partial_\phi^2 \varphi \}
\end{align}
 The solution space is thus characterized by five arbitrary functions of $(u, \phi)$, given by $\beta_0$, $U_0$, $M$, $N$, $\varphi$, with two dynamical constraints given by the time evolution equations of $M$ and $N$. 

Through a similar procedure, the on-shell residual gauge diffeomorphisms and the variations of the solution space are obtained by taking $\Lambda \to 0$ in the expressions (\ref{resBON1}-\ref{resBON3}) and (\ref{var1}-\ref{var2}), respectively. 
The on-shell residual gauge diffeomorphisms are given by
\begin{align}
\xi^u =&~ f,\label{resBON1flat}\\
\xi^\phi =&~ Y - \frac{1}{r} \partial_\phi f\, e^{2\beta_0-2\varphi} ,\label{resBON2flat}\\
\xi^r =&~ - r [ \partial_\phi Y - \omega - U_0 \partial_\phi f + Y \partial_\phi \varphi + f \partial_u \varphi ] \nonumber\\
&+ e^{2\beta_0 - 2 \varphi} (\partial_\phi^2 f - \partial_\phi f \partial_\phi \varphi + 4 \partial_\phi f \partial_\phi \beta_0) - \frac{1}{r} e^{2\beta_0 - 2 \varphi}  \partial_\phi f\, N .\label{resBON3flat}
\end{align}
Under these residual gauge diffeomorphisms, the unconstrained part of the solution space transforms as
\begin{align}
\delta_\xi \varphi =&~ \omega, \\
\delta_\xi \beta_0 =&~ (f \partial_u + Y \partial_\phi)\beta_0 + \left(\frac{1}{2}\partial_u - \frac{1}{2} \partial_u \varphi + U_0 \partial_\phi \right) f - \frac{1}{2}(\partial_\phi Y + Y \partial_\phi \varphi - \omega ), \\
\delta_\xi U_0 =&~ (f \partial_u + Y \partial_\phi - \partial_\phi Y ) U_0 - \left(\partial_u Y + \Lambda e^{4 \beta_0} e^{-2 \varphi} \partial_\phi f\right) + U_0 (\partial_u f +  U_0 \partial_\phi f) ,
\end{align}
while the constrained part transforms as
\begin{align}
\delta_\xi N =&~  (f \partial_u + Y \partial_\phi + 2 \partial_\phi Y +f \partial_u \varphi+  Y \partial_\phi \varphi -\omega-2 U_0 \partial_\phi f )N \nonumber \\ &+ M \partial_\phi f -e^{2\beta_0-2\varphi}[3\partial_\phi^2 f (2\partial_\phi \beta_0 - \partial_\phi \varphi)+ \partial_\phi^3 f\nonumber \\
& + \partial_\phi f( 4 (\partial_\phi \beta_0)^2 - 8 \partial_\phi \beta_0 \partial_\phi \varphi + 2 (\partial_\phi \varphi)^2 + 2 \partial_\phi^2 \beta_0 -\partial_\phi^2 \varphi)  ], \\
\delta_\xi M =&~ (f\pa_u+ \pa_u f+f \pa_u\varphi+ Y \partial_\phi  +\pa_\phi Y+Y\pa_\phi\varphi-\omega)M \nonumber  \\ &-2e^{2\beta_0-2\varphi}\Big[2\pa_\phi^2f\pa_u\beta_0+4\pa_u\pa_\phi f\pa_\phi\beta_0+\pa_u\pa_\phi^2 f +\pa_\phi^2f\pa_\phi U_0+8\pa_\phi^2 f\pa_\phi \beta_0 U_0 \nonumber \\
&+\pa_\phi f\Big((4\pa_\phi\beta_0-\pa_\phi\varphi)(2\pa_u \beta_0-\pa_u \varphi)+4\pa_u\pa_\phi\beta_0+\pa_\phi U_0(8\pa_\phi\beta_0-2\pa_\phi\varphi)\nonumber\\&-\pa_\phi^2 U_0-2\pa_u\pa_\phi\varphi
+U_0(-4\pa_\phi\beta_0\pa_\phi\varphi+8(\pa_\phi\beta_0)^2+4\pa_\phi^2\beta_0+(\pa_\phi\varphi)^2-2\pa_\phi^2\varphi)\Big) \nonumber \\ &-2\pa_\phi^2 f U_0\pa_\phi\varphi+\pa_\phi^3 fU_0 -\pa_u\pa_\phi f\pa_\phi\varphi-\pa_\phi^2 f\pa_u\varphi-2f\pa_\phi\beta_0\pa_u\pa_\phi\varphi \nonumber\\&+2\pa_\phi\beta_0\pa_\phi\omega-2\pa_\phi\beta_0\pa_\phi Y\pa_\phi\varphi-2\pa_\phi\beta_0\pa_\phi^2 Y-2\pa_\phi \beta_0 Y\pa_\phi^2\varphi\Big]. 
\end{align}

\subsection{From asymptotically AdS$_3$ to asymptotically flat spacetime}
\label{From asymptotically AdS to asymptotically flat spacetime}

Here, we discuss how the $\mathfrak{bms}_3$ algebra can be obtained by taking the flat limit of $\mathfrak{diff}(S^1) \oplus \mathfrak{diff}(S^1)$ in asymptotically AdS$_3$ spacetime. 

In asymptotically locally AdS$_3$ spacetime, the Dirichlet boundary conditions are obtained in the Fefferman-Graham gauge by imposing
\begin{equation}
g^{(0)}_{ab} \D x^a \D x^b = \Lambda \D t^2 + \D \phi^2
\label{bound gauge fixing 3d}
\end{equation}  (see definition (AAdS2) given in equation \eqref{BC Dirichlet}). It is a well-known result that the asymptotic symmetry algebra associated with Dirichlet boundary conditions is given by the direct sum between two copies of the Witt algebra, $\mathfrak{diff}(S^1) \oplus \mathfrak{diff}(S^1)$ \cite{Brown:1986nw}. The corresponding surface charges are finite, integrable and form a representation of $\mathfrak{diff}(S^1) \oplus \mathfrak{diff}(S^1)$ with a central extension involving the Brown-Henneaux central charge $c = \frac{3}{2 G |\Lambda|}$.  

The analogous boundary conditions of \eqref{bound gauge fixing 3d} in Bondi gauge can be readily obtained from \eqref{boundary metric in 3d} and are explicitly given by
\begin{equation}
\beta_0 =0 , \qquad U_0 = 0, \qquad \varphi = 0
\label{Dirichltet Bondi suite}
\end{equation}  (see equation \eqref{BC Dirichlet Bondi}). The residual gauge diffeomorphisms preserving these constraints are given by (\ref{resBON1}-\ref{resBON3}), where the parameters satisfy
\begin{equation}
\partial_u f =  \partial_\phi Y , \qquad \partial_u Y = - \Lambda \partial_\phi f , \qquad \omega = 0.
\label{constraint equations Bond Dir}
\end{equation} We express $f$ and $Y$ as $f = \frac{1}{\Lambda} (Y^+ + Y^- )$, $Y = \frac{1}{2} (Y^+ - Y^- )$, where $x^\pm = - \Lambda u \pm \phi$ and $Y^\pm = Y^\pm (x^\pm)$ \cite{Barnich:2012aw}. Using the modified Lie bracket \eqref{modified lie bracket gravity}, one can show that $[\xi (Y^\pm_1) , \xi (Y^\pm_2)]_A = \xi (\hat{Y}^\pm)$, where
\begin{equation}
\hat{Y}^\pm = Y^\pm_1 \partial_\pm Y^\pm_2 - Y^\pm_2 \partial_\pm Y^\pm_1 , 
\label{rel int}
\end{equation} which corresponds to $\mathfrak{diff}(S^1) \oplus \mathfrak{diff}(S^1)$, as it should. Furthermore, on the cylinder, we can expand $Y^\pm$ as $Y^\pm (x^\pm)  = \sum_{m\in \mathbb{Z}} Y_m^\pm e^{i m x^\pm}$, with $\bar{Y}_m^\pm = Y^\pm_{-m}$. Writing $l^+_m = \xi (Y^+ = e^{i m x^+}, Y^- = 0)$ and $l^-_m = \xi (Y^+ =0, Y^- = e^{i m x^-})$, the commutation relations \eqref{rel int} become
\begin{equation}
i[l_m^\pm , l_n^\pm ]_A = (m- n ) l_{m+n}^\pm, \qquad [l_m^\pm, l_n^\mp ]_A = 0 .
\end{equation} 

The flat limit $\Lambda \to 0$ can be readily taken in the Bondi gauge. In this context, the boundary conditions \eqref{Dirichltet Bondi suite} become asymptotically flat boundary conditions (AF3) (equation \eqref{asymp flat 1} together with \eqref{asymp flat 3}). The constraint equations \eqref{constraint equations Bond Dir} reduce to 
\begin{equation}
\partial_u f =  \partial_\phi Y , \qquad \partial_u Y = 0 \qquad \omega = 0.
\end{equation} These equations can be solved as
\begin{equation}
f = T + u \partial_\phi Y, \qquad Y = Y(\phi)
\end{equation} where $T= T(\phi)$ and $Y= Y(\phi)$ are the parameters of supertranslations and superrotations, respectively. Using the modified Lie bracket \eqref{modified lie bracket gravity}, one can show that $[\xi (T_1, Y_1), \xi (T_2, Y_2) ]_A = \xi (\hat{T}, \hat{Y})$, where
\begin{equation}
\begin{split}
\hat{T} &= Y_1 \partial_\phi T_2 + T_1 \partial_\phi Y_2 - Y_2 \partial_\phi T_1 - T_2 \partial_\phi Y_1, \\
\hat{Y} &= Y_1 \partial_\phi Y_2 - Y_2 \partial_\phi Y_1,
\end{split}
\label{rel int 2}
\end{equation} which corresponds to the algebra $\mathfrak{bms}_3 = \mathfrak{diff}(S^1) \loplus_{\text{ad}} \mathfrak{vect}(S^1)$. Expanding $T$ and $Y$ on the circle as $T (\phi) = \sum_{m\in \mathbb{Z}} T_m e^{im \phi}$ and $Y(\phi) = \sum_{m\in \mathbb{Z}} Y_m e^{im \phi}$, and writing $P_m = \xi(T =  e^{im \phi}, Y= 0)$ and $J_m = \xi(T = 0, Y =  e^{im \phi})$, the commutation relations \eqref{rel int 2} reduce to 
\begin{equation}
i[J_m , J_n]_A = (m-n) J_{m+n}, \qquad [P_m, P_n]_A = 0, \qquad i[J_m, P_n]_A = (m-n) P_{m+n} .
\end{equation}

\section{Bondi and Fefferman-Graham gauges in four dimensions}
\label{Relation between Bondi and Fefferman-Graham gauges in four dimensions}
\chaptermark{Bondi and Fefferman-Graham gauges in 4d}

In this section, we repeat the analysis performed in the previous section but for the four-dimensional case. After briefly introducing the Fefferman-Graham gauge in the four-dimensional case, we focus on the Bondi gauge where we derive the most general solution space in asymptotically locally (A)dS$_4$ spacetime (these results were already summarized in an example of subsection \ref{Solution space}). As in the three-dimensional case, the results that we obtain here in four dimensions generalize previous considerations (see e.g. \cite{Barnich:2012aw , Barnich:2013sxa}) by allowing an arbitrary boundary structure encoding the notion of asymptotically locally (A)dS$_4$ spacetime. We also briefly discuss the flat limit of these new results, which allows to find the solution space described in \eqref{parametrization solution space}. Notice that the flat limit process in four dimensions is more subtle and requires a prescription in order to get the right results. Furthermore, we construct the explicit diffeomorphism that maps the Bondi to the Fefferman-Graham gauge, which is the second main result of this section. We finish this section by defining a new set of boundary conditions in asymptotically locally (A)dS$_4$ spacetime whose asymptotic symmetry algebra is the $\Lambda$-BMS$_4$ algebra, written $\mathfrak{bms}^\Lambda_4$. The latter is infinite-dimensional and reduces to $\mathfrak{bms}^\text{gen}_4$ in the flat limit, which was studied in chapter \ref{Generalized BMS and renormalized phase space}. This is the third main result of this section.

\subsection{Fefferman-Graham gauge in 4d}
\label{subsec:Fefferman-Graham gauge in 4d}

\subsubsection{Definition}

We particularize to the four-dimensional case the results discussed in section \ref{Asymptotic symmetries in the gauge fixing approach}. The Fefferman-Graham metric is given by
\begin{equation}
ds^2 =- \frac{3}{\Lambda}\frac{\D \rho^2}{\rho^2} + \gamma_{ab}(\rho,x^c) \D x^a \D x^b \label{FG gauge prime}
\end{equation} (see \eqref{FG gauge}). The infinitesimal diffeomorphisms preserving the Fefferman-Graham gauge are generated by vector fields $\xi^\mu$ satisfying
$\mathcal{L}_\xi g_{\rho \rho} = 0$, $\mathcal{L}_\xi g_{\rho a} = 0$. The first condition leads to the equation $\partial_\rho \xi^\rho = \frac{1}{\rho} \xi^\rho$, which can be solved for $\xi^\rho$ as 
\begin{eqnarray}
\xi^\rho = \sigma (x^a) \rho. \label{AKV 1 prime}
\end{eqnarray}
The second condition leads to the equation $\rho^2 \gamma_{ab} \partial_\rho \xi^b - \frac{3}{\Lambda}\partial_a \xi^\rho =0$, which can be solved for $\xi^a$ as 
\begin{equation}
\xi^a = \xi_0^a (x^b) + \frac{3}{\Lambda}\partial_b \sigma \int_0^\rho \frac{\D \rho'}{\rho'} \gamma^{ab}(\rho', x^c).
\label{AKV 2 prime}
\end{equation} 

\subsubsection{Solution space}

Assuming $\gamma_{ab}= \mathcal{O} (\rho^{-2})$ (see \eqref{preliminary FG}), the general asymptotic expansion that solves Einstein's equations is analytic,
\begin{equation}
\gamma_{ab} = \frac{1}{\rho^2} \, g_{ab}^{(0)} + \frac{1}{\rho} \, g_{ab}^{(1)} + g_{ab}^{(2)} + \rho \, g_{ab}^{(3)} + \mathcal{O}(\rho^2)
\label{preliminary FG}
\end{equation} where $g_{ab}^{(i)}$ are arbitrary functions of $x^a = (t, x^A)$. Following the standard holographic dictionary (see e.g. \cite{deHaro:2000vlm}), we call $g_{ab}^{(0)}$ the boundary metric and 
\begin{equation}
T_{ab} = \frac{\sqrt{3|\Lambda|}}{16\pi G} g_{ab}^{(3)}
\end{equation} 
the energy-momentum tensor. Einstein's equations fix $g^{(1)}_{ab} = 0$ and $g^{(2)}_{ab}$ in terms of $g_{ab}^{(0)}$ while all subleading terms in \eqref{preliminary FG} are determined in terms of the free data $g^{(0)}_{ab}$ and $T_{ab}$ satisfying 
\begin{equation}
D_a^{(0)} T^{ab} = 0 , \quad g^{(0)}_{ab} T^{ab} =0.
\label{condition on energy momentum}
\end{equation} 
Here, $D^{(0)}_a$ is the covariant derivative with respect to $g_{ab}^{(0)}$ and indices are raised with the inverse metric $g^{ab}_{(0)}$. 

\subsubsection{Variation of the solution space}

Expanding the residual gauge diffeomorphisms \eqref{AKV 1 prime} and \eqref{AKV 2 prime} in power of $\rho$ yields
\begin{equation}
\begin{split}
\xi^\rho &= \rho \, \sigma(x^a)\\
\xi^a &= \xi^a_{0} + \frac{3}{2\Lambda} \rho^2 g^{ab}_{(0)} \partial_b \sigma - \frac{3}{4\Lambda} \rho^4 g^{ab}_{(2)} \partial_b \sigma - \frac{3}{5\Lambda} \rho^5 g^{ab}_{(3)} \partial_b \sigma + \mathcal{O}(\rho^6).
\end{split} 
\end{equation} The variations of the data parametrizing the solution space under the residual gauge transformations are given by (see also \cite{Papadimitriou:2010as})
\begin{eqnarray}
\delta_\xi g_{ab}^{(0)} &=& \mathcal{L}_{\xi^c_0} g_{ab}^{(0)} - 2 \sigma\, g_{ab}^{(0)},\\
\delta_\xi T_{ab} &=& \mathcal{L}_{\xi^c_0} T_{ab}+ \sigma\, T_{ab}. \label{eq:TransformationTab}
\end{eqnarray}

\subsubsection{Useful expressions}

Here, we collect some useful formulae needed in the process of holographic renormalization below (see section \ref{Holographic renormalization}), especially the coefficients of the Levi-Civita connection in the Fefferman-Graham gauge. 

The inverse $\gamma^{ab}$ of the $3d$ metric
\begin{equation}
\gamma_{ab} = \frac{1}{\rho^2} \Big( g_{ab}^{(0)} + \rho^2\, g_{ab}^{(2)} + \rho^3 \, g_{ab}^{(3)} + \rho^4 \, g_{ab}^{(4)} + \mathcal{O}(\rho^5)  \Big)
\end{equation}
is given by
\begin{equation}
\gamma^{ab} = \rho^2 \Big[ g^{ab}_{(0)} - \rho^2\, g^{ab}_{(2)} - \rho^3 \, g^{ab}_{(3)} + \rho^4 \, (-g_{(4)}^{ab} + g_{(2)}^{ac} g_{cd}^{(2)} g^{db}_{(0)}) + \mathcal{O}(\rho^5)  \Big]
\end{equation}
We denote $g^{ab}_{(i)} \equiv g_{(0)}^{ac} g_{(0)}^{bd} g^{(i)}_{cd} $. Only $g_{(0)}^{ab}$ is the true inverse matrix of $g^{(0)}_{ab}$; the indices of other fields are simply raised with respect to the boundary metric.

The volume form is given by
\begin{equation}
\sqrt{-g} = \sqrt{\frac{3}{|\Lambda|}} \, \frac{1}{\rho} \, \sqrt{|\gamma|}
\label{eq:DetSplit}
\end{equation}
with
\begin{equation}
\begin{split}
\sqrt{|\gamma|} &= \sqrt{|g_{(0)}|} \, \frac{1}{\rho^3} \Big( 1 + \frac{1}{2} g_{(0)}^{ab} g^{(2)}_{ab} \, \rho^2 + \frac{1}{2} g_{(0)}^{ab} g^{(3)}_{ab}\, \rho^3 + \mathcal{O}(\rho^{4})  \Big) \\
&= \sqrt{|g_{(0)}|} \, \frac{1}{\rho^3} \Big( 1 + \rho^2 \frac{3}{8\Lambda} R_{(0)} + \mathcal{O}(\rho^{4})  \Big).
\end{split} \label{eq:DetPwrSeries}
\end{equation}

We compute the Christoffel symbols for \eqref{FG gauge prime}. Gauge conditions imply directly that
\begin{equation}
\Gamma^\rho_{\rho\rho} = \frac{1}{2} g^{\rho\rho} \partial_\rho g_{\rho\rho} = -\frac{1}{\rho}, \quad \Gamma^\rho_{\rho a} = 0, \quad \Gamma^a_{\rho\rho} = 0.
\end{equation}
Using the power series for $\gamma_{ab}$ and its inverse, we get
\begin{equation}
\begin{split}
\Gamma^\rho_{ab} &= -\frac{1}{2} g^{\rho\rho} \partial_\rho \gamma_{ab} = \frac{\Lambda}{6}\rho^2 \partial_\rho \gamma_{ab} \\
&= - \frac{\Lambda}{3} \frac{1}{\rho} g^{(0)}_{ab} + \frac{\Lambda}{6} \rho^2 g^{(3)}_{ab}  + \mathcal{O}(\rho^{3}),
\end{split}
\label{eq:GammaRhoAB}
\end{equation}
and
\begin{equation}
\begin{split}
\Gamma^a_{\rho b} &= \frac{1}{2} \gamma^{ac} \partial_\rho \gamma_{bc} \\
&= -\frac{1}{\rho} \delta^a_b + \rho \, g_{(0)}^{ac} g^{(2)}_{bc} + \frac{3}{2} \, \rho^2 \,  g_{(0)}^{ac} g^{(3)}_{bc}  + \mathcal{O}(\rho^{3}) .
\end{split}
\label{eq:GammaARhoB}
\end{equation}
Finally,
\begin{equation}
\begin{split}
\Gamma^a_{bc} &= \Gamma^{a}_{bc}[\gamma] \\
&= \Gamma^a_{bc}[g_{(0)}] + \frac{1}{2}\rho^2 g_{(0)}^{ad} (D_b^{(0)} g^{(2)}_{dc} + D_c^{(0)} g^{(2)}_{db} - D_d^{(0)} g^{(2)}_{bc}) \\
& \quad + \frac{1}{2}\rho^3 g_{(0)}^{ad} (D_b^{(0)} g^{(3)}_{dc} + D_c^{(0)} g^{(3)}_{db} - D_d^{(0)} g^{(3)}_{bc}) + \mathcal{O}(\rho^4).
\end{split}
\label{eq:Gammaabc}
\end{equation}

\subsection{Bondi gauge in 4d}
\label{sec2}

We now briefly discuss again the Bondi gauge in four dimensions (see section \ref{Asymptotic symmetries in the gauge fixing approach}). Then we provide a full derivation of the most general solution space in four-dimensional asymtotically locally (A)dS$_4$ spacetime. Throughout this analysis, we discuss the flat limit of these results and relate them to those considered in chapter \ref{Generalized BMS and renormalized phase space}. 

\subsubsection{Definition and residual transformations}

The four-dimensional Bondi metric is given by
\begin{equation}
ds^2 = e^{2\beta} \frac{V}{r} \D u^2 - 2 e^{2\beta}\D u \D r + g_{AB} (\D x^A - U^A \D u)(\D x^B - U^B \D u)
\label{bondi gauge bis}
\end{equation} where $\beta$, $U^A$, $g_{AB}$ and $V$ are arbitrary functions of the coordinates. The $2$-dimensional metric $g_{AB}$ satisfies the determinant condition
\begin{equation}
\partial_r \left(\frac{\det (g_{AB})}{r^4} \right) = 0  \label{eq:DetCond} .
\end{equation} 
Any metric can be written in this gauge. For example, global (A)dS$_4$ is obtained by choosing $\beta = 0$, $U^A = 0$, $V/r = (\Lambda r^2/3)-1$, $g_{AB} = r^2 \mathring q_{AB}$, where $\mathring q_{AB}$ is the unit round-sphere metric. 

Infinitesimal diffeomorphisms preserving the Bondi gauge are generated by vector fields $\xi^\mu$ satisfying
\begin{equation}
\mathcal{L}_\xi g_{rr} = 0, \quad \mathcal{L}_\xi g_{rA} = 0, \quad g^{AB} \mathcal{L}_\xi g_{AB} = 4 \omega(u, x^A).
\label{eq:GaugeConstraints}
\end{equation} The prefactor of $4$ is introduced for convenience. As discussed in appendix \ref{app:detcondBondi}, the last condition is equivalent to the determinant condition \eqref{eq:DetCond}. From \eqref{eq:GaugeConstraints}, we deduce
\begin{equation}
\begin{split}
\xi^u &= f, \\
\xi^A &= Y^A + I^A, \quad I^A = -\partial_B f \int_r^\infty \D r'  (e^{2 \beta} g^{AB}),\\
\xi^r &= - \frac{r}{2} (\mathcal{D}_A Y^A - 2 \omega + \mathcal{D}_A I^A - \partial_B f U^B + \frac{1}{2} f g^{-1} \partial_u g) ,\\
\end{split}
\label{eq:xirpr}
\end{equation} 
where $\partial_r f = 0 = \partial_r Y^A$, and $g= \det (g_{AB})$. The covariant derivative $\mathcal{D}_A$ is associated with the $2$-dimensional metric $g_{AB}$. The residual gauge transformations are parametrized by the 4 functions $\omega$, $f$ and $Y^A$ of $(u,x^A)$.

\subsubsection{Procedure to resolve Einstein's equations}

We solve the Einstein equations $G_{\mu\nu} + \Lambda g_{\mu\nu} = 0$ in Bondi gauge. We follow the integration scheme and notations of \cite{Barnich:2010eb}. In particular, we use the Christoffel symbols that have been derived in this reference.

\textit{Minimal fall-off requirements:} We impose the preliminary boundary condition $g_{AB} = \mathcal{O}(r^2)$ (see \eqref{preliminary boundary condition}) and assume an analytic expansion for $g_{AB}$, namely 
\begin{equation}
g_{AB} = r^2 \, q_{AB}  + r\, C_{AB} + D_{AB} + \frac{1}{r} \, E_{AB} + \frac{1}{r^2} \, F_{AB} + \mathcal{O}(r^{-3}),\label{eq:gABFallOff}
\end{equation} 
where each term involves a symmetric tensor whose components are arbitrary functions of $(u,x^C)$. For $\Lambda \neq 0$, the Fefferman-Graham theorem  \cite{Starobinsky:1982mr,AST_1985__S131__95_0,Skenderis:2002wp,2007arXiv0710.0919F,Papadimitriou:2010as} together with the map between the Fefferman-Graham gauge and Bondi gauge, derived in appendix \ref{app:chgt}, ensures that the expansion \eqref{eq:gABFallOff} leads to the most general solution to the vacuum Einstein equations. For $\Lambda = 0$, the analytic expansion \eqref{eq:gABFallOff} is a hypothesis since additional logarithmic branches might occur \cite{Winicour1985,Chrusciel:1993hx,ValienteKroon:1998vn}. 

This fall-off condition does not impose any constraints on the generators of residual diffeomorphisms \eqref{eq:xirpr}. In the following, upper-case Latin indices are lowered and raised by the $2$-dimensional metric $q_{AB}$ and its inverse. The gauge condition \eqref{eq:DetCond} implies $g^{AB}\p_r g_{AB}=4/r$ which imposes successively that $\det (g_{AB}) = r^4 \det (q_{AB})$, $q^{AB} C_{AB} = 0$ and
\begin{equation}
\begin{split}
&D_{AB} = \frac{1}{4} q_{AB} C^{CD} C_{CD} + \mathcal{D}_{AB} (u,x^C),  \\
&E_{AB} = \frac{1}{2} q_{AB} \mathcal{D}_{CD}C^{CD} + \mathcal{E}_{AB} (u,x^C), \\
&F_{AB} = \frac{1}{2} q_{AB} \Big[ C^{CD}\mathcal{E}_{CD} + \frac{1}{2} \mathcal{D}^{CD}\mathcal{D}_{CD} - \frac{1}{32} (C^{CD}C_{CD})^2 \Big] + \mathcal{F}_{AB}(u,x^C),
\end{split}
\end{equation}
with $q^{AB} \mathcal{D}_{AB} = q^{AB} \mathcal{E}_{AB} = q^{AB} \mathcal{F}_{AB} = 0$. 

\textit{Organization of Einstein's equations:} We organize the equations of motion as follows. First, we solve the equations that do not involve the cosmological constant. The radial constraint $G_{rr} = R_{rr} = 0$ fixes the $r$-dependence of $\beta$, while the cross-term constraint $G_{rA} = R_{rA} = 0$ fixes the $r$-dependence of $U^A$. 

Next, we treat the equations that do depend on $\Lambda$. The equation $G_{ur} + \Lambda g_{ur} = 0$ determines the $r$-dependence of $V/r$ in terms of the previous variables. Noticing that $R = g^{\mu\nu} R_{\mu\nu} = 2 g^{ur} R_{ur} + g^{rr} R_{rr} + 2 g^{rA} R_{rA} + g^{AB} R_{AB}$, and taking into account that $R_{rr} = 0 = R_{rA}$, one gets $G_{ur} + \Lambda g_{ur} = R_{ur} - \frac{1}{2} g_{ur} R + \Lambda g_{ur} = \frac{1}{2} g_{ur} g^{AB} R_{AB} = 0$ so that we can solve equivalently $g^{AB} R_{AB} = 0$. 

Next, we concentrate on the pure angular equation, $G_{AB} + \Lambda g_{AB} = 0$, which can be split into a trace-free part 
\begin{equation}
G_{AB} - \frac{1}{2} g_{AB} \, g^{CD}G_{CD} = 0 \label{eq:GABTF}
\end{equation}
and a pure-trace part
\begin{equation}
g^{CD} G_{CD} + 2\Lambda = 0. \label{eq:GABTFull}
\end{equation}
Consider the Bianchi identities $\nabla_\mu G^{\mu\nu} = 0$ which can be rewritten as
\begin{equation}
2 \sqrt{-g} \nabla_\mu G^\mu_\nu = 2 \partial_\mu (\sqrt{-g}G^\mu_\nu) - \sqrt{-g} G^{\mu\lambda} \partial_\nu g_{\mu\lambda} = 0. \label{rewrite Bianchi}
\end{equation}
Since $\partial_\nu g_{\mu\lambda} = - g_{\mu\alpha}g_{\lambda\beta}\partial_\nu g^{\alpha\beta}$, we have
\begin{equation}
2 \partial_\mu (\sqrt{-g}G^\mu_\nu) + \sqrt{-g} G_{\mu\lambda} \partial_\nu g^{\mu\lambda} = 0.
\end{equation}
Taking $\nu = r$ and noting that $G_{r\alpha} +\Lambda g_{r\alpha} = 0$ have already been solved, one gets
\begin{equation}
G_{AB} \partial_r g^{AB} = \frac{4\Lambda}{r}. \label{eq:int1}
\end{equation}
Recalling that \eqref{eq:GABTF} holds, and that the determinant condition implies that $g^{AB}\partial_r g_{AB} = 4/r$, we see that  \eqref{eq:int1} is equivalent to \eqref{eq:GABTFull}. As a consequence, the equation $G_{AB} + \Lambda g_{AB} = 0$ is completely obeyed if \eqref{eq:GABTF} is solved. Indeed, once the trace-free part \eqref{eq:GABTF} has been set to zero, the tracefull part \eqref{eq:GABTFull} is automatically constrained by the Bianchi identity. Another way to see this is as follows: imposing that $G_{r\alpha} +\Lambda g_{r\alpha} = 0$ holds, \eqref{eq:GABTF} is equivalent to
\begin{equation}
(M^{TF})^A_{\,\,B} \equiv M^A_{\,\,B} - \frac{1}{2} \delta^A_B M^C_{\,\,C} = 0, \quad M^A_{\,\,B} \equiv g^{AC}R_{CB}, \label{eq:MABTF}
\end{equation}
since the trace part of $M^A_{\,\,B}$ has already been set to zero to fix the radial dependence of $V/r$. 

At this stage, Einstein's equations $(r,r)$, $(r,A)$, $(r,u)$ and $(A,B)$ have been solved. The $(u,u)$ and $(u,A)$ components remain to be solved. In doing so, we will derive the evolution equations for the Bondi mass and angular momentum aspects (see section \ref{partII} below). Expressing the $A$ component of the contracted Bianchi identities \eqref{rewrite Bianchi} yields
\begin{equation}
\partial_r \Big[ r^2 \Big( G_{uA} + \Lambda g_{uA} \Big) \Big] = \partial_r \Big[ r^2 \Big( R_{uA} - \Lambda g_{uA} \Big) \Big] = 0. \\
\end{equation}
Therefore, we can isolate the only non-trivial equation to be the $1/r^2$ part of $G_{uA} + \Lambda g_{uA} = 0$. This will determine the evolution of $N^{(\Lambda)}_A(u,x^B)$ related to the Bondi angular momentum aspect. Assuming that $G_{uA} + \Lambda g_{uA} = 0$ is solved, the last Bianchi identity \eqref{rewrite Bianchi} for $\nu = u$ becomes
\begin{equation}
\partial_r \Big[ r^2 \Big( G_{uu} + \Lambda g_{uu} \Big) \Big] = \partial_r \Big[ r^2 \Big( R_{uu} - \Lambda g_{uu} \Big) \Big] = 0,
\end{equation}
and the reasoning is similar. We will solve the $r$-independent part of $r^2 (R_{uu} - \Lambda g_{uu})$, which will uncover the equation governing the time evolution of $M^{(\Lambda)}(u,x^A)$ related to the Bondi mass aspect.

\subsubsection{Solution to the algebraic equations}

We define several auxiliary fields as in \cite{Barnich:2010eb}. Starting from \eqref{eq:gABFallOff}, we can build $k_{AB} = \frac{1}{2} \partial_r g_{AB}$, $l_{AB} = \frac{1}{2} \partial_u g_{AB}$, and $n_A = \frac{1}{2} e^{-2\beta}g_{AB}\partial_r U^B$. The determinant condition \eqref{eq:DetCond} allows us to split the tensors $k_{AB}$ and $l_{AB}$ in leading trace-full parts and subleading trace-free parts as
\begin{equation}
\begin{split}
k^A_B &\equiv g^{AC} k_{BC} = \frac{1}{r} \delta^A_B + \frac{1}{r^2} K^A_B, \qquad K^A_A = 0, \\
l^A_B &\equiv g^{AC} l_{BC} = \frac{1}{2} q^{AC}\partial_u q_{BC} + \frac{1}{r} L^A_B, \qquad L^A_A = 0.
\end{split}
\end{equation}
Note that 
\begin{eqnarray}
l = l^A_A = \frac{1}{2} q^{AB}\partial_u q_{AB} = \partial_u \ln \sqrt{q}.
\end{eqnarray} 
Let us start by solving $R_{rr} = 0$, which leads to
\begin{equation}
\partial_r \beta = -\frac{1}{2r} + \frac{r}{4} k^A_B k^B_A = \frac{1}{4r^3} K^A_B K^B_A.
\end{equation}
Expanding $K^A_B$ in powers of $1/r$, we get
\begin{align}
\beta(u,r,x^A) &= \beta_0 (u,x^A) + \frac{1}{r^2} \Big[ -\frac{1}{32} C^{AB} C_{AB} \Big] + \frac{1}{r^3} \Big[ -\frac{1}{12} C^{AB} \mathcal{D}_{AB} \Big] \label{eq:EOM_beta 2} \\
&\qquad + \frac{1}{r^4}\Big[ - \frac{3}{32} C^{AB}\mathcal{E}_{AB} - \frac{1}{16} \mathcal{D}^{AB}\mathcal{D}_{AB} + \frac{1}{128} (C^{AB}C_{AB})^2 \Big] + \mathcal{O}(r^{-5}). \nonumber
\end{align}
Up to the integration ``constant'' $\beta_0 (u,x^A)$, the condition \eqref{eq:gABFallOff} uniquely determines  $\beta$. In particular, the $1/r$ order is always zero on-shell. This equation also holds for $\Lambda =0$ but standard asymptotic flatness conditions set $\beta_0 = 0$ (see equation \eqref{asymp flat 1}). We keep it arbitrary here. 

Next, we develop $R_{rA} = 0$, which gives
\begin{equation}
\partial_r (r^2 n_A) = r^2 \Big( \partial_r - \frac{2}{r} \Big) \partial_A \beta - \mathcal{D}_B K^B_A.
\end{equation}
We now expand the transverse covariant derivative $\mathcal{D}_A$
\begin{equation}
\Gamma^B_{AC}[g_{AB}] = \Gamma^B_{AC}[q_{AB}] + \frac{1}{r} \Big[ \frac{1}{2} (D_A C^B_C + D_C C^B_A - D^B C_{AC}) \Big] + \mathcal{O}(r^{-2}),
\end{equation}
in terms of the transverse covariant derivative $D_A$ defined with respect to the leading transverse metric $q_{AB}$. This implies in particular that
\begin{equation}
\mathcal{D}_B K^B_A = -\frac{1}{2} D^B C_{AB} + \frac{1}{r} \Big[ -D^B \mathcal{D}_{AB} + \frac{1}{8} \partial_A (C_{BC}C^{BC}) \Big] + \mathcal{O}(r^{-2}).
\end{equation}
Explicitly using \eqref{eq:EOM_beta 2}, we find
\begin{equation}
n_A = -\partial_A \beta_0 + \frac{1}{r}\Big[ \frac{1}{2}D^B C_{AB} \Big] + \frac{1}{r^2} \Big[ \ln r \,  D^B \mathcal{D}_{AB} + N_A \Big] + o(r^{-2})
\end{equation}
where $ N_A$ is a second integration ``constant'' (i.e. $\partial_r  N_A = 0$), which corresponds to the Bondi angular momentum aspect in the asymptotically flat case. After inverting the definition of $n_A$, integrating one time further on $r$ and raising the index $A$, we end up with
\begin{equation}
\begin{split}
U^A = \,\, & U^A_0(u,x^B) +\overset{(1)}{U^A}(u,x^B) \frac{1}{r} + \overset{(2)}{U^A}(u,x^B) \frac{1}{r^2} \\
&+ \overset{(3)}{U^A}(u,x^B) \frac{1}{r^3} + \overset{(\text{L}3)}{U^A}(u,x^B) \frac{\ln r}{r^3} + o(r^{-3})
\end{split} \label{eq:EOM_UA bis}
\end{equation}
with
\begin{eqnarray}
\overset{(1)}{U^A}(u,x^B)\hspace{-6pt} &=&\hspace{-6pt} 2 e^{2\beta_0} \partial^A \beta_0 ,\nonumber \\
\overset{(2)}{U^A}(u,x^B)\hspace{-6pt} &=&\hspace{-6pt} - e^{2\beta_0} \Big[ C^{AB} \partial_B \beta_0 + \frac{1}{2} D_B C^{AB} \Big], \nonumber\\
\overset{(3)}{U^A}(u,x^B)\hspace{-6pt} &=& \hspace{-6pt}- \frac{2}{3} e^{2\beta_0} \Big[ N^A - \frac{1}{2} C^{AB} D^C C_{BC} +   (\partial_B \beta_0 - \frac{1}{3} D_B) \mathcal{D}^{AB} - \frac{3}{16} C_{CD}C^{CD} \partial^A \beta_0  \Big], \nonumber\\
\overset{(\text{L}3)}{U^A}(u,x^B) \hspace{-6pt}&=&\hspace{-6pt} -\frac{2}{3}e^{2\beta_0}D_B \mathcal{D}^{AB}, \label{eq:EOM_UA2 bis}
\end{eqnarray}
where $U^A_0(u,x^B)$  is a new integration ``constant''. Again, this equation also holds if $\Lambda$ is absent, but standard asymptotic flatness sets this additional parameter to zero. As known in standard flat case analysis, the presence of $\mathcal{D}_{AB}$ is responsible for logarithmic terms in the expansion of $U^A$. We will shortly derive that for $\Lambda \neq 0$, $\mathcal{D}_{AB}$ vanishes on-shell.

Given that
\begin{equation}
\begin{split}
M^A_{\,\,B} &= e^{-2\beta} \Big[ (\partial_r + \frac{2}{r}) (l^A_B + k^A_B \frac{V}{r} + \frac{1}{2} \mathcal{D}_B U^A + \frac{1}{2} \mathcal{D}^A U_B) \\
&\qquad\qquad + k^A_C \mathcal{D}_B U^C - k^C_B \mathcal{D}_C U^A + (\partial_u + l)k^A_B + \mathcal{D}_C (U^C k^A_B) \Big] \\
&\qquad + R^A_B[g_{CD}] - 2(\mathcal{D}_B \partial^A \beta + \partial^A \beta \partial_B \beta + n^A n_B),
\end{split}
\end{equation}
we extract the $r$-dependence of $V/r$ thanks to $M^A_{\,\,A} = 0$, which reads as
\begin{equation}
\begin{split}
\partial_r V = & - 2 r (l + D_A U^A) + \\ &e^{2\beta} r^2 \Big[ D_A D^A \beta + (n^A - \partial^A \beta) (n_A - \partial_A \beta) - D_A n^A - \frac{1}{2} R[g_{AB}] + \Lambda \Big] .
\end{split}
\end{equation}
Considering \eqref{eq:gABFallOff}, \eqref{eq:EOM_beta 2} and \eqref{eq:EOM_UA bis}, we get, after integration on $r$
\begin{align}
\frac{V}{r} = &\frac{\Lambda}{3} e^{2\beta_0} r^2 - r (l + D_A U^A_0) \label{eq:EOMVr 2} \\
&- e^{2\beta_0} \Big[ \frac{1}{2}\Big( R[q] + \frac{\Lambda}{8}C_{AB} C^{AB} \Big) + 2 D_A \partial^A \beta_0 + 4 \partial_A \beta_0 \partial^A \beta_0 \Big] - \frac{2  M}{r} + o(r^{-1}) \nonumber 
\end{align}
where $ M(u,x^A)$ is an integration ``constant'' which, in flat asymptotics, is recognized as the Bondi mass aspect. 

Afterwards, we solve \eqref{eq:MABTF} order by order, which provides us with the constraints imposed on each independent order of $g_{AB}$. The leading $\mathcal{O}(r^{-1})$ order of that equation yields
\begin{equation}
\frac{\Lambda}{3} C_{AB} = e^{-2\beta_0} \Big[ (\partial_u - l) q_{AB} + 2 D_{(A} U^0_{B)} - D^C U^0_C q_{AB} \Big].
\label{eq:CAB 2}
\end{equation}
This result shows that there is a discrete bifurcation between the asymptotically flat case and the case $\Lambda \neq 0$. Indeed, when $\Lambda = 0$, the left-hand side vanishes, which leads to a constraint on the time-dependence of $q_{AB}$. Consequently, the field $q_{AB}$ is constrained while $C_{AB}$ is completely free and interpreted as the shear. For (A)dS$_4$ asymptotics, $C_{AB}$ is entirely determined by $q_{AB}$, $\beta_0$ and $U^A_0$, while the boundary metric $q_{AB}=q_{AB}(u,x^A)$ is left completely undetermined by the equations of motion. This is consistent with previous analyses \cite{Ashtekar:2014zfa,He:2015wfa,Saw:2017amv,He:2018ikd,Poole:2018koa}.

Going to $\mathcal{O}(r^{-2})$, we get
\begin{equation}
\frac{\Lambda}{3} \mathcal{D}_{AB} = 0,\label{eq:DAB 2}
\end{equation}
which removes the logarithmic term in \eqref{eq:EOM_UA bis} for $\Lambda \neq 0$, but not for $\Lambda = 0$. The condition at next $\mathcal{O}(r^{-3})$ order
\begin{equation}
\partial_u \mathcal{D}_{AB} + U_0^C D_C \mathcal{D}_{AB} + 2 \mathcal{D}_{C(A} D_{B)}U_0^C = 0,
\end{equation}
is thus trivial for $\Lambda \neq 0$, but reduces to $\partial_u \mathcal{D}_{AB} = 0$ in the flat limit, consistently with previous results. 

Using an iterative argument as in \cite{Poole:2018koa}, we now make the following observation. If we decompose $g_{AB} = r^2 \sum_{n\geq 0} g_{AB}^{(n)} r^{-n}$, we see that the iterative solution of \eqref{eq:MABTF} organizes itself as $\Lambda g_{AB}^{(n)} = \partial_u g_{AB}^{(n-1)} + (...)$ at order $\mathcal{O}(r^{-n})$, $n\in\mathbb{N}_0$. Accordingly, the form of $\mathcal{E}_{AB}$ should have been fixed by the equation found at $\mathcal{O}(r^{-3})$, but it is not the case, since both contributions of $\mathcal{E}_{AB}$ cancel between $G_{AB}$ and $\Lambda g_{AB}$. Moreover, the equation $\Lambda g_{AB}^{(4)} = \partial_u g_{AB}^{(3)} + (...)$ at next order turns out to be a constraint for $g_{AB}^{(4)} \sim \mathcal{F}_{AB}$, determined with other subleading data such as $C_{AB}$ or $\partial_u g_{AB}^{(3)} \sim \partial_u \mathcal{E}_{AB}$. It shows that $\mathcal{E}_{AB}$ is a set of two free data on the boundary, built up from two arbitrary functions of $(u,x^A)$. It shows moreover that there is no more data to be uncovered for $\Lambda \neq 0$. This matches with the number of free data of the solution space in the Fefferman-Graham gauge, as discussed in subsection \ref{sec:FGg}.

As a conclusion, Einstein's equations $(r,r)$, $(r,A)$, $(r,u)$ and $(A,B)$ can be solved iteratively in the asymptotic expansion for $\Lambda \neq 0$. We identified 11 independent functions $\{ \beta_0 (u,x^A)$, $U^A_0 (u,x^B)$, $q_{AB} (u,x^C)$, $ M (u,x^C)$, $ N_A(u,x^C)$, $\mathcal{E}_{AB} (u,x^C)\}$ that determine the asymptotic solution. We see in the following subsection that the remaining equations are equivalent to evolution equations for $ M(u,x^A)$ and $ N_A(u,x^B)$. This contrasts with the asymptotically flat case $\Lambda = 0$ where an infinite series of functions appear in the radial expansion, see e.g. \cite{Barnich:2010eb}.

\subsubsection{Boundary gauge fixing}
\label{sec:bndg}

In this section, we simplify our analysis by imposing a (co-dimension $1$) boundary gauge fixing. The latter can also be interpreted as a partial Dirichlet boundary condition with respect to the bulk spacetime. Let us consider the pullback of the most general Bondi metric satisfying \eqref{eq:gABFallOff} to the boundary $\mathscr{I} \equiv \lbrace r \to \infty \rbrace$, 
\begin{equation}
\left. ds^2 \right|_{\mathscr{I}} = \Big[ \frac{\Lambda}{3}e^{4\beta_0} + U_0^A U^0_A \Big] du^2 - 2 U_A^0 du dx^A + q_{AB} dx^A dx^B.
\end{equation}
We use the boundary gauge freedom to reach the gauge
\begin{equation}
\beta_0 = 0,\quad U^A_0 = 0,\quad \sqrt{q} = \sqrt{\bar q}\label{bndgauge}
\end{equation}
where $\sqrt{\bar q}$ is a fixed area of the two-dimensional transverse space spanned by $x^A$. This gauge is a temporal boundary gauge for $\Lambda < 0$, a radial boundary gauge for $\Lambda > 0$ and a null boundary gauge for $\Lambda = 0$ with $g_{ur} = -1+ \mathcal{O}(r^{-1})$ in \eqref{bondi gauge bis}. 

Intuitively, this amounts to using the gauge freedom at the boundary $\mathscr{I}$, to eliminate three pure-gauge degrees of freedom thanks to a diffeomorphism defined intrinsically on $\mathscr{I}$ and lifted to the bulk in order to preserve the Bondi gauge. Such a transformation also involves a Weyl rescaling of the boundary metric, as can be seen from \eqref{eq:xirpr}, which consists in a redefinition of the coordinate $r$ by an arbitrary factor depending on $(u,x^A)$. We can use this Weyl rescaling to gauge-fix one further quantity in the boundary metric, namely the area of the transverse space. The details are provided below. 
 
Computing the Lie derivative on the Bondi metric on-shell and retaining only the leading $\mathcal{O}(r^2)$ terms, we get the transformation laws of the boundary fields $q_{AB}$, $\beta_0$ and $U^A_0$ under the set of residual gauge transformations \eqref{eq:xirpr}:
\begin{align}
\delta_\xi q_{AB} &= f(\partial_u - l)q_{AB} + (\mathcal{L}_Y - D_C Y^C +2\omega)q_{AB} \nonumber \\
&\quad - 2 (U_{(A}^0 \partial_{B)}f - \frac{1}{2} q_{AB} U^C_0 \partial_C f),\label{eq56} \\
\delta_\xi \beta_0 &= (f \partial_u + \mathcal{L}_Y) \beta_0 + \frac{1}{2}\Big[ \partial_u - \frac{1}{2}l + \frac{3}{2} U^A_0 \partial_A \Big] f - \frac{1}{4} (D_A Y^A  - 2 \omega),\label{eq57} \\
\delta_\xi U^A_0 &= (f\partial_u + \mathcal{L}_Y) U_0^A - \Big[ \partial_u Y^A - \frac{1}{\ell^2} e^{4\beta_0} q^{AB} \partial_B f \Big] + U_0^A (\partial_u f + U_0^B \partial_B f).\label{eq58}
\end{align}
The first equation implies that $q^{AB}\delta_\xi q_{AB} = 4\omega$. We can therefore adjust the Weyl generator $\omega$ in order to reach the gauge $ \sqrt{q} = \sqrt{\bar q}$. The form of the infinitesimal transformations \eqref{eq57}-\eqref{eq58} involves $\p_u f$ and $\p_u Y^A$. This ensures that a finite gauge transformation labelled by $f,Y^A$ can be found by integration over $u$ to reach $\beta_0=0$, $U^A_0=0$, at least in a local patch. As a result, the conditions \eqref{bndgauge} can be reached by gauge fixing, at least locally. The vanishing of the inhomogeneous contributions in the transformation laws \eqref{eq57}-\eqref{eq58} constrains parameters $f,Y^A$ and reduces the set of allowed vectors among \eqref{eq:xirpr}. The remaining residual transformations are studied in subsection \ref{The Lambda-BMS group and its flat limit}.

\subsubsection{Constraint equations as Bondi evolution equations}
\label{partII}

Assuming the gauge fixing conditions \eqref{bndgauge}, we are now ready to present the evolution equations that follow from the remaining Einstein equations. Moreover, we suppose that $\mathcal{D}_{AB} = 0$ in the case $\Lambda = 0$ to simplify our computation. As justified before, the $\mathcal{O}(r^0)$ part of $r^2 (R_{uA} - \Lambda g_{uA}) = 0$ will fix the temporal evolution of $N_A$. From the Christoffel symbols, we can develop the first term as
\begin{align}
R_{uA} = &-(\partial_u - l) \partial_A \beta - \partial_A l - (\partial_u + l) n_A  \\
&+ n_B \mathcal{D}^B U_A - \partial_B \beta \mathcal{D}_A U^B + 2 U^B (\partial_A \beta \partial_B \beta + n_A n_B) \nonumber \\
&+ \mathcal{D}_B \Big[ l^B_A + \frac{1}{2} ( \mathcal{D}^B U_A - \mathcal{D}_A U^B ) + U^B (\partial_A \beta - n_A) \Big] + 2 n_B l^B_A \nonumber \\
&- \frac{1}{2}(\partial_r + 2\partial_r \beta + \frac{2}{r})  \partial_A \frac{V}{r} - \frac{V}{r} (\partial_r + \frac{2}{r})n_A + k_A^B (\partial_B \frac{V}{r} + 2 \frac{V}{r} n_B) \nonumber \\
&- e^{-2\beta} (\partial_r + \frac{2}{r}) \Big[ U^B(l_{AB} + \frac{V}{r} k_{AB} + \mathcal{D}_{(A}U_{B)}) \Big] \nonumber \\
&- e^{-2\beta} U^B \Big[ (\partial_u + l) k_{AB} - 4 l^C_{(A} k_{B)C} - 2 k^C_A k_{BC} \frac{V}{r} + \mathcal{D}_C (k_{AB}U^C) - 2 k_{C(A}\mathcal{D}^C U_{B)} \Big]. \nonumber
\end{align}
Let us emphasize that the $r$-dependence of the fields is not yet explicit in this expression, so the upper case Latin indices are lowered and raised by the full metric $g_{AB}$ and its inverse. Expanding all the fields in power series of $1/r$ in $R_{uA}$ and $\Lambda g_{uA}$ and selecting the $1/r^2$ terms yields
\begin{equation}
(\partial_u + l)  N_A^{(\Lambda)} - \partial_A  M^{(\Lambda)} - \frac{\Lambda}{2} D^B  J_{AB} = 0. \label{eq:EvolutionNA}
\end{equation}
Here, we defined with hindsight the Bondi mass and angular momentum aspects for $\Lambda \neq 0$ as
\begin{align}
M^{(\Lambda)} &=  M + \frac{1}{16} (\partial_u + l)(C_{CD}C^{CD}), \label{eq:hatM} \\
N^{(\Lambda)}_A &=  N_A - \frac{3}{2\Lambda} D^B (N_{AB} - \frac{1}{2} l C_{AB}) - \frac{3}{4} \partial_A (\frac{1}{\Lambda} R[q] - \frac{3}{8}  C_{CD}C^{CD}), \label{eq:hatNA}
\end{align}
and the traceless symmetric tensor $J_{AB}$ ($q^{AB} J_{AB} = 0$) as
\begin{align}
J_{AB} = &-\mathcal{E}_{AB} -\frac{3}{\Lambda^2} \Big[ \partial_u (N_{AB} - \frac{1}{2} lC_{AB})  -\frac{\Lambda}{2} q_{AB} C^{CD}(N_{CD} - \frac{1}{2} l C_{CD}) \Big] \nonumber \\
&\quad +\frac{3}{\Lambda^2} (D_A D_B l - \frac{1}{2} q_{AB} D_C D^C l) \nonumber \\
&\quad -\frac{1}{\Lambda} (D_{(A}D^C C_{B)C} - \frac{1}{2} q_{AB} D^C D^D C_{CD}) \nonumber \\
&\quad +C_{AB} \Big[ \frac{5}{16} C_{CD}C^{CD} + \frac{1}{2\Lambda}R[q]\Big] . \label{eq:hatJAB}
\end{align}
We used the notation $N_{AB} \equiv \partial_u C_{AB}$. This tensor is symmetric and obeys $q^{AB}N_{AB} = \frac{\Lambda}{3} C^{AB}C_{AB}$. When $\Lambda = 0$, $N_{AB}$ is thus traceless and represents the Bondi news tensor. 

We will justify the definitions of Bondi mass and angular momentum aspects in section \ref{sec:FGg}. Note that $\partial_u q_{AB}$ has been eliminated using \eqref{eq:CAB 2}. The transformations of these fields under the residual gauge symmetries $\xi$ preserving the Bondi gauge \eqref{bondi gauge bis} and the boundary gauge \eqref{bndgauge} are given by
\begin{align}
\delta_\xi M^{(\Lambda)} &= [f\partial_u + \mathcal{L}_Y + \frac{3}{2}(D_A Y^A + f l - 2 \omega)]M^{(\Lambda)} - \frac{\Lambda}{3} N_A^{(\Lambda)} \partial^A f, \label{eq:VarM} \\
\delta_\xi N_A^{(\Lambda)} &= [f\partial_u + \mathcal{L}_Y + D_B Y^B + f l - 2 \omega] N_A^{(\Lambda)} + 3 M^{(\Lambda)} \partial_A f + \frac{\Lambda}{2} J_{AB} \partial^B f, \label{eq:VarNA} \\
\delta_\xi J_{AB} &= [f\partial_u + \mathcal{L}_Y + \frac{1}{2}(D_C Y^C + f l - 2 \omega)] J_{AB} \nonumber \\
&\quad\, - \frac{4}{3} (N_{(A}^{(\Lambda)}\partial_{B)}f - \frac{1}{2} N_C^{(\Lambda)} \partial^C f q_{AB}). \label{eq:VarJAB}
\end{align}

The asymptotically flat limit is not trivial in the equation \eqref{eq:EvolutionNA} due to terms $\sim \Lambda^{-1}$ above which we collect here:
\begin{equation}
\begin{split}
-\frac{3}{2\Lambda} \Big[ &(\partial_u + l) D^B (\partial_u C_{AB} - \frac{1}{2} l C_{AB}) - D^B \partial_u (\partial_u C_{AB} - \frac{1}{2} l C_{AB}) \\
&\quad + \frac{1}{2} (\partial_u + l) \partial_A R[q] + D^B (D_A D_B l - \frac{1}{2} D_C D^C l q_{AB}) \Big]. \label{eq:ProblematicTerms}
\end{split}
\end{equation}
There are two subtle steps here needed to massage the evolution equation before taking the limit $\Lambda\to 0$. First, we must develop the remaining $u$-derivatives acting on covariant derivatives and taking the constraint \eqref{eq:CAB 2} into account to highlight $\Lambda$ factors. Next, we can extract the trace of $N_{AB}$, which also contains a residual contribution $\sim \Lambda$. We end up with
\begin{equation}
\begin{split}
\eqref{eq:ProblematicTerms} = \,\, &\frac{1}{2} D_C (N_{AB}^{TF} C^{BC}) + \frac{1}{4} N_{BC}^{TF} D_A C^{BC} - \frac{1}{4} D_A D_B D_C C^{BC} \\
&+ \frac{1}{8} C^B_C C^C_B \partial_A l - \frac{3}{16} l \partial_A (C^B_C C^C_B)
\end{split}
\label{eq:Probl2}
\end{equation}
where $N_{AB}^{TF}$ denotes the trace-free part of $N_{AB}$. The following identities turn out to be useful for the computation:
\begin{equation}
\begin{split}
(\partial_u +l) H^{AB} &= q^{AC}\partial_u H_{CD}q^{BD} - l H^{AB} - \frac{2\Lambda}{3} C^{C(A}H^{B)}_C,   \\
(\partial_u+l) (D^B H_{AB}) &= D^B \partial_u H_{AB} - \frac{1}{2} q^{CD}H_{CD}\partial_A l   \\
& \quad\, - \frac{\Lambda}{3}\Big[ D_C (H_{AB}C^{BC}) + \frac{1}{2} H^{BC}D_A C_{BC}\Big], \\
(\partial_u+l) C^{AB}C_{AB} &= 2 N^{AB}C_{AB} - l C_{AB}C^{AB}, \\
(\partial_u + l) \partial_A R[q] &= - (D^B D_B + \frac{1}{2} R[q]) \partial_A l + \frac{\Lambda}{3} D_A D_B D_C C^{BC}
\end{split}
\end{equation}
where $H_{AB}(u,x^C)$ is any symmetric rank 2 transverse tensor. We note that $N_{AB}^{TF} C^{BC} + C_{AB} N^{BC}_{TF} = \delta_A^C C_{BD}N^{BD}_{TF}$, thanks to which the first term of \eqref{eq:Probl2} can be rewritten as
\begin{equation}
\frac{1}{2} D_C (N_{AB}^{TF} C^{BC}) = \frac{1}{4} D_B (N_{AC}^{TF} C^{BC}-C_{AC} N^{BC}_{TF}) + \frac{1}{4} \partial_A (C_{BD}N^{BD}_{TF}).
\end{equation}
We can now present \eqref{eq:EvolutionNA} in a way that makes terms in $\Lambda$ explicit:
\begin{align}
(\partial_u + l)  N_A &- \partial_A  M - \frac{1}{4} C_{AB} \partial^B R[q] - \frac{1}{16} \partial_A (N_{BC}^{TF} C^{BC}) \label{eq:EvolutionNA_Explicit} \\
&- \frac{1}{32} l \partial_A (C_{BC}C^{BC}) +\frac{1}{4} N_{BC}^{TF} D_A C^{BC} + \frac{1}{4} D_B (C^{BC} N_{AC}^{TF} - N^{BC}_{TF} C_{AC}) \nonumber \\
&+\frac{1}{4} D_B (D^B D^C C_{AC} - D_A D_C C^{BC}) + \frac{\Lambda}{2} D^B(\mathcal{E}_{AB} - \frac{7}{96} C^C_D C^D_C C_{AB}) = 0.\nonumber 
\end{align}
As a result, the asymptotically flat limit can be safely taken and \eqref{eq:EvolutionNA_Explicit} reduces to
\begin{align}
(\partial_u + l)  N_A &- \partial_A  M - \frac{1}{4} C_{AB} \partial^B R[q] - \frac{1}{16} \partial_A (N_{BC}C^{BC}) \nonumber \\
&- \frac{1}{32} l \partial_A (C_{BC}C^{BC}) +\frac{1}{4} N_{BC} D_A C^{BC} + \frac{1}{4} D_B (C^{BC} N_{AC} - N^{BC} C_{AC}) \nonumber \\
&+\frac{1}{4} D_B (D^B D^C C_{AC} - D_A D_C C^{BC}) = 0,
\end{align}
which fully agrees with (4.49) of \cite{Barnich:2010eb} after a change of conventions\footnote{The Bondi news tensor is defined in \cite{Barnich:2010eb} as $N_{AB}^{\text{there}} = \partial_u C_{AB} - l C_{AB}$ while we define $N_{AB}^{\text{here}} = \partial_u C_{AB}$.}. It must be mentioned that $N_{AB} = N_{AB}^{TF}$ when $\Lambda = 0$.

We now derive the temporal evolution of $ M$, encoded in the $r$-independent part of $r^2 (R_{uu} - \Lambda g_{uu})=0$. The first term is worked out to be
\begin{equation}
\begin{split}
R_{uu} = \hspace{0.2cm} &(\partial_u + 2 \partial_u \beta + l) \Gamma^u_{uu} + (\partial_r + 2\partial_r \beta + \frac{2}{r}) \Gamma^r_{uu} + (\mathcal{D}_A + 2\partial_A \beta) \Gamma^A_{uu}  \\
&- 2\partial_u^2 \beta - \partial_u l - (\Gamma^u_{uu})^2 - 2 \Gamma^u_{uA}\Gamma^A_{uu} - (\Gamma^r_{ur})^2 -2\Gamma^r_{uA}\Gamma^A_{ur} - \Gamma^A_{uB}\Gamma^B_{uA}
\end{split}
\end{equation}
where all Christoffel symbols can be found on page 26 of \cite{Barnich:2010eb}. We finally get
\begin{equation}
(\partial_u + \frac{3}{2}l) M^{(\Lambda)} + \frac{\Lambda}{6} D^A  N^{(\Lambda)}_A + \frac{\Lambda^2}{24} C_{AB} J^{AB} = 0. \label{eq:EvolutionM}
\end{equation}
Here, the asymptotically flat limit is straightforward and gives
\begin{equation}
\begin{split}
(\partial_u + \frac{3}{2}l )  M +\frac{1}{8} N_{AB} N^{AB} - \frac{1}{8} l N_{AB} C^{AB} + \frac{1}{32} l^2 C_{AB} C^{AB} - \frac{1}{8} D_A D^A R[q] \\
- \frac{1}{4} D_A D_B N^{AB} + \frac{1}{4} C^{AB} D_A D_B l + \frac{1}{4} \partial_{(A} l D_{B)} C^{AB} + \frac{1}{8} l D_A D_B C^{AB} = 0,
\end{split}\label{duM prime}
\end{equation}
in agreement with (4.50) of \cite{Barnich:2010eb}. 
As a conclusion, in Bondi gauge \eqref{bondi gauge bis} with fall-off condition \eqref{eq:gABFallOff} and boundary gauge fixing \eqref{bndgauge}, the general solution to Einstein's equations is entirely determined by the seven free functions of $(u,x^A)$ for the case $\Lambda \neq 0$: $q_{AB}$ with fixed area $\sqrt{\bar q}$, $ M$, $N_A$ and trace-free $J_{AB}$ where $M$ and $N_A$ are constrained by the evolution equations \eqref{eq:EvolutionM} and \eqref{eq:EvolutionNA}. This contrasts with the asymptotically flat case $\Lambda = 0$ where an infinite series of functions
appearing in the radial expansion of $g_{AB}$ have to be specified to parametrize the solution (see e.g. \cite{Barnich:2010eb}).

\subsection{Dictionary between Fefferman-Graham and Bondi gauges}
\label{sec:FGg}

In appendix \ref{app:chgt}, we establish a coordinate transformation between Fefferman-Graham and Bondi gauges, which extends the procedure used in \cite{Poole:2018koa} to a generic spacetime metric. The boundary metric in the Fefferman-Graham gauge is related to the functions in the Bondi gauge through
\begin{equation}
g_{tt}^{(0)} = \frac{\Lambda}{3} e^{4 \beta_0} + U_0^C U_{C}^0 , \qquad g_{tA}^{(0)} = - U_A^0, \qquad g_{AB}^{(0)} =   q_{AB},
\label{g0 in term of Bondi}
\end{equation} 
where all functions on the right-hand sides are now evaluated as functions of $(t,x^A)$. 

The parameters $\lbrace \sigma,\xi^t_0, \xi^A_0 \rbrace$ of the residual gauge diffeomorphisms in the Fefferman-Graham gauge \eqref{AKV 1 prime} and \eqref{AKV 2 prime} can be related to those of the Bondi gauge appearing in \eqref{eq:xirpr} through
\begin{equation}
\begin{split}
\xi^t_0 &= f ,\\
\xi^A_0 &= Y^A, \\
\sigma &= \frac{1}{2} (D_A Y^A + f l - U_0^A \partial_A f - 2\omega),
\end{split}
\label{translation parameters}
\end{equation} 
where all functions on the right-hand sides are also evaluated as functions of $(t,x^A)$. 

The boundary gauge fixing \eqref{bndgauge} described in section \ref{sec:bndg} can now be understood as a gauge fixation of the boundary metric to
\begin{eqnarray}
g^{(0)}_{tt} = \frac{\Lambda}{3},\qquad g^{(0)}_{tA} =0,\qquad \text{det}(g_{(0)})=  \frac{\Lambda}{3} \bar q.\label{bndgauge2}
\end{eqnarray}
For $\Lambda < 0$ (resp. $\Lambda > 0$), this is exactly the temporal (resp. radial) gauge for the boundary metric, with a fixed area form for the 2-dimensional transverse space. 

Let us develop the constraint equations \eqref{condition on energy momentum} after boundary gauge fixing. First, the tracelessness condition determines the trace of $T_{AB}$ to be 
\begin{equation}
q^{AB}T_{AB} = -\frac{3}{\Lambda} T_{tt}.
\end{equation}
We define $T^{TF}_{AB}$ as the trace-free part of $T_{AB}$, i.e. $T_{AB} = T^{TF}_{AB} -\frac{3}{2\Lambda} T_{tt} q_{AB}$. The conservation equation $D_a^{(0)} T^{ab} = 0$ reads as
\begin{equation}
\begin{split}
(\partial_t + \frac{3}{2} l) T_{tt} + \frac{\Lambda}{3} D^A T_{tA} - \frac{\Lambda}{6} \partial_t q_{AB} T^{AB}_{TF} &= 0, \\
(\partial_t + l) T_{tA} - \frac{1}{2} \partial_A T_{tt} + \frac{\Lambda}{3} D^B T^{TF}_{AB} &= 0.
\end{split}
\label{EOM FG}
\end{equation}
Pursuing the change of coordinates to the Fefferman-Graham gauge up to fourth order in $\rho$, it can be shown that the stress tensor is given, in terms of Bondi variables, by
\begin{equation}
T_{ab} = \frac{\sqrt{3 |\Lambda|}}{16\pi G} \left[
\begin{array}{cc}
-\frac{4}{3} M^{(\Lambda)} & -\frac{2}{3}  N^{(\Lambda)}_B \\ 
-\frac{2}{3}  N^{(\Lambda)}_A &  J_{AB} + \frac{2}{\Lambda} M^{(\Lambda)} q_{AB}
\end{array} 
\right], \label{eq:RefiningTab}
\end{equation}
where $M^{(\Lambda)} (t,x^A)$ and $N^{(\Lambda)}_A (t,x^B)$ are the boundary fields defined as \eqref{eq:hatM}-\eqref{eq:hatNA} and $J_{AB}$ is precisely the tensor \eqref{eq:hatJAB}, all evaluated as functions of $t$ instead of $u$.  The conservation equations \eqref{EOM FG} are, in fact, equivalent to \eqref{eq:EvolutionM} and \eqref{eq:EvolutionNA} after using the dictionary \eqref{eq:RefiningTab} and solving $\partial_t q_{AB}$ in terms of $C_{AB}$ using \eqref{eq:CAB 2}. Moreover, we checked that the transformation laws \eqref{eq:VarM}-\eqref{eq:VarJAB} are equivalent to \eqref{eq:TransformationTab}. We therefore identified the Bondi mass aspect $M^{(\Lambda)}$ and the Bondi angular momentum aspect $N^{(\Lambda)}_A$ as the components $T_{tt}$ and $T_{tA}$ of the holographic stress-tensor, up to a normalization constant.

\subsection{Symmetries and flat limit}
\label{Symmetries and flat limit}

In contrast to the three-dimensional case discussed in subsection \ref{From asymptotically AdS to asymptotically flat spacetime}, the BMS group in four dimensions is not readily obtained by taking the flat limit of the asymptotic symmetry group associated with Dirichlet boundary conditions in AdS. In the following, we discuss the technical issue of finding a version of BMS in AdS, which reduces to the BMS group in the flat limit. Then, we present our new set of boundary conditions in asymptotically locally (A)dS$_4$ spacetime that leads to the $\Lambda$-BMS$_4$ algebra $\mathfrak{bms}_4^\Lambda$. We show that in the flat limit, this reduces to $\mathfrak{bms}_4^{\text{gen}}$.

\subsubsection{The problem to obtain BMS in the flat limit}
\label{The problem to obtain BMS in the flat limit}

In this subsection, mimicking the three-dimensional case discussed in subsection \ref{From asymptotically AdS to asymptotically flat spacetime}, we consider Dirichlet boundary conditions defining asymptotically AdS$_4$ spacetimes in the Fefferman-Graham gauge:
\begin{equation}
g_{ab}^{(0)} \D x^a \D x^b  = \frac{\Lambda}{3} \D t^2 + \mathring{q}_{AB} \D x^A \D x^B ,
\label{dirichlet 4d}
\end{equation} where $\mathring{q}_{AB}$ is the unit sphere metric (see definition (AAdS2) given in equation \eqref{BC Dirichlet}). It has been shown in \cite{Henneaux:1985tv} that the asymptotic symmetry algebra is given by $\mathfrak{so}(3,2)$ and the associated charges are finite, integrable, and form a representation of $\mathfrak{so}(3,2)$ without central extension. Therefore, we obtain a finite-dimensional algebra, which already ends any hope of obtaining BMS in the flat limit. 

Using \eqref{g0 in term of Bondi}, we can translate the Dirichlet boundary condition \eqref{dirichlet 4d} into the Bondi gauge as
\begin{equation}
\beta_0 = 0, \qquad U^0_A = 0, \qquad q_{AB} = \mathring{q}_{AB}
\label{cond asymp flat}
\end{equation} (this is the four-dimensional analogue of \eqref{Dirichltet Bondi suite}). The residual gauge diffeomorphisms preserving these constraints are given by \eqref{eq:xirpr}, where the parameters satisfy
\begin{equation}
\begin{split}
&\partial_u f =  \half D_A Y^A , \qquad \partial_u Y^A = - \frac{\Lambda}{3} \mathring{q}^{AB} \partial_B f , \qquad \omega = 0, \\
&\mathcal{L}_Y \mathring{q}_{AB} = ( D_C Y^C ) \mathring{q}_{AB} , 
\end{split}
\label{constraint equations Bond Dir prime}
\end{equation} where the last equation tells us that $Y^A$ is a conformal Killing vector of the unit 2-sphere metric. One can show that these conditions imply that $f$ and $Y^A$ are the parameters of the $\mathfrak{so}(3,2)$ asymptotic symmetry algebra, as it should be \cite{Barnich:2013sxa , Compere:2019bua}. Therefore, we conclude that, despite the conditions being interpreted as asymptotically flat boundary conditions (AF3) (see equation \eqref{asymp flat 1} together with \eqref{asymp flat 3}) in the flat limit $\Lambda \to 0$, we recover the Poincaré group instead of the BMS group \cite{Barnich:2013sxa , Compere:2019bua}. In particular, the supertranslations cannot be recovered through this process.

\subsubsection{The $\Lambda$-BMS group and its flat limit}
\label{The Lambda-BMS group and its flat limit}

We now circumvent this issue by proposing a new set of boundary condition in asymptotically locally (A)dS$_4$. We require that
\begin{equation}
\beta_0 = 0, \qquad U^0_A = 0, \qquad \sqrt{q} = \sqrt{\mathring{q}}, 
\label{new boundary conditions}
\end{equation} where $\mathring{q}$ is the determinant of the unit sphere metric (this last condition leads to $\delta \sqrt{q} =0$). Several comments can be made about these boundary conditions. They are very similar to \eqref{cond asymp flat}, except that we allow some fluctuations of the two-dimensional boundary metric $q_{AB}$ with fixed determinant. These boundary conditions are inspired by those investigated in the asymptotically flat context in chapter \ref{Generalized BMS and renormalized phase space} (see equation \eqref{repeated BC}). Furthermore, the boundary conditions \eqref{new boundary conditions} are precisely the conditions imposed in the boundary gauge fixing \eqref{bndgauge} (with $\bar{q} = \mathring{q}$) to write the evolution equations of the Bondi mass and the angular momentum aspect with respect to the $u$ coordinate. Finally, we notice that the boundary conditions \eqref{new boundary conditions} are valid for both $\Lambda > 0$ and $\Lambda < 0$. Indeed, as discussed around equation \eqref{bndgauge}, every solution written in the Bondi gauge and satisfying the preliminary boundary conditions $g_{AB} = \mathcal{O}(r^2)$, can be transformed through a diffeomorphism to satisfy \eqref{new boundary conditions}. Therefore, this does not constrain the Cauchy problem in dS$_4$. This contrasts with the Dirichlet boundary conditions \eqref{cond asymp flat} that do not make sense to impose in dS$_4$ since they would strongly constrain the Cauchy problem.  

The residual gauge diffeomorphisms \eqref{eq:xirpr} preserving the boundary conditions \eqref{new boundary conditions} have the following constraints on their parameters:
\begin{equation}
\partial_u  f = \frac{1}{2} D_A Y^A , \qquad \partial_u Y^A = - \frac{\Lambda}{3} q^{AB} \partial_B f, \qquad \omega = 0 .
\label{lambda bms conditions} 
\end{equation} Note that the solutions of these equations admit three integration ``constants'' $S(x^A)$, $V^A(x^B)$, though these are difficult to solve explicitly for an arbitrary transverse metric $q_{AB}$ in terms of these functions (see appendix C of \cite{newCompere} for an explicit solution in the case $q_{AB} = \mathring{q}_{AB}$). We call the vectors generated by $S(x^A)$ and $V^A(x^B) $ the supertranslation and superrotation generators, respectively. The use of this terminology will be justified below. In the Fefferman-Graham notation, the equations in \eqref{lambda bms conditions} are equivalent to 
\begin{equation}
\begin{split}
\sigma &= \frac{1}{2}D_A^{(0)} \xi^A_0 , \\
\partial_t  \xi^t_0 &= \frac{1}{2} D_A^{(0)}  \xi^A_0, \quad \partial_t \xi^A_0 = -\frac{\Lambda}{3}g^{AB}_{(0)}\partial_B \xi^t_{(0)}.
\label{sig}
\end{split}
\end{equation}

As already discussed in one of the examples in subsection \ref{Asymptotic symmetry algebra}, the asymptotic Killing vectors satisfy the following commutation relations with the modified Lie bracket \eqref{modified lie bracket gravity}:
\begin{equation}
[\xi (f_1, Y^A_1 ) , \xi (f_2, Y^A_2)]_A = \xi (\hat{f}, \hat{Y}^A )  ,
\end{equation} where 
\begin{align}
\hat{f} &= Y_1^A \partial_A f_2 + \frac{1}{2} f_1 D_A Y_2^A - \delta_{\xi (f_1, Y^A_1 )} f_2 - (1 \leftrightarrow 2) \label{bms like algebra 1}, \\
\hat{Y}^A &= Y^B_1 \partial_B Y_2^A - \frac{\Lambda}{3} f_1 q^{AB} \partial_B f_2 -   \delta_{\xi(f_1, Y_1^A)}Y^A_2  - (1 \leftrightarrow 2).
\label{bms like algebra 2}
\end{align} 
In the asymptotically flat limit $\Lambda = 0$, the functions $Y^A$, $f$ reduce to $Y^A = Y^A(x^B)$, $f = T(x^A) + \frac{u}{2} D_A V^A$ and the structure constants reduce to the ones of the generalized BMS$_4$ algebra $\mathfrak{bms}_4^\text{gen}$ (see equations \eqref{fAndY} and \eqref{struct const gen}). For $\Lambda \neq 0$, the supertranslations do not commute and the structure constants depend explicitly on $q_{AB}$. We therefore find the structure of a Lie algebroid \cite{Crainic,Lyakhovich:2004kr,Barnich:2017ubf,Barnich:2010xq}\footnote{The existence of the $\Lambda$-BMS$_4$ Lie algebroid is not in contradiction with recent no-go results \cite{Safari:2019zmc} that were obtained for Lie algebra deformations. Here, we have a field-dependent Lie algebroid deformation of the BMS Lie algebra in asymptotically locally (A)dS$_4$ spacetimes.}. We call it the $\Lambda$-BMS$_4$ algebra and we write it $\mathfrak{bms}_4^\Lambda$. This algebroid gives an infinite-dimensional algebra at each point of the solution space. Indeed, it always contains the area preserving diffeomorphisms given by $\xi = \xi ( f = 0, Y^A = Y^A (x^B))$, where $D_A Y^A = 0$\footnote{These vectors are called area preserving diffeomorphisms since, for a diffeomorphism on a two-dimensional Riemannian manifold with metric $q_{AB}$, the determinant transforms infinitesimally as $\delta_Y \sqrt{q} = D_A Y^A$. Therefore, the divergence-free vectors fields $Y^A$ generate diffeomorphisms that preserve the area.}.

When the transverse metric $q_{AB}$ is equal to the unit round sphere metric $\mathring{q}_{AB}$, we are back to the Dirchlet boundary conditions \eqref{cond asymp flat} and, therefore, $\mathfrak{bms}_4^\Lambda$ reduces to  $\mathfrak{so}(3,2)$ for $\Lambda <0$ and the $\mathfrak{so}(1,4)$ algebra for $\Lambda > 0$ (see \cite{Barnich:2013sxa} and appendix A of \cite{Compere:2019bua}).

\section{Holographic renormalization and surface charges}
\label{Holographic renormalization}

In this section, we reproduce the holographic renormalization in asymptotically locally (A)dS$_4$ spacetime in the Fefferman-Graham gauge \cite{deHaro:2000vlm , Skenderis:2002wp , Compere:2008us}. In this process, we assume only the preliminary boundary condition $\gamma_{ab} = \mathcal{O}(\rho^{-2})$ (see \eqref{preliminary FG}). This allows us to obtain the renormalized presymplectic form, from which we extract the charges for the most general solution space of asymptotically locally (A)dS$_4$ spacetime in the Fefferman-Graham gauge. Then, we compute the charge algebra and show that it closes under the modified bracket, without central extension.

\subsection{Presymplectic structure and its ambiguities}

In what follows, we will see that the counter-terms brought to the presymplectic form by the holographic renormalization process can be interpreted as ambiguities from the point of view of the Iyer-Wald procedure discussed in subsection \ref{Relation between Barnich-Brandt and Iyer-Wald procedures}. Let us mention two possible sources of ambiguities in the procedure that will appear in this process. 

A first ambiguity is the one discussed in subsection \ref{Relation between Barnich-Brandt and Iyer-Wald procedures} (see equation \eqref{ambiguity presymplectic potential IW} and the discussion that follows) and allows us to shift the presymplectic potential $\boldsymbol{\theta}[g; \delta g]$ by an exact $(n-1)$-form as
\begin{equation}
\boldsymbol{\theta}[g; \delta g] \to \boldsymbol{\theta}[g; \delta g] - \D \mathbf{Y}[g; \delta g] .
\end{equation} This leads to the following shift in the presymplectic form
\begin{equation}
\boldsymbol{\omega}[g; \delta g, \delta g] \to \boldsymbol{\omega}[g. \delta g, \delta g] - \delta \D \mathbf{Y} [g; \delta g]. 
\end{equation} In particular, this ambiguity has already been used in section \ref{sec:phasespace} to renormalize the symplectic structure in asymptotically flat spacetime.

Another freedom that we have is to modify the Lagrangian $\mathbf{L}[g]$ of the theory by boundary terms, 
\begin{equation}
\bm L_{EH}[g] \to \bm L_{EH}[g] + \D \bm A[g] .
\label{ambigu 2}
\end{equation} This shifts the presymplectic potential by an exact term 
\begin{equation}
\bm \theta [g ; \delta g] \to \bm \theta [g ; \delta g] + \delta \bm A[g]
\label{ambigu 1}
\end{equation} but leaves the presymplectic form invariant ($\delta^2 = 0$). Therefore, this freedom does not lead to further ambiguity in the symplectic structure. 

For a non-vanishing cosmological constant, the Einstein-Hilbert Lagrangian is
\begin{equation}
\bm L_{EH}[g] = \frac{1}{16\pi G}\left( R[g] - 2\Lambda\right)\sqrt{-g}\, \D^4 x . \label{eq:LEH}
\end{equation}
The associated canonical presymplectic potential is given by  
\begin{equation}
\bm \theta_{EH} [g;\delta g] = \frac{\sqrt{-g}}{16\pi G} \Big(\nabla_\nu (\delta g)^{\mu\nu} - \nabla^\mu (\delta g)^\nu_{\phantom{\nu}\nu}\Big) (\D^3 x)_{\mu}
\end{equation}
where $(\delta g)^{\mu\nu} = g^{\mu\alpha}g^{\nu\beta}\delta g_{\alpha\beta}$ (see equation \eqref{canonical presymplectic potential}). 

The radial component of the presymplectic potential can be computed as follows:
\begin{equation}
\begin{split}
\theta^\rho_{EH} [g;\delta g] &= \frac{\sqrt{-g}}{16\pi G}  \Big(\nabla_\alpha (\delta g)^{\rho \alpha} - g^{\rho\rho}\partial_{\rho} (\delta g)^\alpha_{\phantom{\alpha}\alpha}\Big)\\
&= \frac{\sqrt{-g}}{16\pi G}  \Big(\Gamma^\rho_{ab} \gamma^{ac}\gamma^{bd} \delta \gamma_{cd} -  g^{\rho\rho} \partial_\rho (\gamma^{ab} \delta \gamma_{ab})\Big).
\end{split}
\end{equation}
Expanding the metric $\gamma_{ab}(\rho,x^c)$ in powers of $\rho$, we get
\begin{align}
\theta^\rho_{EH} [g; \delta g] &=  \sqrt{\frac{3}{|\Lambda|}} \left[ -\frac{1}{\rho^3}\frac{2\Lambda}{3}\frac{\delta\sqrt{|g_{(0)}|}}{16\pi G} +\frac{1}{\rho}\left(-\frac{3}{4}\delta L_{EH,(0)} + \partial_a \theta^a_{EH,(0)}\right)\right] \nonumber\\
&+\frac{1}{2}\text{sgn}(\Lambda)\sqrt{|g_{(0)}|}\, T^{ab}\delta g_{ab}^{(0)} + \mathcal O(\rho).
\end{align}
We denoted by $L_{EH,(0)} = \frac{1}{16\pi G} R_{(0)}\sqrt{|g_{(0)}|}$ the Einstein-Hilbert Lagrangian density for the boundary metric field $g_{ab}^{(0)}$ and $\theta^a_{EH,(0)}$ the canonical boundary term in the variation $\delta L_{EH,(0)}$. We observe that the presymplectic potential is radially divergent as we approach the boundary $\mathscr I \equiv \{\rho = 0 \}$, so we need a renormalization procedure to obtain a well-defined symplectic structure at $\mathscr I$, allowing us to compute the surface charges. The precise form of the divergence suggests that there is a boundary-covariant way to subtract these divergences by refining the action principle of pure gravity in asymptotically locally (A)dS$_4$ spacetimes: this is the point of the \textit{holographic renormalization} \cite{deHaro:2000vlm , Skenderis:2002wp , Compere:2008us} that we review in the next section.

\subsection{Holographic renormalization}

The action for general relativity in asymptotically locally (A)dS$_4$ spacetimes is given by \cite{deHaro:2000vlm , Skenderis:2002wp , Compere:2008us}
\begin{equation}
S = \frac{1}{16\pi G} \int_{\mathscr M} \text d^4 x \, \sqrt{|g|} \, (R[g]-2\Lambda) + \int_{\mathscr{I}} \D^3 x\, L_{GHY} + \int_{\mathscr I} \D^3 x\, L_{ct} .
\label{total action}
\end{equation} Here, $\mathscr M$ denotes the bulk spacetime and $\mathscr{I} = \partial \mathscr M$ its boundary. We impose that $\int_{\mathscr M} \text d^4 x = \int_0^\infty \text d\rho' \int_{\rho = \rho'} \text d^3 x$. Remark that this convention sets the \textit{lower} bound of the radial integral to be the boundary. The integration measure $d^3 x$ should be understood as a measure on the hypersurface at fixed $\rho = \rho'$. In particular, consistently with the notations of appendix \ref{Useful results}, we have
\begin{equation}
(\D^3x)^{n=3} = \frac{1}{3!} \epsilon^{n=3}_{abc} \D x^a \wedge \D x^b \wedge \D x^c \equiv \frac{1}{1!3!} \epsilon^{n=4}_{\rho abc} \D x^a \wedge \D x^b \wedge \D x^c = (\D^3x)^{n=4}_\rho .
\end{equation} This allows us to interpret the top form on the hypersurface $\rho = \rho'$ as co-dimension $1$ with respect to the four-dimensional spacetime (for example, $L_{GHY} (\D^3x)^{n=3} |_{\rho = \rho'} \equiv L^\rho_{GHY} (\D^3x)_\rho^{n=4} |_{\rho = \rho'}$). The first term in \eqref{total action} is the bare Einstein-Hilbert action $S_{EH}$, the second term is the Gibbons-Hawking-York term $S_{GHY}$ and the third term is the counter-term action $S_{ct}$. 

Let us describe the additional boundary terms in \eqref{total action} and justify their presence. To have a well-defined variational principle for Dirichlet boundary conditions, i.e. when all induced fields at the boundary are kept fixed ($\delta \gamma_{ab}|_{\mathscr I} = 0$), the action must be completed by the usual Gibbons-Hawking-York boundary term $S_{GHY} = \int_{\partial\mathscr M} \D^3 x\, L_{GHY}$. Let us denote the outward normal unit vector by $\bm n = n^\mu \partial_\mu$, such that $n^\mu n_\mu = \eta$, where $\eta = - \text{sgn}(\Lambda)$. Here, ``outward'' means that the vector points from the inside of the enclosed region to the outside. Recall that $\mathscr I$ is defined as the set of roots of the scalar field $f(\rho,x^a) = \rho$. Hence $n_\mu$ is collinear to $\partial_\mu f$ and differs only by a normalization factor and a relative sign. Since the coordinate $\rho$ increases inwards, $n^\mu$ must point in the direction of decreasing $f$, such that $n^\mu \partial_\mu f < 0$, independently of $\eta$. We get $n_{\mu} = - \eta \sqrt{|g_{\rho\rho}|} \delta^\rho_\mu$. The knowledge of this unit normal vector allows us to define the extrinsic curvature as the trace of the second fundamental form $K = \gamma^{ab}K_{ab} = \frac{1}{2}\gamma^{ab}\mathcal L_{\bm n}\gamma_{ab}$, and build the Gibbons-Hawking-York piece
\begin{equation}
S_{GHY} = \frac{1}{8\pi G} \eta \int_{\mathscr I} \text d^3 x \sqrt{|\gamma|} K = \frac{1}{8\pi G}\sqrt{\frac{3}{|\Lambda|}} \int_{\mathscr I} \D^3 x \, \frac{\Lambda}{3}  \, \rho\partial_\rho \sqrt{|\gamma|}.
\end{equation} 
An important observation is that the on-shell action $S_{EH} + S_{GHY}$ is divergent. In order to deal with these divergences, we introduce an infrared cut-off $\varepsilon > 0$ (called the \textit{regulator}) sufficiently small so that the Fefferman-Graham expansion is still valid around $\lbrace\rho=\varepsilon\rbrace$. The regulated variational principle
\begin{equation}
S^{\text{reg}} = \frac{1}{16\pi G}\int_{\varepsilon}^{\infty} \text d\rho' \int_{\rho=\rho'} \text d^3 x (R[g]-2\Lambda)\sqrt{|g|} + \frac{1}{8\pi G} \, \eta \int_{\rho=\varepsilon} \text d^3 x  \sqrt{|\gamma|} K
\end{equation}
possesses two divergent pieces on-shell
\begin{equation}
S^{\text{reg}} = \frac{1}{16\pi G} \sqrt{\frac{3}{|\Lambda|}} \int_{\rho=\varepsilon} \text d^3 x \left[ -\frac{4\Lambda}{3}\sqrt{|g_{(0)}|}\frac{1}{\varepsilon^3} + \frac{1}{2}R_{(0)}\sqrt{|g_{(0)}|} \frac{1}{\varepsilon} + \mathcal{O}(\varepsilon) \right]. \label{eq:Sreg}
\end{equation}
The holographic renormalization procedure amounts to supplying the regulated variation principle with a second counterterm $S_{ct} = \int_{\rho=\varepsilon} \D^3 x\, L_{ct}$ which must obey several requirements: $S_{ct}$ is a boundary action constructed from a Lagrangian $L_{ct}$ considered as a top-form living on the regulated hypersurface $\lbrace \rho=\varepsilon\rbrace$. The latter is built up from covariant objects living on $\lbrace \rho=\varepsilon\rbrace$, but is not required to be covariant with respect to the bulk geometry. In particular, it will involve the metric $\gamma_{ab}(\varepsilon,x^c)$ only. The renormalization requirement imposes that $S^{\text{reg}} + S_{ct} = \mathcal O(\varepsilon^0)$ after expanding in power series of $\varepsilon$. The working counterterm has been prescribed in \cite{deHaro:2000vlm , Skenderis:2002wp} and is given by
\begin{equation}
S_{ct} = \int_{\rho=\varepsilon} \text d^3 x\, L_{ct}[\gamma],\quad L_{ct}[\gamma] = \frac{1}{16\pi G}\sqrt{\frac{3}{\Lambda}} \left[\frac{4\Lambda}{3}\sqrt{|\gamma|}-R[\gamma]\sqrt{|\gamma|}\right].
\end{equation}
Indeed, it evidently satisfies the first two requirements, and we also check the last one by expanding $L_{ct}$ in $\varepsilon$,
\begin{equation}
L_{ct} = \frac{1}{16\pi G} \sqrt{\frac{3}{|\Lambda|}} \left[ \frac{4\Lambda}{3}\sqrt{|g_{(0)}|}\frac{1}{\varepsilon^3} - \frac{1}{2}R_{(0)}\sqrt{|g_{(0)}|} \frac{1}{\varepsilon} + \mathcal{O}(\varepsilon) \right],
\end{equation}
hence $S^{\text{reg}} + S_{ct} = \mathcal O(\varepsilon)$. For later purposes, we define the presymplectic potential associated with $L_{ct}$ as $\delta {L}_{ct} = \frac{\delta {L}_{ct}}{\delta \gamma^{ab}} \delta \gamma^{ab} + \partial_a {\theta}^a_{ct} [\gamma ; \delta \gamma]$. It is given explicitly by 
\begin{equation}
\begin{split}
{\theta}^a_{ct}[\gamma ; \delta \gamma] &= -\frac{1}{16 \pi G} \sqrt{\frac{3}{|\Lambda|}} \sqrt{|\gamma|} \left[D_b (\delta \gamma )^{ab} - \gamma^{ab} D_b {(\delta \gamma)^c}_c \right] \\
&= -\frac{1}{16 \pi G} \rho \sqrt{|g|} \left[D_b (\delta \gamma )^{ab} - \gamma^{ab} D_b {(\delta \gamma)^c}_c \right] \\
&= - \rho \theta^a_{EH} [\gamma; \delta \gamma],
\end{split}
\end{equation} where $D_a$ denotes the Levi-Civita connection with respect to $\gamma_{ab}$. Therefore, 
\begin{equation}
{\theta}^a_{ct}[\gamma ; \delta \gamma]\Big|_{\rho=\varepsilon} = -\sqrt{\frac{3}{\Lambda}} \Theta^a_{EH,(0)}[g_{(0)};\delta g_{(0)}]\frac{1}{\varepsilon} + \mathcal O(\varepsilon). 
\end{equation}

Now we can concentrate on the renormalization of the presyplectic potential. On-shell, we have
\begin{equation}
\begin{split}
\delta S_{EH}^{\text{reg}} &=  \int_{\rho\geq \varepsilon} \text d^4 x \, \partial_\mu \theta_{EH}^\mu [g; \delta g] = - \int_{\rho=\varepsilon} \text d^3 x \, \theta^\rho_{EH}[g ; \delta g], \label{eq:DeltaSEHreg}
\end{split}
\end{equation} where the minus sign in the last equality is due to the fact that we integrate on $\rho$ from the boundary to the bulk, which gives the negative orientation to the Stokes formula. The resulting integrand is only the $\rho$ component of $\boldsymbol{\theta}_{EH}$ since the outward normal to the regulating surface is collinear to $\partial_\rho$. Therefore, we can prove by a straightforward computation that the renormalization of the presymplectic potential works as follows \cite{Compere:2008us}:
\begin{equation}
\theta^\rho_{\text{ren}} [g; \delta g] = \theta^{\rho}_{EH} - \delta L_{GHY} - \delta L_{ct}+ \partial_a \theta^a_{ct} = -\frac{1}{2}\eta \sqrt{|g_{(0)}|}T^{ab}\delta g^{(0)}_{ab} + \mathcal{O}(\rho). \label{eq:RenormalisationOfTheta}
\end{equation}
We deduce from \eqref{eq:RenormalisationOfTheta} that the modification brought to the presymplectic potential by the holographic renormalization is two-fold. The contribution of the exact terms with respect to $\delta$ are top-forms on the regularized boundary that can be promoted as bulk co-dimension $1$-forms, collectively denoted by $\bm A = A^\rho (\text d^4 x)_\rho$, with $A^\rho \equiv -(L_{GHY}+L_{ct})$, such that $\bm \theta_{EH} \to \bm\theta' = \bm \theta_{EH}+\delta\bm A$ after renormalization (see \eqref{ambigu 1}). The contribution of the Iyer-Wald ambiguity appears here thanks to $\theta^a_{ct}$, which is a co-dimension $1$-form on the regularized boundary. Again we can promote it as a co-dimension $2$-form on the bulk geometry $\bm Y = Y^{\rho a}(\text d^2 x)_{\rho a}$, and $Y^{\rho a}[g ; \delta g] \equiv \theta^a_{ct}[\gamma;\delta\gamma]$. The potential will be modified as $\bm\theta'\to\bm\theta' - \text d\bm Y$ (see \eqref{ambigu 2}), or in components,
\begin{align}
\theta'^\rho \to \theta'^\rho +\partial_a Y^{\rho a} = \theta'^\rho + \partial_a \theta^a_{ct}, \label{eq:YmodThetaRho} \\
\theta'^a \to \theta'^a + \partial_\rho Y^{a \rho} = \theta'^a - \partial_\rho \theta^a_{ct},
\label{eq:YmodThetaA}
\end{align}
In particular, \eqref{eq:YmodThetaRho} is consistent with \eqref{eq:RenormalisationOfTheta}, and it can be shown that \eqref{eq:YmodThetaA} renormalizes the tangent components of the presymplectic potential as well. Finally, the renormalized presymplectic current is given by
\begin{equation}
\omega_{\text{ren}}^\rho [g;\delta g,\delta g] = - \frac{1}{2}\eta \delta \left(\sqrt{|g_{(0)}|}T^{ab}\right) \wedge \delta g^{(0)}_{ab} + \mathcal{O}(\rho).
\label{presymplectic form prime}
\end{equation}

\subsection{Surface charges}

Once the expression for the renormalized presymplectic potential $\bm \theta_{\text{ren}}[g;\delta g]$ is established, one can compute the Iyer-Wald co-dimension 2 form as $\bm k_{\xi, \text{ren}}[g;\delta g] = -\delta \bm Q_{\xi, \text{ren}}[g] + \bm Q_{\delta \xi, \text{ren}}[g]  + i_\xi \bm\theta_{\text{ren}} [g;\delta g]$ (see equation \eqref{Iyer-Wald}). In the present context, instead of directly computing this expression, we propose an ansatz for the co-dimension 2 form inspired by the results of \cite{Compere:2008us} that were obtained for a sub-case ($\sigma = 0$, $\delta \xi^a = 0$, $\Lambda < 0$). Our ansatz for the components $\rho a$ of the co-dimension 2 form associated with the most general asymptotically locally (A)dS$_4$ spacetime in Fefferman-Graham gauge is given by
\begin{equation}
k_{\xi,\text{ren}}^{\rho a}[g;\delta g] = \eta \delta \left( \sqrt{|g_{(0)}|} {T^a}_b \right)\xi^b_0 -\frac{1}{2}\eta \sqrt{|g_{(0)}|}\,\xi_0^a\,T^{bc}\delta g^{(0)}_{bc} +\mathcal{O}(\rho). \label{eq:FinalCharges}
\end{equation} To confirm that this proposal is the correct one, we check that it satisfies the conservation law \eqref{breaking in the conservation 2}, namely
\begin{equation}
\text d \bm k_{\xi,\text{ren}}[g; \delta g] = \bm \omega_{\text{ren}}[g;\mathcal L_\xi g , \delta g] \quad\Rightarrow\quad \partial_a k_{\xi,\text{ren}}^{\rho a}[g; \delta g] = \omega^\rho_{\text{ren}} [g; \mathcal L_\xi g , \delta g] . \label{eq:FundamentalThm}
\end{equation} The detailed computation can be found in appendix \ref{Check of the conservation law}. Since the co-dimension 2 form $k_{\xi,\text{ren}}[g; \delta g]$ is defined up to an exact co-dimension 2 form, we are certain that the proposal \eqref{eq:FinalCharges} is the right one for the renormalized presymplectic form \eqref{presymplectic form prime}.

\subsection{Charge algebra}

\subsubsection{Modified Lie bracket for residual gauge diffeomorphisms}

Let us denote by $\xi$ and $\chi$ two arbitrary residual gauge diffeomorphisms of the Fefferman-Graham expansion which are of the form \eqref{AKV 1 prime} and \eqref{AKV 2 prime}. We recall that the modified Lie bracket is given by
\begin{equation}
[\xi,\chi]_A = [\xi,\chi] - \delta_\xi \chi + \delta_\chi \xi, \label{eq:ModLieBracket}
\end{equation}
with $\delta_\xi g_{\mu\nu} \equiv \mathcal L_\xi g_{\mu\nu}$ (see \eqref{modified lie bracket gravity}). We now provide an explicit computation of this bracket for the present case. Since $\xi$ and $\chi$ preserve the Fefferman-Graham gauge, they satisfy
\begin{equation}
\left\lbrace
\begin{split}
\xi^\rho &= \rho \sigma_\xi(x^a), \quad \partial_\rho \xi^a = \frac{3}{\Lambda}\frac{1}{\rho}\gamma^{ab}\partial_b \sigma_\xi, \quad \lim_{\rho\to 0} \xi^a = \xi^a_0(x^b), \\
\chi^\rho &= \rho \sigma_\xi(x^a), \quad \partial_\rho \chi^a = \frac{3}{\Lambda}\frac{1}{\rho}\gamma^{ab}\partial_b \sigma_\chi, \quad \lim_{\rho\to 0} \chi^a = \chi^a_0(x^b). \\
\end{split} \right. \label{eq:RefVec}
\end{equation}
As a result, the computation of $[\xi,\chi]^\rho_A$ is straightforward and gives
\begin{equation}
\frac{1}{\rho}[\xi,\chi]^\rho_A = \left( \xi^a\partial_a\sigma_\chi - \chi^a\partial_a\sigma_\xi\right) - \delta_\xi\sigma_\chi + \delta_\chi\sigma_\xi.
\end{equation}
Taking a derivative with respect to $\rho$, and again using $\rho^2 \gamma_{ab} \partial_\rho \xi^b - \frac{3}{\Lambda}\partial_a \xi^\rho =0$, we get
\begin{equation}
\partial_\rho \left( \frac{1}{\rho}[\xi,\chi]^\rho_A \right) = \partial_\rho \xi^a \partial_a \sigma_\chi -\partial_\rho \chi^a\partial_a\sigma_\xi = 0,
\end{equation}
which shows that $[\xi,\chi]^\rho_\star = \rho \hat \sigma$, and
\begin{equation}
\hat \sigma = \frac{1}{\rho}[\xi,\chi]^\rho_A \Big|_{\rho=0} =  \xi_0^a\partial_a \sigma_\chi - \chi^a_0\partial_a \sigma_\xi - \delta_\xi\sigma_\chi + \delta_\chi\sigma_\xi.
\end{equation}
Let us now consider the transverse components. By evaluating the commutator at leading order, we derive that
\begin{equation}
\hat \xi_0^a = \lim_{\rho\to 0} [\xi,\chi]^a_A = [\xi_0,\chi_0]^a - \delta_\xi\chi_0^a + \delta_\chi\xi_0^a.
\end{equation}
Recalling that $\delta_\xi \gamma^{ab} = \mathcal L_\xi \gamma^{ab} = \rho\sigma_\xi \partial_\rho \gamma^{ab} + \xi^c \partial_c \gamma^{ab} - 2 \gamma^{c(a}\partial_c \xi^{b)}$ and explicitly using \eqref{eq:RefVec} to express $\partial_\rho \xi^a$ and $\partial_\rho \chi^b$ in terms of $\sigma_\xi$ and $\sigma_\chi$, respectively, a direct computation yields
\begin{equation}
\partial_\rho \left( [\xi,\chi]^a_A \right) = \frac{3}{\Lambda}\frac{1}{\rho}\gamma^{ab} \partial_b \hat \sigma.
\end{equation}
We have just proven that residual gauge diffeomorphisms $\xi$ and $\chi$ of the Fefferman-Graham gauge satisfy 
\begin{equation}
[\xi(\sigma_\xi,\xi_0),\chi(\sigma_\chi,\chi_0)]_A =  \xi (\hat \sigma,\hat \zeta_0), \label{eq:VectorAlgebra1}
\end{equation}
where
\begin{equation}
\begin{split}
\hat \sigma &= \xi_0^a\partial_a \sigma_\chi - \chi^a_0\partial_a \sigma_\xi - \delta_\xi\sigma_\chi + \delta_\chi\sigma_\xi, \\
\hat \zeta^a_0 &=  \xi^b_0\partial_b \chi^a_0 - \chi^b_0\partial_b \xi^a_0 - \delta_\xi\chi_0^a + \delta_\chi\xi_0^a.
\end{split} \label{eq:VectorAlgebra}
\end{equation}

\subsubsection{Charge algebra with modified bracket}

The co-dimension 2 form derived in \eqref{eq:FinalCharges} is generically non-integrable for the most general asymptotically locally (A)dS$_4$ boundary conditions in the Fefferman-Graham gauge. Therefore, the representation theorem \eqref{charge algebra integrable} does not hold and one has to consider the modified bracket for the charges.  

The leading term of \eqref{eq:FinalCharges} ($\sim \rho^0$) can be written as
\begin{equation}
k_{\xi, \text{ren}}^{\rho a}[g ; \delta g] \Big|_{\rho = 0} = \delta J^a_\xi[g] + \Xi^a_\xi[g ; \delta g] ,
\end{equation}
where the integrable part is taken as
\begin{equation}
J^a_\xi[g] = \eta \sqrt{|g_{(0)}|}\,g_{(0)}^{ac}T_{bc}\xi_0^b
\label{integrable current}
\end{equation}
and the corresponding non-integrable part 
\begin{equation}
\Xi^a_\xi[g; \delta g] = -\frac{1}{2}\eta\sqrt{|g_{(0)}|} \xi_0^a \left( T^{bc}\delta g_{bc}^{(0)}\right) - J^a_{\delta\xi}.
\label{non integrble current}
\end{equation}
Defining the modified bracket as
\begin{equation}
\lbrace J_\xi[g],J_\chi[g] \rbrace^a_* = \delta_\chi J_\xi^a[g] + \Xi_\chi^a[g;\delta_\xi g]
\label{modified bracket AAdS}
\end{equation} (this is the analogue of \eqref{modified bracket charges}), we show in appendix \ref{sec:Charge algebra} that 
\begin{equation}
\lbrace J_\xi[g],J_\chi[g] \rbrace^a_* = J^a_{[\xi,\chi]_A}[g] + \partial_b \ell^{ab}[g]. 
\label{charge algebra AAdS}
\end{equation} where $\partial_b \ell^{ab}[g]$ (see equation \eqref{boundary term algebra}) is a term that will disappear once integrated on $S^2_\infty$. Therefore, when considering the modified bracket \eqref{modified bracket AAdS}, we conclude that the currents satisfy a consistent algebra without central extension. This contrasts with the asymptotically flat case where a $2$-cocycle appears in the right-hand side (compare \eqref{charge algebra AAdS} with \eqref{algebra modified}). Of course, as discussed around equation \eqref{modif central charge}, modifying the split \eqref{integrable current}-\eqref{non integrble current} between integrable and non-integrable parts will bring a trivial 2-cocycle in the algebra \eqref{charge algebra AAdS}. This situation is similar to the case studied in \cite{Adami:2020amw} where the 2-cocycle could be absorbed by choosing an appropriate split.

\section{$\Lambda$-BMS$_4$ phase space and its flat limit}
\label{phase space and its flat limit}

In the previous section, we obtained through the holographic renormalization process the renormalized co-dimension 2 form associated with the most general asymptotically locally (A)dS$_4$ boundary conditions in the Fefferman-Graham gauge. We saw that the associated currents satisfy an algebra by using the modified bracket. In subsection \ref{Lambda phase space in Fefferman-Graham gauge}, we particularize the analysis for the boundary conditions \eqref{new boundary conditions} and obtain the symplectic structure and the surface charges associated with the $\mathfrak{bms}_4^\Lambda$ asymptotic symmetry algebra. In subsection \ref{Translation into the Bondi gauge}, we express the symplectic structure in terms of the Bondi gauge variables and perform the diffeomorphism discussed in appendix \ref{app:chgt}. As discussed in section \ref{sec2}, the solution space with boundary conditions \eqref{new boundary conditions} reduces to the solution space considered in chapter \ref{Generalized BMS and renormalized phase space}. Similarly, in section \ref{Symmetries and flat limit}, we showed that $\mathfrak{bms}^\Lambda_4$ reduces to $\mathfrak{bms}^\text{gen}_4$ in the flat limit. In subsection \ref{Renormalization in Lambda and flat limit}, after renormalization of $\sim 1/\Lambda$ divergences, we prove that this limit also holds at the level of the phase space. 

\subsection{$\Lambda$-BMS$_4$ phase space in Fefferman-Graham gauge}
\label{Lambda phase space in Fefferman-Graham gauge}

Using \eqref{g0 in term of Bondi}, the boundary conditions \eqref{new boundary conditions} can be expressed in the Fefferman-Graham gauge as
\begin{equation}
g_{tt}^{(0)} = \frac{\Lambda}{3}, \quad g_{tA}^{(0)} = 0, \quad \sqrt{|g_{(0)}|} = \sqrt{\frac{|\Lambda|}{3}}\sqrt{\mathring q}, \label{eq:LBMSConditions}
\end{equation} where $\mathring q$ is the determinant of the unit sphere metric. The asymptotic Killing vectors are the residual gauge diffeomorphisms given in \eqref{AKV 1 prime} and \eqref{AKV 2 prime}, whose parameters satisfy \eqref{sig}. Inserting the conditions \eqref{eq:LBMSConditions} into the renormalized presymplectic potential \eqref{eq:RenormalisationOfTheta}, we obtain
\begin{equation}
\theta^\rho_{\Lambda\text{-BMS}} [g; \delta g ] = - \frac{\sqrt{\mathring{q}}}{2} \eta \sqrt{\frac{|\Lambda|}{3}} g^{AB}_{(0)} \delta T_{AB}^{TF} + \mathcal{O} (\rho) , 
\label{theta in FG}
\end{equation} where $T_{AB}^{TF} = T_{AB} - \frac{1}{2} g_{AB}^{(0)} (g^{CD}_{(0)} T_{CD})$. Similarly, the presymplectic form \eqref{presymplectic form prime} reduces to 
\begin{equation}
\omega^\rho_{\Lambda\text{-BMS}} [g; \delta g ] =  - \frac{\sqrt{\mathring{q}}}{2} \eta  \sqrt{\frac{|\Lambda|}{3}} \delta g^{AB}_{(0)} \wedge \delta T_{AB}^{TF} + \mathcal{O} (\rho) 
\end{equation} From \eqref{eq:FinalCharges}, we deduce that the $\Lambda$-BMS$_4$ surface charges are given by
\begin{equation}
\ndelta H_\xi^{\Lambda\text{-BMS}} [g] =  \int_{S^2_\infty} 2 (\D^2 x)_{\rho t} k^{\rho t}_{\xi, \Lambda\text{-BMS}} [g; \delta g] = \delta H_\xi^{\Lambda\text{-BMS}} [g] + \Theta_\xi^{\Lambda\text{-BMS}} [g;\delta g] , \label{Lambda-BMS charges 1}
\end{equation} where we performed a split between integrable and non-integrable parts as in \eqref{integrable current} and \eqref{non integrble current}:
\begin{equation}
\begin{split}
H_\xi^{\Lambda\text{-BMS}}[g] &= -\sqrt{\frac{3}{|\Lambda|}} \int_{S^2_\infty} \D^2 \Omega ~\left[ \xi^t_0 T_{tt} + \xi_0^B T_{tB}   \right], \\
\Theta_\xi^{\Lambda\text{-BMS}} [g;\delta g] &= \sqrt{\frac{3}{|\Lambda|}} \int_{S^2_\infty} \D^2 \Omega ~\left[ \frac{\Lambda}{6}\xi_0^t g_{(0)}^{AB}\delta T_{AB} \right] - H_{\delta\xi}^{\Lambda\text{-BMS}}[g] .
\end{split}\label{Lambda-BMS charges 2}
\end{equation} Here, $\D^2 \Omega = 2 \sqrt{\mathring q} (\D^2 x)_{\rho t}$ denotes the measure on $S^2_\infty$. As a corollary of \eqref{charge algebra AAdS}, they satisfy an algebra for the modified bracket 
\begin{equation}
\lbrace H^{\Lambda\text{-BMS}}_\xi[g], H^{\Lambda\text{-BMS}}_\chi[g] \rbrace_\star = H_{[\xi,\chi]_A}^{\Lambda\text{-BMS}}[g] , \label{eq:AlgebraLBMS}
\end{equation} with $\lbrace H^{\Lambda\text{-BMS}}_\xi[g], H^{\Lambda\text{-BMS}}_\chi[g] \rbrace_\star = \delta_{\chi} H^{\Lambda\text{-BMS}}_\xi[g] + \Theta_\chi^{\Lambda\text{-BMS}} [g;\delta_\xi g]$, and $[\xi,\chi]_A$ given by\begin{equation}
[\xi(\xi^t_{0} , \xi^A_{0}), \chi(\chi^t_{0} , \chi^A_{0})]_A = \hat \xi(\hat{\xi}^t_{0} , \hat{\xi}^A_{0}) ,
\label{struc const}
\end{equation} where
\begin{equation}
\left\lbrace \,\,
\begin{split}
{\hat{\xi}}_0^t &= \xi^A_{0} \partial_A \chi_{0}^t + \frac{1}{2} \xi_{0}^t  D_A^{(0)} \chi^A_{0} - \delta_{\xi(\xi^t_{0} , \xi^A_{0})} \chi^t_{0} - (\xi \leftrightarrow \chi ), \\
{\hat{\xi}}_0^A &= {\xi}_{0}^B \partial_B \chi^A_{0} - \frac{\Lambda}{3} \xi_{0}^t g^{AB}_{(0)} \partial_B \chi_{0}^t  - \delta_{\xi(\xi^t_{0} , \xi^A_{0})} \chi^A_{0} - (\xi \leftrightarrow \chi ). \label{structure constant Lambda BMSSSS}
\end{split}
\right.
\end{equation} This is a corollary of \eqref{eq:VectorAlgebra} and \eqref{eq:VectorAlgebra1}. Alternatively, one can obtain the commutation relations \eqref{structure constant Lambda BMSSSS} from those written in the Bondi variables \eqref{bms like algebra 2}, using the dictionary \eqref{translation parameters}.

\subsection{Translation into the Bondi gauge}
\label{Translation into the Bondi gauge}

The next step is to perform the change of coordinates described in appendix \ref{app:chgt} between Fefferman-Graham and Bondi gauges and deduce the transformation of the presymplectic potential. Starting from Fefferman-Graham coordinates, we first go to tortoise coordinates $(r_\star,x^a_\star)$. The presymplectic potential reads as $\bm \theta_{\Lambda\text{-BMS}} = \theta^{\rho}_{\Lambda\text{-BMS}} (\D^3 x)_{\rho} + \theta_{\Lambda\text{-BMS}}^a (\D^3 x)_{a}$. At leading order, $\rho = -\frac{\Lambda}{3}r_\star + \mathcal O(r_\star^2)$ and $x^a = x^a_\star + \mathcal O(r_\star)$, hence $\theta^{r_\star}_{\Lambda\text{-BMS}}= \theta^{\rho}_{\Lambda\text{-BMS}} + \mathcal O(r_\star)$. Therefore, the leading order of the radial component of the presymplectic potential is not affected. Now we can reach the Bondi gauge $(u,r,x^A)$ by a second change of coordinates
\begin{equation}
t_\star = u + \frac{3}{\Lambda}\frac{1}{r} + \mathcal O\left(\frac{1}{r^3}\right), \quad
r_\star = \frac{3}{\Lambda}\frac{1}{r} + \mathcal O\left(\frac{1}{r^3}\right), \quad x^A_\star = x^A.
\label{diffeo-stage2}
\end{equation}
We obtain $\theta^{r}_{\Lambda\text{-BMS}} = \theta^{r_\star}_{\Lambda\text{-BMS}} + \mathcal O(r)$\footnote{Since the field-dependence involved in the diffeomorphism \eqref{diffeo-stage2} appears only at subleading orders in $r$, we assumed that the variational operator $\delta$ is not affected at leading order and therefore does not bring any contribution to the leading order in the transformation of $\theta^\rho_{\Lambda\text{-BMS}}$.}. Therefore, using \eqref{eq:RefiningTab}, the renormalized presymplectic potential \eqref{theta in FG} is expressed in Bondi gauge as
\begin{equation}
\theta^r_{\Lambda\text{-BMS}}[g;\delta g] = \frac{\Lambda}{2}\frac{\sqrt{\mathring q}}{16\pi G} \, q^{AB}\delta J_{AB} + \mathcal{O}(r^{-1}).
\label{thetabondi ren}
\end{equation} The associated presymplectic form is readily obtained
\begin{equation}
\omega^r_{\Lambda\text{-BMS}}[g;\delta g] = \frac{\Lambda}{2}\frac{\sqrt{\mathring q}}{16\pi G} \, \delta q^{AB} \wedge \delta J_{AB} + \mathcal{O}(r^{-1}).
\label{lambda bms symplectic form}
\end{equation} As discussed in \cite{Compere:2019bua}, by analogy with the flat case (see e.g. \eqref{non conservation BMS}), this last expression allows us to identify the Bondi news functions in (A)dS$_4$ as the symplectic couple $(q^{AB}, J_{AB})$.

\subsection{Renormalization in $\Lambda$ and flat limit}
\label{Renormalization in Lambda and flat limit}

The Bondi gauge admitting a well-defined flat limit, nothing prevents us from considering the flat limit of the above symplectic structure. Let us recall the prescription followed in subsection \ref{sec2} when discussing the flat limit of the solution space:

\begin{enumerate}
\item In the Bondi gauge, we explicitly identify the dependence in $\Lambda$ in the different expressions by using the relations in section \ref{sec2} (see e.g. \eqref{eq:hatM}, \eqref{eq:hatNA}, \eqref{eq:hatJAB}), until obtaining only the functions $C_{AB}$, $q_{AB}$, $N_{AB}^{TF}$, $M$, $N_{A}$ and their derivatives with respect to the angles, where $N_{AB}^{TF}$ designates the trace-free part of $N_{AB} = \partial_u C_{AB}$. These functions are those that admit a well-defined interpretation in the flat limit, which contrasts with $J_{AB}$, $M^{(\Lambda)}$ and $N_A^{(\Lambda)}$. Furthermore, the relation \eqref{eq:CAB 2} is extensively used in order to exchange terms $\sim \partial_u q_{AB}$ for terms $\sim \Lambda C_{AB}$. The following identities turn out to be useful for the computation:
\begin{equation}
\begin{split}
\partial_u H^{AB} &= q^{AC}\partial_u H_{CD}q^{BD} - \frac{2\Lambda}{3} C^{C(A}H^{B)}_C,   \\
\partial_u (D^B H_{AB}) &= D^B \partial_u H_{AB} - \frac{\Lambda}{3}\Big[ D_C (H_{AB}C^{BC}) + \frac{1}{2} H^{BC}D_A C_{BC}\Big], \\
\partial_u C^{AB}C_{AB} &= 2 N^{AB}C_{AB} , \\
\partial_u \partial_A R[q] &=  \frac{\Lambda}{3} D_A D_B D_C C^{BC} ,
\end{split}
\end{equation}
where $H_{AB}(u,x^C)$ is any symmetric rank 2 transverse tensor. 
\item Once this procedure is achieved, we take the flat limit by putting $\Lambda \to 0$. As we later explain, this limit may require a renormalization procedure to remove the divergences $\sim \frac{1}{\Lambda}$. 
\end{enumerate}

Let us apply the first step of the procedure to the presymplectic potential. Starting from its expression in the Bondi gauge \eqref{thetabondi ren} and using \eqref{eq:hatJAB}, we get 
\begin{equation}
\begin{split}
\theta^r_{\Lambda\text{-BMS}} [g;\delta g] &= \frac{\sqrt{\mathring q}}{16\pi G} \Big[ \frac{3}{2\Lambda}\partial_u (N_{AB}^{TF}\delta q^{AB}) + \frac{1}{2}\Big( N^{AB}_{TF} + \frac{1}{2} R[q]q^{AB}\Big) \delta C_{AB}  \\
&\qquad\qquad\quad + \frac{1}{2} D_A D^C C_{BC}\delta q^{AB}\Big] + \mathcal O(\Lambda; r^{-1} ),
\end{split}
\end{equation} where the notation $\mathcal{O}(\Lambda, r^{-1})$ designates terms $\mathcal{O}(r^{-1})$ and/or $\mathcal{O}(\Lambda)$. A striking observation is that we have a term $\sim \Lambda^{-1}$ in this expression, which does not allow us to go to the second stage of the flat limit procedure. We have to suppress this divergence before taking $\Lambda \to 0$. A way to proceed is to use the ambiguity allowed by the covariant phase space formalism \eqref{ambiguity presymplectic potential IW}. Indeed, noticing that the term $\sim \Lambda^{-1}$ can be expressed as
\begin{equation}
\frac{\sqrt{\mathring q}}{16\pi G}\frac{3}{2\Lambda} \partial_u (N_{AB}^{TF}\delta q^{AB}) = \frac{1}{16\pi G} \frac{3}{\Lambda}\partial_u \left[\frac{1}{2}\partial_u(\sqrt{\mathring q}\,  C_{AB}\delta q^{AB})\right] + \frac{\sqrt{\mathring q}}{16\pi G} \, \delta(C^{AB}N_{AB}^{TF}) , 
\end{equation} we define 
\begin{equation}
Y^{ru}_{(\Lambda)}[g;\delta g] = -\frac{1}{16\pi G} \frac{3}{\Lambda}\frac{1}{2}\partial_u(\sqrt{\mathring q}\,  C_{AB}\delta q^{AB}), 
\end{equation} and $Y^{rA} = 0$. As discussed in \cite{newCompere}, the presence of this Iyer-Wald ambiguity can be justified by adding corner terms in the variational principle \eqref{total action}. The presymplectic potential is renormalized as
\begin{equation}
\begin{split}
&\theta^r_{\text{ren}(\Lambda)}[g;\delta g] = \theta^r_{\Lambda\text{-BMS}} [g;\delta g] + \partial_u Y^{ru}_{(\Lambda )}[g;\delta g] \\
&\quad= \frac{\sqrt{\mathring q}}{16\pi G} \Big[ \frac{1}{2}\Big( N^{AB}_{TF} + \frac{1}{2} R[q]q^{AB}\Big) \delta C_{AB} + \frac{1}{2} \Big(D_A D^C C_{BC}\Big) \delta q^{AB} + \delta (C^{AB}N^{TF}_{AB}) \Big] \\
&\qquad + \mathcal{O}(\Lambda; r^{-1} ) \label{eq:ThetaWellDefinedFlatLimit}
\end{split}
\end{equation} and is finite in the limit $\Lambda \to 0$. The associated symplectic potential $\omega^r_{\text{ren}(\Lambda)}$ is explicitly given by 
\begin{equation}
\begin{split}
&\omega^r_{\text{ren}(\Lambda)} [g;\delta_1 g,\delta_2 g] \\
&\quad =  \frac{\sqrt{\mathring{q}}}{16\pi G} \left[ \frac{1}{2}\delta_1 \left(N^{AB}_{TF} + \frac{1}{2}R[q]q^{AB}\right) \wedge \delta_2 C_{AB} + \frac{1}{2} \delta_1 \left( D_A D^C C_{BC} \right) \wedge \delta_2 q^{AB}\right]\\
&\qquad+ \mathcal{O}(\Lambda; r^{-1} ).
\end{split}
\end{equation} Finally, taking the flat limit $\Lambda \to 0$, we obtain 
\begin{equation}
\begin{split}
&\omega^r_{\text{ren}(\Lambda)} [g;\delta_1 g,\delta_2 g] \\
&\quad =  \frac{\sqrt{\mathring q}}{16\pi G} \left[ \frac{1}{2}\delta_1 \left(N^{AB}_{TF} + \frac{1}{2}R[q]q^{AB}\right) \wedge \delta_2 C_{AB} + \frac{1}{2} \delta_1 \left( D_A D^C C_{BC} \right) \wedge \delta_2 q^{AB}\right] \\
&\qquad+ \mathcal{O}( r^{-1} ).
\end{split}
\end{equation} This result precisely corresponds to the presymplectic form \eqref{omegegar} obtained in asymptotically flat spacetime. Therefore, we have shown that through an appropriate renormalization process, the flat limit of the $\mathfrak{bms}^\Lambda_4$ symplectic structure yields the $\mathfrak{bms}^\text{gen}_4$ symplectic structure in asymptotically flat spacetime. 

An interesting observation is that, to obtain \eqref{omegegar}, we had to renormalize the symplectic structure using the Iyer-Wald ambiguity \eqref{ambiguity presymplectic potential IW} with a term $Y^{ru} = -r \frac{1}{2} \frac{\sqrt{\mathring q}}{16 \pi G} C_{AB} \delta q^{AB}$ to remove the $\sim r$ divergences. To take the flat limit in the present context, we also had to renormalize the symplectic structure with a term $Y^{ru}_{(\Lambda ) } = -\frac{1}{16\pi G} \frac{3}{\Lambda}\frac{1}{2}\partial_u(\sqrt{\mathring q}\,  C_{AB}\delta q^{AB})$ to remove the $\sim \Lambda$ divergences. Therefore, even if the nature of the divergences is different in both contexts, the expressions are astonishingly very similar and may rely on deeper reasons.

\section{New boundary conditions for asymptotically locally AdS$_4$ spacetime}
\label{New boundary conditions for asymptotically locally}
\chaptermark{New boundary conditions in AdS$_4$}

We now particularize our discussion to the case $\Lambda < 0$. The presymplectic form \eqref{presymplectic form prime} obtained through the holographic renormalization procedure is generically non-vanishing for asymptotically locally AdS$_4$ spacetimes. Allowing some flux at infinity leads to an ill-defined Cauchy problem \cite{Ishibashi:2004wx}. Depending on the physical context, one may be interested in studying open systems allowing flux at infinity (see e.g. \cite{Jana:2020vyx}) or isolated systems with a well-defined dynamics (see e.g. \cite{Ishibashi:2004wx}). In this section, we propose a new set of boundary conditions for which the symplectic flux vanishes. The associated phase space admits the Schwarzschild-AdS$_4$ black hole and a stationary rotating solution distinct from the Kerr-AdS$_4$ black hole. The asymptotic symmetry algebra is shown to be a subalgebra of $\mathfrak{bms}_4^\Lambda$ consisting of time translations and area-preserving diffeomorphisms.

\subsection{Mixed boundary conditions}

We start from the expression of the presymplectic form \eqref{presymplectic form prime} that we repeat here for $\Lambda < 0$ ($\eta = 1$)
\begin{equation}
\omega_{\text{ren}}^\rho [g; \delta g,\delta g] = - \frac{1}{2} \delta \left(\sqrt{|g_{(0)}|}T^{ab}\right) \wedge \delta g^{(0)}_{ab} + \mathcal{O}(\rho).
\end{equation} In the literature, both Dirichlet and Neumann boundary conditions have been studied to set this presymplectic form to zero. On the one hand, Dirichlet boundary conditions  \cite{Henneaux:1985tv} amount to freezing the components of the boundary metric $g^{(0)}_{ab}$ to the ones of the unit cylinder while leaving the holographic stress-tensor $T^{ab}$ free. The resulting asymptotic symmetry algebra is the algebra of exact symmetries of global AdS$_4$, namely $\mathfrak{so}(3,2)$. On the other hand, Neumann boundary conditions \cite{Compere:2008us} freeze the components of $T^{ab}$ while leaving the boundary metric $g^{(0)}_{ab}$ free. The resulting asymptotic symmetry group is empty: all residual gauge transformations have vanishing charges. 

We now present new mixed Dirichlet-Neumann boundary conditions. We first impose the boundary conditions \eqref{eq:LBMSConditions}, which leads to $\mathfrak{bms}_4^\Lambda$. This is a Dirichlet boundary condition on a part of the boundary metric, which is reachable locally by a choice of boundary gauge. The symplectic flux at the spatial boundary is then given by \eqref{lambda bms symplectic form}, which we repeat here: 
\begin{equation}
\omega^\rho_{\Lambda\text{-BMS}}[g;\delta g] = \frac{\Lambda}{2}\frac{\sqrt{\mathring q}}{16\pi G} \, \delta q^{AB} \wedge \delta J_{AB} + \mathcal{O}(r^{-1}).
\end{equation} We now further impose the Neumann boundary conditions 
\begin{equation}
J_{AB} = 0.  \label{BCads}
\end{equation}
This cancels the symplectic flux, as required. The boundary condition \eqref{BCads} restricts the solution space.

\subsection{Asymptotic symmetry algebra}

Let us now derive both the asymptotic symmetries preserving the boundary conditions and the associated charge algebra. 

The boundary gauge fixing \eqref{eq:LBMSConditions} is preserved by the $\mathfrak{bms}_4^\Lambda$ asymptotic symmetry algebra of residual gauge transformations as derived in section \ref{Symmetries and flat limit} (see equation \eqref{sig}). We now show that the boundary condition \eqref{BCads} further reduces $\mathfrak{bms}_4^\Lambda$ to the direct sum $\mathbb{R} \oplus \mathcal{A}$, where $\mathbb{R}$ are the time translations and $\mathcal{A}$ is the algebra of two-dimensional area-preserving diffeomorphisms. We further show that the charges associated with this asymptotic symmetry algebra are finite, integrable, conserved and generically non-vanishing on the phase space.  

The variation of $J_{AB}$ is given by 
\begin{equation}
\begin{split}
\delta_\xi {J}_{AB} &= (\xi_0^t \partial_t + \mathcal{L}_{\xi_0^C} + \sigma) {J}_{AB} - \frac{4}{3} \Big[  N_{(A} \partial_{B)}\xi_0^t - \frac{1}{2} N_{C} \partial^C \xi_0^t q_{AB} \Big] \\
&\overset{\eqref{sig}}{=} \Big[ \xi_0^t \partial_t + \mathcal{L}_{\xi_0^C} + \frac{1}{2} D_A \xi^A_0 \Big] {J}_{AB} - \frac{4}{3} \Big[  N_{(A} \partial_{B)}\xi_0^t - \frac{1}{2} N_{C} \partial^C \xi_0^t q_{AB} \Big].
\end{split}
\end{equation} 
We recall that  $D_A$ is the covariant derivative with respect to the transverse metric $g^{(0)}_{AB} = q_{AB}$. Imposing $\delta_\xi J_{AB} = 0$ leads to the following constraint on the $\mathfrak{bms}_4^\Lambda$ asymptotic Killing vectors:
\begin{equation}
\partial_A \xi_0^t =0.
\label{angle independance}
\end{equation} Therefore, the asymptotic symmetry generators satisfy the relations
\begin{equation}
\partial_t \xi^t_0 = \frac{1}{2} D_A \xi^A_{0} , \qquad \partial_t \xi^A_0 = 0.
\end{equation} The second equation implies $\xi^A_0 = V^A(x^B)$, while the first gives
\begin{equation}
\xi^t_0 = S + \frac{t}{2} D_A V^A
\end{equation} where $S$ is a constant by virtue of \eqref{angle independance}, and $D_A V^A \equiv c$, where $c$ is also a constant. Using Helmholtz's theorem, the vector $V^A$ can be decomposed into a divergence-free and a curl-free part as $V^A = \epsilon^{AB} \p_B \Phi + q^{AB} \p_B \Psi$, where $\Psi$ and $\Phi$ are functions of $x^C$. Injecting this expression for $V^A$ into this equation gives $D_A D^A \Psi = c$. This equation admits a solution if and only if $c=0$, which is given by $\Psi = 0$. Therefore, the asymptotic symmetry generators are given by 
\begin{equation}
\xi^t_0 = S, \qquad \xi^A_0 =  \epsilon^{AB} \p_B \Phi(x^C)
\label{new ads symmetry vectors}
\end{equation} where $S$ is a constant and $\Phi(x^C)$ is arbitrary. Writing $\xi = \xi (S, \Phi)$, the residual gauge diffeomorphisms, the commutation relations \eqref{struc const} and \eqref{structure constant Lambda BMSSSS} reduce to $[\xi (S_1, \Phi_1),  \xi (S_2, \Phi_2) ]_A = \xi(\hat S, \hat \Phi)$, where
\begin{equation}
\hat{S} = 0, \qquad \hat{\Phi} = \epsilon^{AB}\p_A \Phi_2 \p_B \Phi_1. 
\label{algebra are preserving}
\end{equation} Hence, after imposing the boundary condition \eqref{BCads}, $\mathfrak{bms}_4^\Lambda$ reduces to the $\mathbb{R} \oplus \mathcal{A}$ algebra, where $\mathbb{R}$ denotes the abelian time translations and $\mathcal{A}$ is the algebra of two-dimensional area-preserving diffeomorphisms. The latter symmetries are an infinite-dimensional extension of the $\mathfrak{so}(3)$ rotations. 

Let us now study the associated surface charges. Starting from the $\mathfrak{bms}_4^\Lambda$ surface charges given in \eqref{Lambda-BMS charges 1} and \eqref{Lambda-BMS charges 2}, and imposing the boundary condition \eqref{BCads}, one sees that the charges are integrable. The integrated charges reduce to 
\begin{equation}
H_\xi^{\text{AdS}}[g] = - \sqrt{\frac{3}{|\Lambda|}}\int_{S^2_\infty} d^2 \Omega \,  [ S T_{tt} + T_{tA} \epsilon^{AB} \p_B \Phi ] .
\label{charge expression}
\end{equation} From this expression, we see that the charges associated with the symmetry $\mathbb{R} \oplus \mathcal{A}$ are generically non-vanishing. Taking $S = 1$ and $\Phi = 0$ gives the energy. The first harmonic modes of $\Phi$ give the angular momenta, while the higher modes give an infinite tower of charges. Using \eqref{algebra are preserving} and \eqref{EOM FG}, a simple computation shows that the charges \eqref{charge expression} satisfy the algebra
\begin{equation}
\delta_{\xi (S_2, \Phi_2)} H^{\text{AdS}}_{\xi (S_1, \Phi_1)} = H^{\text{AdS}}_{\xi (\hat S,\hat \Phi)} .
\end{equation} The charges form a representation of $\mathbb{R} \oplus \mathcal{A}$ without central extension. This result is also a direct consequence of \eqref{eq:AlgebraLBMS} when taking \eqref{BCads} into account.  

\subsection{Stationary solutions}

Here, we study the stationary sector of the phase space associated with the boundary conditions. In this subsection, we write $\Lambda  = - 3 / \ell^2$. The Schwarzschild-AdS$_4$ solution is included in the phase space. Indeed, Schwarzschild-AdS$_4$ can be set in the Fefferman-Graham gauge, which allows to identify $q_{AB} = \mathring q_{AB}$ the unit metric on the sphere, as well as $T_{tt} = - \frac{M}{4 \pi G \ell}$, $T_{tA} = 0$ and $J_{AB} = 0$. 

The boundary metric and holographic stress-tensor of Kerr-AdS$_4$ are given in the conformally flat frame by \cite{Awad:1999xx,Bhattacharyya:2007vs,Bhattacharyya:2008ji}
\begin{eqnarray}
g^{(0)}_{ab}dx^a dx^b &=& -\ell^{-2}dt^2 + d\theta^2 +  \sin^2\theta d\phi^2,\label{flat} \\
T^{ab} &=& T_{\text{Kerr}}^{ab}  \equiv - \frac{m \gamma^3\ell }{8 \pi} (3 u^a u^b + g_{(0)}^{ab}),
\end{eqnarray}
where $\Xi = 1-a^2 \ell^{-2}$ and 
\begin{eqnarray}
u^a \p_a = \gamma \ell (\p_t +\frac{a}{\ell^2} \p_\phi),\qquad \gamma^{-1} \equiv \sqrt{1-\frac{a^2}{\ell^2} \sin^2\theta}. 
\end{eqnarray} 
The mass and angular momentum are $M = -\ell \int \D^2 \Omega \, T_{tt} = \frac{m}{\Xi^2}$,  $J = Ma = -\ell \int \D^2  T_{t\phi} = \frac{ma}{\Xi^2}$. We observe that $J_{AB} \neq 0$. Therefore, the Kerr-AdS$_4$ solution is not included in the phase space. However, it is possible to obtain a stationary axisymmetric solution with $J_{AB} = 0$ as follows. The most general diagonal, traceless, divergence-free, stationary and axisymmetric $T^{ab}$ is given by 
\begin{equation}
\begin{split}
&T^{tt}_{\text{corr}} = \ell^2 [2 T^{\theta\theta}(\theta) + \tan \theta ~{T^{\theta \theta}}^\prime  (\theta)], \quad T^{\theta \theta}_{\text{corr}}  = T^{\theta \theta}(\theta), \\
&T^{\phi\phi}_{\text{corr}}  = \frac{1}{\sin^2 (\theta)} [T^{\theta\theta}(\theta) + \tan \theta ~{T^{\theta\theta}}^\prime(\theta) ]
\end{split}
\end{equation} and the other components are set to zero. We consider the sum of $T_{\text{Kerr}}+T_{\text{corr}}$. We solve for $T^{\theta\theta}(\theta)$ to set $J^{AB} = 0$. The regular solution at $\mathscr I$ is unique and given by
\begin{eqnarray}
T^{tt}= -\frac{m \ell^3}{4\pi},\qquad T^{t\phi} = -\frac{3a m \ell \gamma^5}{8 \pi }, \qquad T^{AB} = -\frac{m \ell}{8 \pi} q^{AB}. 
\end{eqnarray}
The mass and angular momentum are $M  =- \ell \int \D^2 \Omega \, T_{tt}= m$,  $J= - \ell \int \D^2 \Omega T_{t\phi}= \frac{ma}{\Xi^2}$. It would be interesting to know whether this solution is regular in the bulk of spacetime. 

From the conservation of the stress-energy tensor $T^{ab}$ given by the first equation of \eqref{condition on energy momentum}, the most general stationary solution with flat boundary metric \eqref{flat} is only constrained by the following conditions:
\begin{equation}
D_A N^A = 0 \Leftrightarrow N^A = \epsilon^{AB} D_B \alpha (x^C), \quad \partial_A M = 0 , \label{soll}
\end{equation} where $\alpha (x^C)$ is an arbitrary function of $x^C$. To obtain these expressions, we also used equations \eqref{eq:RefiningTab} and \eqref{BCads}. Therefore, even for stationary solutions, we see that the charges associated with the area-preserving diffeomorphisms are generically non-vanishing. It would be interesting to study the regularity of the general solutions \eqref{soll} in the bulk of spacetime.

\chapter{Conclusion}

The discovery of the global BMS symmetry group at null infinity came as a surprise in 1962. This infinite-dimensional endowing of the Poincaré group was, however, necessary to include radiative spacetimes in the four-dimensional analysis. Since then, the extensions of the BMS group have highlighted the richness of the asymptotic structure of the gravitational field. 

In this thesis, we have explored the extensions of the BMS group and their implications for the phase space of the theory. Furthermore, we have established new relations between those symmetries and the gravitational memory effects. We have also elaborated on the covariant phase space methods, allowing us to compute the gravitational surface charges in a first order framework. 

Before concluding this manuscript, we would like to mention some current or future research directions that are suggested by the present work. 

As discussed in chapter \ref{ch:Asymptotic symmetries and surface charges}, we have always adopted the gauge fixing approach throughout this thesis \cite{Ruzziconi:2019pzd}. This approach allows us to eliminate the arbitrary functions of the gauge transformations and therefore fix the dependence of the residual gauge diffeomorphisms at all orders in the expansion parameter. However, even if one can always reach a gauge by definition, the gauge transformations that are necessary to reach a particular gauge might be large gauge transformations, namely, they could be associated with non-vanishing surface charges \cite{Grumiller:2016pqb , Grumiller:2017sjh}. Therefore, it would be interesting to study asymptotic symmetries by considering only partial gauge fixings. For example, we showed in \cite{newMarios, newMarios2} that the Bondi gauge can be embedded in the Derivative expansion, which is a partial gauge fixing admitting additional parameters in its solution space. Another example is the extended Fefferman-Graham gauge considered, for example, in \cite{Grumiller:2016pqb , Ciambelli:2019bzz}, where the Weyl transformations preserving the radial foliation are well defined.

Furthermore, it has been noted that the solution space of three-dimensional general relativity transforms in the coadjoint representation of the asymptotic symmetry group \cite{Barnich:2015uva}. It would be insightful to investigate how the coadjoint patterns appear in the transformation of the solution space of four-dimensional gravity. Based on the results established in section \ref{Application to asymptotically flat 4d gravity}, we have already found that only a subsector of the solution space transforms in the coadjoint representation of BMS$_4$. This subsector couples with the radiation to form the full four-dimensional gravitational theory. Furthermore, it has been shown that three-dimensional gravity could be described by a geometric action defined on coadjoint orbits \cite{Barnich:2017jgw}. It would be fascinating to have the same construction for the coadjoint subsector of four-dimensional gravity. These questions are part of our current research.  

Moreover, we saw in this thesis that some gravitational memory effects could be related to the BMS symmetries in asymptotically flat spacetimes. Since, in chapter \ref{ch:LambdaBMS group}, we found the analogue of the BMS group in asymptotically locally (A)dS$_4$ spacetimes, it would be worth investigating if similar relations exist with memory effects in these kinds of asymptotics (see e.g. \cite{Tolish:2016ggo, Hamada:2017gdg}).

The AdS/CFT correspondence and the associated holographic dictionary are now clearly stated and have been checked in many situations. Surprisingly, though, the analogous holographic correspondence in asymptotically flat spacetimes is poorly understood. However, all the ingredients needed to clearly state the holographic duality in flat space and its associated dictionary are now present. Indeed, from the point of view of the bulk theory, the Bondi expansion of the metric enables us to approach the spacetime boundary in the flat case, as the Fefferman-Graham does in the AdS case. Furthermore, as discussed in much detail in chapter \ref{ch:LambdaBMS group}, the Bondi gauge also exists in asymptotically AdS spacetimes and has been related to the Fefferman-Graham gauge \cite{Poole:2018koa , Compere:2019bua}. Therefore, many results and interpretations of the bulk spacetime metric obtained in AdS can be directly imported into flat space by taking the well-defined flat limit in the Bondi gauge. For example, the process of holographic renormalization could be adapted for asymptotically flat spacetimes at null infinity. From the point of view of the dual boundary theory, using the geometric action construction mentioned above which is based on coadjoint methods, one could construct a boundary action invariant under the BMS$_4$ symmetry. This would be the effective action of the theory dual to the coadjoint subsector of four-dimensional asymptotically flat gravity. Adding Hamiltonians and source terms to this action would lead to an effective dual description of the full asymptotically flat gravity theory.

\appendix

\chapter{Useful results and conventions}
\label{Useful results}

In this appendix, we establish some important frameworks and conventions. The aim of this formalism is to manipulate some local expressions, as this is convenient in field theory. We closely follow \cite{Compere:2019qed, Compere:2007az , Ruzziconi:2019pzd}. 

\section{Jet bundles}
\label{Jet bundles}

Let $M$ be the $n$-dimensional spacetime with local coordinates $x^\mu$ ($\mu = 0, \ldots , n-1$). The fields, written as $\phi = (\phi^i)$, are supposed to be Grassmann even. The \textit{jet space} $J$ consists in the fields and the symmetrized derivatives of the fields $(\phi, \phi_\mu , \phi_{\mu\nu}, \ldots)$, where $\phi^i_{\mu_1 \ldots \mu_k} = \frac{\partial}{\partial x^{\mu_1}} \ldots  \frac{\partial}{\partial x^{\mu_k}} \phi^i$. The \textit{symmetrized derivative} is defined as
\begin{equation}
\frac{\partial \phi^i_{\nu_1 \ldots \nu_k}}{\partial \phi^j_{\mu_1 \ldots \mu_k}}  = \delta_{(\mu_1}^{\nu_1} \ldots \delta_{\mu_k)}^{\nu_k} \delta^{i}_j .
\end{equation} In the jet space, the cotangent space at a point is generated by the variations of the fields and their derivatives at that point, namely $(\delta \phi, \delta \phi_{\mu}, \delta \phi_{\mu\nu}, \ldots)$. The \textit{variational operator} is defined as
\begin{equation}
\delta = \sum_{k \ge 0} \delta \phi^i_{\mu_1 \ldots  \mu_k} \frac{\partial}{\partial \phi^i_{\mu_1 \ldots \mu_k}} .
\end{equation} We choose all the $\delta \phi$, $\delta \phi_\mu$, $\delta \phi_{\mu\nu}, \ldots$ to be Grassmann odd, which implies that $\delta^2 = 0$. Hence, $\delta$ is seen as an exterior derivative on the jet space. 

Now, we define the \textit{jet bundle} as the fiber bundle with local trivialization $(x^\mu, \phi,\phi_\mu, \phi_{\mu\nu} ,\ldots)$. Locally, the total space of the jet bundle looks like $M \times J$. A section of this fiber bundle is a map $x \to (\phi (x), \phi_\mu (x) , \phi_{\mu\nu}(x), \ldots )$. The \textit{horizontal derivative} is defined as
\begin{equation}
\mathrm{d} = \mathrm{d}x^\mu \partial_\mu , \quad \text{where} \quad \partial_\mu = \frac{\partial}{\partial x^\mu} + \sum_{k \ge 0} \left( \phi^i_{\mu \nu_1 \ldots \nu_k} \frac{\partial}{\partial  \phi^i_{\nu_1 \ldots \nu_k}} + \delta \phi^j_{\mu \nu_1 \ldots \nu_k} \frac{\partial}{\partial  \delta \phi^j_{\nu_1 \ldots \nu_k}} \right) .
\label{horizontal}
\end{equation} In this perspective, the variational operator can also be seen as the \textit{vertical derivative}, i.e. the derivative along the fibers. The exterior derivative on the total space can be defined as $\mathrm{d}_{Tot} = \mathrm{d} + \delta$. Notice that both $\mathrm{d}$ and $\delta$ are Grassmann odd and they anti-commute, namely
\begin{equation}
\mathrm{d} \delta = - \delta \mathrm{d} .
\end{equation} On the jet bundle, we write $\Omega^{p,q}$ for the set of functions that are $p$-forms with respect to the spacetime and $q$-forms with respect to the jet space\footnote{One often refers to a \textit{variational bicomplex} structure}. 

\section{Some operators}

In this subsection, we introduce additional operators used in the text and discuss their properties. 

The \textit{Euler-Lagrange} derivative of a \textit{local function} $f$, i.e. a function on the total space of the jet bundle $f = f[x,\phi,\phi_\mu, \phi_{\mu\nu}, \ldots ]$, is defined as
\begin{equation}
\frac{\delta f}{\delta \phi^i} = \sum_{k\ge 0} (-1)^k \partial_{\mu_1} \ldots \partial_{\mu_k} \left(\frac{\partial f}{\partial \phi^i_{\mu_1 \ldots \mu_k}} \right) .
\label{Euler lagrange def}
\end{equation} This operator satisfies
\begin{equation}
\frac{\delta f}{\delta \phi^i} = 0 \quad \Leftrightarrow \quad f = \partial_\mu j^\mu ,
\label{euler lagrange and total}
\end{equation} where $j^\mu$ is a local function (for a proof, see e.g. section 1.2 of \cite{Barnich:2018gdh}). 

The \textit{variation under a transformation of characteristic $Q$} (i.e. $\delta_Q \phi^i = Q^i$) is given by
\begin{equation}
\delta_Q f = \sum_{k\ge 0} (\partial_{\mu_1} \ldots \partial_{\mu_k} Q^i ) \frac{\partial f}{\partial \phi^i_{\mu_1 \ldots \mu_k}} + (\partial_{\mu_1} \ldots \partial_{\mu_k} \delta Q^j ) \frac{\partial f}{\partial \delta \phi^j_{\mu_1 \ldots \mu_k}} .
\label{def variation}
\end{equation}  The Lie bracket of characteristics is defined by $[Q_1,Q_2] = \delta_{Q_1} Q_2 - \delta_{Q_2} Q_1$ and satisfies $[\delta_{Q_1}, \delta_{Q_2}] = \delta_{[Q_1,Q_2]}$. A contracted variation of this type is Grassmann even and we have
\begin{equation}
\delta_Q  \mathrm{d} = \mathrm{d} \delta_Q , \quad \delta \delta_Q = \delta_Q \delta .
\label{commutation relation delta Q}
\end{equation} We also have the following relation between the variation under a transformation of characteristic $Q$ and the Euler-Lagrange derivative:
\begin{equation}
\delta_Q \frac{\delta f}{\delta \phi^i} = \frac{\delta }{\delta \phi^i} (\delta_Q f ) - \sum_{k \ge 0} (-1)^k \partial_{\mu_1} \ldots \partial_{\mu_k} \left( \frac{\partial Q^j}{\partial \phi^i_{\mu_1\ldots\mu_k}} \frac{\delta f}{\delta \phi^j} \right) .
\label{variation and euler lagrange}
\end{equation}

Let $\boldsymbol{\alpha} \in \Omega^{n-k, q}$. We use the notation
\begin{equation}
\boldsymbol{\alpha} = \alpha^{\mu_1 \ldots \mu_{k}} (\mathrm{d}^{n-k}x)_{\mu_1 \ldots \mu_k} ,
\end{equation} where
\begin{equation}
(\mathrm{d}^{n-k}x)_{\mu_1 \ldots \mu_k} = \frac{1}{k!(n-k)!} \epsilon_{\mu_1 \ldots \mu_k \nu_1 \ldots \nu_{n-k}} \mathrm{d}x^{\nu_1} \wedge \ldots \wedge \mathrm{d}x^{\nu_{n-k}}
\end{equation} and where $\epsilon_{\mu_1 \ldots \mu_n}$ is completely antisymmetric and $\epsilon_{01\ldots n-1} = 1$. We can check that
\begin{equation}
\mathrm{d} \boldsymbol{\alpha} = (-1)^q \partial_\sigma {\alpha}^{[\mu_1 \ldots \mu_{k-1} \sigma]} (\mathrm{d}^{n-k+1}x)_{\mu_1 \ldots \mu_{k-1}} .
\end{equation} The \textit{interior product of a spacetime form} with respect to a vector field $\xi$ is defined as
\begin{equation}
\iota_\xi = \xi^\mu \frac{\partial}{\partial \mathrm{d}x^\mu} .
\end{equation} We can also define the \textit{interior product of a jet space form} with respect to a characteristic $Q$ as
\begin{equation}
i_Q = \sum_{k\ge 0} ( \partial_{\mu_1} \ldots \partial_{\mu_k} Q^i ) \frac{\partial}{\partial \delta \phi^i_{\mu_1 \ldots \mu_k}} .
\end{equation} It satisfies
\begin{equation}
i_Q \delta + \delta i_Q = \delta_Q, \quad i_{Q_1} \delta_{Q_2} - \delta_{Q_2} i_{Q_1} = i_{[Q_1,Q_2]} .
\end{equation}

The \textit{homotopy operator} $I_{\delta \phi}^p : \Omega^{p,q} \mapsto \Omega^{p-1, q+ 1}$ is defined as
\begin{equation}
I_{\delta \phi}^p \boldsymbol{\alpha} = \sum_{k\ge 0} \frac{k + 1}{n-p+ k +1} \partial_{\mu_1} \ldots \partial_{\mu_k} \left( \delta \phi^i \frac{\delta}{\delta \phi^i_{\mu_1\ldots\mu_k \nu}} \frac{\partial \boldsymbol{\alpha}}{\partial \mathrm{d}x^\nu} \right)
\label{homotopy operator}
\end{equation} for $\boldsymbol{\alpha} \in \Omega^{p,q}$. This operator satisfies the following relations
\begin{align}
\delta &= \delta \phi^i \frac{\delta}{\delta \phi^i} - \mathrm{d} I^n_{\delta \phi} \quad \text{when acting on spacetime $n$-forms} , \\
\delta &= I^{p+1}_{\delta \phi} \mathrm{d} - \mathrm{d} I_{\delta \phi}^p \quad \text{when acting on spacetime $p$-forms ($p<n$)} .
\label{commutation homotopy op}
\end{align} Furthermore,
\begin{equation}
\delta I^p_{\delta \phi} = I^p_{\delta \phi} \delta .
\end{equation} Notice that the homotopy operator is used to prove the algebraic Poincar\'e lemma \eqref{Poincare lemma}.

Similarly, the \textit{homotopy operator with respect to gauge parameters $f=(f^\alpha)$} is defined as $I_f^p : \Omega^{p,q} \mapsto \Omega^{p-1,q}$, where
\begin{equation}
I^p_f \boldsymbol{\alpha} = \sum_{k\ge 0} \frac{k + 1}{n-p+ k +1} \partial_{\mu_1} \ldots \partial_{\mu_k} \left(f^\alpha \frac{\delta}{\delta f^\alpha_{\mu_1\ldots\mu_k \nu}} \frac{\partial \boldsymbol{\alpha}}{\partial \mathrm{d} x^\nu} \right) .
\end{equation} It satisfies
\begin{equation}
I^{p+1}_f \mathrm{d} + \mathrm{d} I^p_f  = 1 .
\end{equation}

\renewcommand{\theequation}{\thechapter.\arabic{equation}}

\chapter{Determinant condition in Bondi gauge}
\label{app:detcondBondi}

In this appendix, we discuss the determinant condition used to define the Bondi gauge in equation \eqref{Bondi gauge} and repeated here:
\begin{equation}
\partial_r \left( \frac{\det g_{AB}}{r^{2(n-2)}} \right) = 0.
\label{det cond disc}
\end{equation} Let us emphasize that this condition is weaker than the historical one given by $\det (g_{AB}) = r^{2(n-2)} \det (\mathring{q}_{AB})$, where $\mathring{q}_{AB}$ is the unit sphere metric \cite{Bondi:1962px , Sachs:1962wk , Sachs:1962zza}. The relaxed determinant condition \eqref{det cond disc} is inspired by \cite{Barnich:2010eb} and is essential if one wants to consider Weyl rescalings of the transverse boundary metric.

To illustrate this claim, we derive the implication of the determinant condition in the derivation of the residual gauge diffeomorphisms. The equation \eqref{det cond disc} is equivalent to 
\begin{equation}
\det (g_{AB}) = r^{2(n-2)} \chi (u, x^C), 
\label{int condi}
\end{equation} where $\chi$ is an arbitrary function of $(u, x^C)$. When the preliminary boundary condition $g_{AB} = \mathcal{O}(r^2) ~\Leftrightarrow~g_{AB} = r^2 q_{AB} + o (r^2)$ is imposed, we have $\chi (u, x^C) = \det (q_{AB})$. From \eqref{int condi}, we obtain
\begin{equation}
\delta_\xi \ln [\det (g_{AB})] = g^{AB} \mathcal{L}_\xi g_{AB} = \delta_\xi \ln \chi \equiv 2(n-2) \omega ,
\label{introduction of omega}
\end{equation} where we introduced the parameter $\omega (u,x^A)$. We deduce
\begin{equation}
\xi^r = -\frac{r}{(n-2)} \left[ \mathcal{D}_A \xi^A - U^A \partial_A \xi^u + \frac{1}{2} \xi^u \partial_u \ln g - (n-2) \omega \right] , 
\label{xir in Bondi gauge}
\end{equation} where $\mathcal{D}_A$ is the covariant derivative with respect to $g_{AB}$ and $g = \det (g_{AB})$. Indices are lowered and raised by $g_{AB}$ and its inverse. 

In this derivation, the introduction of the parameter $\omega$ in \eqref{introduction of omega} is somewhat peculiar and may seem artificial. This way of introducing the parameters of the residual gauge diffeomorphisms \textit{by hand} was also used in \cite{newMarios, newMarios2} to define the additional parameters in the Derivative expansion compared to the Bondi gauge. Let us show here that \eqref{xir in Bondi gauge} can be deduced from the determinant condition without forcing $\omega$ in \eqref{introduction of omega}. The condition \eqref{introduction of omega} tells us that
\begin{equation}
\delta_\xi \ln [\det (g_{AB})] \sim \text{order $r^0$} .
\end{equation} Working out the left-hand side yields
\begin{equation}
\frac{(n-2)}{r} \xi^r + \mathcal{D}_A \xi^A + \frac{1}{2} \xi^u \partial_u \ln g -  U^A \partial_A \xi^u \sim \text{order $r^0$}
\end{equation} Taking into account the other gauge conditions in \eqref{Bondi gauge}, the preliminary boundary conditions $g_{AB} = \mathcal{O}(r^2)$ and the associated fall-offs imposed by the Einstein equations, we obtain that $\xi^r$ is determined at all orders, except at leading order $\sim r$. In other words, writing $\xi^r = r R(u, x^A) + o(r)$, the remaining free parameter is $R (u,x^A)$ and
\begin{equation}
\xi^r = r R(u, x^A) + \left[ -\frac{r}{(n-2)} \left( \mathcal{D}_A I^A - U^A \partial_A f \right) \right]\Big|_{\mathcal{O}(r^{n<1})},
\end{equation} where $I^A$ is defined in \eqref{eq:xir} and the notation $\mathcal{O}(r^{n<1})$ means that the expression inside the brackets is truncated for terms of order $\sim r$ or higher. Finally, doing the following field-dependent redefinition of the free-parameter:
\begin{equation}
R (u, x^A)= \omega (u, x^A) - \frac{1}{(n-2)}  \left[ \mathcal{D}_A Y^A  + \frac{1}{2} \xi^u \partial_u \ln g  - U_0^A \partial_A f  \right],
\end{equation} we recover the original result \eqref{xir in Bondi gauge}.

\chapter{Further results in Newman-Penrose formalism}

\section{Newman-Unti solution space in NP formalism}
\label{NP solution}

When conditions \eqref{gauge conditions} supplemented by the fall-off
conditions \eqref{fall-off conditions} are imposed, the asymptotic
expansion of on-shell spin coefficients, tetrads and the associated
components of the Weyl tensor can be determined. All the coefficients
in the expansions are functions of the three coordinates
$u,\zeta, \bar{\zeta}$. In this approach to the characteristic initial
value problem, freely specifiable initial data at fixed $u_0$ is given
by $\Psi_0 (u_0, r, \zeta, \bar{\zeta})$ in the bulk with the
fall-offs given below and by
$(\Psi^0_2 + \bar{\Psi}^0_2)(u_0, \zeta, \bar{\zeta})$,
$\Psi^0_1 (u_0, \zeta, \bar{\zeta})$ at $\mathscr{I}^+$. The
asymptotic shear $\sigma^0(u, \zeta, \bar{\zeta})$ and the conformal
factor $P(u, \zeta, \bar{\zeta})$ are free data at $\mathscr{I}^+$ for
all $u$.

Explicitly,
\begin{equation*}
\begin{split}
&\Psi_0=\frac{\Psi_0^0}{r^5} + \frac{\Psi_0^1}{r^6} + \frac{\Psi_0^2}{r^7} + \cO(r^{-8}),\\
&\Psi_1=\frac{\Psi_1^0}{r^4}-\frac{\xbar \eth
  \Psi_0^0}{r^5}+\frac{2\sigma^0\xbar\sigma^0\Psi_1^0 + \frac52 \eth\xbar\sigma^0 \Psi_0^0 + \frac12 \xbar\sigma^0 \eth \Psi_0^0 - \frac12 \xbar\eth \Psi_0^1}{r^6} + \cO(r^{-7}),\\
&\Psi_2=\frac{\Psi_2^0}{r^3}-\frac{\xbar \eth \Psi_1^0}{r^4} + \frac{2\eth\xbar\sigma^0 + \frac12 \lambda^0\Psi_0^0 + \frac32 \sigma^0\xbar\sigma^0\Psi_2^0 + \frac12 \xbar\sigma^0\eth\Psi_1^9 + \frac12 \xbar\eth^2\Psi_0^0}{r^5}+ \cO(r^{-6}),\\
&\Psi_3=\frac{\Psi_3^0}{r^2}-\frac{\xbar
  \eth\Psi_2^0}{r^3}+\cO(r^{-4}),\quad
\Psi_4=\frac{\Psi_4^0}{r}-\frac{\xbar \eth \Psi_3^0}{r^2}+\cO(r^{-3}),\\
\end{split}
\end{equation*}
\begin{equation*}
\begin{split}
&\rho=-\frac{1}{r}-\frac{\sigma^0\xbar\sigma^0}{r^3}+\cO(r^{-5}),\
\sigma=\frac{\sigma^0}{r^2}+ \frac{\bar{\sigma}^0 \sigma^0 \sigma^0 - \frac{1}{2} \Psi^0_0}{r^4}+\cO(r^{-5}),\\
&\tau=-\frac{\Psi^0_1}{2r^3}+ \frac{\frac{1}{2} \sigma^0
  \bar{\Psi}^0_1 + \bar{\eth} \Psi^0_0}{3r^4}+ \cO(r^{-5}),\
\alpha=\frac{\alpha^0}{r} +\frac{\xbar\sigma^0\xbar\alpha^0}{r^2}
+\frac{\sigma^0\xbar\sigma^0\alpha^0}{r^3}+\cO(r^{-4}),
\\
&\beta=-\frac{\xbar\alpha^0}{r}-\frac{\sigma^0\alpha^0}{r^2}
-\frac{\sigma^0\xbar\sigma^0\xbar\alpha^0+\half
  \Psi^0_1}{r^3}+\cO(r^{-4}),\
\gamma=\gamma^0-\frac{\Psi^0_2}{2r^2}+\frac{2 \bar{\eth} \Psi^0_1 +
  \alpha^0 \Psi^0_1 - \bar{\alpha}^0
  \bar{\Psi}^0_1}{6r^3}+\cO(r^{-4}), \\
&
 \mu=\frac{\mu^0}{r} -
 \frac{\sigma^0\lambda^0+\Psi^0_2}{r^2}+\frac{\sigma^0 \bar{\sigma}^0
   \mu^0 + \frac{1}{2}\bar{\eth} \Psi^0_1}{r^3}+\cO(r^{-4}),\
 \nu=\nu^0-\frac{\Psi^0_3}{r}+\frac{\xbar \eth
  \Psi^0_2}{2r^2}+\cO(r^{-3}),\\
&
\lambda=\frac{\lambda^0}{r}-\frac{\xbar\sigma^0
  \mu^0}{r^2}+\frac{\sigma^0 \bar{\sigma}^0\lambda^0 + \frac{1}{2}
  \bar{\sigma}^0 \Psi^0_2}{r^3}+\cO(r^{-4}),\\
\end{split}
\end{equation*}
\begin{equation*}
\begin{split}
&X^\zeta = \overline{X^{\bar{\zeta}}}= \frac{\bar{P}\Psi^0_1}{6 r^3}
+\cO(r^{-4}), \quad
\omega=\frac{\xbar \eth \sigma^0}{r} -\frac{\sigma^0\eth \xbar\sigma^0
  +\half \Psi^0_1}{r^2}+\cO(r^{-3}),\\
&U=-r(\gamma^0+\xbar\gamma^0) + \mu^0-\frac{\Psi^0_2 + \xbar
  \Psi^0_2}{2r}+ \frac{\bar{\eth} \Psi^0_1 + \eth
  \bar{\Psi}^0_1}{6r^2}+ \cO(r^{-3}),\\
&L^\zeta=\overline{{\bar{L}}^{\bar{\zeta}}}=-\frac{\sigma^0
  \bP}{r^2}+\cO(r^{-4}), \quad
L^{\bar{\zeta}}=\overline{{\bar{L}}^\zeta}=\frac{P}{r}+\frac{\sigma^0
  \xbar\sigma^0 P}{r^3}+\cO(r^{-4}),\\
\end{split}
\end{equation*} where
\begin{equation*}
\begin{split}
&\alpha^0=\half \bP \p \ln P,\quad
\gamma^0=-\half \p_u \ln \bP,\quad \nu^0=\xbar \eth (\gamma^0+\xbar\gamma^0),\\
&\mu^0=-\half P \bP \p\xbar\p \ln P\bP = -\half \bar{\eth}\eth \ln
P\bP= -\frac{R}{4} ,\quad
\lambda^0= \dot{\xbar\sigma^0} + \xbar \sigma^0 (3\gamma^0 - \xbar \gamma^0),\\
&\Psi^0_2 - \bar{\Psi}^0_2 = \bar{\eth}^2 \sigma^0 - \eth^2
\bar{\sigma}^0 + \bar{\sigma}^0 \bar{\lambda}^0 - \sigma^0 \lambda^0
\\
&\Psi^0_3 = - \eth \lambda^0 + \bar{\eth} \mu^0 , \\
&\Psi^0_4 = \bar{\eth} \nu^0 - (\partial_u + 4 \gamma^0) \lambda^0,
\end{split}
\end{equation*}
and
\begin{align*}
&\p_u\Psi^0_0 + (\gamma^0 + 5 \xbar \gamma^0)\Psi^0_0=\eth\Psi^0_1+3\sigma^0\Psi^0_2,\\
&\p_u\Psi^0_1 + 2 (\gamma^0 + 2 \xbar \gamma^0)\Psi^0_1=\eth\Psi^0_2+2\sigma^0\Psi^0_3,\\
&\p_u\Psi^0_2 + 3 (\gamma^0 + \xbar \gamma^0)\Psi^0_2=\eth\Psi^0_3 + \sigma^0\Psi^0_4,\\
&\p_u\Psi^0_3 + 2 (2 \gamma^0 + \xbar \gamma^0)\Psi^0_3=\eth\Psi^0_4,\\
&\p_u
 \mu^0=-2(\gamma^0+\xbar\gamma^0)\mu^0+\xbar\eth\eth(\gamma^0+\xbar\gamma^0),\\
&\p_u\alpha^0=-2\gamma^0\alpha^0-\xbar\eth\xbar\gamma^0,
\end{align*}
\begin{align*}
\p_u \Psi_0^1 + (2\gamma^0 + 6\xbar\gamma^0) \Psi_0^1 = - \xbar\eth (\eth\Psi_0^0 + 4 \sigma^0 \Psi_1^0),
\end{align*}
\begin{multline*}
\p_u \Psi_0^2 + (3\gamma^0 + 7\xbar\gamma^0) \Psi_0^2 = -
\frac12\xbar\eth \eth\Psi_0^1 + 3 \mu^0 \Psi_0^1 + 5 (\Psi_1^0
\Psi_1^0 - \Psi_0^0 \Psi_2^0 - \frac12 \Psi_0^0 \xbar\Psi_2^0)\\
+ 5 \xbar\eth \sigma^0 \xbar\eth\Psi_0^0 + 3
\eth\xbar\sigma^0\eth\Psi_0^0 + \frac52 \sigma^0\xbar\eth^2\Psi_0^0 +
\frac52 \eth^2\xbar\sigma^0 \Psi_0^0 + \frac12 \xbar\sigma^0
\eth^2\Psi_0^0 + \frac92 \sigma^0\xbar\sigma^0 \eth\Psi_1^0\\
+12\sigma^0\eth\xbar\sigma^0 \Psi_1^0 +
2\xbar\sigma^0\eth\sigma^0\Psi_1^0 +
\frac{15}{2}\xbar\sigma^0(\sigma^0)^2\Psi_2^0 +
\frac52\sigma^0\lambda^0\Psi_0^0.
\end{multline*}

\section{Parameters of residual gauge transformations}\label{ASG}

For computational purposes, it turns out to be more convenient to
determine the parameters of residual gauge transformations by using
the generating set given in \eqref{eq:15num2} rather than the one in
\eqref{eq:23}.

Asking that conditions \eqref{gauge conditions} be preserved on-shell yields
\begin{itemize}
\item
  $0=\delta_{\xi,\omega}\; e_1^u=-\p_r \xi^u \Longrightarrow
  \xi^u=f(u,\zeta,\bar \zeta)$.
\item
  $0=\delta_{\xi,\omega}\; e_2^u=-e_2^\alpha \p_\alpha f + \omega^{12}
  \Longrightarrow \omega^{12}=\p_u f + X^A \p_A f$.
\item
  $0=\delta_{\xi,\omega}\; e_3^u=-e_3^\alpha \p_\alpha f + \omega^{42}
  \Longrightarrow \omega^{24}= L^A \p_A f$.
\item
  $0=\delta_{\xi,\omega}\; e_4^u=-e_4^\alpha \p_\alpha f + \omega^{32}
  \Longrightarrow \omega^{23}= \bL^A \p_A f$.
\item
  $0=\delta_{\xi,\omega}\; e_1^r=-e_1^\alpha \p_\alpha \xi^r +
  \omega^{2a}e_a^r \Longrightarrow \xi^r=-\p_u f r + Z(u,\zeta,\bar \zeta) - \p_A
  f \int^{+\infty}_r dr[\omega \bL^A + \bomega L^A + X^A]$.
\item
  $0=\delta_{\xi,\omega}\; e_1^A=-e_1^\alpha \p_\alpha \xi^A +
  \omega^{2a}e_a^A \Longrightarrow \xi^A=Y^A(u,\zeta,\bar \zeta) - \p_B f
  \int^{+\infty}_r dr[L^A \bL^B + \bL^AL^B ]$.
\item
  $\delta_{\xi,\omega}\;\bar\pi=0\iff 0=\delta_{\xi,\omega}\;
  \Gamma_{321}=l^\mu \p_\mu \omega^{41} + \Gamma_{32a} \omega^{2a}
  \Longrightarrow \omega^{14}=\omega^{14}_0(u,\zeta,\bar \zeta) + \p_A f
  \int^{+\infty}_r dr[\bar{\lambda} \bL^A + \bar{\mu} L^A]$.
\item $\delta_{\xi,\omega}\;\pi=0\iff 0=\delta_{\xi,\omega}\;
  \Gamma_{421}=l^\mu \p_\mu \omega^{31} + \Gamma_{42a} \omega^{2a}
  \Longrightarrow
\omega^{13}=\omega^{13}_0(u,\zeta,\bar \zeta) + \p_A f \int^{+\infty}_r
dr[\lambda L^A + \mu \bL^A]$.
\item
  $\delta_{\xi,\omega}\;(\epsilon-\bar\epsilon)=0\iff
  0=\delta_{\xi,\omega}\; \Gamma_{431}=l^\mu \p_\mu \omega^{43} +
  \Gamma_{43a} \omega^{2a} \Longrightarrow
  \omega^{34}=\omega^{34}_0(u,\zeta,\bar \zeta) - \p_A f \int^{+\infty}_r
  dr[(\bar{\alpha}-\beta) \bL^A + (\bar{\beta}-\alpha) L^A]$.

\item $\epsilon+\bar\epsilon=0=\kappa=\bar\kappa$ is equivalent to
  $\Gamma_{211}= \Gamma_{311}=\Gamma_{411}=0$, $\rho-\bar\rho=0$ is
  equivalent to $\Gamma_{314}-\Gamma_{413}=0$ while
  $\tau-\bar\alpha-\beta=0$ is equivalent to
  $\Gamma_{213}-\Gamma_{312}=0$. On-shell, i.e., in the absence of
  torsion, these conditions on spin coefficients hold as a consequence
  of the tetrad conditions imposed in \eqref{gauge conditions}. It
  follows that requiring these conditions to be preserved on-shell by
  gauge transformations does not give rise to new conditions on the
  parameters. This can also be checked by direct computation.
\end{itemize}
Asking that the fall-off conditions \eqref{fall-off conditions} be
preserved on-shell yields
\begin{itemize}
\item $\delta_{\xi,\omega}\; e_2^A=\cO(r^{-1}) \Longrightarrow \p_u Y^A=0$.
\item $\delta_{\xi,\omega}\; g_{\zeta\zeta}=\cO(r^{-1})
  \Longrightarrow \bar{\partial} Y^{\zeta}=0\iff Y^\zeta = Y(\zeta)$.
\item $\delta_{\xi,\omega}\; g_{\bar{\zeta}\bar{\zeta}}=\cO(r^{-1})
  \Longrightarrow \partial Y^{\bar{\zeta}}=0\iff Y^{\bar{\zeta}} = \bar{Y} (\bar{\zeta})$.
\item
  $\delta_{\xi,\omega} \; \Gamma_{314} = \mathcal{O}(r^{-3})
  \Longrightarrow Z =\half \bDelta f$.
\item
  $\delta_{\xi,\omega} \; \Gamma_{312} =\mathcal{O}(r^{-2})
  \Longrightarrow \omega^{14}_0 = (\gamma^0 + \bar{\gamma}^0)P
  \bar{\partial} f - P \partial_u \bar{\partial}f$.
\item
  $\delta_{\xi,\omega} \; \Gamma_{412} =\mathcal{O}(r^{-2})
  \Longrightarrow \omega^{13}_0 = (\gamma^0 + \bar{\gamma}^0)\bar{P}
  {\partial} f - \bar{P} \partial_u {\partial}f$.
\item $\delta_{\xi,\omega}\; \Psi_0 = \mathcal{O}(r^{-5})$ does not
  impose further constraints.
\end{itemize}

\section{Action on solution space: original parametrization}
\label{sec:oracti}

Besides \eqref{conformal factors}, if $s_o=(Y,\bar Y,f,\omega_0)$, one finds
\begin{equation}
\begin{split}
&\delta_{s_o} \sigma^0=[Y \p +\bY \bp + f \p_u + \p_u f +
2\omega^{34}_0]\sigma^0 - \eth^2 f,\\
&\delta_{s_o} \Psi^0_0=[Y \p +\bY \bp + f \p_u + 3\p_u f +
2\omega^{34}_0]\Psi^0_0 + 4 \Psi^0_1 \eth f,\\
&\delta_{s_o} \Psi^0_1=[Y \p +\bY \bp + f \p_u + 3\p_u f +
\omega^{34}_0]\Psi^0_1 + 3 \Psi^0_2 \eth f,\\
&\hspace{-1cm} \delta_{s_o} \left(\frac{\Psi^0_2 +
    \bar{\Psi}^0_2}{2}\right)=[Y \p +\bY \bp + f \p_u + 3\p_u f
]\left(\frac{\Psi^0_2 + \bar{\Psi}^0_2}{2}\right)
+  \Psi^0_3 \eth f +\bar{\Psi}^0_3 \overline{\eth} f.
\end{split}
\label{transfo solution space 1}
\end{equation}
When $\Psi_0$ can be expanded in powers of $1/r$, $
\Psi_0 = \sum_{n=0}^\infty \frac{\Psi_0^n}{r^{n+5}}$,
one also has
\begin{multline}\label{transfo solution space 2}
\delta_{s_o} \Psi^1_0 = [Y \p +\bY \bp + f \p_u + 4\p_u f + 2\omega^{34}_0] \Psi^1_0
\\ + [- \frac{5}{2} \bDelta f - 5 \eth f \overline{\eth} - \overline{\eth}f \eth] \Psi_0^0
- 4 \sigma^0 \overline{\eth} f \Psi^0_1,
\end{multline}
\begin{multline}\label{transfo solution space 3}
\delta_{s_o} \Psi^2_0 = [Y \p +\bY \bp + f \p_u + 5\p_u f
+ 2\omega^{34}_0] \Psi^2_0
+ [- 3 \bDelta f -  3\eth f \overline{\eth} - \overline{\eth}f \eth] \Psi_0^1
\\ + [5 \overline{\eth} \sigma^0 \overline{\eth}f + 15 \eth
\overline{\sigma}^0 \eth f+ 5 \sigma^0 \overline{\eth}f
\overline{\eth} + 3 \bar{\sigma}^0 \eth f \eth] \Psi^0_0
+ 12 \sigma^0 {\bar{\sigma}}^0 \eth f \Psi^0_1,
\end{multline}
\begin{multline}\label{transfo solution space 4}
 \delta_{s_o} \Psi^n_0 = [Y \p +\bY \bp + f \p_u + (n+3)\p_u f + 2\omega^{34}_0] \Psi^n_0
\\+ (\text{inhomogeneous terms}).
\end{multline}
For later purposes, we also give the variations of composite
quantities in terms of free data,
\begin{equation}
\begin{split}
&\delta_{s_o} \lambda^0=[Y \p +\bY \bp + f \p_u +  2\p_u f
- 2\omega^{34}_0]\lambda^0 - \p_u\xbar\eth^2 f + (\xbar \gamma^0 - 3
\gamma^0) \xbar\eth^2 f,\\
&\delta_{s_o} \Psi^0_2=[Y \p +\bY \bp + f \p_u + 3\p_u f
]\Psi^0_2 + 2 \Psi^0_3 \eth f,\\
&\delta_{s_o} \Psi^0_3=[Y \p +\bY \bp + f \p_u + 3\p_u f -
\omega^{34}_0]\Psi^0_3 + \Psi^0_4 \eth f,\\
&\delta_{s_o} \Psi^0_4=[Y \p +\bY \bp + f \p_u + 3\p_u f -
2\omega^{34}_0]\Psi^0_4.
\end{split}
\label{transfo solution space}
\end{equation}

\section{Useful relations}
\label{Useful relations}

Some useful relations for the computation of the current algebra are
summarized here.
\begin{align*}
&\p_u f=\frac12 (\eth\cY+\xbar\eth\xbar\cY) + f(\gamma^0+\xbar\gamma^0),\\
&\hat f =\frac12 f_1 (\eth\cY_2 + \xbar\eth\xbar\cY_2) + \cY_1 \eth
  f_2 + \xbar\cY_1 \xbar\eth f_2 - (1\leftrightarrow2),\\
&\hat \cY = \cY_1 \eth^2 \cY_2 - \cY_2 \eth^2 \cY_1,\quad \hat
  {\xbar\cY} = \xbar\cY_1 \xbar\eth^2 \xbar\cY_2 - \xbar\cY_2
    \xbar\eth^2 \xbar\cY_1,\\
  &\eth^2\hat\cY=\eth\cY_1 \eth^2\cY_2 + \cY_1 \eth^3\cY_2
    - (1\leftrightarrow2),\quad \eth\xbar\eth\hat{\xbar\cY}
    =\xbar\cY_1\eth\xbar\eth^2\xbar\cY_2 - (1\leftrightarrow2),\\
  &\eth^3\hat\cY=2\eth\cY_1 \eth^3\cY_2 + \cY_1 \eth^4\cY_2
    - (1\leftrightarrow2),\quad \eth^2\xbar\eth\hat{\xbar\cY}
    =\xbar\cY_1\eth^2\xbar\eth^2\xbar\cY_2 - (1\leftrightarrow2),\\
  &\xbar\eth\eth^3\cY=2\cY\eth^2\mu^0
    + 4\eth\mu^0\eth\cY,\quad\xbar\eth^2\eth^2\cY=2\xbar\eth\eth\mu^0
    \cY
    + 2\xbar\eth\mu^0\eth\cY + 4(\mu^0)^2\cY, \\
&\eth\hat f=\frac12 f_1 \eth(\eth\cY_2+\xbar\eth\xbar\cY_2) + \cY_1
   \eth^2 f_2 + \xbar\cY_1\eth\xbar\eth f_2 + \frac12(\eth\cY_1
 - \xbar\eth\xbar\cY_1)\eth f_2 - (1\leftrightarrow2),\\
&\eth\xbar\eth \xbar\cY=2\mu^0\xbar\cY,\quad
  \xbar\eth\eth\cY=2\mu^0\cY,
  \quad \p_u\eth\cY=2\xbar\nu^0\cY,\\
  &\p_u \eth f=\frac12\eth(\eth\cY+\xbar\eth\xbar\cY)
    +\eth f(\gamma^0-\xbar\gamma^0)+f \xbar\nu^0,\\
&\p_u\eth^2\cY=2\eth\xbar\nu^0\cY + 2 \xbar\nu^0\eth\cY - 2\xbar\gamma^0\eth^2\cY,\\
  &\p_u\eth\xbar\eth\xbar\cY=2\eth\nu^0\xbar\cY
    - 2 \xbar\gamma^0\eth\xbar\eth\xbar\cY,\\
  &\p_u\eth^2 f=\frac12\eth^2(\eth\cY+\xbar\eth\xbar\cY)
    + \eth^2 f (\gamma^0-3\xbar\gamma^0) + f\eth\xbar\nu^0,\\
  &\p_u\eth\xbar\eth f=\frac12\eth\xbar\eth(\eth\cY+\xbar\eth\xbar\cY)
    - \eth\xbar\eth f (\gamma^0 + \xbar\gamma^0)
    + \xbar\eth f \xbar\nu^0   + \eth f \nu^0 + f\eth\nu^0,\\
  &\p_u\eth\xbar\sigma^0=\eth\lambda^0 + \xbar\nu^0\xbar\sigma^0
    - (\xbar\gamma^0+3\gamma^0)\eth\xbar\sigma^0,\\
  &\p_u \eth\mu^0=\xbar\eth\xbar\nu^0 - 2\mu^0\xbar\nu^0
    - 2(\gamma^0+2\xbar\gamma^0)\eth\mu^0,\\
&\xbar\eth\eth\xbar\nu^0=\eth^2\nu^0-2\mu^0\xbar\nu^0.
\end{align*}
If one wants to compute the current algebra from the expressions
derived in the standard Cartan formalism \cite{Barnich:2016rwk},
one needs to transform the spin coefficients into a Lorentz
connection with a space-time index in NU gauge. Using
the notations of subsection
\ref{sec:relat-newm-penr}, together with the gauge choice for the
tetrads \eqref{gauge
  conditions} (and thus also
\eqref{cotetrad 4d case}), we have
\begin{align*}
  \Gamma_{12u} &= -(\gamma + \bar{\gamma}) -\tau  X^A \bar{L}_A  -
\bar{\tau} X^A L_A, &\Gamma_{12A} = \tau \bar{L}_A + \bar{\tau} L_A,\\
  \Gamma_{13u} &= - \tau - \sigma X^A \bar{L}_A - \rho X^A L_A,
&\Gamma_{13A} = \sigma \bar{L}_A + \rho L_A, \\
  \Gamma_{14u} &= -\bar{\tau} - \bar{\sigma} X^A L_A - \rho X^A
                 \bar{L}_A,
&\Gamma_{14A} = \rho \bar{L}_A + \bar{\sigma} L_A, \\
  \Gamma_{23u} &= \bar{\nu} + \bar{\lambda} X^A \bar{L}_A + \bar{\mu}
                 X^A L_A,
&\Gamma_{23A} = - \bar{\lambda} \bar{L}_A - \bar{\mu} L_A ,\\
  \Gamma_{24u} &= \nu + \mu X^A \bar{L}_A + \lambda X^A L_A , &\Gamma_{24A} = - \mu \bar{L}_A - \lambda L_A ,\\
  \Gamma_{34u} &= (\gamma - \bar{\gamma}) + (\beta - \bar{\alpha}) X^A
                 \bar{L}_A + ( \alpha - \bar{\beta} ) X^A L_A ,
&\Gamma_{34A} = (\bar{\alpha} - \beta ) \bar{L}_A + (\bar{\beta} -\alpha   ) L_A ,\\
  \Gamma_{abr} &= 0.
\end{align*}

\chapter{Map from Bondi to Fefferman-Graham gauge}
\label{app:chgt}

In this section, we find the explicit change of coordinates that maps a general vacuum asymptotically locally (A)dS$_4$ spacetime ($\Lambda \neq 0$) in Bondi gauge to Fefferman-Graham gauge \cite{Compere:2019bua}. This procedure will lead to the  explicit map between the free functions defined in Bondi gauge $ \lbrace q_{AB}, \beta_0,U^A_0,\mathcal{E}_{AB},M,N_A\rbrace $ and the holographic functions defined in Fefferman-Graham gauge, namely the boundary metric $g_{ab}^{(0)}$ and the boundary stress-tensor encoded in $g^{(3)}_{ab}$. 

We follow and further develop the procedure introduced in \cite{Poole:2018koa}. We first note that one can map the (A)dS$_4$ vacuum metric in retarded coordinates 
\begin{equation}
ds^2 = \left(\frac{\Lambda r^2}{3}-1 \right) du^2 - 2 du dr + r^2 \mathring{q}_{AB} dx^A dx^B ,
\label{global ads metric}
\end{equation} to the global patch 
\begin{equation}
ds^2 = -\Big(1 -\frac{\Lambda r^2}{3} \Big)dt^2 + \Big(1 -\frac{\Lambda r^2}{3} \Big)^{-1} dr^2 +r^2 \mathring q_{AB} dx^A dx^B
\end{equation}
by using $u = t - r_\star$, where the tortoise coordinate is $r_\star \equiv \sqrt{-\frac{3}{\Lambda}} \left[\arctan \left(r \sqrt{-\frac{\Lambda}{3}} \right)  -\frac{\pi}{2} \right]$ for $\Lambda <0$ and $r_\star \equiv \sqrt{\frac{3}{\Lambda}} \left[\text{arcoth} \left(r \sqrt{\frac{\Lambda}{3}} \right) \right]$ for $\Lambda > 0$. The next step is to transform the radial coordinate $r$ into the tortoise coordinate $r_\star$ which maps $r=\infty$ to $r_\star=0$. The change of coordinates from $(t_\star ,r_\star,x^A_\star)$ to Fefferman-Graham gauge $(t,\rho,x^A)$ can then be performed perturbatively in series of $\rho$ around $\rho = 0$, identified with $r_\star = 0$. 

The general algorithm is then the following:
\begin{enumerate}
\item Starting from any asymptotically locally (A)dS$_4$ solution formulated in Bondi gauge $(u,r,x^A)$, we perform the preliminary change to the tortoise  coordinate system,
\begin{equation}
\begin{split}
u &\to t_\star -r_\star, \quad x^A \to x^A_\star, \\
r &\to     \left\{
    \begin{array}{ll}
        \sqrt{-\frac{3}{\Lambda}} \tan \left( r_\star \sqrt{-\frac{\Lambda}{3}}  +\frac{\pi}{2} \right) & \mbox{if } \Lambda < 0 \\
       \sqrt{\frac{3}{\Lambda}} \text{coth} \left( r_\star \sqrt{\frac{\Lambda}{3}} \right)  & \mbox{if } \Lambda > 0
    \end{array}
\right.      =    
  \frac{3}{\Lambda r_\star} + \frac{r_\star}{3} - \frac{\Lambda r_\star^3}{135} + \mathcal{O}(r_\star^{5}).
\end{split}
\end{equation}
\item We reach the Fefferman-Graham gauge at order $N \geq 0$ perturbatively,
\begin{equation}
g_{\rho\rho} = -\frac{3}{\Lambda}\frac{1}{\rho^2} \Big( 1 + \mathcal{O}(\rho^{N+1}) \Big), \quad g_{\rho t} = \frac{1}{\rho^2}\mathcal{O}(\rho^{N+1}), \quad g_{\rho A} = \frac{1}{\rho^2}\mathcal{O}(\rho^{N+1}), \label{eq:FGgaugecond}
\end{equation} thanks to a second change of coordinates,
\begin{equation}
\begin{split}
r_\star &\to \sum_{n=1}^{N+1} R_n (t,x^A)\rho^n, \\
t_\star &\to t + \sum_{n=1}^{N+1} T_n (t,x^A)\rho^n, \\
x^A_\star &\to x^A + \sum_{n=1}^{N+1} X^A_n (t,x^B)\rho^n.
\end{split}
\end{equation}
\end{enumerate}
To obtain all the free functions in $\gamma_{ab}$, we must proceed up to order $N=3$. For each $n$, each gauge condition \eqref{eq:FGgaugecond} can be solved separately and will algebraically determine $R_n$, $T_n$ and $X_n^A$ respectively. Only the function $R_1(t,x^A)$ remains unconstrained by these conditions, since it represents a Weyl transformation on the boundary metric that is allowed within Fefferman-Graham gauge. We fix this freedom by requiring the normalization $g_{AB}^{(0)}=q_{AB}$.

We use the following shorthand notations for subleading fields in Bondi gauge:
\begin{equation}
\begin{split}
\frac{V}{r} &= \frac{\Lambda}{3}r^2 + r \ V_{(1)} (t,x^A) + V_{(0)} (t,x^A) + \frac{2M}{r} + \mathcal{O}(r^{-2}), \\
U^A &= U_{0}^A(t,x^B) + \frac{1}{r}U_{(1)}^A(t,x^B) + \frac{1}{r^2} U_{(2)}^A(t,x^B) + \frac{1}{r^3} U_{(3)}^A(t,x^B) + \mathcal{O}(r^{-4}), \\
\beta &= \beta_{0}(t,x^A) + \frac{1}{r^2}\beta_{(2)}(t,x^A)+ \mathcal{O}(r^{-4}).
\end{split}
\end{equation}
whose explicit on-shell values can be read off in \eqref{eq:EOMVr 2} and \eqref{eq:EOM_UA2 bis}. That will state the equations in a more compact way. All the fields are now evaluated on $(t,x^A)$ since the time coordinate on the boundary can be defined as $t$ as well as $u$. We also define  some recurrent structures appearing in the diffeomorphism as differential operators on boundary scalar fields $f(t,x^A)$:
\begin{equation}
\begin{split}
P[f] &= \frac{1}{2} e^{-4\beta_0} (\partial_t f + U_0^A \partial_A f), \\
Q[f;g] &= P[f] - 2  P[g] f, \\
B_A[f] &= \frac{1}{2} e^{-2\beta_0} (\partial_A - 2\partial_A \beta_0) f.
\end{split}
\end{equation}
$P^n[f]$ denotes $n$ applications of $P$ on $f$, for example $P^2[f] \equiv P[P[f]]$. Now we can write down the perturbative change of coordinate to Fefferman-Graham gauge:

\begin{align*}
R_1 (t,x^A) &= -\frac{3}{\Lambda}, \\
R_2 (t,x^A) &= \frac{9}{2\Lambda^2} e^{-2\beta_0} V_{(1)}, \\
R_3 (t,x^A) &= \frac{3}{2\Lambda} \beta_{(2)} - \frac{3}{\Lambda^2} \Big( 1 + \frac{3}{4} e^{-2\beta_0} V_{(0)} \Big) +\frac{27}{2\Lambda^3}  \Big( Q[V_{(1)};\beta_0] - \frac{3}{8} e^{-4\beta_0} V_{(1)}^2 \Big), \\
R_4 (t,x^A) &= \frac{3}{\Lambda^2} e^{-2\beta_0} \Big( M + 2 e^{4\beta_0} P[\beta_{(2)}] - \frac{5}{2} V_{(1)} \beta_{(2)}  \Big) \\
	&\hspace{-40pt}\quad -\frac{9}{\Lambda^3} \Big\lbrace Q[V_{(0)};\beta_0] + \frac{1}{4} e^{-4\beta_0} \Big[ U^A_{(1)} \partial_A V_{(1)} - 2 V_{(1)} U^A_{(1)} \partial_A\beta_0 - 3 V_{(1)} (2 e^{2\beta_0} + V_{(0)}) \Big] \Big\rbrace \\
	&\hspace{-40pt}\quad + \frac{27}{\Lambda^4} e^{2\beta_0} \Big[ P^2[V_{(1)}] - 2 V_{(1)} \Big( P^2[\beta_0]+ \frac{1}{2}e^{-4\beta_0} Q[V_{(1)};\beta_0] - \frac{3}{32} e^{-8\beta_0} V_{(1)}^2 \Big) - 2 P[\beta_0]P[V_{(1)}] \Big], \\
	&\ \\
T_1 (t,x^A) &= (1 - e^{-2\beta_0}) R_1(t,x^A),\\
T_2 (t,x^A) &=  (1 - e^{-2\beta_0}) R_2 (t,x^A) -\frac{18}{\Lambda^2} \Big( P[\beta_0] - \frac{1}{4}e^{-4\beta_0} V_{(1)} \Big),\\
T_3 (t,x^A) &=  (1- e^{-2\beta_0})R_3(t, x^A) -\frac{3}{\Lambda^2} e^{-2\beta_0} (1 + e^{-2\beta_0} V_{(0)} - 2 \partial^A \beta_0 \partial_A \beta_0) \\
	&\quad + \frac{9}{\Lambda^3} e^{-2\beta_0} \Big( Q[V_{(1)};		\beta_0] - 4 e^{4\beta_0} P^2[\beta_0]- \frac{1}{2}		e^{-4\beta_0} V_{(1)}^2 \Big), \\
T_4 (t,x^A) &= (1- e^{-2\beta_0})R_4(t, x^A) \\
& \quad+ \frac{9}{2\Lambda^2} \Big[ e^{-4\beta_0} \Big( M - \beta_{(2)} V_{(1)} - \frac{1}{3} U_{(2)}^A \partial_A \beta_0 \Big) -\frac{1}{2} (P[\beta_{(2)}] - 8 \beta_{(2)} P[\beta_0]) \Big] \\
	&\quad - \frac{27}{\Lambda^3} \Big\lbrace \frac{1}{8} e^{-2\beta_0} \Big( 3 Q[V_{(0)};\beta_0] - \frac{8}{3} P[\beta_0]V_{(0)} - 2 e^{-4\beta_0} V_{(1)}V_{(0)} \Big) \\
		&\qquad\qquad + \frac{1}{3} e^{-2\beta_0} \Big( P[U_{(1)}^A]\partial_A\beta_0 + \frac{3}{2} U_{(1)}^A \partial_A P[\beta_0] \Big) \\
		&\qquad\qquad - \frac{1}{12}e^{-4\beta_0} \Big[ U_{(1)}^A B_A [V_{(1)}] + 6 V_{(1)} - 2 (V_{(1)}\partial_A \beta_0 + 2 \partial_B \beta_0 \partial_A U_0^B)\partial^A \beta_0 \Big] \Big\rbrace \\
	&\quad + \frac{81}{\Lambda^4} \Big\lbrace -\frac{2}{3} e^{4\beta_0} \Big( P^3[\beta_0] + 2 P[\beta_0]P^2[\beta_0] \Big) + \frac{1}{4} \Big(P^2[V_{(1)}] - 2 V_{(1)} P^2[\beta_0]\Big) \\
	&\qquad\qquad +\frac{1}{6} P[\beta_0] \Big[\Big(\frac{13}{4}e^{-4\beta_0} V_{(1)} - 8 P[\beta_0] \Big) V_{(1)} + P[V_{(1)}] \Big] \\
	&\qquad\qquad - \frac{1}{16} e^{-4\beta_0} V_{(1)} \Big(5P[V_{(1)}]- e^{-4\beta_0} V_{(1)}^2\Big) \Big\rbrace, \\
&\ \\
X_1^A (t,x^B) &= (T_1 - R_1) U_0^A,\\
X_2^A (t,x^B) &= (T_2 - R_2) U_0^A - \frac{3}{2\Lambda} e^{-2\beta_0} U_{(1)}^A + \frac{9}{\Lambda^2} P[U_0^A],\\
X_3^A (t,x^B) &= (T_3 - R_3) U_0^A + \frac{1}{\Lambda} e^{-2\beta_0} U_{(2)}^A \\
	&\quad - \frac{6}{\Lambda^2} \Big[ Q[U_{(1)}^A;\beta_0] + \frac{1}{2} B^A[V_{(1)}] + \frac{1}{4} e^{-4\beta_0} (U_{(1)}^B \partial_B U^A_0 - V_{(1)}U_{(1)}^A) \Big] \\
		&\quad + \frac{18}{\Lambda^3} e^{2\beta_0} Q[P[U_0^A];\beta_0], \\
X_4^A (t,x^B) &= (T_4 - R_4) U_0^A - \frac{3}{4\Lambda} e^{-2\beta_0} \Big[ U_{(3)}^A + \frac{1}{2}e^{2\beta_0} (\partial^A \beta_{(2)} - 8 \beta_{(2)}\partial^A\beta_0) \Big] \\
&\quad + \frac{9}{2\Lambda^2} \Big[ Q[U_{(2)}^A;\beta_0] - \frac{1}{2} e^{-4\beta_0} \Big( V_{(1)} U_{(2)}^A - \frac{1}{3} U_{(2)}^B \partial_B U_{(0)}^A \Big) - 2 \beta_{(2)} P[U_0^A] \\
		&\qquad\qquad + \frac{1}{4} B^A[V_{(0)}] + \frac{1}{2} C^{AC} B_C[V_{(1)}] + \frac{1}{2} e^{-2\beta_0} U_{(1)}^C B_C[ U^A_{(1)} ] \Big] \\
		&\quad -\frac{27}{\Lambda^3} \Big\lbrace  e^{2\beta_0} \Big( P[Q[U_{(1)}^A;\beta_0]] + P[B^A[V_{(1)}]] - \frac{1}{2} q^{AC} P[B_C[V_{(1)}]] \Big) \\
		&\qquad\qquad - \frac{1}{2}e^{-2\beta_0} \Big[ V_{(1)} P[U_{(1)}^A] - \frac{2}{3}\Big( P[U_{(1)}^C] + \frac{1}{2} B^C[V_{(1)}] - 5 P[\beta_0]U_{(1)}^C \Big)\partial_C U_0^A \\
		&\qquad\qquad + \frac{1}{2} P[V_{(1)}]U_{(1)}^A - \frac{2}{3}(V_{(0)} - 8 e^{2\beta_0} \partial^B \beta_0 \partial_B \beta_0)P[U_0^A] - U_{(1)}^B P[\partial_B U_0^A] \\
		&\qquad\qquad + \frac{1}{2} (\partial^A U_0^C)B_C[V_{(1)}] \Big] + 3 P[\beta_0] V_{(1)}\partial^A \beta_0 \\
		&\qquad\qquad - e^{-4\beta_0} \Big[ \frac{3}{32} \Big(\partial^A (V_{(1)}^2) - \frac{20}{3} \partial^A \beta_0 V_{(1)}^2 \Big) + \frac{1}{6} (V_{(1)}  \partial^B \beta_0 - \partial^C\beta_0 \partial_C U_0^B) \partial_B U_0^A \Big] \Big\rbrace. \\ 
		&\quad +\frac{81}{\Lambda^4} \Big\lbrace \frac{1}{3} e^{4\beta_0} P^3[U_0^A] + \Big[ \frac{1}{4}e^{-4\beta_0} V_{(1)}^2 - \frac{1}{3} Q[V_{(1)};\beta_0] - \frac{4}{3} e^{4\beta_0} (P^2[\beta_0] + P[\beta_0]^2)  \Big] P[U_0^A] \Big\rbrace. \nonumber
\end{align*}

Several consistency checks can be performed at each stage of the computation. The boundary metric in Fefferman-Graham gauge must be equivalent to the pulled-back metric on the hypersurface $\lbrace r\to \infty \rbrace$ in Bondi gauge, up to the usual replacement $u\to t$:
\begin{equation}
g_{ab}^{(0)} = \left[
\begin{array}{cc}
\frac{\Lambda}{3} e^{4\beta_0} + U_0^C U^0_C & -U_B \\ 
-U_A & q_{AB}
\end{array} 
\right].
\end{equation}
At subleading orders, $g^{(1)}_{ab}$ and $g^{(2)}_{ab}$ must be algebraically determined by $g_{ab}^{(0)}$ and its first and second derivatives, which turns out to be the case. The constraint \eqref{eq:CAB 2} forces $g^{(1)}_{ab} = 0$ while the annulation of $\mathcal{D}_{AB}(t,x^C)$ \eqref{eq:DAB 2} results in
\begin{equation}
g^{(2)}_{ab} = \frac{3}{\Lambda} \Big[ R^{(0)}_{ab} - \frac{1}{4} R_{(0)} g^{(0)}_{ab} \Big].
\end{equation}
We do not give the full general form of $g^{(3)}_{ab}$, but it can be proven that this tensor is traceless with respect to $g^{(0)}_{ab}$, and that the equations of motion in Bondi gauge are necessary and sufficient to show its conservation $D^{(0)}_a g_{(3)}^{ab} = 0$, as we argued in the main text.

After boundary gauge fixing $\beta_0 = 0$, $U_0^A = 0$, the expressions of each coefficient in the diffeomorphism simplify drastically:
\begin{align*}
R_1 (t,x^A) &= -\frac{3}{\Lambda}, \\
R_2 (t,x^A) &= \frac{9}{2\Lambda^2} V_{(1)}, \\
R_3 (t,x^A) &= \frac{3}{2\Lambda} \beta_{(2)} - \frac{3}{\Lambda^2} \Big( 1 + \frac{3}{4} V_{(0)} \Big) +\frac{27}{2\Lambda^3}  \Big( \frac{1}{2} \partial_t V_{(1)} - \frac{3}{8} V_{(1)}^2 \Big), \\
R_4 (t,x^A) &= \frac{3}{\Lambda^2} \Big( M + \partial_t \beta_{(2)} - \frac{5}{2} V_{(1)} \beta_{(2)} \Big) -\frac{9}{\Lambda^3} \Big[ \frac{1}{2} \partial_t V_{(0)} - \frac{3}{4} V_{(1)} (2 + V_{(0)}) \Big] \\
	&\quad + \frac{27}{\Lambda^4} \Big( \frac{1}{4}\partial_t^2[V_{(1)}] - \frac{1}{4}\partial_t V_{(1)}^2 + \frac{6}{32} V_{(1)}^3\Big). \\
	&\ \\
T_1 (t,x^A) &= 0,\\
T_2 (t,x^A) &=  \frac{9}{2\Lambda^2} V_{(1)},\\
T_3 (t,x^A) &=  -\frac{3}{\Lambda^2} (1 + V_{(0)}) + \frac{9}{2\Lambda^3} \Big( \partial_t V_{(1)} - V_{(1)}^2 \Big), \\
T_4 (t,x^A) &= \frac{9}{2\Lambda^2} \Big[ M - \frac{1}{4} (\partial_t \beta_{(2)} + 4 V_{(1)}\beta_{(2)}) \Big] + \frac{9}{\Lambda^3} \Big[ - \frac{9}{16} \partial_t V_{(0)} + \frac{3}{2} V_{(1)} (1+\frac{1}{2} V_{(0)}) \Big] \\
	&\quad + \frac{81}{\Lambda^4} \Big(  \frac{1}{16} \partial_t^2 V_{(1)}  - \frac{5}{64} \partial_t V_{(1)}^2 + \frac{1}{16}V_{(1)}^3 \Big). \\
	&\ \\	
X_1^A (t,x^B) &= X_2^A (t,x^B) = 0, \\
X_3^A (t,x^B) &= \frac{1}{\Lambda} U_{(2)}^A - \frac{3}{2\Lambda^2} \partial^A V_{(1)}, \\
X_4^A (t,x^B) &= - \frac{3}{4\Lambda} \Big( U_{(3)}^A + \frac{1}{2}\partial^A \beta_{(2)} \Big) + \frac{9}{2\Lambda^2} \Big( \frac{1}{2}\partial_t U_{(2)}^A - \frac{1}{2}  V_{(1)} U_{(2)}^A + \frac{1}{8} \partial^A V_{(0)} \Big) \\
		&\quad -\frac{27}{16\Lambda^3}  \ q^{AB} \Big( \partial_t \partial_B V_{(1)} + \frac{1}{2} V_{(1)} \partial_B V_{(1)} \Big).
\end{align*}

\renewcommand{\theequation}{\thesection.\arabic{equation}}

\chapter{Detailed computations in Fefferman-Graham gauge}

\section{Check of the conservation law}
\label{Check of the conservation law}

We explicitly check that the ansatz \eqref{eq:FinalCharges} satisfies \eqref{eq:FundamentalThm}. We start by computing the right-hand side of \eqref{eq:FundamentalThm}. The variations $\delta_\xi \sqrt{|g_{(0)}|}$ and $\delta_\xi T^{ab}$ are given by
\begin{align}
\delta_\xi \sqrt{|g_{(0)}|} &= \frac{1}{2}\sqrt{|g_{(0)}|} \, g^{ab}_{(0)}\delta_\xi g_{ab}^{(0)} = \sqrt{|g_{(0)}|} ( D_a^{(0)}\xi_0^a - 3 \sigma_\xi ), \label{eq:deltaG0} \\
\delta_\xi T^{ab} &= \mathcal L_{\xi_0} T^{ab} + 5 \sigma_\xi T^{ab}. \label{eq:deltaTab}
\end{align}
Recalling that $T_{ab}$ obeys ${T^a}_a = 0$ and $D_a^{(0)}T^{ab} =0$ on-shell, we get
\begin{align}
\delta \theta^\rho_{\text{ren}}[g; \mathcal L_\xi g] &= -\eta \delta\left(\sqrt{|g_{(0)}|} T^{ab}\right)D_a^{(0)}\xi^0_b - \eta \sqrt{|g_{(0)}|} T^{ab} \delta \left(D_a^{(0)}\xi^0_b\right),\\
-\delta_\xi \theta^\rho_{\text{ren}}[g ; \delta g] &=\frac{1}{2}\eta \sqrt{|g_{(0)}|} \left( D^{(0)}_c \xi^c_0 T^{ab} + \mathcal L_{\xi_0} T^{ab}\right)\delta g_{ab}^{(0)} + \eta \sqrt{|g_{(0)}|} T^{ab}\delta (D_a^{(0)}\xi^0_b).
\end{align}
The left-hand side reads as
\begin{equation}
\begin{split}
\partial_a k^{\rho a}_\xi[g ; \delta g] &= -\eta \delta \left(\sqrt{|g_{(0)}|}T^{ab}\right)D_a^{(0)} \xi^0_b - \eta \sqrt{|g_{(0)}|} T^{ab} \delta g_{bc}^{(0)} D_a^{(0)}\xi^c_0 \\
&\quad + \frac{1}{2}\eta \sqrt{|g_{(0)}|} D_a^{(0)}\xi^a_0 T^{bc} \delta g_{bc}^{(0)} + \frac{1}{2}\eta \sqrt{|g_{(0)}|} \xi^a_0 D_a^{(0)} T^{bc}\delta g_{bc}^{(0)}.
\end{split}
\end{equation}
Using $\mathcal L_{\xi_0} (T^{ab}) = \xi_0^c D^{(0)}_c T^{ab} - 2 T^{c(a}D_c^{(0)} \xi_0^{b)}$, we have
\begin{equation}
\begin{split}
&\partial_a k^{\rho a}_\xi[g; \delta g]+\omega^\rho_{\text{ren}} [g; \delta g,\mathcal L_\xi g]\\
&\qquad\qquad = \frac{1}{2}\eta\sqrt{|g_{(0)}|} \left(\mathcal L_{\xi_0} T^{ab}\delta g_{ab}^{(0)} + 2 T^{ab}\delta g_{bc}^{(0)}D_a^{(0)} \xi^c_0 - \xi^c_0 D_c^{(0)}T^{ab} \delta g_{ab}^{(0)}\right) = 0,
\end{split}
\end{equation}
which finishes the verification. 

\section{Charge algebra}
\label{sec:Charge algebra}

Here, we write the explicit computations leading to the result \eqref{charge algebra AAdS}. 

The computation is on-shell, so in particular $g_{(0)}^{ab}T_{ab}=0$ and $D_a^{(0)}T^{ab} = 0$. Let us start by computing $\delta_\chi J^a_\xi[g]$. The computation is direct and takes benefit of \eqref{eq:deltaG0} and \eqref{eq:deltaTab}:
\begin{equation}
\begin{split}
\delta_\chi J^a_\xi[g] &= \eta \sqrt{|g_{(0)}|} D_d^{(0)}(\chi^d_0 T^{ab}) g^{(0)}_{bc}\xi^c_0 - \eta \sqrt{|g_{(0)}|} T^{bd}D_d^{(0)}\chi^a_0 g_{bc}^{(0)} \xi^c_0 \\
&\quad + \eta \sqrt{|g_{(0)}|} T^{ab} (D_c^{(0)}\chi_b^0) \xi^c_0 + J^a_{\delta_\chi\xi}[g].
\end{split}
\end{equation}
To obtain the second term is just a matter of replacement, so
\begin{equation}
\Xi_\chi^a[\delta_\xi g,g] = -\eta \sqrt{|g_{(0)}|} \chi^a_0 T^{bc} D_b^{(0)}\xi_c^0 - J^a_{\delta_\xi\chi}[g].
\end{equation}
Summing both contributions and using the fact that $T^{ab}$ is divergence-free, we get
\begin{equation}
\begin{split}
&\delta_\chi J^a_\xi[g] + \Xi_\chi^a[\delta_\xi g,g] \\
&\quad = \eta \sqrt{|g_{(0)}|} \, {T^a}_b \left(\xi^c D_c^{(0)}\chi^b_0 - \chi^c_0 D_c^{(0)}\xi_0^b\right) + J^a_{\delta_\chi\xi} - J^a_{\delta_\xi\chi} - 2\eta \partial_b \left( \sqrt{|g_{(0)}|}\chi^{[a}_0 {T^{b]}}_c\xi^c_0 \right) \\
&\quad = \eta \sqrt{|g_{(0)}|} \, {T^a}_b [\xi,\chi]^b + J^a_{\delta_\chi\xi} - J^a_{\delta_\xi\chi} - 2\eta \partial_b \left( \sqrt{|g_{(0)}|}\chi^{[a}_0 {T^{b]}}_c\xi^c_0 \right) \\
&\quad = J^a_{[\xi,\chi]} + J^a_{\delta_\chi\xi} - J^a_{\delta_\xi\chi} - 2\eta \partial_b \left( \sqrt{|g_{(0)}|}\chi^{[a}_0 {T^{b]}}_c\xi^c_0 \right) \\
&\quad = J^a_{[\xi , \chi]_A} - 2\eta \partial_b \left( \sqrt{|g_{(0)}|}\chi^{[a}_0 {T^{b]}}_c\xi^c_0 \right).
\end{split}
\end{equation}
The last term exhibits the exterior derivative of a 2-form,
\begin{equation}
\bm \ell = \ell^{ab}(\text d^2 x)_{ab}, \quad \ell^{ab} = - 2\eta \sqrt{|g_{(0)}|}\chi^{[a}_0 {T^{b]}}_c\xi^c_0 . \label{boundary term algebra}
\end{equation}
Therefore, we have shown that the charge algebra represents the vector algebra without additional 2-cocycle,
\begin{equation}
\lbrace J_\xi[g],J_\chi[g] \rbrace^a_\star = J^a_{[\xi,\chi]_A}[g] + \partial_b \ell^{ab}[g]. 
\end{equation}

\addcontentsline{toc}{chapter}{References}

\providecommand{\href}[2]{#2}\begingroup\raggedright\endgroup

\end{document}